This article is published here:
PNAS November 30, 2021 118 (48) e2111213118;
https://doi.org/10.1073/pnas.2111213118

The dataset and SI files are published here:
https://doi.org/10.5281/zenodo.4943234

There are only minor editing differences between the preprint below
and the final published article.

# Seeds of imperfection rule the
# mesocrystalline disorder in natural anhydrite single crystals

Tomasz M. Stawski*[1], Glen J. Smales[1], Ernesto Scoppola[2], Diwaker Jha[3],
Luiz F. G. Morales[4,5], Alicia Moya[6], Richard Wirth[7], Brian R. Pauw[1], Franziska Emmerling[1, 8],
and Alexander E. S. Van Driessche**[6]

* tomasz.stawski@bam.de; **Alexander.Van-Driessche@univ-grenoble-alpes.fr

1. Federal Institute for Materials Research and Testing (BAM), 12489, Berlin, Germany;
2. Max Planck Institute of Colloids and Interfaces, 14476 Potsdam, Germany;
3. HyperSpecs.de, Berlin, Germany
4. Scientific Center for Optical and Electron Microscopy (ScopeM), ETH Zürich, 8093, Zürich;
5. Geological Institute, Department of Earth Sciences, ETH Zürich, 8092 Zurich, Switzerland;
6. Université Grenoble Alpes, Université Savoie Mont Blanc, CNRS, IRD, IFSTTAR, ISTerre, F-38000 Grenoble, France;
7. German Research Centre for Geosciences, GFZ, Interface Geochemistry, Telegrafenberg, 14473 Potsdam, Germany;
8. Humboldt-Universität zu Berlin, Department of Chemistry, 12489, Berlin, Germany

Orcid:
TMS: https://orcid.org/0000-0002-0881-5808
GJS: https://orcid.org/0000-0002-8654-9867
ES: https://orcid.org/0000-0002-6390-052X
DJ:
LFGM: https://orcid.org/0000-0002-8352-850X
AM: https://orcid.org/0000-0002-0328-012X
RW:
BRP: https://orcid.org/0000-0002-8605-838X
FE: https://orcid.org/0000-0001-8528-0301
AESVD: https://orcid.org/0000-0003-2528-3425

**Keywords:**
Calcium sulfate; anhydrite; Naica; single crystal; crystallisation; nucleation; particle-mediated

**Competing Interest Statement:**
The authors declare no competing interest.

**Significance statement**

Many people perceive crystals as the embodiment of perfect order, though in reality, it is well-understood that monocrystals possess imperfections. By considering large anhydrite crystals from the famous Naica Mine ("*Cueva de los cristales*"), an extended picture begins to emerge, revealing a suite of correlated self-similar void defects spanning multiple length-scales. These flaws, in the macroscopic crystal, likely stem from "seeds of imperfection" originating from a particle-mediated nucleation pathway. Hence, building a crystal could be viewed as Nature stacking blocks in a game of Tetris, whilst slowly forgetting the games core concept and failing to fill rows completely.

**Abstract**

In recent years, we have come to appreciate the astounding intricacies associated with the formation of minerals from ions in aqueous solutions. In this context, a number of studies have revealed that the nucleation of calcium sulfate systems occurs non-classically, involving the aggregation and reorganization of nanosized prenucleation species. In recent work, we have shown that this particle-mediated nucleation pathway is actually imprinted in the resultant micron-sized $CaSO_4$ crystals. This property of $CaSO_4$ minerals provides us with the unique opportunity to search for evidence of non-classical nucleation pathways in geological environments. In particular, we focused on large anhydrite crystals extracted from the Naica mine in Mexico. We were able to shed light on this mineral's growth history by mapping defects at different length scales. Based on this, we argue that the nano-scale misalignment of the structural sub-units, observed in the initial calcium sulfate crystal seeds, propagate through different length-scales both in morphological, as well as strictly crystallographic aspects, eventually causing the formation of large mesostructured single crystals of anhydrite. Hence, the non-classical nucleation mechanism introduces a "seed of imperfection", which leads to a macroscopic "single" crystal, in which its fragments do not fit together at different length-scales in a self-similar manner. Consequently, anisotropic voids of various sizes are formed with very well-defined walls/edges. Though, at the same time, the material retains in part its single crystal nature.

**Introduction**

The formation, transformation and dissolution of minerals in aqueous solutions plays a key role in the natural and engineering evolution of the Earth's surface. These factors control geological processes as diverse as the mass transfers within the lithosphere, elemental cycling, natural water composition, soil formation and biomineralization in living organisms, sequestration of $CO_2$, (sea)water desalination, geological nuclear waste storage, the setting of cement, and the synthesis of advanced functional materials, to name just a few. Although there are long standing theories to explain these mineral processes, in recent times a vast amount of new evidence has surfaced challenging these traditional views. In the particular case of nucleation, a number of precursor and intermediate solute/solid species, both stable and metastable, have been identified. The observation of these prenucleation phases extends the classical view of one step nucleation towards multi-step "non-classical" models(1–3). Although these recent insights have significantly expanded our view of mineral formation, only a reduced matrix of physicochemical parameters have been explored in the lab, which may not be fully representative of conditions prevalent in natural or even engineered environments. Consequently, it remains uncertain whether the multistep pathways observed in a laboratory setting are truly universal or only incidental.

To evaluate the general applicability of these so-called non-classical mechanisms to out-of-the lab environments, mineral formation should ideally be observed in situ. Nevertheless, this is a daunting task taking into account the inherent stochastic nature of the nucleation process, combined with the close to equilibrium conditions usually prevailing in natural environments. One way to circumvent these difficulties is to identify evidence/remnants of the non-classical nucleation pathways within the subsequent crystals, similar to what has been reported for several biominerals(4). Recently, we have shown that the particle-mediated nucleation pathway of $CaSO_4$(5–7) is imprinted in the resultant single crystals(8), which are almost universally mesocrystalline(9–11) in nature. This property of calcium sulfate minerals provides us with a unique tool to search for evidence of non-classical nucleation pathways in different geological processes spanning time-scales of 1000s of years and beyond(12). In particular, we focused on single crystals of natural anhydrite (i.e. anhydrite AII phase). Together with gypsum, anhydrite is commonly encountered in evaporitic, and also hydrothermal, environments on the Earth's surface(7). Although there are different polymorphs(13–15) (AI, AII and AIII; see SI: Supplementary Note 1 and Fig. S1), only anhydrite AII crystallizes from aqueous solutions(16–18), either directly or via a dissolution and reprecipitation process from gypsum. In the Naica Mine (Cueva de los cristales or Cave of the Crystals or Giant Crystal Cave) in Mexico, anhydrite samples exceeding 10 cm in length(19–21) can be found (see SI: Figs. S2 and S3), with these large natural crystals of anhydrite AII containing unique fingerprints of the growth processes that have taken place over a period of millenia.

We used a multitechnique approach to extract the growth history of an anhydrite single crystal from the Naica mine to better understand its internal structure at different length scales and correlate it with the particle-mediated crystallization model of calcium sulfate(22–24). In particular, we argue that the nano-scale misalignment of the structural sub-units, observed in the initial calcium sulfate crystal seed, propagates through different length scales both in morphological, as well as, strictly structural aspects, eventually leading to the formation of large, mesostructured, single crystals of anhydrite.

**Results**

*Local heterogeneities observed in "single" anhydrite crystals at the nanoscale*

Fig. 1 shows high-resolution transmission electron microscopy (HRTEM) lattice fringe images of two focused ion beam (FIB) foils obtained from arbitrary locations, several µm apart from each other, on a single anhydrite crystal, (Figs. 1A&B). Both images display the same crystal orientation and their near-identical fast Fourier transforms (FFTs), showing only individual maxima (FFT-calculated diffraction spots), confirming that the material is indeed single crystal in nature (inset I in Fig. 1A and II in Fig. 1B). The relatively long distance between both fields of view, further indicates that we are dealing with a continuous single crystal, which is in agreement with X-ray diffraction measurements (see Methods). However, the individual spots in the FFTs are not circular in shape, but elliptical. This directly implies mosaicity at the length-scales corresponding to the field of view of the HRTEM images (i.e. nanoscale < 100 nm). To emphasize any structural variations we used Fourier filtering(8) to calculate the inverse FFT real-space images corresponding to the HRTEM micrographs (Figs. 1C&D). In addition, a false-colour map was applied to highlight the structural features.

The obtained images indicate that the HRTEM regions shown in Fig. 1A and B are composed of areas with lattice fringes aligned in the same manner (areas marked with pink rectangles in complementary representations in Fig. 1). However, the extent of order within the crystal lattice continuously fluctuates, with better ordered domains separated by less ordered ones. If we trace selected fringes across the field of view (straight lines in Fig. 1C), it becomes apparent that some domains are slightly misaligned by <1° (see arrows in inset III in Fig. 1C). Small disordered areas act as transitional parts between the ordered areas (see arrows in inset IV in Fig. 1D). Based on HRTEM, these order-disorder structural modulations occur typically at a length-scale of ~10-20 nm.

We also collected selected-area electron diffraction (SAED) patterns from the FIB foils using an effective aperture of ~1 μm (SI: Fig. S4A). The obtained data confirms the single crystalline nature of the anhydrite specimen, while the elliptical diffraction spots indicate strong mosaicity in the [0 1 0] direction. This indicates that the observed order-disorder regions are anisotropic, with shorter length-scales perpendicular to the [0 1 0] plane and longer-ones parallel. This is also directly visible in real-space (Fig. 1), but the HRTEM images represent significantly smaller areas of the crystal than that probed for SAED measurements. Therefore, the latter indicates that the alignment of the anisotropic defects extends over a length-scale of at least 1 μm. It is also important to note, the SAED also contains a significant (i.e. above the noise background) diffuse scattering contribution, e.g. a raster of streaks parallel to the [1 0 0] and [0 1 0] directions passing through the diffraction spots. Fig. S4A in SI shows that scattering in the [0 1 0] direction has higher intensity than that in the [100] direction. Although it might be difficult to unequivocally interpret the origin of the diffuse scattering without in-depth structural modelling (25–30), it appears reasonable to conclude that the observed streaks are related to the presence of the aforementioned order-disorder modulation regions (Fig. 1). In Fig. S4B we show a dark-field image collected with one of the diffracted beams from Fig. S4A. Such a low-magnification dark-field image from a near-perfect single crystal should exhibit a uniform contrast(8). This is not the case here, with the image highlighting an order-disorder modulation over a distance of hundreds of nanometers. The visible domains are also oriented in the same way as that seen in Fig. 1, though Fig. S4B shows that this extends over a greater length-scale.

Finally, it should be noted that FIB-prepared TEM lamellae, in general, may have up to 30 nm of amorphous material on both sides (31–34). With a sample thickness of approximately 150 nm, the scattering volume of the undisturbed crystal is substantially larger than the amorphous or FIB affected layer on both sides of the foil. Furthermore, this effect should reveal itself as a decrease in the diffraction contrast because the amorphous layer is perpendicular to the beam during TEM imaging (34). Conversely, in our thin sections the order-disorder modulation is parallel to the beam, as observed in transmission, which indicates that it is highly unlikely to be a result of the FIB thinning. Furthermore, the observed structural effects correlate well to those obtained from other techniques as we show.

*Structural heterogeneity at the nanoscale - a global perspective*
The internal homogeneity of an anhydrite single crystal was also probed by means of transmission X-ray scattering measurements at scattering vector q-ranges corresponding to small-angle X-ray scattering (SAXS; ~1-70 nm) and wide-angle X-ray scattering (WAXS; <1 nm). In contrast to HRTEM, the signal measured by scattering originates from a large sample volume of an order of 1 $mm^3$, and as such it provides statistically global information about the nanostructure. The general considerations on how an idealised single crystal ought to scatter in SAXS and WAXS are summarised in the SI: Supplementary Note 2. Fig. 2 presents 2D SAXS patterns for an anhydrite crystal in position S at 0° and 21° tilt respectively (see also SI: Fig. S3). For both of the crystal orientations, distinctly different anisotropic scattering patterns were observed. The patterns show that the anhydrite crystal structure is heterogeneous at a length-scale < ~70 nm ($q > 0.1$ $nm^{-1}$), and that these structural heterogeneities are orientated. The fact that we can observe such structural features in the first place means that their average electron density is different from the one of the surrounding matrix. It can be either higher, which is unlikely since anhydrite is the densest phase of $CaSO_4$, or lower if there is disorder and/or empty voids/pores. In Fig. 2A, the direction of the high-intensity scattering pattern is approximately parallel to the apparent vertical long axis of the crystal (inset in Fig. 2A, SI Fig. S3A z-axis). This suggests that the nanosized structural features, from which this pattern originates, have their longer dimensions aligned approximately perpendicularly to the crystal vertical long axis (and thus their shorter dimensions parallel to the long axis, SI Fig. S3B x-axis). When the crystal is tilted the resulting cross-shaped scattering patterns represent cross-sectional components of the anisotropic objects. The obtained cross-sections present a more complex scattering pattern (Fig. 2B). The intensity in Fig. 2B does not extend as far towards high-$q$ as that seen in Fig. 2A (compare $q_x$ and $q_y$ scales). Hence, the scattering in Fig. 2B originates from larger scattering features, than the one in Fig. 2A. Such a cross-shaped pattern for the tilted orientation, together with the one from Fig. 2A, implies that the scattering objects are rod- or platelet-like in nature, with the long-axis of these objects aligned perpendicular to the long axis of the crystal. Furthermore, the scattering profile in the direction perpendicular to the long axis of the crystal exhibits small-angle diffraction peaks (indicated with arrows in Fig. 2B). Such peaks most likely originate from a regular arrangement of mesoscale features in the crystal along this direction. A more complete interpretation is presented further in the text in the context of the microtomography data.

In order to compare intensities, 2D patterns were converted to polar coordinates (SI: Fig. S5A&B), based on which we calculated 1D scattering profiles (see SI: Supplementary Note 3) from the selected directions (dotted lines in Figs. 2A and 2B). The resulting 1D scattering profiles (Fig. 2C) have the form of straight lines in a log-log representation, with both converging to background scattering at ~1000 cts. Both profiles scale proportionally to $I(q) \propto q^{<-3}$, but the exponents are higher than the typical smooth interface dependence of $I(q) \propto q^{-4}$ (Porod interface (35, 36)). These relatively feature-poor forms indicate that scattering arises from objects extending beyond the available $q$-range and/or exhibit high polydispersity, meaning that the objects are >~70 nm. In addition, the scattering exponents between -3 and -4 point to rough interfacial, surface-fractal-like scattering (37), but due to the limited $q$-range this cannot be unequivocally confirmed. The high-intensity profile (black curve, Fig. 2C) converges to background scattering at $q$ of ~1 $nm^{-1}$, whereas the low-intensity scattering (purple curve, Fig. 2C) converges at ~0.3 $nm^{-1}$. This implies that the length-scale aspect ratio between the perpendicular scattering features is >3:1. For the tilted orientation, shown in Fig. 2D, the perpendicular scattering patterns (I & II) are similar in terms of intensity and their characteristic length-scales. In both cases the dominant interfacial scattering exponent is <-3 and the patterns converge to the background level at ~0.6 $nm^{-1}$. However, scattering profiles in II, in addition to an interfacial-type scattering profile (i.e. a straight line), also exhibit relatively well-pronounced small-angle diffraction peaks corresponding to $d$-spacings of 22, 19, 13 and 12 nm. Profile III in Fig. 2D is calculated from a "streak" in Fig. 2B in the direction of 0° (see also SI: Fig. S5B and Supplementary Note 3) and contains a straight-line interfacial component, but its scaling follows a less steep dependence of $I(q) \propto q^{-2.5}$, and small-angle diffraction peaks at 13, 9 and 7 nm. The presence of the peaks in II and III indicates the contribution of a scattering structure factor $S(q)$, which describes the regular, paracrystalline arrangement of the scattering objects(38, 39) , where the normalised structure factor is defined as S(q) → 1 for q →∞. The occurrence of such a paracrystalline structure factor has two major implications: (1) it potentially explains why the observed interfacial scattering exponents are less steep than the expected -4; (2) $S(q)$ is direction-dependent and the contributing scattering features are closely-spaced/correlated only along the directions where the peaks are observed.

In summary, the SAXS data tells us that (1) the single crystal is structurally heterogeneous at the mesoscale; (2) the heterogeneities are highly anisotropic and preferentially orientated, with their longer dimensions perpendicular to the apparent long axis of the crystal; (3) the tilt of the crystal reveals the presence of a paracrystalline structure factor only in some cross-sectional directions. Although the exact crystallographic alignment of $S(q)$, with respect to the crystal structure, is not known, its presence could be explained by the fact that the anisotropic scattering features are, to a certain degree, regularly arranged in the plane perpendicular to the long axis, where the actual scattering profiles with peaks (II and III in Fig. 2D) originate from cross-sections (between 0 and 90°) of such anisotropic superstructure.

Fig. 3A shows a four-panel composite WAXS diffraction pattern in polar coordinates from an anhydrite crystal at 0° relative tilt. The patterns were measured for four orientations (SI: Fig. S3B) resulting from a rotation of the crystal around the goniometer's vertical axis, Z. Consecutive panels correspond to positions: S (starting), N (180° clockwise), W (90° clockwise), E (270° clockwise). The pattern consists of individual diffraction spots, which again confirms that we are dealing with a single-crystalline material. The diffraction spots are, in general, very similar in shape and broadening, and at first glance they do not reveal any obvious structural defects, such as strong mosaicity. However, in our WAXS measurements, we probed only a very limited set of crystal orientations. Furthermore, there are some very apparent exceptions from this trend, most noticeable for the peaks at $q$ ~28 nm$^{-1}$ (marked with arrows in Fig. 3A). The strongest of these reflections is at 166° in panel S, and has the form of a cross (dashed white rectangle in Fig. 3A). The reflection appears to be broadened both in $q$ and the azimuthal-angle direction, whilst also being composed of several sub-reflections (Fig. 3B). This cross also has an asymmetric counterpart for the same $q$ at 76° in the same panel (i.e. 90° apart in the azimuthal angle), and weaker analogs at 5° and -85° in panel N, due to the rotation of a crystal by 180° around the Z-axis (SI: Fig. S3B). The reflections in panel S are also accompanied by long streaked lines extending at 90° in directions parallel to the cross in Cartesian coordinates (Fig. 3C), which makes them appear as curves in polar coordinates (Fig. 3B). For a given peak, the two considered broadening values in polar coordinates are essentially independent from each other, because they are related to different structural effects. The increased broadening in the $q$-direction is correlated with structural effects, which typically affect the lattice $d$-spacing e.g. caused by strain. The broadening in the azimuthal direction expresses the structural coherence of a crystal and is a measure of mosaicity, as is the case for the reflections at $q$ ~

28 nm$^{-1}$. The observed occurrence of such mosaic effects only for a single group of diffraction spots implies that the potential defects in the crystal structure are strongly anisotropic. Furthermore, the observed streaks are indicative of diffuse scattering, which signifies the presence of anisotropic disordered features at the mesoscale, associated with the structural defects. Both effects (strong mosaicity and diffuse scattering) are in agreement with the interpretation of the SAXS and TEM data presented earlier.

*Structural heterogeneity at the micron-scale and beyond*

We further evaluated the structure of a single crystal at the micrometer-length scale using X-ray microtomography (µCT). This technique allows for the 3D visualisation of a structure, based on the absorption contrast i.e. differences in the electron density at a resolution of 550x550x550 nm$^3$ voxel-size. In Fig. 4A, two selected 3D projections of a reconstructed crystal derived from ring-uncorrected slices are shown (see Methods). These images reveal the presence of voids within the crystal volume. In SI: Videos S1 and S2 complete 360° rotations of the crystal are shown: around the Z-axis in the XY-plane, and around the Y-axis in the XZ-plane, respectively. These renderings show that the crystal volume contains objects of different sizes, spanning from several tens of microns, down to objects at the voxel-size resolution limit. Fig. 4B shows two of the, ring-corrected and cropped, XY slices of the crystal (the Z direction is out-of-plane ) that contain a number of these objects (in blue), surrounded by a relatively homogenous matrix (in brown). Considering the origin of contrast in µCT, the objects have significantly lower absorption than that of the surrounding crystal matrix, and are thus attributed to voids/pores within the crystal. These voids exhibit very straight and well-defined edges in the XY-plane (as indicated by the intensity profile function in the inset in Fig. 4B). Fig. 4C shows a selected projection of a segmentation highlighting the voids, with a full-rotation of these objects around the Z-axis in the XY-plane can be found in SI: Video S3. Overall, voids are clustered into channel-like features parallel to the Z-axis (long-axis) of the crystal. Along the Z-axis, the largest voids (~25 µm in XY) form continuous regular structures, whereas the smaller ones, although not necessarily connected with each other, are grouped in pillar-like arrangements. To further evaluate the properties of the void arrangements we performed statistical analysis on all slices (Fig. 4D, see Methods for details). This analysis highlights what is directly observable in Fig. 4C, and in SI: Video S3., i.e. on average the features are anisotropic, with a width <~5 µm and a length >~10 µm in XY, and <~5 µm for width and >~5 µm for length in YZ. In addition, they exhibit preferred orientation in the Y direction of the XY-plane (at 90° to X, hence parallel to Y), and in the Z-direction in the ZY-plane (at 0° to Z, hence normal to Y), as can be seen in Fig. 4E. The void walls thus appear to approximate morphologically preferred faces of anhydrite, i.e. {0 1 0} and {1 0 0}.

*Bridging length scales: nanometres to microns*

We further characterised the voids exposed at the crystal facets using atomic force microscopy (AFM; Fig. 5). Detailed observation of different crystallographic faces revealed that on the top face (Fig. 5A) nano- to macroscopic sized porosity is observed (Figs. 5B&C), while on the side faces (Fig. 5A) no such porosity is found (Fig. 5D). This multi-scale surface imaging reveals that surface voids are rectangular/regular in shape and in the size range of nano- to micrometers (~50 nm to 25 µm). Although, with AFM the depth of these pores cannot be probed, the obtained surface images directly complement the observations made at different length scales by SAXS and µCT. A quantitative analysis of the topographical AFM images reveals a porosity of 7.4 ± 1.6 %, obtained as the area of the pores divided by the total area of the AFM images. Furthermore, we measured the dimensions of the pores directly from the height profiles of a wide range of AFM images (from 1 to 120 µm field-of-view), which were acquired at different locations of the same sample. Fig. 5E shows the as-obtained pore/void size distribution, which covers a wide range of pore sizes up to ~60 µm. However, pores larger than 10 µm only constitute a minor contribution to the total porosity. In fact, the highest contribution corresponds to the pores smaller than 0.5 µm. The pore size distribution at the nanoscale highlights that a dominant pore size exists at ~85 nm as shown in Fig. 5F. The AFM study is intrinsically limited to the external crystal surfaces, however considering the observations from SAXS and µCT, it is inferred that the void size distribution, within the crystal, is analogous to that at the surface.

We also probed the surface of the studied single crystal at the intermediate length-scales using electron backscatter diffraction (EBSD), which bridges findings between X-ray scattering and microtomography by providing crystallographic and lattice distortion information. In Fig. 6A, we show an EBSD orientation map of one of the crystal facets of the studied single crystal. This map represents an area of 51 x 51 µm$^2$, has a *quasi*-uniform cyan color and corroborates the idea that, at the macroscale, the studied sample is a single crystal. Crystallographic orientation plots indicate that the EBSD mapped facet is parallel to the anhydrite (9 5 3) crystal plane. However, when we calculate the kernel average misorientation (KAM) assuming a maximum misorientation of 2.5°, it becomes apparent that the probed area is composed of heterogeneous regions (Fig. 6B). This is further evidenced by low-angle grain boundaries, with many of them completely closed and looking like small "cells" that are neighboured by other cells that have slightly mismatched angles between each other down to the 100 nm resolution used in the map (Fig. 6C). The degree of their misorientation is very small, as seen in the histogram of Fig. 6D, with a mismatch between blue-to-green regions in the range of 0.3° to <1°, which is in agreement with TEM data. The mean misorientation is 0.35° with a standard deviation of 0.22°. Furthermore, the observed misorientation does not have a random character. In Figs. 6B&D one can observe that the upper half of the image (mean misorientation 0.40°, std. dev. 0.26°) contains more pronounced disorder than the lower half (mean 0.30°, std. dev. 0.15°).

## Discussion

*Anhydrite: single, poly- or mesocrystalline?*

Although an ideal single crystal is assumed to have a continuous, perfectly ordered structure, real-life crystalline materials often contain point, line and/or planar defects. If the amount and/or extension of defects are significant enough to physically separate crystalline domains by forming grain boundaries, the material is considered to be polycrystalline. When no long-range order is observed anymore, the solid phase is referred to as amorphous. Surprisingly, our anhydrite samples cannot be strictly classified into either of these groups. On the one hand, the diffraction measurements and the external appearance of the investigated mineral sample seem to indicate that we are dealing with a regular single crystal. Accordingly, its structure can be solved using standard methods of single crystal X-ray diffraction (see Methods and SI: anhydrite.cif for the solved/refined structure). On the other hand, our detailed structural characterisation at different length-scales revealed that the anhydrite crystal contains numerous types of intercorrelated structural defects, which separate co-aligned crystalline areas identifiable at the nanoscale. This corresponds with the general definition provided for so-called mesocrystals(9, 11, 40)**,** which is a relatively novel concept, originally derived from experimental data and observations focusing primarily on biomineralization. Importantly, most mesocrystals identified so far formed in the presence of templating (macro)molecules, and are linked to bioinspired mineral systems. Consequently, these mesocrystals are composites of crystalline material and templates. However, we observe for anhydrite crystals a segmented structure that appears to have formed in the absence of a template; as no template remnants are found in the crystal nor are organic templates expected considering the formation environment(41). Moreover, we previously observed the same phenomenon for micron-sized "single" crystals of gypsum and bassanite[8] formed under controlled laboratory conditions (i.e. in the absence of templating molecules). Hence, in the case of calcium sulfate, it seems to further blur the already murky distinction between single- and meso-crystals(42).

*Seeds of imperfection: an origin of the mesocrystallinity?*
At the nanoscale, i.e. <~100 nm, HRTEM imaging reveals order-disorder modulation with crystallographic domains 10-20 nm in size. This modulation exhibits smooth transitions among the neighbouring regions, and can be interpreted as a reminiscence of the particle mediated nucleation/crystallization. We proposed that tiny misalignments (<1º) between the crystalline regions, in combination with the disordered areas, act as "seeds of imperfection" for the further crystal growth, resulting in the formation of a macroscopic mesocrystal. The misorientations between the crystalline areas, of the order of the 1° observed here, were also reported for other non-classical systems (e.g. bismuth or magnetite) growing via particle-mediated processes, such as orientated attachment(23, 43). Micron-sized anhydrite mesocrystals were also found to be a byproduct of the bacterial dehydration of gypsum under very dry conditions(44), where particle-mediated nucleation is suggested as the reason behind its mesocrystallinity. Our earlier studies(5, 6, 45) on the nucleation of gypsum and bassanite, as well as those conducted by other groups(46–53), show that the crystalline phases of $CaSO_4$ nucleate and form within a micron-sized amorphous matrix of aggregated primary species(45, 52), several nanometers in size (for gypsum sub-3 nm). This constitutes the first stage of crystallisation, which produces the initial micron-sized meso crystalline seeds. Importantly, the re-structuring processes within the amorphous aggregates do not continue until a near-perfect, homogeneous single crystal is obtained, but instead comes to a halt during the observation window, because any mass transport processes inside such aggregates must be subject to slow diffusion when compared to those associated with ion transportation through the bulk aqueous solution. This early-stage crystallization leads to the formation of several-micron-sized single-crystalline seeds. What happens afterwards is unclear, but the particle-mediated stage may be followed by ion-by-ion growth, which is the prevailing mechanism under thermodynamic conditions close to equilibrium. During this secondary growth stage, the initial mesostructured single crystals would act as the aforementioned seeds of imperfection. In fact, in situ AFM experiments conducted in a controlled laboratory environment using chemically pure solutions have shown so far that ion-by-ion growth can take place on the {1 0 0} and {0 1 0} facets at room temperature(54) and on the {1 0 0} facet between 70-120ºC(55).  It is generally assumed that this growth mode should lead to rather continuous crystal structures. But, this does not correspond with what we observe in our anhydrite samples. This leaves open two scenarios: (1) the anhydrite samples examined by us did not experience significant ion-by-ion growth or (2) this growth mode can replicate the preexisting, or underlying, mesocrystalline matrix.

*Voids are a key feature to keep the crystal structure together*
Irrespective of the prevailing growth mechanism occurring during the post-nucleation stage, the inherent misalignment and disorder in the crystalline matrix has to be compensated for if a crystal is to grow to macroscopic dimensions. We hypothesize that voids (SI: Fig. S8) are formed to compensate for the lattice strain/internal stress during growth. This would explain why voids are observed at all length scales, as discussed above. The fact that these voids, over time, develop straight facets to minimize their surface free energy, and thus lower the overall excess energy of the crystals seems to corroborate our hypothesis.

The scattering behaviour observed in SAXS can be explained by the presence of voids at the length-scale of ~100 nm. Tilt-dependent SAXS patterns actually correspond to structural features, which could very well look exactly like those in μCT, if it was not for the fact that CT probes length scales > 500 nm, whereas SAXS in our configuration is limited to << 500 nm, where the largest structures (>70 nm) could only be partially observed, if at all. This implies that the void topology extends to significantly smaller length-scales than those measured with μCT, in a self-similar manner. This is confirmed by our AFM images (Fig. 5), which show that the voids can be <100 nm, with an average size of 85 nm. HRTEM revealed orientated disordered regions that can be correlated with the diffuse scattering/streaks observed in SAED and WAXS. In this regard, SAXS is likely to produce structural features from both the voids and the disordered nano-sized regions, since they would all exhibit lower electron density than the crystalline anhydrite matrix, and therefore contribute to the scattering contrast. The orientated and anisotropic character of the voids indicates, in this context, that they form as semi-regular errors in the replication/growth processes in a similar manner as the order-disorder modulation is both regular and anisotropic. Such a behaviour also explains the presence of the orientation-dependent small-angle diffraction peaks from semi-periodic structures. Indeed, we see large semi-periodic structures in μCT, and thus we infer that they have analogs at shorter length-scales. Finally, the EBSD results bridge our interpretation of the scattering data with what we observe in microtomography and other methods. This technique accesses the intermediate length-scales of several tens to several hundreds of nanometers, and highlights a heterogenous crystallographic character of the single crystal. Hence, the original seeds of imperfection are expressed through the length-scales both in morphological (e.g. voids), as well as strictly structural (e.g. crystallographic) aspects of the actual single crystal.

In the context of the discussion above, it is also worth considering whether the observed voids might simply constitute dissolution etch pits(56). The voids observed here extend into the volume of the crystal whereas etch pits are typically limited to the surfaces of crystals. The deepening of etch pits leads to the formation of inverse pyramid dissolution patterns(56, 57) and not vertical wells/channels as those observed in our anhydrite samples. Moreover, etch pits should be filled up again when the crystal surface continues to grow, so it is highly unlikely that the remnants of a large number of etch pits should be preserved within the bulk of the crystal.

**Conclusions and outlook**

Ongoing investigations into the nucleation of various mineral systems seem to insinuate that non-classical pathways are much more prevalent than was considered a decade or two ago. In fact, it is even possible that some materials can nucleate both classically and non-classically, a phenomenon that was recently demonstrated for calcium oxalate (58). Here, we show that the intrinsic mesocrystallinity, seen in anhydrite(8, 44) is imprinted evidence of non-classical, particle-mediated nucleation. This nucleation mechanism introduces seeds of imperfection, which still leads to the formation of a macroscopic single crystal, though its fragments do not fit together at different length-scales in a self-similar manner. This results in the formation of anisotropic voids of various sizes with very well-defined walls/edges, which approximately follow the anhydrite faces with the lowest surface free energy. This resembles nature playing a game of Tetris, which in some ways it is losing.

## Materials and Methods
### Anhydrite single crystals
Macroscopic well-formed translucent anhydrite samples were obtained from the Naica mine, Chihuahua, Mexico (municipality of Saucillo, the mine is owned by Industrias Peñoles). This mining area is located on the northern side of the Sierra de Naica (59), and constitutes one of the main lead and silver deposits in the world. Hydrothermal fluid circulation associated with Tertiary dikes formed these Ag-Pb-Zn deposits(60). During the late hydrothermal stage sulfuric acid formed by oxidation of the underlying sulfides and reacted with the available limestone to form calcium-sulfate–rich waters that eventually precipitated anhydrite masses(21). These specimens (Figs. S2 and S3) are famous for their high purity, light blue color and large size (single crystals can easily reach >10 cm). Almost all the experiments were performed on the same selected single crystal shard shown in SI: Fig. S3, which was chipped off from a bigger body of crystals similar to the one shown in SI: Fig. S2. The only exception was in the case of the atomic force microscopy characterisation (AFM, see below), for which we used another pristine shard from the same group.

### Single Crystal X-ray Diffraction
Single crystal X-ray diffraction experiments were performed on a Bruker D8 Venture system with graphite-monochromatic Mo-Kα radiation (λ = 0.71073 Å). For the diffraction experiments the crystal (SI: Fig. S3) was not modified in any way, to ensure sample preservation for further analyses. Data reduction was performed with Bruker AXS SAINT(61) and SADABS(62) packages. The structure was solved in the space group Cmcm using direct methods and completed using differential Fourier maps calculated with SHELXL 2018(63). Full matrix least-squares refinements were performed on $F^2$ using SHELXL 2018(63) with anisotropic displacement parameters for all atoms. All programs were run under the WinGX (v. 1.80) system(64). VESTA (v. 3.5.7) was used for structure visualization(65). The resulting crystal information file (CIF) is included as a part of the SI (anhydrite.cif). Diffraction unequivocally confirmed the single-crystalline character of the investigated anhydrite sample.

### Scattering Methods

Small- and wide-angle X-ray scattering (SAXS/WAXS) measurements were conducted using the MOUSE instrument(66) . X-rays were generated from a microfocus X-ray tube, followed by multilayer optics to parallelize and to monochromatise the X-ray beams to wavelength of Mo Kα (λ = 0.71073 Å). Scattered radiation was detected on an in-vacuum Eiger 1M detector (Dectris, Switzerland), which was placed at multiple distances between 52 - 2354 mm from the sample. Beam parameters were kept consistent for all sample-to-detector distances used with a spot size of 643 $\mu$m (fwhm).

The anhydrite single crystal was placed on a goniometer for data collection to allow multiple orientations to be probed in SAXS and WAXS (SI: Fig. S3). The initial crystal orientation was arbitrary, and the goniometer was set to null positions. This starting orientation with a XZ-tilt of 0° constituted position S (SI: Fig. S3A). WAXS was measured starting from S, and at further positions N-W-E, 90° apart from each other, corresponding to a rotation of the crystal around its axis in the XY plane (SI: Fig. S3B). For SAXS, the crystal was returned to S (SI: Fig. S3A), and measured in a second step in position S with an additional XZ-tilt of 21° (SI: Fig. S3C).

The resulting data was processed using the DAWN software package (v. 2.20) in a standardized complete 2D correction pipeline with uncertainty propagation(67, 68). These included, among other steps, essential corrections for sample transmission and the instrument background subtraction. For SAXS, in order to compare the intensities in different directions the 2D patterns were also converted to polar coordinates ("cake" plots). Such a representation allows for an easy integration of the direction dependent-scattering intensities to 1D scattering curves. The azimuthal positions of the intensity directions of interest, as well as their angular widths are directly obtained from the mean intensity profiles.

In the case of WAXS, such a polar representation was the only one used, due to the fact that version 2.20 of DAWN applies a small-angle approximation to scale the $q_x$- and $q_y$-axis in 2D patterns. This issue does not affect SAXS, but at higher scattering angles the resulting scales are incorrect for 2D images in Cartesian coordinates. However, the small-angle approximation is not utilised for calculating the "cake" plots, hence they are rendered correctly for all angular ranges. In those cases when 2D WAXS in Cartesian coordinates were required, we back-calculated them from the "cake" plots rather than use outputs from DAWN, so that $q_x$ and $q_y$ were expressed correctly. Further processing and analysis of reduced 2D scattering datasets was performed in Python using NumPy, SciPy and Pandas(69–72). The dataset is deposited at Zenodo(73).

*Transmission Electron Microscopy*

In order to analyse single crystals under a transmission electron microscope (TEM), we prepared ~15 $\mu$m x 4 $\mu$m thin foils (~100 - 150 nm thickness) using the focused ion beam technique (FIB, FEI FIB200) following a standard procedure (31, 32). Neither did the crystal show any signs of alteration under the vacuum of the instrument during cutting/milling, nor did the foils when imaged in TEM. For TEM imaging and selected-area electron diffraction (SAED), a Tecnai F20 XTWIN TEM was used at 200 kV, equipped with a field-emission gun electron source. SAED patterns were collected using an aperture with an effective diameter of ~1 μm and the diffraction plates were developed in a high-dynamic range Ditabis Micron scanner. To correctly interpret any preferred orientation or texture-related effects in the TEM images, the objective stigmatism of the electron beam was corrected by ensuring the fast Fourier transform (FFT) was circular over the amorphous carbon film.

*X-ray Microtomography*

Microtomography (µCT) was performed with an EASYTOM (RX Solutions) equipped with a LaB$_6$ filament. The final resolution was set to 0.55 µm obtained by applying a voltage of 100 kV and a current of 100 µA and by collecting 2816 sinograms over a 360° rotation. With these settings, the collection time was set to ~38 h. 3D reconstruction was performed by the software provided by the manufacturer (RX Solutions). The collected 2816 sinograms were reconstructed using the Back Projection Algorithm into a 3D tomogram, where the YZ plane was 1447 x 1718 pixels$^2$, and the YX plane was 1447 x 1716 pixels$^2$. The reconstructed tomogram constituted raw data, which were processed by reslicing the dataset into the XY and XZ planes. For quantitative analysis, the 3D data was further processed by applying artifact correction and data restoration algorithms and scripts described in refs. (74–76). The images in the XY plane revealed typical ring artifacts from the reconstruction processes, which were partially suppressed following the referenced method (74). In the next step, the images were corrected for an illumination drift using histogram matching through the Z axis in the stack. A median filter with a 3-pixel 2D kernel together with a non-local means filter to suppress noise were applied which was necessary to perform feature analysis, because the rings and noise were hindering the deduction of the solid-void threshold parameter. Such as-calculated images were trimmed in order to remove the edge/background parts of the slices leaving only the measured crystal. We calculated gradients distribution of features that measured along which axes the defects were dominant, to see if the defects exhibited any anisotropy. This was performed separately for the XY and YZ plane because of the memory limitations, and demonstrated that the features were strongly anisotropic. Finally, an ellipse fit on the features was performed to further characterize their dimensions, which showed features were longer than wider. This was also readily visible from a visual inspection of the 3D projections. An example of uncorrected and corrected images is shown in SI: Fig. S6. The dataset and the uncompressed Videos S1-S3 are deposited at Zenodo(73).

*Electron Backscatter Diffraction (EBSD)*

We used electron backscatter diffraction (EBSD) in a scanning electron microscope (SEM)To characterize the general orientation and the intracrystalline distortion of the studied anhydrite single crystal. This was done on a FEI Quanta200 F SEM with EDAX EBSD/EDS detectors and Team/OIM Analysis software. The orientation map was collected from one of the single crystal facets, which exhibited very high apparent smoothness, and hence did not require any special sample preparation. The crystal was orientated in the vacuum chamber in a way that the elongated direction of the single crystal (~Z in SI: Fig. S3) was horizontal in the EBSD orientation maps and in the pole figures. The SEM operating conditions included an accelerating voltage of 20 kV, beam current of 8 nA, working distance of 15 mm and a step size of 100 nm, in an uncoated sample, with the SEM working under low vacuum (30 Pa $H_2O$). Post-acquisition processing included confidence index (CI) standardization with a grain tolerance angle of 5°, followed by one iteration of CI neighbour correlation considering only grains with CI >0.1. Afterwards, we removed all the pixels with CI <0.2 and image quality below 25% to ensure that the orientations presented here were correct. From the orientation map, we then calculated the kernel average misorientation (KAM) map calculated in relation to a fixed distance between neighbours, which showed the average misorientation of a given pixel in the map in comparison to all its neighbours. The as-obtained map had a threshold misorientation angle of 2.5° and was calculated in relation to the 1st neighbour pixel. The crystal orientation data was plotted in the upper hemisphere of an equal-angle stereographic projection (SI: Fig. S7). Here, we plotted the three main crystal directions of anhydrite ([1 0 0],[0 1 0] and [0 0 1]) plus the direction <5 8 6>, which had one of the 4 symmetrically equivalent directions ([5 8 6],[-5 -8 6],[-5 8 -6] and [5 -8 -6]) plotted right in the middle of the pole figure and thus indicate which crystal direction is approximately normal to the reader when looking at the orientation map. That indicated that this direction was normal to the facet, which was equivalent to a crystal plane within the  {9 5 3} group ((9 5 3), (-9 -5 3), (-9 5 -3) or (9 -5 -3)).

*Atomic Force Microscopy*
Topographical features of the studied anhydrite sample (see also Methods: Anhydrite single crystals*)* were evaluated using Atomic Force Microscopy (AFM) operating in contact mode using a MFP-3D microscope from Asylum Research (Santa Barbara, USA). The maximum range of the piezo scanner was 120 µm in the planar direction (XY) and 15 µm in the vertical direction (Z). All the AFM images were acquired by using triangular silicon nitride cantilevers (PNP-TR from NanoWorld) with a nominal spring constant of 0.08 N·m$^{-1}$. Before each experiment the used cantilever was routinely calibrated using the thermal method. All the obtained images were processed using the AR and WSxM software (77). Images collected at a field-of-view length-scale of ~120 µm were post-processed by applying a thresholding algorithm that filtered out all topographical features above 20 nm (set as a threshold value). As a result, filtered AFM images were obtained showing almost exclusively the pores/voids, which allowed us to perform quantitative calculations of the volume, surface and perimeter of the pores.

## Acknowledgments
We acknowledge Max Planck Institute of Colloids and Interfaces (MPI) for granting access to the microtomography instrument. We thank Daniel Werner for assisting with the µCT measurements.

# Figures

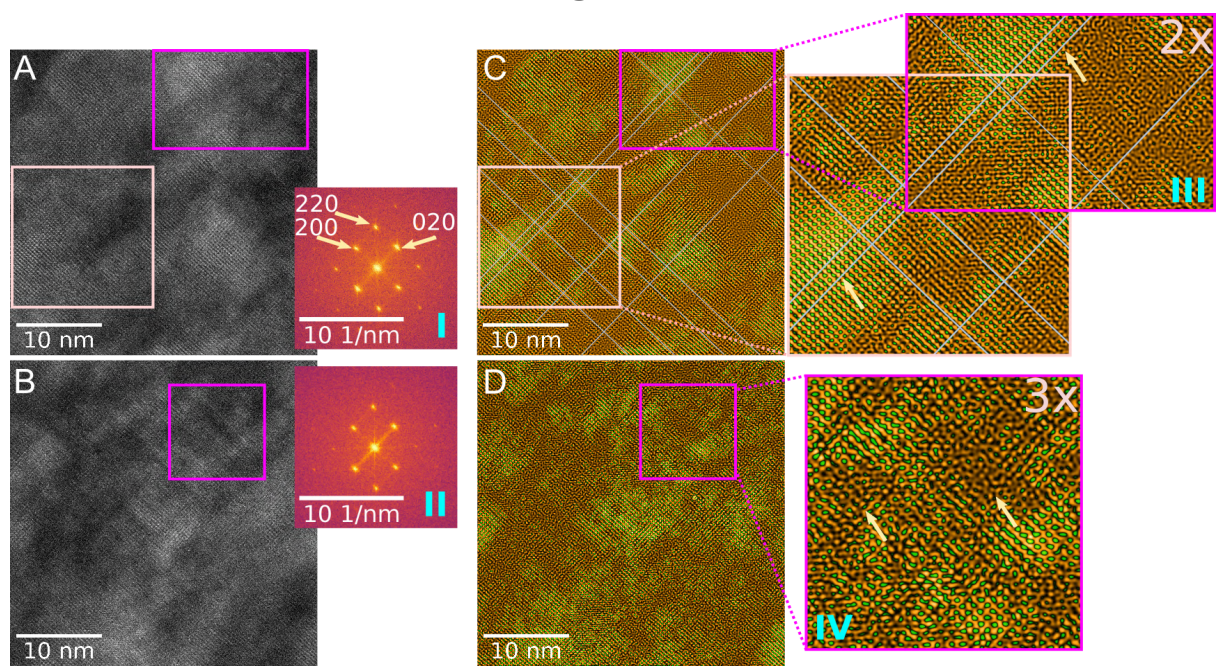

Fig. 1. HRTEM analysis of a FIB foil from the anhydrite single crystal; A) and B) show two similar HRTEM images of the small sections of the foil, which are >10 µm apart from each other. Inset I in (A), and II in (B) show FFTs calculated to the respective images; selected reflections in FFTs are indexed for anhydrite; flux: ~8x10⁵ e⁻Å⁻²s⁻¹, estimated received fluence ~1x10²⁷ e⁻m⁻²; Pink rectangles in (A) and (B) correspond to the same regions-of-interest (ROIs) in (C) and (D), respectively; C) Inverse FFT and filtered image, which highlights the order-disorder in (A) with a fake-colour palette applied; light-blue lines trace the lattice fringes in two directions; inset III shows 2x magnified overlap between the ROIs marked by the pink rectangles, so that the light-blue lines remain continuous; the arrows point to an apparent lattice fringe shift (*pseudo*-dislocation) along the selected light-blue line; D) Inverse FFT and filtered image, which highlights the order-disorder in (B) with a fake-colour palette applied; inset IV shows the ROI contained in a pink rectangle and 3x magnified; the arrows in IV point to highly-disordered regions in the crystal.

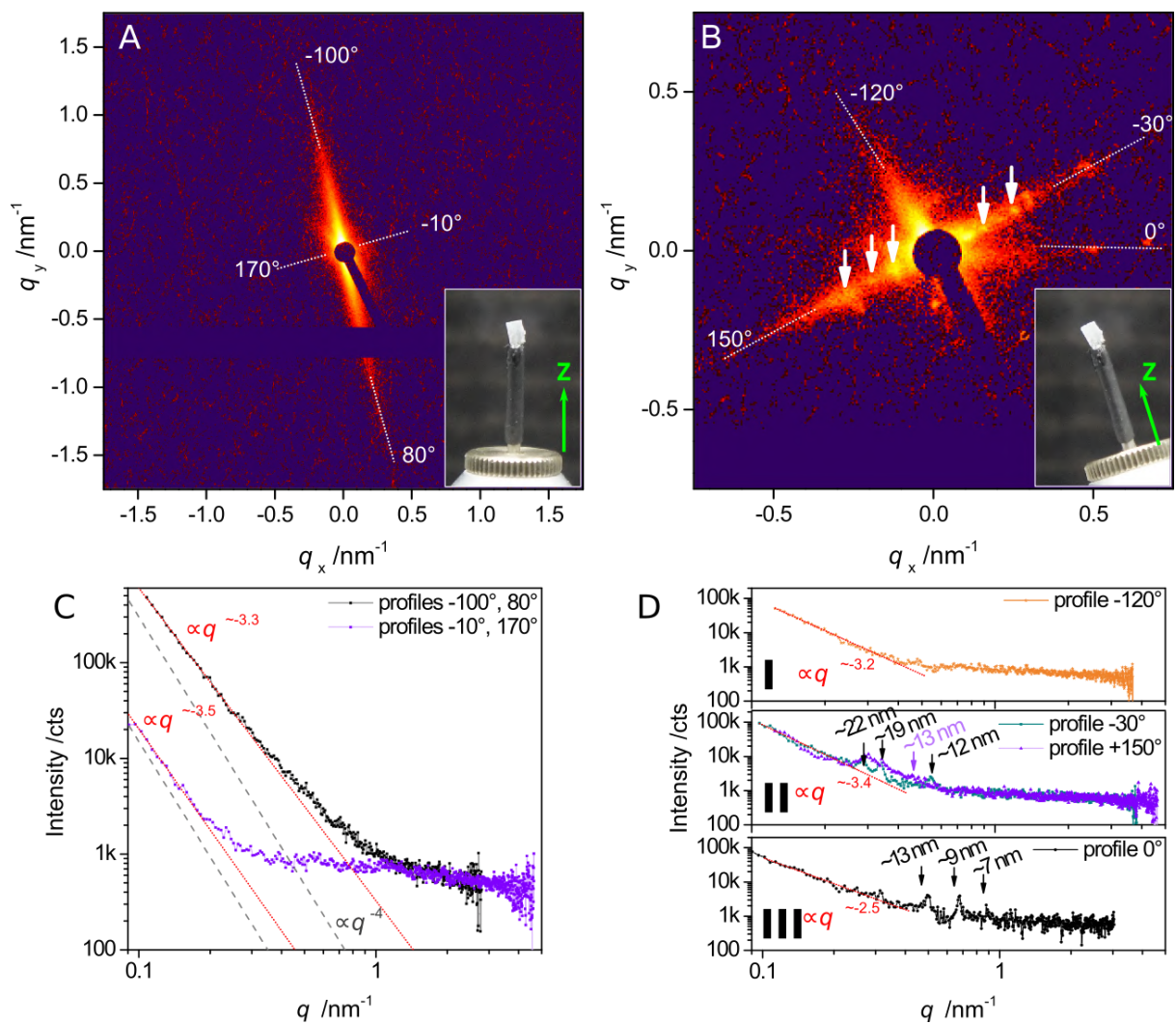

Fig. 2. SAXS scattering patterns from the anhydrite crystal. A) 2D SAXS (left) pattern of the anhydrite crystal in Position S at 0° (see also SI: Fig. S3A); B) 2D SAXS (left) pattern of the anhydrite crystal in Position S at 21° tilt (see also SI: Fig. S3C); In (A) and (B) The angular directions marked with dotted lines indicate integration directions based on the polar coordinate representations shown in SI: Fig. S5; the insets show the two orientations of the single crystal in accordance with SI: Fig. S5; approximate long z-axis direction of the crystal is indicated with green arrows; C) and D) Direction-dependent scattering curves integrated from (A) and (B) respectively (see also SI: Fig. S5). Fitted scattering dependencies in a form of $I(q) \propto q^{-a}$, where $-a$ is a scattering exponent, are indicated with dotted red lines; the Porod-scattering (smooth interface) $I(q) \propto q^{-4}$ is shown with dashed black lines; C) Position S at 0°, averaged high-intensity direction (black), and low-intensity direction (purple); curves are obtained by integrating intensity profiles with centroids of azimuthal angles as written in the legends, and based on (A) and II in SI: Fig. S5A; each of two symmetric profiles are averaged together; D) Position S at 21° tilt; three characteristic scattering directions are shown I (orange), II (cyan and purple), III (black); curves are obtained by integrating intensity profiles with centroids of azimuthal angles as written in the legends, and based on (B) and II in SI: Fig. S5B.

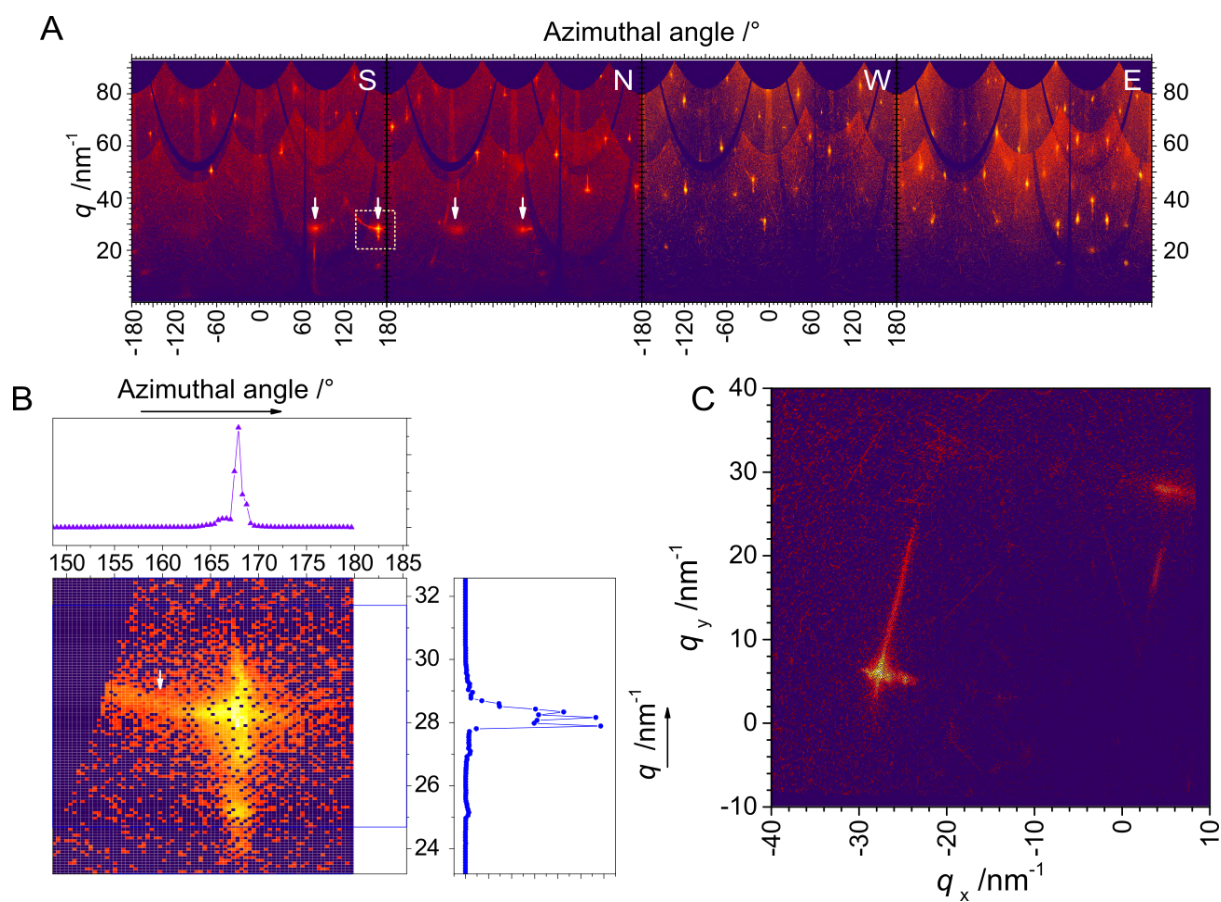

Fig. 3. WAXS diffraction/scattering patterns from the anhydrite crystal. A) Composite WAXS 2D diffraction pattern, which consists of four panels. The data are plotted in polar coordinates (the "cake plot"). The diffraction patterns were measured for four orientations resulting from a rotation of a crystal around the goniometer's vertical axis, Z, where consecutive panels correspond to positions in Fig. 1B: I (starting), II (180° clockwise), III (90° clockwise), IV (270° clockwise). Furthermore, each of the panels comprises five sub-patterns obtained by moving the detector in a plane perpendicular to the beam i.e. emulating a larger area detector. For the overlapping pixels among such sub-patterns the intensities were averaged out. The discussed peaks at $q \sim 28$ nm$^{-1}$ are indicated with arrows; the cross-shaped reflection is marked with a dashed rectangle; B) close-up 2D WAXS in polar coordinates, and profile plots of the reflection marked in (A); an arrow points to a diffuse scattering streak; C) 2D WAXS in Cartesian coordinates of the two peaks at $q \sim 28$ nm$^{-1}$ from the 1st panel in (A); the diffuse scattering streaks are well-pronounced; the Cartesian-coordinates representation is re-calculated from the polar coordinates.

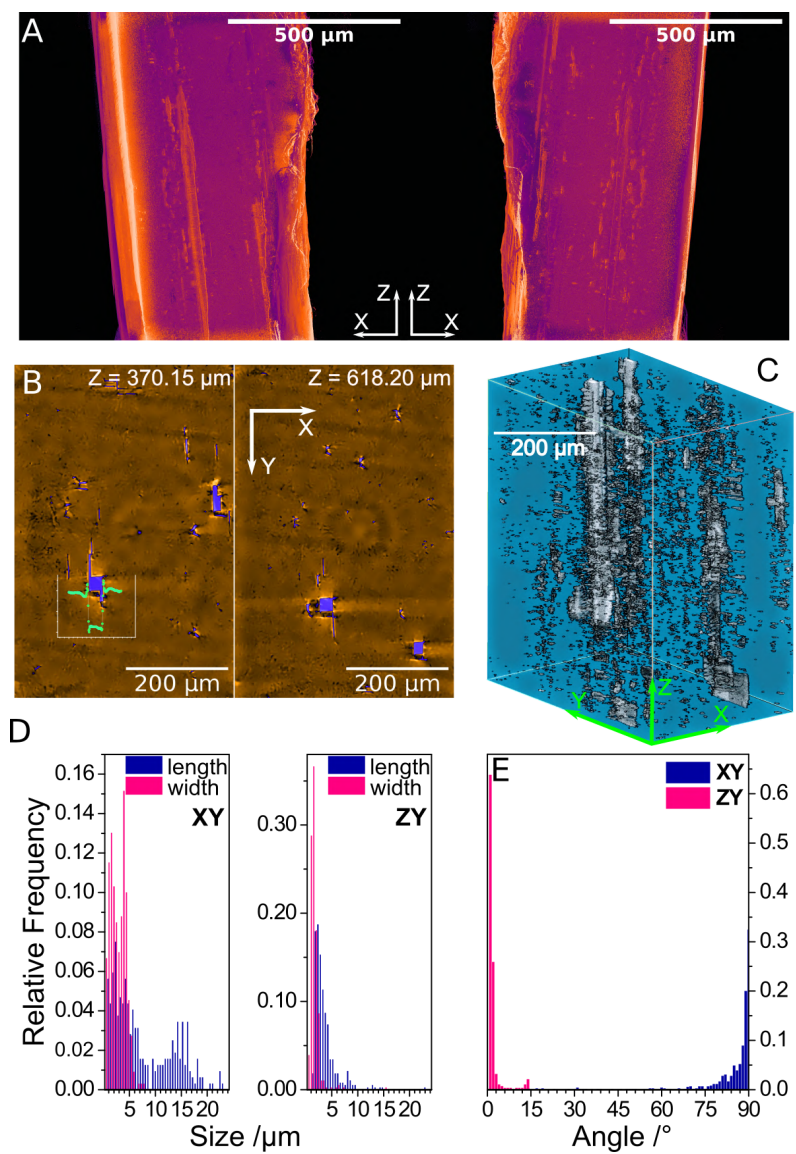

Fig. 4. Microtomographic reconstructions and analysis. A) Selected projections from a 3D reconstruction of an uncorrected (see Methods) microtomography data set collected for the anhydrite single crystal; the projections show the two sides of the crystal (left and right) and highlight internal defect structure; see also SI: Video S1 - overview 360° rotation in the XY-plane around Z, and SI: Video S2 - overview 360° rotation in the XZ-plane around Y; B) Selected processed (see Methods) cross-sections in the XY plane for two arbitrary Z-values; the voids are shown in blue; an inset graph on the left shows the overall abruptness of the contrast transition between the void and the surrounding crystal matrix; the co-centric rings are a typical artifact of the reconstruction processes, and in the image they are already partially suppressed (see Methods); C) a projection of a segmentation which shows the void structure within the crystal, derived from the processed data such as those in (B); see also SI: Video S3 - 360° rotation in the XY-plane around Z which highlights the void structure; D) distribution of voids' dimensions in XY- and ZY-planes calculated from the corrected data, which demonstrates the anisotropic character of the defects; E) distribution of voids' orientations in XY- and ZY-planes calculated from the corrected data, which demonstrates the preferred  orientation of the defects along Z and Y.

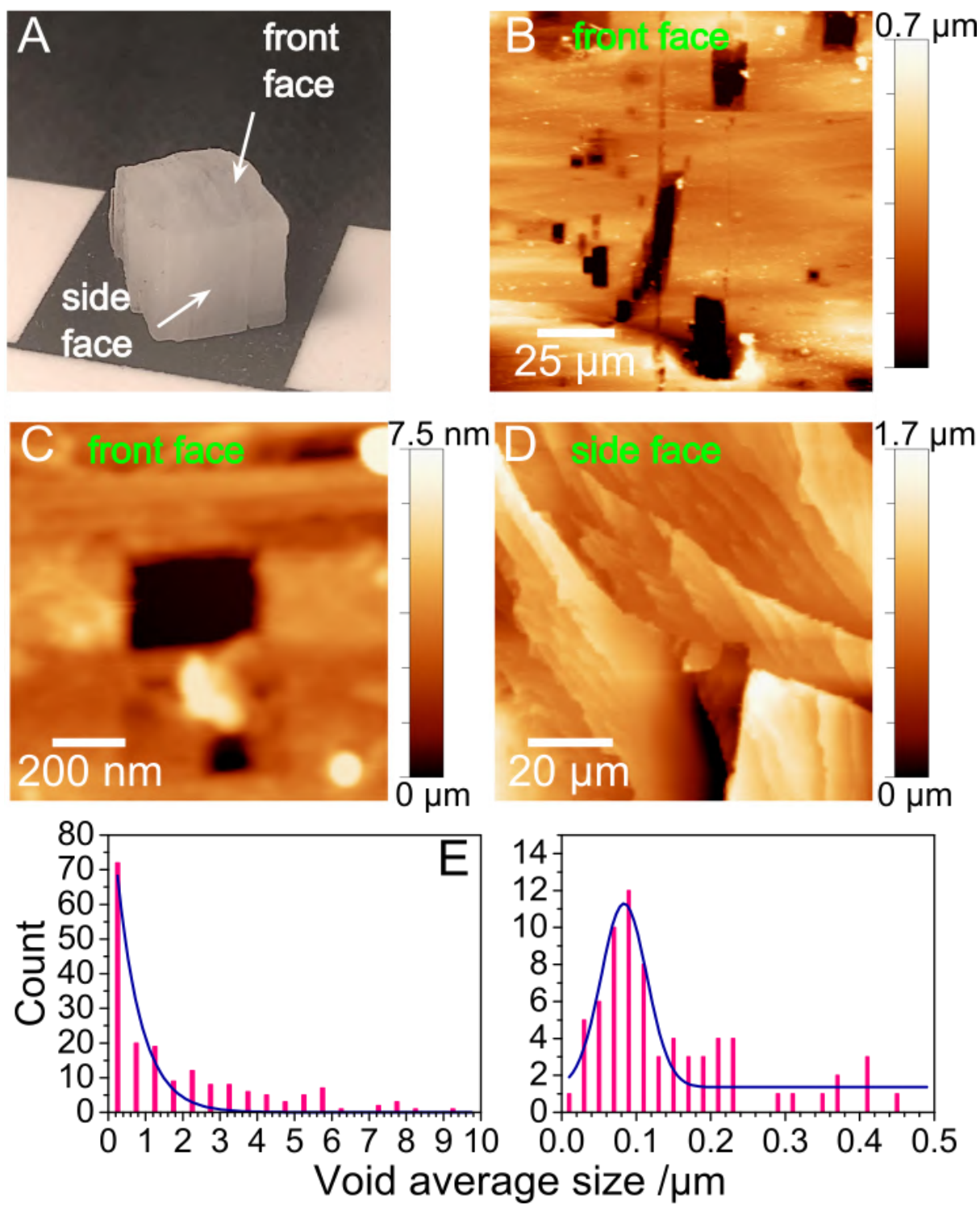

Fig. 5. AFM characterisation of the crystal facets. A) View of a typical hydrothermal anhydrite sample from Naica; a side of the black square, on which the crystal is laid, is 1 cm. AFM images of the selected sides of the anhydrite crystal, which show topographical details of B) the top face at low magnification; C) the top face at high magnification; D) the side face at low magnification. The smallest observed voids exhibit well-defined edges and are <100 nm in size. E) the pore/void size distribution on the anhydrite surface obtained from the quantitative analysis of the AFM topographical images; F) the pore/void size distribution at the nanometre-length scale, where a maximum can be observed at ~85 nm; an average void size is calculated as (length+width)/2.

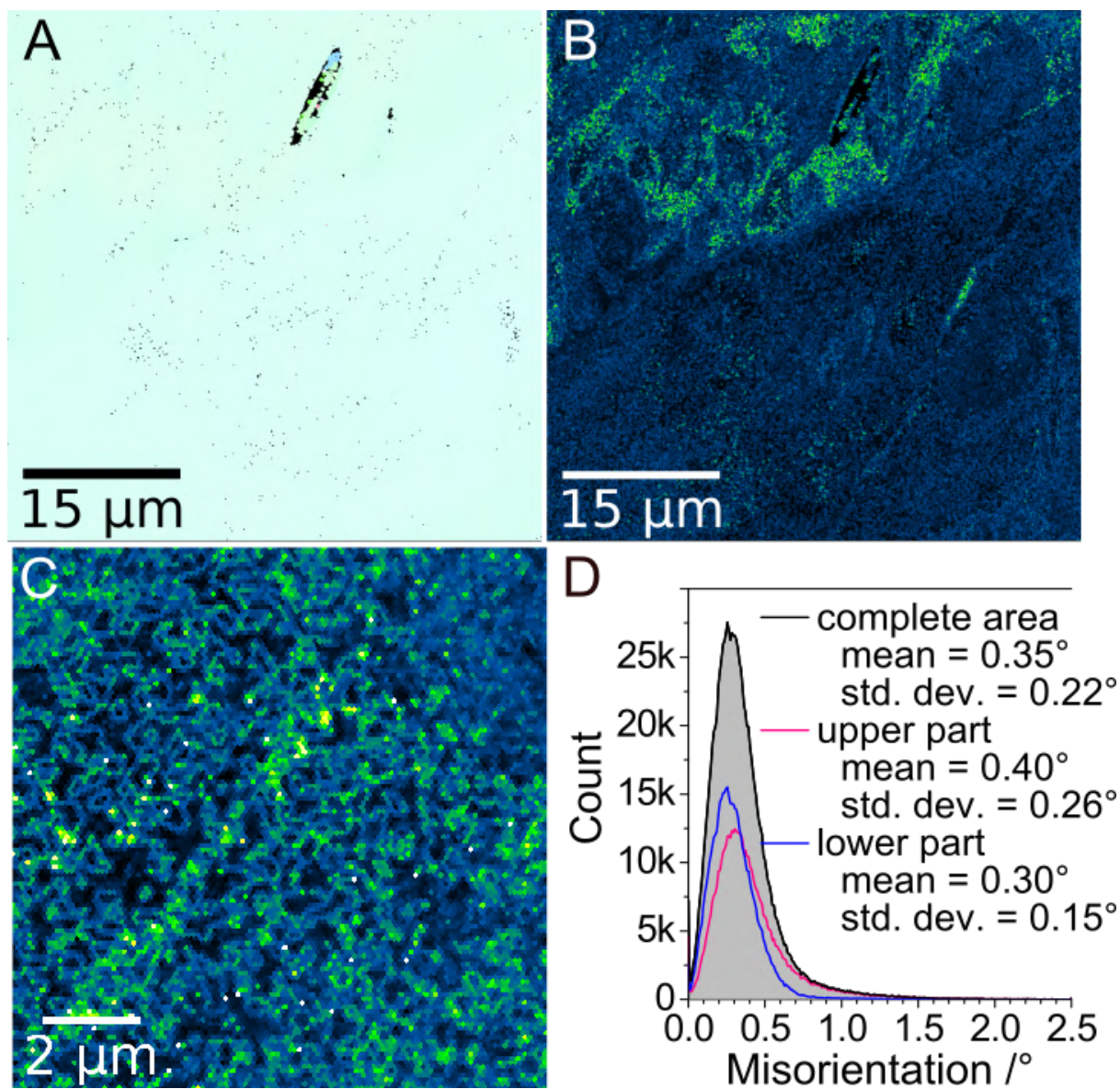

Fig. 6. EBSD maps of the selected crystal facet. A) Orientation map from the (9 5 6) facet parallel to Z in SI: Fig. S3. The corresponding pole figure is shown in SI: Fig. S7. The total viewed area is 51 x 51 µm$^2$ at a resolution of 100 nm. A *quasi*-uniform cyan colour indicates that the studied sample is a single crystal; B) Kernel average misorientation (KAM) map for the 1st neighbours with a 2.5° threshold, which highlights a heterogeneous character of (A); C) A cropped image from (B) which further illustrates the intrinsic disorder down to ~100 nm; D) distribution of orientations in (B) for the complete area (black), the upper half of the image (pink) and the lower half of the image (blue); in (B) and (C) the intensity is expressed using a Green-Fire-Blue palette (black-blue-green-yellow) from ImageJ2(78) , where green codes misorientation of ~ 0.9° and black of 0°.

# Supplementary Information for:

## Seeds of imperfection rule the mesocrystalline disorder in natural anhydrite single crystals

Tomasz M. Stawski*, Glen J. Smales, Ernesto Scoppola, Diwaker Jha, Luiz F. G. Morales, Alicia Moya, Richard Wirth, Brian R. Pauw, Franziska Emmerling, and Alexander E. S. Van Driessche**

* tomasz.stawski@bam.de; **alexander.van-driessche@univ-grenoble-alpes.fr

## Table of Contents:

### *Supplementary Figures*

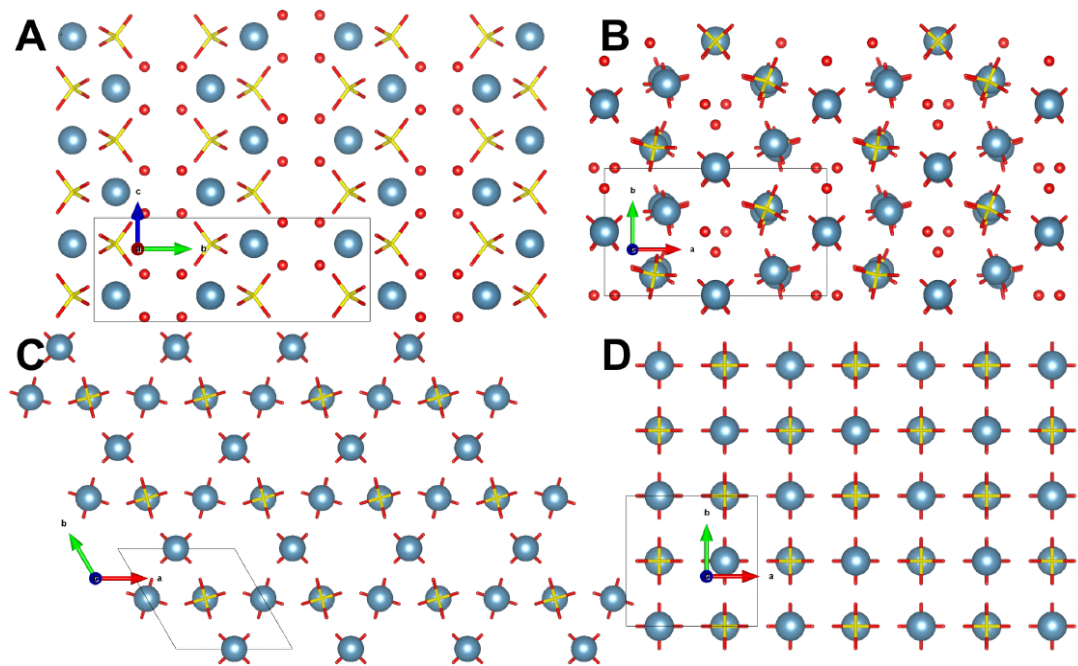

Fig. S1. Visualisation of different crystal structures of CaSO$_4$; Blue spheres - Ca$^{2+}$, yellow-red stick tetrahedra - SO$_4^{2-}$, red spheres - O in H$_2$O; The crystal orientations/projections are expressed by three vectors *a* - red, *b* - green, *c* - blue; A) CaSO$_4$·2H$_2$O, gypsum; B) CaSO$_4$·0.5H$_2$O, bassanite; C) γ-CaSO$_4$, AIII anhydrite; D) β-CaSO$_4$, AII insoluble anhydrite. Prepared in VESTA(1).

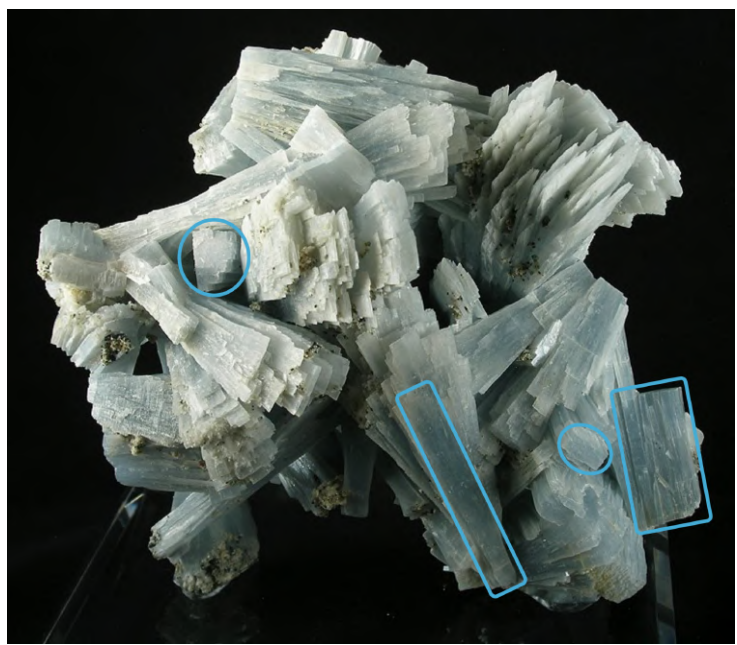

Fig. S2. Centimetric hydrothermal anhydrite single crystals from the Naica Mine, Mexico. Specimen size is 16.8 cm x 15.4 cm x 10.8 cm, and the largest anhydrite "fans" are up to 10.5 cm long(2). Blue squares indicate top views of single crystals, while blue rectangles indicate side views of single crystals. (Image: R. M. Lavinsky, CC-BY-SA-3.0)

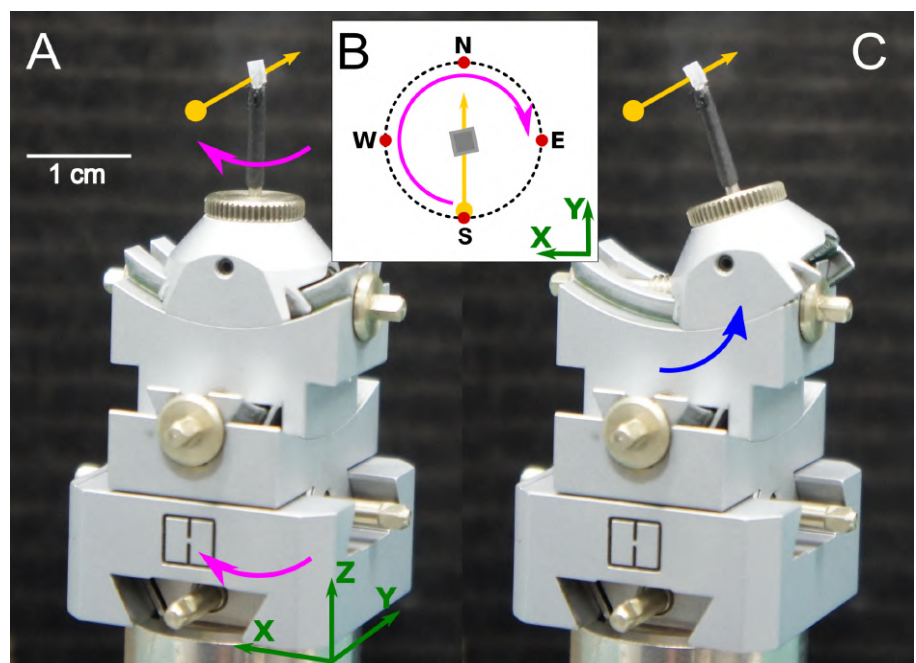

Fig. S3. Anhydrite crystal orientations during WAXS and SAXS measurements. The initial crystal orientation was arbitrary, and the goniometer was set to null positions. The XYZ cardinal directions are indicated in green; the X-ray beam direction is marked with a yellow arrow, where the beam is parallel to Y; the magenta arrow indicates the direction of rotation in the XY plane; the blue arrow indicates a tilt direction in XZ plane; A) Initial crystal orientation, S, used for the WAXS and SAXS measurements; XZ-tilt was 0°; B) Further crystal positions S-E, 90° apart from each other, used only for WAXS corresponding to a rotation of the crystal around its axis in the XY plane; C) Crystal in position S with an additional XZ-tilt of 21° used for SAXS.

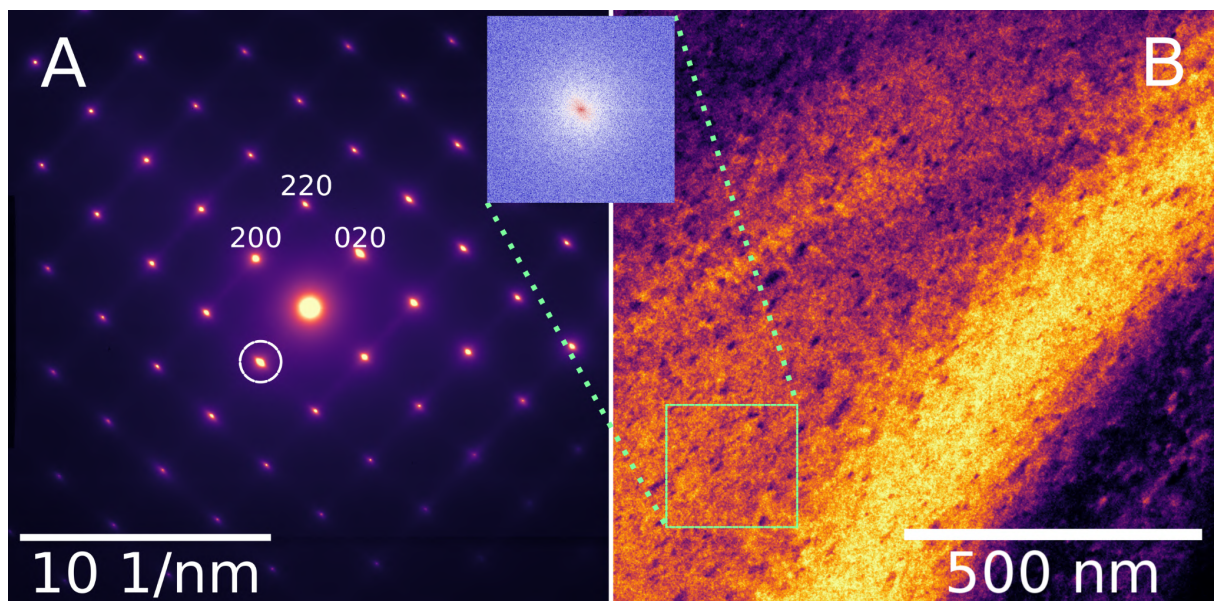

Fig. S4. A) SAED pattern from a FIB foil, which corresponds to a field of view of under 2 μm x 2 μm, and measured with a 1 μm aperture. Characteristic reflections of anhydrite are indexed. Diffuse scattering streaks are visible. They are parallel to 100 (weaker) and 010 (stronger) directions, and passing through all diffraction spots; the diffracted beam used for dark-field imaging is marked with a circle; B) dark-field TEM image with a field of view of 1.07 μm x 1.07 μm collected with a diffracted beam marked in (A); an inset shows a FFT from the area marked with a green rectangle; the the dark-field image highlights the crystallographic heterogeneities in the imaged foil, where a uniform crystal lattice should exhibit homogenous contrast; the local FFT shows that the objects within the rectangle are anisotropic and are orientated. Further analysis of this data set can be also found in our earlier work regarding mesocrystallinity in calcium sulfate(3).

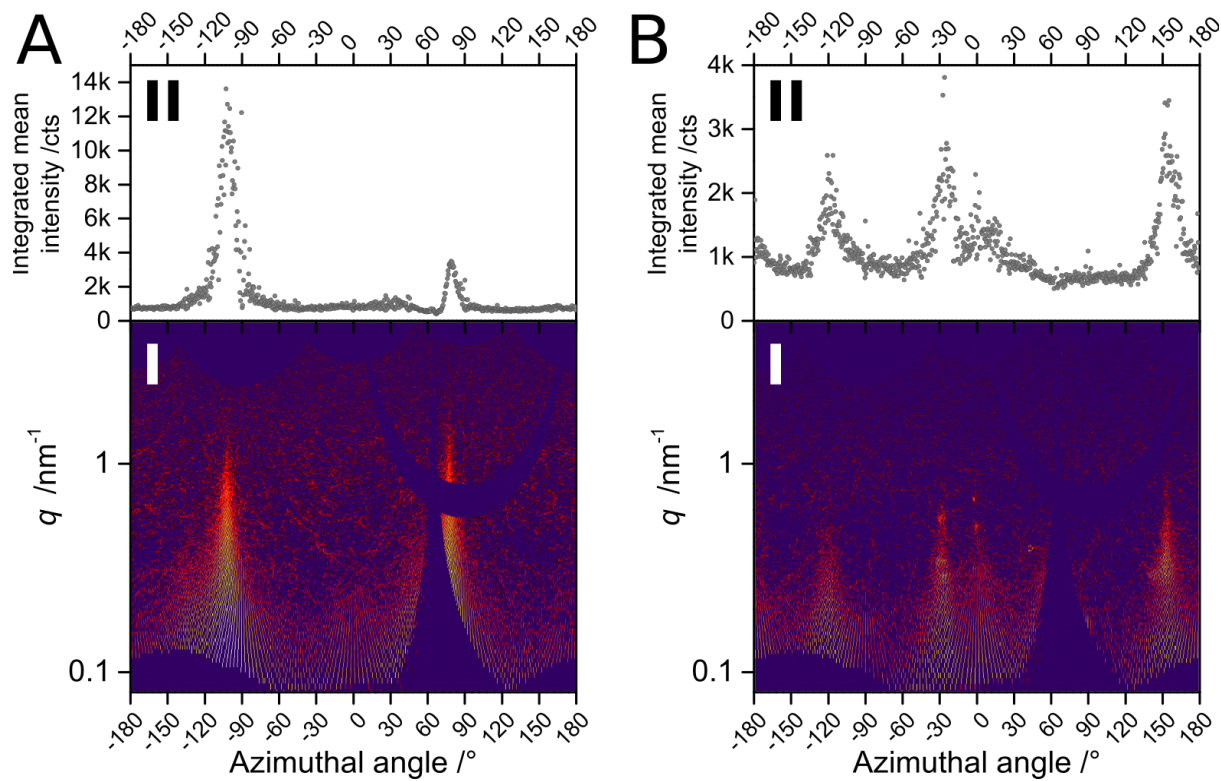

Fig. S5. Polar-coordinate representation of the SAXS data in Fig. 2 in the main text, where a cake plot is in I, and the integrated mean intensity from I is plotted in II. A) Transformed SAXS from Fig. 2A of the crystal in Position S at 0°; B) Transformed SAXS from Fig. 2B of the crystal in Position S at 21° tilt.

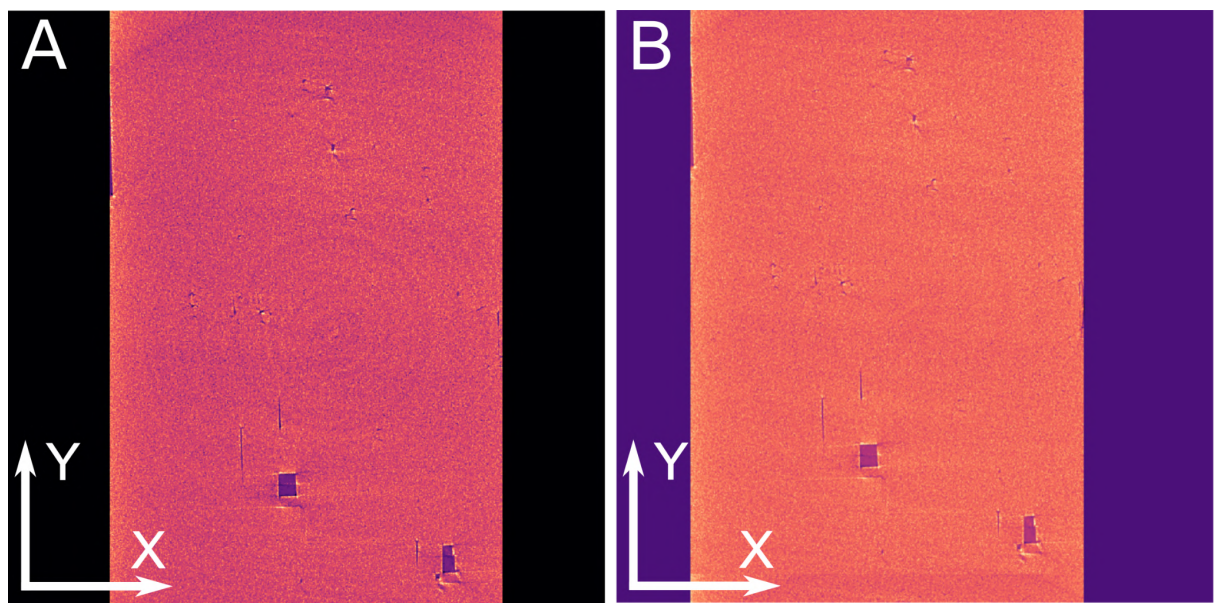

Fig. S6. An example of reconstructed images in the XY-plane from our tomography dataset, before (A) and after (B) corrections following ref.(4). In (A) the artifactual reconstruction co-centric rings are clearly visible, whereas in (B) they are considerably suppressed, although still present.

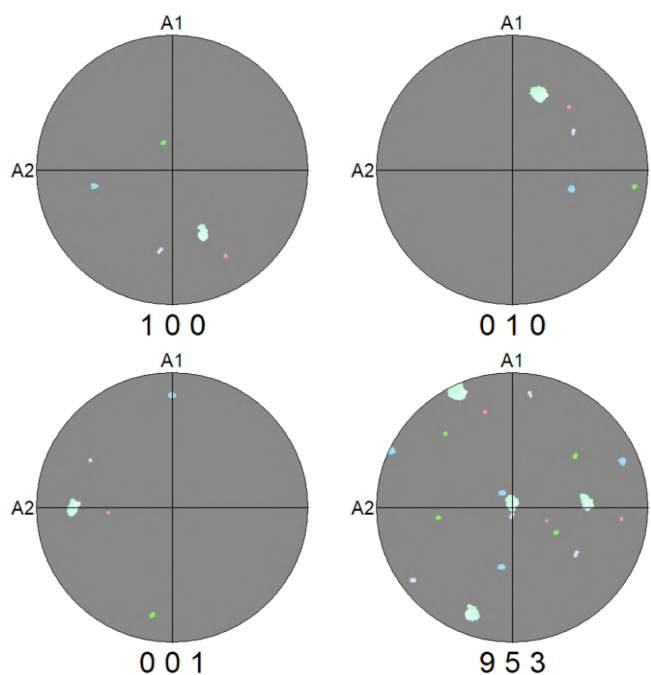

Fig. S7. Pole figures of the crystal orientation data plotted in the upper hemisphere of an equal-angle stereographic projection with the three main crystal directions of anhydrite ([1 0 0], [0 1 0], [0 0 1]) and the poles to (9 5 3), one of which plots right in the middle of the pole figure and indicates that the mapped facet is parallel to this anhydrite crystal plane.

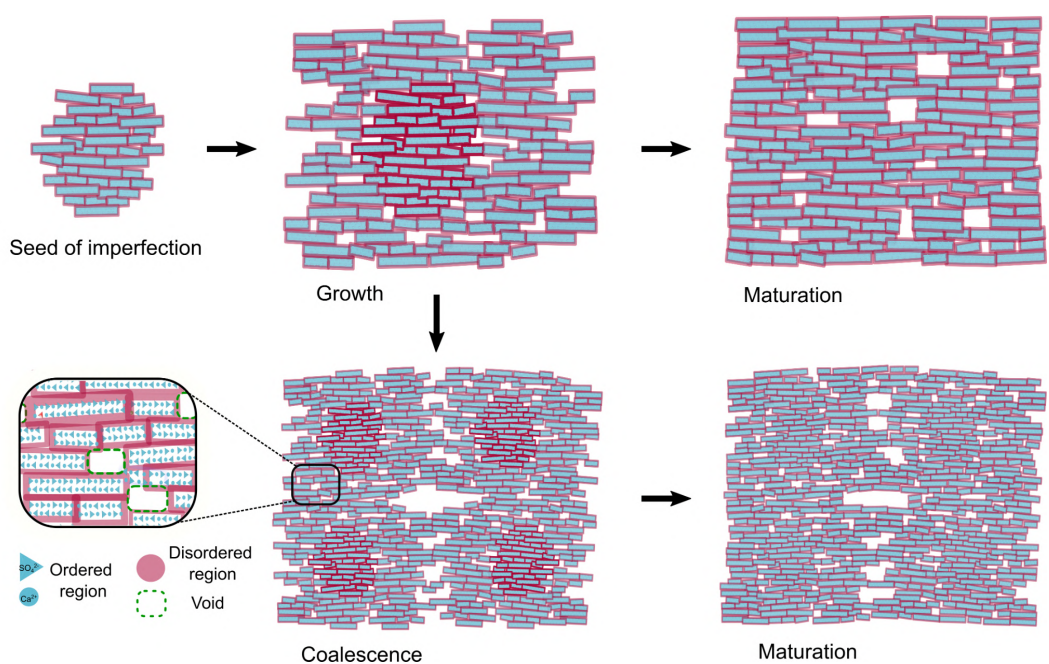

Fig. S8. Schematic representation of a tentative void formation mechanism. Through a nonclassical particle-mediated nucleation pathway(3, 5),a so-called "seed of imperfection" is formed. Such seeds will continue to grow through ion-by-ion addition and/or coalescence, during which lattice mismatches pile up. These disordered areas need to be compensated for growth to continue, which is achieved by the creation of voids that reduce the crystal structure strain. Over time, i.e. during maturation, these voids adopt shapes dominated by the more stable crystallographic directions to minimize their surface free energy.

# *Supplementary Notes*

*Supplementary Note 1: Anhydrite and other phases of CaSO$_4$*

Anhydrite is one of the three major phases of calcium sulfate. Its geological name directly implies that it is anhydrous (CaSO$_4$) in contrast to dihydrate gypsum (CaSO$_4$·2H$_2$O) and hemihydrate bassanite (CaSO$_4$·0.5H$_2$O). In terms of crystallography and physicochemical properties anhydrite at least three major (sub)phases ought to be considered designated as AI, AII and AIII. They differ in terms of structure, stability, formation pathway and natural occurrence. In the conceptually easiest case, anhydrite is formed in the course of dehydration of gypsum by heat treatment under wet and/or dry conditions. For the temperatures upto ~200 °C, gypsum converts first to bassanite: an alpha-form for dry calcination in air, and a beta-form for heating in water, aqueous solutions, organic solvents e.g. alcohols or in the presence of water vapour(5). The structural differences between the α- and β- hemihydrates are rather minor (although present) and appear to be more relevant at a morphological level. In general, the layered structure of gypsum (with alternating water and CaSO$_4$ in parallel sheets, Fig. S1A) is converted into a distinct channel structure of bassanite  (Fig. S1B). The channels are formed by CaSO$_4$ chains aligned along the *c*-axis, and are filled with water molecules. Upon prolonged and slow exposure to high temperatures (upto ~200 ºC) water is lost gradually, which converts bassanite to an AIII anhydrite (γ-CaSO$_4$) phase. AIII anhydrite is crystallographically distinct from bassanite, but the two compounds are closely related, and often difficult to unequivocally recognise by diffraction methods(6). In bassanite, along the *c*-axis direction parallel to the channel  (Fig. S1B), a channel is formed/surrounded by three Ca$^{2+}$ ions and three sulfate tetrahedra arranged into an imperfect hexagonal pattern, while in AIII anhydrite an analogous hexagonal arrangement appears perfectly ordered (Fig. S1C). Importantly, AIII anhydrite is metastable and will react with even trace amounts of water, and convert back to bassanite through, as it appears, a simple crystal lattice rearrangement. Therefore, the AIII phase is also known as soluble anhydrite, and is found practically only in engineered environments. In this regard, the further rehydration of bassanite back to gypsum is far more complex as it occurs through a dissolution-reprecipitation crystallisation mechanism(7). An ongoing heat-treatment beyond ~200, and up to 1200 °C, converts the metastable AIII form into an orthorhombic AII phase (insoluble anhydrite, β-CaSO$_4$), which has either two or three pseudo-phases depending on the actual calcination temperature (Fig. S1D). These sub-phases are crystallographically identical and the differences are morphological, which still affects e.g. reactivity with water. In air AII is stable up to 1200 °C and in water AII is the thermodynamic stable form from(5, 7) ~42-58 °C also up to 1200 °C, above which a high-temperature AI form is found(6, 8) (α-CaSO$_4$). As it is explained in the main text, anhydrite AII is commonly encountered in evaporitic environments on the Earth's surface and it is also the only natural anhydrite phase (at least on Earth), which can crystallise directly from aqueous solutions. Hence, a natural AII anhydrite phase can be obtained through two non-related pathways/mechanisms: high-temperature calcination(8, 9) or aqueous crystallisation(10). Although the final material is identical in terms of the molecular structure, at the microstructural level the resulting solid phases are clearly different in terms of morphology, crystallite size and habit etc. In particular, a slow growth from aqueous solution may potentially yield macroscopic single crystals of AII anhydrite, which are not conceivable in the course of thermal treatment of hydrated calcium sulfate phases.

*Supplementary Note 2: Scattering from idealised single crystals*

Here, we first consider how an idealised macroscopic single crystal, i.e. infinitely large and continuous from the point of probed length-scales, should scatter in both SAXS and WAXS regions. Typically, for a crystal of a simple inorganic compound one would expect that at decreasing scattering angles in SAXS, the probed characteristic distances increase, and thus at the mesoscale the considered single crystal would appear to be a homogenous and a continuous object in terms of electron density contrast. Effectively, SAXS should yield a featureless flat signal, *i.e.* $I(q) \propto q^0$, proportional to the square of the electron density of the material. Moreover, such a SAXS pattern would be isotropic and independent from the crystal orientation. Mesoscale and atomic-scale defects in the crystal structure, such as point defects, dislocations or stacking faults, at the length-scales probed by SAXS would not contribute sufficiently to electron density contrast variations in the material. On the other hand, for the considered case WAXS directly should correspond to a typical single crystal diffraction measurement, in which sharp diffraction spots are observed. The intensity and the position/arrangement of single spots would depend on the actual structure and symmetry, as well as on the relative crystal orientation with respect to the beam and the detector. Thus, in general the recorded scattering signal should be anisotropic in WAXS.

*Supplementary Note 3: Analysis of direction-dependent SAXS profiles*

In Fig. 2A in the main text, two regions are considered: (1) a high-intensity profile with a centroid at -100° together with its symmetric counterpart at +80° integrated for azimuthal angles ±22.5°, and (2) a low-intensity profile at -10° and 170° integrated in the same way. The two resulting curves are shown in Fig. 2C. In comparison, the cross-shaped scattering pattern in Fig. 2B contains more direction-dependent components. Based on representation II in Fig. 2B, the first high-intensity direction has a centroid at -120°, but its symmetric counterpart at +60° is nearly fully covered by the beamstop. The resulting integrated curve I is shown Fig. 2D. The second high-intensity direction has a pair of centroids at -30° and +150° i.e. perpendicular to the 1st direction. In this case the two curves (II in Fig. 2D) were not averaged together to highlight better the weak small-angle diffraction peaks, which are stronger for the +150° centroid side. In addition, the pattern in Fig. 2B shows a narrow streak at ~30° diagonally from the main cross-shaped pattern at 0°. This streak nominally should also have a +180° counterpart, but it is less pronounced and noisy, and hence was not included in the integration to a final curve III in Fig. 2D. The observed slight asymmetries in the complementary scattering profiles most likely stem from sample thickness variations and hence different absorption values across the sample.

# Supplementary Videos

*Supplementary Video 1:* 3D projection of a reconstructed crystal derived from the X-ray microtomography (µCT) measurements. 360° rotation of the crystal around the Z-axis in the XY-plane.

*Supplementary Video 2:* 3D projection of a reconstructed crystal derived from the X-ray microtomography (µCT) measurements. 360° rotation of the crystal around the Y-axis in the XZ-plane.

*Supplementary Video 3:* 3D projection of a reconstructed crystal derived from the X-ray microtomography (µCT) measurements with a segmentation highlighting the voids. 360° rotation of the crystal around the Z-axis in the XY-plane.

# Supplementary Files

*Anhydrite structural file:* Solved anhydrite structure from the analysed single crystal sample. Solution from single-crystal X-ray diffraction. Please note: for technical reasons the file is provided in a TXT format, and the extension should be simply changed to CIF.

# Supplementary References

```
_audit_creation_method            SHELXL-2013
_chemical_name_systematic
;
 ?
;
_chemical_name_common             ?
_chemical_melting_point           ?
_chemical_formula_moiety          'Ca O4 S'
_chemical_formula_sum
 'Ca O4 S'
_chemical_formula_weight          136.14

 _atom_type_symbol
 _atom_type_description
 _atom_type_scat_dispersion_real
 _atom_type_scat_dispersion_imag
 _atom_type_scat_source
 'O'  'O'   0.0106   0.0060
 'International Tables Vol C Tables 4.2.6.8 and 6.1.1.4'
 'S'  'S'   0.1246   0.1234
 'International Tables Vol C Tables 4.2.6.8 and 6.1.1.4'
 'Ca'  'Ca'   0.2262   0.3064
 'International Tables Vol C Tables 4.2.6.8 and 6.1.1.4'

_space_group_crystal_system       orthorhombic
_space_group_IT_number            63
_space_group_name_H-M_alt         'C m c m'
_space_group_name_Hall            '-C 2c 2'

_shelx_space_group_comment
;
The symmetry employed for this shelxl refinement is uniquely defined
by the following loop, which should always be used as a source of
symmetry information in preference to the above space-group names.
They are only intended as comments.
;

 _space_group_symop_operation_xyz
 'x, y, z'
 'x, -y, -z'
 '-x, y, -z+1/2'
 '-x, -y, z+1/2'
 'x+1/2, y+1/2, z'
 'x+1/2, -y+1/2, -z'
 '-x+1/2, y+1/2, -z+1/2'
 '-x+1/2, -y+1/2, z+1/2'
 '-x, -y, -z'
 '-x, y, z'
 'x, -y, z-1/2'
 'x, y, -z-1/2'
 '-x+1/2, -y+1/2, -z'
 '-x+1/2, y+1/2, z'
 'x+1/2, -y+1/2, z-1/2'
 'x+1/2, y+1/2, -z-1/2'

_cell_length_a                    6.992(2)
_cell_length_b                    6.2281(17)
_cell_length_c                    7.017(2)
_cell_angle_alpha                 90
_cell_angle_beta                  90
_cell_angle_gamma                 90
_cell_volume                      305.59(15)
_cell_formula_units_Z             4
_cell_measurement_temperature     298(2)
```

```
_cell_measurement_reflns_used      66
_cell_measurement_theta_min        2.70
_cell_measurement_theta_max        29.14

_exptl_crystal_description         plate
_exptl_crystal_colour              colourless
_exptl_crystal_density_meas        ?
_exptl_crystal_density_method      ?
_exptl_crystal_density_diffrn      2.959
_exptl_crystal_F_000               272
_exptl_transmission_factor_min
0.004
_exptl_transmission_factor_max
0.07
_exptl_crystal_size_max            2.5
_exptl_crystal_size_mid            2
_exptl_crystal_size_min            1
_exptl_absorpt_coefficient_mu      2.555

_exptl_absorpt_correction_type     'multi-scan'
_exptl_absorpt_correction_T_min    ?
_exptl_absorpt_correction_T_max    ?
_exptl_absorpt_process_details     '(SADABS; Sheldrick, 1996)'

_exptl_special_details
;
 ?
;

_diffrn_ambient_temperature        298(2)
_diffrn_radiation_wavelength       0.71073
_diffrn_radiation_type             MoK\a
_diffrn_radiation_source           'fine-focus sealed tube'
_diffrn_radiation_monochromator    graphite
_diffrn_measurement_device_type    'Bruker APEX-II CCD'
_diffrn_measurement_method         '\f and \w scans'
_diffrn_detector_area_resol_mean   ?
_diffrn_reflns_number              1727
_diffrn_reflns_av_unetI/netI       0.0780
_diffrn_reflns_av_R_equivalents    0.2273
_diffrn_reflns_limit_h_min         -8
_diffrn_reflns_limit_h_max         9
_diffrn_reflns_limit_k_min         -8
_diffrn_reflns_limit_k_max         8
_diffrn_reflns_limit_l_min         -9
_diffrn_reflns_limit_l_max         9
_diffrn_reflns_theta_min           4.382
_diffrn_reflns_theta_max           27.468
_diffrn_reflns_theta_full          25.242
_diffrn_measured_fraction_theta_max   0.990
_diffrn_measured_fraction_theta_full  0.988
_diffrn_reflns_Laue_measured_fraction_max   0.990
_diffrn_reflns_Laue_measured_fraction_full  0.988
_diffrn_reflns_point_group_measured_fraction_max   0.990
_diffrn_reflns_point_group_measured_fraction_full  0.988
_reflns_number_total               201
_reflns_number_gt                  200
_reflns_threshold_expression       'I > 2\s(I)'
_reflns_Friedel_coverage           0.000
_reflns_Friedel_fraction_max       .
_reflns_Friedel_fraction_full      .

_reflns_special_details
;
 Reflections were merged by SHELXL according to the crystal
 class for the calculation of statistics and refinement.

 _reflns_Friedel_fraction is defined as the number of unique
 Friedel pairs measured divided by the number that would be
```

possible theoretically, ignoring centric projections and
systematic absences.
;

_computing_data_collection       'Bruker APEX2'
_computing_cell_refinement       'Bruker SAINT'
_computing_data_reduction        'Bruker SAINT'
_computing_structure_solution    'SHELXS-97 (Sheldrick 2008)'
_computing_structure_refinement  'SHELXL-2014 (Sheldrick 2014)'
_computing_molecular_graphics    'Bruker SHELXTL'
_computing_publication_material  'Bruker SHELXTL'

_refine_special_details
;
 ?
;
_refine_ls_structure_factor_coef  Fsqd
_refine_ls_matrix_type            full
_refine_ls_weighting_scheme       calc
_refine_ls_weighting_details
'w=1/[\s^2^(Fo^2^)+(0.0387P)^2^+0.9886P] where P=(Fo^2^+2Fc^2^)/3'
_atom_sites_solution_primary      ?
_atom_sites_solution_secondary    ?
_atom_sites_solution_hydrogens    .
_refine_ls_hydrogen_treatment     undef
_refine_ls_extinction_method      SHELXL
_refine_ls_extinction_coef        0.097(15)
_refine_ls_extinction_expression
'Fc^*^=kFc[1+0.001xFc^2^\l^3^/sin(2\q)]^-1/4^'
_refine_ls_number_reflns          201
_refine_ls_number_parameters      22
_refine_ls_number_restraints      0
_refine_ls_R_factor_all           0.0460
_refine_ls_R_factor_gt            0.0457
_refine_ls_wR_factor_ref          0.1091
_refine_ls_wR_factor_gt           0.1087
_refine_ls_goodness_of_fit_ref    1.190
_refine_ls_restrained_S_all       1.190
_refine_ls_shift/su_max           0.000
_refine_ls_shift/su_mean          0.000

 _atom_site_label
 _atom_site_type_symbol
 _atom_site_fract_x
 _atom_site_fract_y
 _atom_site_fract_z
 _atom_site_U_iso_or_equiv
 _atom_site_adp_type
 _atom_site_occupancy
 _atom_site_site_symmetry_order
 _atom_site_calc_flag
 _atom_site_refinement_flags_posn
 _atom_site_refinement_flags_adp
 _atom_site_refinement_flags_occupancy
 _atom_site_disorder_assembly
 _atom_site_disorder_group
Ca1 Ca 0.5000 0.8479(2) 0.7500 0.0110(7) Uani 1 4 d S T P . .
S1 S 0.5000 0.6549(3) 0.2500 0.0097(7) Uani 1 4 d S T P . .
O1 O 0.3299(4) 0.5155(5) 0.2500 0.0157(9) Uani 1 2 d S T P . .
O2 O 0.5000 0.7970(5) 0.4177(4) 0.0147(9) Uani 1 2 d S T P . .

 _atom_site_aniso_label
 _atom_site_aniso_U_11
 _atom_site_aniso_U_22
 _atom_site_aniso_U_33
 _atom_site_aniso_U_23

```
 _atom_site_aniso_U_13
 _atom_site_aniso_U_12
Ca1 0.0127(8) 0.0096(11) 0.0108(8) 0.000 0.000 0.000
S1 0.0099(9) 0.0091(13) 0.0101(8) 0.000 0.000 0.000
O1 0.0118(16) 0.0114(18) 0.0240(16) 0.000 0.000 -0.0023(11)
O2 0.0193(16) 0.0115(18) 0.0132(16) -0.0009(10) 0.000 0.000

_geom_special_details
;
 All esds (except the esd in the dihedral angle between two l.s. planes)
 are estimated using the full covariance matrix.  The cell esds are taken
 into account individually in the estimation of esds in distances, angles
 and torsion angles; correlations between esds in cell parameters are only
 used when they are defined by crystal symmetry.  An approximate (isotropic)
 treatment of cell esds is used for estimating esds involving l.s. planes.
;

 _geom_bond_atom_site_label_1
 _geom_bond_atom_site_label_2
 _geom_bond_distance
 _geom_bond_site_symmetry_2
 _geom_bond_publ_flag
Ca1 O2 2.353(3) . ?
Ca1 O2 2.353(3) 3_656 ?
Ca1 O1 2.459(3) 6_566 ?
Ca1 O1 2.459(3) 13_566 ?
Ca1 O2 2.506(3) 11_576 ?
Ca1 O2 2.506(3) 9_676 ?
Ca1 O1 2.557(4) 2_566 ?
Ca1 O1 2.557(4) 9_666 ?
Ca1 S1 3.0966(17) 9_676 ?
Ca1 S1 3.1315(17) 9_666 ?
Ca1 S1 3.4962(10) 13_666 ?
Ca1 S1 3.4962(10) 13_566 ?
S1 O2 1.473(3) 3_655 ?
S1 O2 1.473(3) . ?
S1 O1 1.473(3) 10_655 ?
S1 O1 1.473(3) . ?
S1 Ca1 3.0966(17) 9_676 ?
S1 Ca1 3.1315(17) 9_666 ?
S1 Ca1 3.4962(10) 13_566 ?
S1 Ca1 3.4962(10) 13_666 ?
O1 Ca1 2.459(3) 13_566 ?
O1 Ca1 2.557(4) 9_666 ?
O2 Ca1 2.506(3) 9_676 ?

 _geom_angle_atom_site_label_1
 _geom_angle_atom_site_label_2
 _geom_angle_atom_site_label_3
 _geom_angle
 _geom_angle_site_symmetry_1
 _geom_angle_site_symmetry_3
 _geom_angle_publ_flag
O2 Ca1 O2 164.51(16) . 3_656 ?
O2 Ca1 O1 92.67(3) . 6_566 ?
O2 Ca1 O1 92.67(3) 3_656 6_566 ?
O2 Ca1 O1 92.67(3) . 13_566 ?
O2 Ca1 O1 92.67(3) 3_656 13_566 ?
O1 Ca1 O1 139.50(16) 6_566 13_566 ?
O2 Ca1 O2 125.77(9) . 11_576 ?
O2 Ca1 O2 69.72(12) 3_656 11_576 ?
O1 Ca1 O2 72.21(7) 6_566 11_576 ?
O1 Ca1 O2 72.21(7) 13_566 11_576 ?
O2 Ca1 O2 69.72(12) . 9_676 ?
O2 Ca1 O2 125.77(9) 3_656 9_676 ?
O1 Ca1 O2 72.21(7) 6_566 9_676 ?
O1 Ca1 O2 72.21(7) 13_566 9_676 ?
```

O2 Ca1 O2 56.04(13) 11_576 9_676 ?
O2 Ca1 O1 83.15(7) . 2_566 ?
O2 Ca1 O1 83.15(7) 3_656 2_566 ?
O1 Ca1 O1 137.97(14) 6_566 2_566 ?
O1 Ca1 O1 82.52(6) 13_566 2_566 ?
O2 Ca1 O1 141.39(6) 11_576 2_566 ?
O2 Ca1 O1 141.39(6) 9_676 2_566 ?
O2 Ca1 O1 83.15(7) . 9_666 ?
O2 Ca1 O1 83.15(7) 3_656 9_666 ?
O1 Ca1 O1 82.52(6) 6_566 9_666 ?
O1 Ca1 O1 137.97(14) 13_566 9_666 ?
O2 Ca1 O1 141.39(6) 11_576 9_666 ?
O2 Ca1 O1 141.39(6) 9_676 9_666 ?
O1 Ca1 O1 55.45(14) 2_566 9_666 ?
O2 Ca1 S1 97.75(8) . 9_676 ?
O2 Ca1 S1 97.74(8) 3_656 9_676 ?
O1 Ca1 S1 69.75(8) 6_566 9_676 ?
O1 Ca1 S1 69.75(8) 13_566 9_676 ?
O2 Ca1 S1 28.02(6) 11_576 9_676 ?
O2 Ca1 S1 28.02(7) 9_676 9_676 ?
O1 Ca1 S1 152.28(7) 2_566 9_676 ?
O1 Ca1 S1 152.28(7) 9_666 9_676 ?
O2 Ca1 S1 82.25(8) . 9_666 ?
O2 Ca1 S1 82.26(8) 3_656 9_666 ?
O1 Ca1 S1 110.25(8) 6_566 9_666 ?
O1 Ca1 S1 110.25(8) 13_566 9_666 ?
O2 Ca1 S1 151.98(7) 11_576 9_666 ?
O2 Ca1 S1 151.98(7) 9_676 9_666 ?
O1 Ca1 S1 27.72(7) 2_566 9_666 ?
O1 Ca1 S1 27.72(7) 9_666 9_666 ?
S1 Ca1 S1 180.0 9_676 9_666 ?
O2 Ca1 S1 89.961(4) . 13_666 ?
O2 Ca1 S1 89.961(4) 3_656 13_666 ?
O1 Ca1 S1 20.54(8) 6_566 13_666 ?
O1 Ca1 S1 160.04(9) 13_566 13_666 ?
O2 Ca1 S1 90.25(2) 11_576 13_666 ?
O2 Ca1 S1 90.25(2) 9_676 13_666 ?
O1 Ca1 S1 117.44(8) 2_566 13_666 ?
O1 Ca1 S1 61.99(7) 9_666 13_666 ?
S1 Ca1 S1 90.29(2) 9_676 13_666 ?
S1 Ca1 S1 89.71(2) 9_666 13_666 ?
O2 Ca1 S1 89.961(4) . 13_566 ?
O2 Ca1 S1 89.961(4) 3_656 13_566 ?
O1 Ca1 S1 160.04(9) 6_566 13_566 ?
O1 Ca1 S1 20.54(8) 13_566 13_566 ?
O2 Ca1 S1 90.25(2) 11_576 13_566 ?
O2 Ca1 S1 90.25(2) 9_676 13_566 ?
O1 Ca1 S1 61.99(7) 2_566 13_566 ?
O1 Ca1 S1 117.44(8) 9_666 13_566 ?
S1 Ca1 S1 90.29(2) 9_676 13_566 ?
S1 Ca1 S1 89.71(2) 9_666 13_566 ?
S1 Ca1 S1 179.43(5) 13_666 13_566 ?
O2 S1 O2 106.1(3) 3_655 . ?
O2 S1 O1 110.75(9) 3_655 10_655 ?
O2 S1 O1 110.75(9) . 10_655 ?
O2 S1 O1 110.75(9) 3_655 . ?
O2 S1 O1 110.75(9) . . ?
O1 S1 O1 107.7(3) 10_655 . ?
O2 S1 Ca1 53.07(13) 3_655 9_676 ?
O2 S1 Ca1 53.07(13) . 9_676 ?
O1 S1 Ca1 126.14(14) 10_655 9_676 ?
O1 S1 Ca1 126.14(14) . 9_676 ?
O2 S1 Ca1 126.93(13) 3_655 9_666 ?
O2 S1 Ca1 126.93(13) . 9_666 ?
O1 S1 Ca1 53.86(14) 10_655 9_666 ?
O1 S1 Ca1 53.86(14) . 9_666 ?
Ca1 S1 Ca1 180.0 9_676 9_666 ?
O2 S1 Ca1 90.172(15) 3_655 13_566 ?
O2 S1 Ca1 90.172(15) . 13_566 ?

O1 S1 Ca1 143.58(14) 10_655 13_566 ?
O1 S1 Ca1 35.85(13) . 13_566 ?
Ca1 S1 Ca1 90.29(2) 9_676 13_566 ?
Ca1 S1 Ca1 89.71(2) 9_666 13_566 ?
O2 S1 Ca1 90.172(15) 3_655 13_666 ?
O2 S1 Ca1 90.172(15) . 13_666 ?
O1 S1 Ca1 35.85(13) 10_655 13_666 ?
O1 S1 Ca1 143.58(14) . 13_666 ?
Ca1 S1 Ca1 90.29(2) 9_676 13_666 ?
Ca1 S1 Ca1 89.71(2) 9_666 13_666 ?
Ca1 S1 Ca1 179.43(5) 13_566 13_666 ?
S1 O1 Ca1 123.6(2) . 13_566 ?
S1 O1 Ca1 98.41(16) . 9_666 ?
Ca1 O1 Ca1 137.97(14) 13_566 9_666 ?
S1 O2 Ca1 150.82(19) . . ?
S1 O2 Ca1 98.91(14) . 9_676 ?
Ca1 O2 Ca1 110.28(12) . 9_676 ?

O2 S1 O1 Ca1 -58.74(13) 3_655 . . 13_566 ?
O2 S1 O1 Ca1 58.74(13) . . . 13_566 ?
O1 S1 O1 Ca1 180.0 10_655 . . 13_566 ?
Ca1 S1 O1 Ca1 0.0 9_676 . . 13_566 ?
Ca1 S1 O1 Ca1 180.0 9_666 . . 13_566 ?
Ca1 S1 O1 Ca1 180.0 13_666 . . 13_566 ?
O2 S1 O1 Ca1 121.26(13) 3_655 . . 9_666 ?
O2 S1 O1 Ca1 -121.26(13) . . . 9_666 ?
O1 S1 O1 Ca1 0.0 10_655 . . 9_666 ?
Ca1 S1 O1 Ca1 180.0 9_676 . . 9_666 ?
Ca1 S1 O1 Ca1 180.0 13_566 . . 9_666 ?
Ca1 S1 O1 Ca1 0.0 13_666 . . 9_666 ?
O2 S1 O2 Ca1 180.000(1) 3_655 . . . ?
O1 S1 O2 Ca1 -59.73(14) 10_655 . . . ?
O1 S1 O2 Ca1 59.73(14) . . . . ?
Ca1 S1 O2 Ca1 180.000(1) 9_676 . . . ?
Ca1 S1 O2 Ca1 0.000(1) 9_666 . . . ?
Ca1 S1 O2 Ca1 89.771(19) 13_566 . . . ?
Ca1 S1 O2 Ca1 -89.77(2) 13_666 . . . ?
O2 S1 O2 Ca1 0.000(0) 3_655 . . 9_676 ?
O1 S1 O2 Ca1 120.27(14) 10_655 . . 9_676 ?
O1 S1 O2 Ca1 -120.27(14) . . . 9_676 ?
Ca1 S1 O2 Ca1 180.000(0) 9_666 . . 9_676 ?
Ca1 S1 O2 Ca1 -90.23(2) 13_566 . . 9_676 ?
Ca1 S1 O2 Ca1 90.23(2) 13_666 . . 9_676 ?

TITL fe575_a.res in Cmcm
CELL 0.71073 6.9923 6.2281 7.0171 90.000 90.000 90.000
ZERR 4.00 0.0020 0.0017 0.0020 0.000 0.000 0.000
LATT 7
SYMM x, -y, -z
SYMM -x, y, -z+1/2

```
SYMM -x, -y, z+1/2
SFAC O S CA
UNIT 16 4 4
LIST 6 ! automatically inserted. Change 6 to 4 for CHECKCIF!!
TEMP 24.850
ACTA
L.S. 4
FMAP 2
PLAN -0 0 0.00
HTAB
OMIT -3 55
BOND $H
CONF
WGHT     0.038700    0.988600
EXTI     0.097406
FVAR     0.69578
CA1   3    0.500000    0.847886    0.750000    10.25000    0.01267    0.00955 =
           0.01080    0.00000    0.00000    0.00000
S1    2    0.500000    0.654919    0.250000    10.25000    0.00992    0.00911 =
           0.01011    0.00000    0.00000    0.00000
O1    1    0.329905    0.515478    0.250000    10.50000    0.01176    0.01140 =
           0.02400    0.00000    0.00000   -0.00234
O2    1    0.500000    0.796973    0.417750    10.50000    0.01929    0.01151 =
           0.01318   -0.00090    0.00000    0.00000
REM <hkl>
REM D:\frames\FE\fe575\work\autostructure_private\fe575_a.hkl
REM </hkl>
HKLF 4 1 1 0 0 0 1 0 0 0 1

REM  fe575_a.res in Cmcm
REM R1 =  0.0457 for     200 Fo > 4sig(Fo)  and  0.0460 for all     201 data
REM     22 parameters refined using      0 restraints

END

WGHT      0.0387      0.9891

REM No hydrogen bonds found for HTAB generation

REM Highest difference peak  0.880,  deepest hole -0.607,  1-sigma level  0.171
Q1    1    0.5000  0.6666  0.3865  10.50000  0.05    0.88
Q2    1    0.5000  0.9970  0.7500  10.25000  0.05    0.70
Q3    1    0.5000  0.5007  0.2500  10.25000  0.05    0.69
Q4    1    0.5000  0.8436  0.6294  10.50000  0.05    0.66
Q5    1    0.5000  0.8182  0.2500  10.25000  0.05    0.63
Q6    1    0.4274  0.3910  0.4164  11.00000  0.05    0.45
Q7    1    0.2675  0.3577  0.2500  10.50000  0.05    0.44
Q8    1    0.2602  0.5452  0.3825  11.00000  0.05    0.44
Q9    1    0.5000  0.7569  0.4915  10.50000  0.05    0.42
Q10   1    0.5626  0.9141  0.4216  11.00000  0.05    0.39
Q11   1    0.2267  0.5417  0.2500  10.50000  0.05    0.38
Q12   1    0.5876  0.8312  0.3820  11.00000  0.05    0.33
Q13   1    0.2017  0.6662  0.3387  11.00000  0.05    0.33
Q14   1    0.7658  0.7623  0.3985  11.00000  0.05    0.28
Q15   1    0.5000  0.6971  0.7500  10.25000  0.05    0.28
Q16   1    0.5000  0.5000  0.5000  10.25000  0.05    0.27
Q17   1    0.1596  0.4315  0.3248  11.00000  0.05    0.20
Q18   1    0.6636  0.8870  0.5192  11.00000  0.05    0.15
Q19   1    0.3318  0.5930  0.4937  11.00000  0.05    0.11
Q20   1    0.5000  0.9473  0.4996  10.50000  0.05    0.09
;
_shelx_res_checksum   79090

_shelx_hkl_file
;
   0   0   1    0.70    0.08
   0   0   1    1.14    0.12
   0   0  -1   -0.03    0.06
   0   0  -1    0.40    0.06
```

| | | | | |
|---|---|---|---|---|
| 0 | 0 | -1 | 1.55 | 0.12 |
| 0 | 0 | 3 | 0.29 | 0.19 |
| 0 | 0 | -3 | 14.22 | 0.51 |
| 0 | 0 | -3 | 0.42 | 0.23 |
| 0 | 0 | -4 | 2843.14 | 7.76 |
| 0 | 0 | 5 | 0.57 | 0.32 |
| 0 | 0 | -5 | 0.90 | 0.23 |
| 0 | 0 | 6 | 3318.62 | 9.80 |
| 0 | 0 | -6 | 3111.85 | 8.41 |
| 0 | 0 | -6 | 2989.13 | 10.35 |
| 0 | 0 | 7 | 1.15 | 1.17 |
| 0 | 0 | -7 | 2.72 | 0.46 |
| 0 | 0 | -7 | 3.46 | 0.98 |
| 0 | 0 | 8 | 1056.98 | 6.74 |
| 0 | 0 | -8 | 932.11 | 6.34 |
| 0 | 0 | -8 | 1000.55 | 7.34 |
| 0 | 0 | 9 | 0.53 | 1.02 |
| 0 | 0 | -9 | 5.50 | 1.32 |
| 0 | 0 | -9 | 2.41 | 0.89 |
| 0 | 0 | 10 | 636.19 | 6.36 |
| 0 | 0 | -10 | 968.24 | 6.87 |
| 0 | 0 | -10 | 784.15 | 7.76 |
| 0 | 0 | 11 | 2.14 | 2.02 |
| 0 | 0 | -11 | -1.62 | 1.55 |
| 0 | 0 | -11 | 2.64 | 1.48 |
| 0 | 0 | 12 | 455.37 | 7.22 |
| 0 | 0 | -12 | 1117.70 | 10.57 |
| 0 | 0 | 13 | -4.61 | 3.76 |
| 0 | 0 | -13 | 1.16 | 2.22 |
| 0 | 0 | 14 | 59.47 | 7.10 |
| 0 | 0 | -14 | 338.13 | 7.26 |
| -2 | 0 | 1 | 5.89 | 0.54 |
| -2 | 0 | -1 | 0.67 | 0.17 |
| -2 | 0 | -1 | 0.47 | 0.11 |
| 2 | 0 | 1 | 0.44 | 0.13 |
| 2 | 0 | 1 | 3.04 | 0.23 |
| 2 | 0 | -1 | 0.91 | 0.22 |
| -2 | 0 | 2 | 773.14 | 3.94 |
| -2 | 0 | 2 | 1193.22 | 2.91 |
| -2 | 0 | -2 | 938.35 | 3.30 |
| -2 | 0 | -2 | 1102.88 | 3.07 |
| 2 | 0 | 2 | 1124.64 | 3.11 |
| 2 | 0 | -2 | 1774.73 | 6.34 |
| 2 | 0 | -2 | 1104.70 | 2.80 |
| -2 | 0 | 3 | 0.55 | 0.19 |
| -2 | 0 | -3 | 0.59 | 0.17 |
| 2 | 0 | 3 | 0.40 | 0.24 |
| 2 | 0 | -3 | 1.22 | 0.44 |
| 2 | 0 | -3 | 3.20 | 0.41 |
| -2 | 0 | 4 | 1162.06 | 4.31 |
| -2 | 0 | -4 | 775.37 | 3.98 |
| 2 | 0 | 4 | 1012.68 | 4.29 |
| 2 | 0 | -4 | 1260.13 | 4.73 |
| 2 | 0 | -4 | 1053.88 | 5.88 |
| -2 | 0 | 5 | 0.79 | 0.30 |
| -2 | 0 | -5 | 1.19 | 0.32 |
| 2 | 0 | 5 | 0.72 | 0.44 |
| 2 | 0 | -5 | 11.83 | 0.92 |
| 2 | 0 | -5 | 3.25 | 0.53 |
| -2 | 0 | 6 | 1874.58 | 7.15 |
| -2 | 0 | -6 | 1674.35 | 7.52 |
| 2 | 0 | 6 | 2033.27 | 7.84 |
| 2 | 0 | -6 | 2084.64 | 7.22 |
| 2 | 0 | -6 | 1804.62 | 9.12 |
| -2 | 0 | 7 | 0.45 | 0.66 |
| -2 | 0 | -7 | 3.25 | 0.46 |
| -2 | 0 | -7 | 7.51 | 0.91 |
| 2 | 0 | 7 | 1.91 | 0.70 |
| 2 | 0 | -7 | 18.05 | 1.33 |
| 2 | 0 | -7 | 3.73 | 1.52 |
| -2 | 0 | 8 | 484.07 | 4.50 |
| -2 | 0 | -8 | 467.64 | 4.32 |
| -2 | 0 | -8 | 457.15 | 4.87 |
| 2 | 0 | 8 | 524.55 | 5.09 |
| 2 | 0 | -8 | 569.05 | 6.08 |
| -2 | 0 | 9 | 0.39 | 0.96 |
| -2 | 0 | -9 | 0.76 | 0.76 |
| -2 | 0 | -9 | 21.45 | 1.74 |
| 2 | 0 | 9 | -1.41 | 1.18 |
| 2 | 0 | -9 | -0.71 | 1.10 |
| -2 | 0 | 10 | 359.81 | 4.80 |
| -2 | 0 | -10 | 427.25 | 4.85 |
| -2 | 0 | -10 | 416.08 | 5.70 |
| 2 | 0 | 10 | 358.28 | 5.25 |
| 2 | 0 | -10 | 759.60 | 7.40 |
| 2 | 0 | -10 | 502.49 | 6.65 |
| -2 | 0 | 11 | 0.30 | 1.90 |
| -2 | 0 | -11 | 5.25 | 1.43 |
| -2 | 0 | -11 | 4.74 | 1.58 |
| 2 | 0 | 11 | 1.88 | 2.45 |
| 2 | 0 | -11 | 5.37 | 2.03 |
| 2 | 0 | -11 | 13.32 | 1.95 |
| -2 | 0 | 12 | 238.82 | 6.00 |
| -2 | 0 | -12 | 789.41 | 7.69 |
| -2 | 0 | -12 | 779.20 | 13.44 |
| 2 | 0 | 12 | 279.02 | 6.78 |
| 2 | 0 | -12 | 823.99 | 10.44 |
| -2 | 0 | 13 | 1.00 | 3.40 |
| -2 | 0 | -13 | 1.39 | 2.38 |
| 2 | 0 | 13 | -5.52 | 4.72 |
| 2 | 0 | -13 | 4.45 | 2.46 |
| -2 | 0 | 14 | 65.64 | 6.52 |
| -2 | 0 | -14 | 210.01 | 6.37 |
| 2 | 0 | 14 | 45.25 | 9.91 |
| 4 | 0 | 0 | 3982.91 | 9.37 |
| -4 | 0 | -1 | 2.56 | 0.31 |
| 4 | 0 | 1 | 1.50 | 0.26 |
| -4 | 0 | 2 | 1856.05 | 6.26 |
| -4 | 0 | 2 | 962.97 | 2.66 |
| -4 | 0 | -2 | 1013.43 | 4.53 |
| -4 | 0 | -2 | 918.42 | 3.04 |
| 4 | 0 | 2 | 801.71 | 4.15 |
| 4 | 0 | -2 | 1434.86 | 6.42 |
| -4 | 0 | 3 | 0.80 | 0.24 |
| -4 | 0 | -3 | 1.63 | 0.27 |
| -4 | 0 | -3 | 0.34 | 0.49 |
| 4 | 0 | 3 | 0.31 | 0.47 |
| 4 | 0 | -3 | -0.03 | 0.44 |
| -4 | 0 | 4 | 864.66 | 3.74 |
| -4 | 0 | -4 | 785.68 | 4.59 |
| 4 | 0 | 4 | 907.83 | 4.14 |
| 4 | 0 | -4 | 1221.37 | 6.96 |
| -4 | 0 | -5 | 1.75 | 0.42 |
| 4 | 0 | 5 | 0.18 | 0.65 |
| 4 | 0 | -5 | 4.27 | 0.74 |
| -4 | 0 | 6 | 1517.74 | 6.50 |
| -4 | 0 | -6 | 1516.84 | 7.73 |
| 4 | 0 | 6 | 1591.52 | 7.09 |
| 4 | 0 | -6 | 1834.63 | 9.57 |
| -4 | 0 | 7 | 0.01 | 0.64 |
| -4 | 0 | -7 | 2.48 | 0.74 |
| 4 | 0 | 7 | 1.51 | 1.12 |
| 4 | 0 | -7 | 2.51 | 1.01 |
| -4 | 0 | 8 | 412.11 | 4.17 |
| -4 | 0 | -8 | 459.57 | 4.70 |
| -4 | 0 | -8 | 558.74 | 5.62 |
| 4 | 0 | 8 | 349.17 | 7.46 |
| 4 | 0 | -8 | 534.36 | 6.02 |
| -4 | 0 | 9 | 1.66 | 1.07 |
| -4 | 0 | -9 | -0.67 | 0.85 |
| -4 | 0 | -9 | 5.05 | 1.04 |
| 4 | 0 | 9 | -3.81 | 1.98 |
| 4 | 0 | -9 | -1.17 | 1.38 |
| -4 | 0 | 10 | 287.63 | 4.41 |
| -4 | 0 | -10 | 376.19 | 5.06 |
| -4 | 0 | -10 | 378.09 | 5.60 |
| 4 | 0 | 10 | 182.69 | 4.89 |
| 4 | 0 | -10 | 490.24 | 5.82 |
| -4 | 0 | 11 | 1.23 | 2.02 |
| -4 | 0 | -11 | -1.85 | 1.51 |
| -4 | 0 | -11 | 1.59 | 1.56 |
| 4 | 0 | 11 | -10.00 | 3.50 |
| -4 | 0 | 12 | 252.41 | 5.53 |
| -4 | 0 | -12 | 593.27 | 7.24 |
| -4 | 0 | -12 | 579.37 | 8.03 |
| -4 | 0 | 13 | -5.25 | 3.34 |
| -4 | 0 | -13 | 4.92 | 3.05 |
| -4 | 0 | -13 | -2.62 | 2.40 |
| 6 | 0 | 0 | 3630.80 | 11.33 |
| 6 | 0 | 1 | 0.93 | 0.49 |
| 6 | 0 | -1 | 2.78 | 0.55 |
| -6 | 0 | -2 | 1791.88 | 4.83 |
| 6 | 0 | 2 | 1447.54 | 7.14 |
| 6 | 0 | -2 | 2051.19 | 8.86 |
| -6 | 0 | 3 | 3.20 | 0.46 |
| -6 | 0 | -3 | 28.26 | 1.19 |
| -6 | 0 | -3 | 1.37 | 0.77 |
| 6 | 0 | -3 | 2.20 | 0.63 |
| -6 | 0 | 4 | 1461.14 | 5.98 |
| -6 | 0 | -4 | 1457.23 | 7.70 |
| 6 | 0 | -4 | 1827.88 | 10.65 |
| -6 | 0 | 5 | 3.18 | 0.82 |
| -6 | 0 | -5 | 2.17 | 0.63 |
| 6 | 0 | -5 | 3.01 | 0.83 |
| -6 | 0 | 6 | 1722.69 | 7.15 |
| -6 | 0 | -6 | 1857.49 | 9.55 |
| 6 | 0 | 6 | 973.21 | 5.54 |
| 6 | 0 | -6 | 1661.61 | 11.69 |
| -6 | 0 | 7 | 3.12 | 1.00 |
| -6 | 0 | -7 | 0.66 | 0.86 |
| 6 | 0 | 7 | -1.79 | 2.43 |
| 6 | 0 | -7 | 1.68 | 1.25 |
| -6 | 0 | 8 | 589.51 | 5.21 |
| -6 | 0 | -8 | 701.35 | 6.64 |
| -6 | 0 | -8 | 717.03 | 6.92 |
| 6 | 0 | -8 | 704.00 | 6.27 |
| -6 | 0 | 9 | 1.64 | 1.57 |
| -6 | 0 | -9 | -0.63 | 1.13 |
| -6 | 0 | -9 | 2.17 | 1.22 |
| 6 | 0 | -9 | -2.75 | 1.57 |
| -6 | 0 | 10 | 323.51 | 5.05 |
| -6 | 0 | -10 | 486.72 | 6.39 |
| -6 | 0 | -10 | 559.72 | 7.22 |
| -6 | 0 | 11 | -3.08 | 2.54 |
| -6 | 0 | -11 | -1.04 | 2.04 |
| -6 | 0 | -11 | 7.36 | 2.15 |
| -6 | 0 | 12 | 257.57 | 5.95 |
| -6 | 0 | -12 | 613.13 | 8.00 |
| -6 | 0 | -12 | 617.91 | 8.59 |
| -6 | 0 | -13 | 0.72 | 3.56 |
| 8 | 0 | 0 | 874.48 | 6.62 |
| 8 | 0 | 1 | 0.94 | 0.73 |
| 8 | 0 | -1 | 3.28 | 0.82 |
| -8 | 0 | -2 | 478.39 | 4.83 |
| 8 | 0 | 2 | 355.03 | 4.35 |
| 8 | 0 | -2 | 427.51 | 4.74 |
| -8 | 0 | -3 | 6.29 | 0.90 |
| 8 | 0 | -3 | 2.61 | 0.95 |
| -8 | 0 | 4 | 242.39 | 3.29 |
| -8 | 0 | -4 | 416.57 | 4.69 |
| 8 | 0 | -4 | 404.35 | 4.70 |
| -8 | 0 | -5 | 2.27 | 0.99 |
| 8 | 0 | -5 | 5.97 | 1.18 |
| -8 | 0 | 6 | 645.61 | 4.94 |
| -8 | 0 | -6 | 682.76 | 6.64 |
| 8 | 0 | -6 | 635.48 | 5.87 |
| -8 | 0 | 7 | 4.89 | 1.62 |
| -8 | 0 | -7 | 4.04 | 1.33 |
| 8 | 0 | -7 | -1.95 | 1.30 |
| -8 | 0 | 8 | 245.17 | 4.16 |
| -8 | 0 | -8 | 246.74 | 4.82 |
| -8 | 0 | -8 | 262.63 | 4.98 |
| 8 | 0 | -8 | 213.97 | 3.87 |
| -8 | 0 | 9 | 6.83 | 2.47 |
| -8 | 0 | -9 | 1.17 | 2.19 |
| -8 | 0 | -9 | -0.07 | 2.05 |
| -8 | 0 | 10 | 126.89 | 4.66 |
| -8 | 0 | -10 | 211.63 | 5.30 |
| -8 | 0 | -10 | 220.70 | 5.63 |
| -8 | 0 | 11 | 5.59 | 3.33 |
| -8 | 0 | -11 | 1.73 | 2.47 |
| -8 | 0 | -11 | 4.02 | 3.17 |
| 10 | 0 | 0 | 509.05 | 5.93 |
| 10 | 0 | -1 | 3.11 | 1.46 |
| 10 | 0 | 1 | 2.27 | 1.18 |
| 10 | 0 | -1 | 1.65 | 1.15 |
| -10 | 0 | -2 | 337.07 | 4.73 |
| 10 | 0 | 2 | 269.40 | 4.57 |
| 10 | 0 | -2 | 286.30 | 4.46 |
| -10 | 0 | -3 | 0.59 | 1.39 |
| 10 | 0 | -3 | 8.87 | 1.38 |
| -10 | 0 | -4 | 322.64 | 4.95 |
| 10 | 0 | -4 | 247.00 | 4.17 |
| -10 | 0 | -5 | 1.83 | 1.58 |
| 10 | 0 | -5 | 0.72 | 1.46 |
| -10 | 0 | -6 | 459.66 | 6.36 |
| 10 | 0 | -6 | 360.43 | 4.91 |
| -10 | 0 | -7 | 1.66 | 2.02 |
| -10 | 0 | -7 | 4.98 | 3.26 |
| 10 | 0 | -7 | 2.42 | 2.33 |
| -10 | 0 | -8 | 194.71 | 5.28 |
| -10 | 0 | -8 | 207.41 | 5.54 |
| -10 | 0 | 9 | 2.01 | 4.51 |
| -10 | 0 | -9 | -0.24 | 2.45 |
| -10 | 0 | -9 | 1.79 | 2.79 |
| -10 | 0 | -10 | 133.53 | 5.88 |
| -10 | 0 | -10 | 164.44 | 6.46 |
| -12 | 0 | 0 | 488.57 | 6.45 |
| 12 | 0 | 0 | 486.00 | 6.72 |
| -12 | 0 | 1 | -6.44 | 2.50 |
| -12 | 0 | -1 | 0.95 | 2.37 |
| 12 | 0 | -1 | 5.37 | 1.99 |
| -12 | 0 | -2 | 353.10 | 5.92 |
| 12 | 0 | -2 | 296.61 | 5.34 |
| -12 | 0 | -3 | 3.88 | 2.44 |
| 12 | 0 | -3 | 2.11 | 2.66 |
| -12 | 0 | -4 | 323.02 | 5.92 |
| -12 | 0 | -4 | 315.54 | 6.11 |
| 12 | 0 | -4 | 252.27 | 5.03 |
| -12 | 0 | -5 | 6.86 | 2.64 |
| -12 | 0 | -5 | 7.76 | 2.78 |
| 12 | 0 | -5 | 0.57 | 2.49 |
| -12 | 0 | -6 | 371.03 | 6.82 |
| -12 | 0 | -6 | 375.17 | 7.09 |
| -12 | 0 | -7 | -3.85 | 2.90 |
| -12 | 0 | -7 | -3.74 | 3.42 |
| -1 | -1 | 0 | 2.39 | 0.19 |
| -1 | -1 | 0 | 10.06 | 0.31 |
| 1 | -1 | 0 | 5.03 | 0.36 |
| 1 | -1 | 0 | 6.53 | 0.38 |
| 1 | -1 | 0 | 5.91 | 0.29 |
| 1 | -1 | 0 | 1.07 | 0.18 |
| -1 | 1 | 0 | 5.45 | 0.42 |
| -1 | 1 | 0 | 11.51 | 0.46 |
| -1 | 1 | 0 | 7.67 | 0.33 |
| -1 | 1 | 0 | 12.03 | 0.39 |
| 1 | 1 | 0 | 4.59 | 0.20 |
| 1 | 1 | 0 | 2.06 | 0.19 |
| -1 | -1 | 1 | 129.22 | 1.49 |
| -1 | -1 | -1 | 144.35 | 1.10 |
| -1 | -1 | -1 | 123.36 | 0.88 |
| -1 | -1 | -1 | 160.62 | 1.15 |
| 1 | -1 | 1 | 150.57 | 1.08 |
| 1 | -1 | 1 | 131.15 | 1.23 |
| 1 | -1 | 1 | 115.02 | 0.96 |
| 1 | -1 | 1 | 171.04 | 1.26 |
| 1 | -1 | -1 | 181.03 | 1.84 |
| 1 | -1 | -1 | 154.44 | 1.39 |
| 1 | -1 | -1 | 152.99 | 1.08 |
| -1 | 1 | 1 | 169.67 | 1.77 |
| -1 | 1 | 1 | 81.54 | 1.30 |
| -1 | 1 | 1 | 106.61 | 0.94 |
| -1 | 1 | 1 | 136.55 | 1.08 |
| -1 | 1 | -1 | 189.43 | 1.22 |
| -1 | 1 | -1 | 162.56 | 1.36 |
| -1 | 1 | -1 | 163.13 | 1.17 |
| -1 | 1 | -1 | 212.80 | 1.40 |
| 1 | 1 | 1 | 111.53 | 0.83 |
| 1 | 1 | 1 | 110.02 | 0.95 |
| 1 | 1 | -1 | 197.69 | 1.80 |
| -1 | -1 | 2 | 182.65 | 1.11 |
| -1 | -1 | 2 | 297.78 | 2.21 |
| -1 | -1 | 2 | 218.43 | 1.51 |
| -1 | -1 | -2 | 213.00 | 1.55 |
| -1 | -1 | -2 | 181.85 | 1.41 |
| -1 | -1 | -2 | 221.93 | 1.54 |
| 1 | -1 | 2 | 153.68 | 1.09 |
| 1 | -1 | 2 | 222.25 | 1.78 |
| 1 | -1 | 2 | 216.05 | 1.57 |
| 1 | -1 | -2 | 246.81 | 2.19 |
| 1 | -1 | -2 | 199.94 | 2.12 |
| -1 | 1 | 2 | 196.23 | 1.84 |
| -1 | 1 | 2 | 223.69 | 1.53 |
| -1 | 1 | -2 | 221.74 | 1.29 |
| -1 | 1 | -2 | 178.81 | 1.35 |
| -1 | 1 | -2 | 249.39 | 1.68 |
| 1 | 1 | 2 | 145.57 | 1.30 |
| 1 | 1 | -2 | 196.43 | 1.35 |
| 1 | 1 | -2 | 384.66 | 2.71 |
| 1 | 1 | -2 | 306.50 | 1.96 |
| -1 | -1 | 3 | 1021.53 | 3.96 |
| -1 | -1 | 3 | 887.27 | 3.49 |
| -1 | -1 | -3 | 878.24 | 3.58 |
| -1 | -1 | -3 | 781.37 | 3.56 |
| 1 | -1 | 3 | 915.87 | 3.97 |
| 1 | -1 | 3 | 909.04 | 3.70 |
| 1 | -1 | -3 | 1001.51 | 4.51 |
| 1 | -1 | -3 | 866.55 | 4.73 |
| -1 | 1 | 3 | 930.82 | 3.58 |
| -1 | 1 | -3 | 985.24 | 3.87 |
| -1 | 1 | -3 | 681.41 | 3.07 |
| 1 | 1 | 3 | 944.20 | 3.67 |
| 1 | 1 | -3 | 809.18 | 3.46 |
| 1 | 1 | -3 | 1469.80 | 5.61 |
| -1 | -1 | 4 | 69.71 | 1.07 |
| -1 | -1 | 4 | 64.76 | 1.12 |
| -1 | -1 | -4 | 63.27 | 1.09 |
| -1 | -1 | -4 | 57.71 | 1.14 |
| 1 | -1 | 4 | 70.40 | 1.20 |
| 1 | -1 | -4 | 66.96 | 1.28 |
| 1 | -1 | -4 | 65.53 | 1.46 |
| 1 | -1 | -4 | 61.73 | 0.88 |
| -1 | 1 | 4 | 69.32 | 1.15 |
| -1 | 1 | -4 | 50.95 | 1.00 |
| 1 | 1 | 4 | 73.88 | 1.22 |
| 1 | 1 | -4 | 54.39 | 0.69 |
| 1 | 1 | -4 | 61.36 | 1.15 |
| -1 | -1 | 5 | 20.50 | 0.80 |
| -1 | -1 | -5 | 19.63 | 0.74 |
| -1 | -1 | -5 | 18.22 | 0.83 |
| 1 | -1 | 5 | 23.89 | 0.87 |
| 1 | -1 | -5 | 32.19 | 1.02 |
| 1 | -1 | -5 | 20.91 | 1.06 |
| -1 | 1 | 5 | 23.41 | 0.84 |
| -1 | 1 | -5 | 17.03 | 0.76 |
| 1 | 1 | 5 | 22.63 | 0.90 |
| 1 | 1 | -5 | 17.99 | 0.57 |
| 1 | 1 | -5 | 18.24 | 0.87 |
| -1 | -1 | 6 | 0.81 | 0.63 |
| -1 | -1 | -6 | 1.37 | 0.40 |
| -1 | -1 | -6 | 1.02 | 0.49 |
| 1 | -1 | 6 | 5.91 | 0.63 |
| 1 | -1 | -6 | 0.95 | 0.49 |
| -1 | 1 | 6 | 1.05 | 0.64 |
| -1 | 1 | -6 | 1.56 | 0.44 |
| 1 | 1 | 6 | 11.56 | 0.98 |
| 1 | 1 | -6 | 2.91 | 0.38 |
| 1 | 1 | -6 | 2.77 | 0.51 |
| -1 | -1 | 7 | 32.42 | 1.51 |
| -1 | -1 | -7 | 4.59 | 0.60 |
| -1 | -1 | -7 | 4.11 | 0.81 |
| 1 | -1 | -7 | 4.49 | 0.68 |
| 1 | -1 | -7 | 5.42 | 1.01 |
| -1 | 1 | 7 | 4.36 | 0.81 |
| -1 | 1 | -7 | 4.86 | 0.54 |
| -1 | 1 | -7 | 6.06 | 0.88 |
| 1 | 1 | 7 | 7.51 | 0.74 |
| 1 | 1 | -7 | 5.35 | 0.52 |
| 1 | 1 | -7 | 4.26 | 0.65 |
| -1 | -1 | 8 | 7.16 | 0.91 |
| -1 | -1 | -8 | 6.46 | 1.00 |
| -1 | -1 | -8 | 4.43 | 0.86 |
| 1 | -1 | 8 | 14.67 | 1.60 |
| 1 | -1 | -8 | 5.19 | 1.23 |
| 1 | -1 | -8 | 7.37 | 1.03 |
| -1 | 1 | 8 | 5.29 | 0.87 |
| -1 | 1 | -8 | 9.69 | 0.96 |
| -1 | 1 | -8 | 5.28 | 0.81 |
| 1 | 1 | 8 | 8.95 | 1.17 |
| 1 | 1 | -8 | 10.72 | 1.14 |
| 1 | 1 | -8 | 9.30 | 1.14 |
| -1 | -1 | 9 | 70.67 | 2.36 |
| -1 | -1 | -9 | 82.22 | 2.36 |
| -1 | -1 | -9 | 82.53 | 2.65 |
| 1 | -1 | 9 | 80.16 | 2.63 |
| 1 | -1 | -9 | 136.04 | 4.04 |
| -1 | 1 | 9 | 67.37 | 2.38 |
| -1 | 1 | -9 | 78.24 | 2.21 |
| -1 | 1 | -9 | 79.35 | 2.49 |
| 1 | 1 | 9 | 56.78 | 2.42 |
| 1 | 1 | -9 | 98.96 | 2.21 |
| 1 | 1 | -9 | 87.96 | 2.80 |
| -1 | -1 | 10 | 2.29 | 1.30 |
| -1 | -1 | -10 | 0.99 | 1.11 |
| -1 | -1 | -10 | 2.28 | 1.26 |
| 1 | -1 | 10 | 0.53 | 1.48 |
| 1 | -1 | -10 | 1.70 | 1.41 |
| 1 | -1 | -10 | 2.43 | 1.38 |
| -1 | 1 | 10 | 0.62 | 1.39 |
| -1 | 1 | -10 | 4.15 | 1.03 |
| -1 | 1 | -10 | 2.22 | 1.15 |
| 1 | 1 | 10 | 6.47 | 1.70 |
| 1 | 1 | -10 | 2.84 | 1.29 |
| -1 | -1 | 11 | 0.81 | 1.72 |
| -1 | -1 | -11 | 1.03 | 1.51 |
| -1 | -1 | -11 | 1.95 | 1.59 |
| 1 | -1 | 11 | 1.81 | 2.06 |
| 1 | -1 | -11 | 4.18 | 1.64 |
| -1 | 1 | 11 | 0.04 | 1.86 |
| -1 | 1 | -11 | 2.13 | 1.48 |
| -1 | 1 | -11 | 2.18 | 1.50 |
| 1 | 1 | 11 | 2.90 | 2.26 |
| 1 | 1 | -11 | 2.96 | 1.67 |
| -1 | -1 | -12 | 1.05 | 2.13 |
| 1 | -1 | 12 | 1.19 | 2.88 |
| 1 | -1 | -12 | 3.71 | 2.96 |
| -1 | 1 | 12 | -5.28 | 4.96 |
| -1 | 1 | -12 | 6.47 | 2.26 |
| 1 | 1 | 12 | 2.51 | 3.24 |
| 1 | 1 | -12 | 7.40 | 2.06 |
| -1 | -1 | 13 | 0.14 | 3.38 |
| -1 | -1 | -13 | -2.95 | 2.18 |
| 1 | -1 | 13 | 2.22 | 4.05 |
| 1 | -1 | -13 | -0.31 | 2.08 |
| -1 | 1 | 13 | 4.03 | 3.63 |
| -1 | 1 | -13 | 1.75 | 2.39 |
| 1 | 1 | -13 | 7.10 | 2.46 |
| -1 | -1 | 14 | 0.10 | 4.77 |
| -1 | -1 | -14 | -5.38 | 2.76 |
| 1 | -1 | 14 | -1.30 | 5.88 |
| 1 | -1 | -14 | 2.46 | 3.40 |
| -1 | 1 | 14 | -15.01 | 5.49 |
| -1 | 1 | -14 | -6.51 | 2.97 |
| 1 | 1 | 14 | -20.11 | 6.71 |
| 1 | 1 | -14 | 0.65 | 2.77 |
| 3 | -1 | 0 | 1924.32 | 6.73 |
| -3 | 1 | 0 | 1383.91 | 3.85 |
| -3 | 1 | 0 | 769.78 | 3.09 |
| 3 | 1 | 0 | 1381.52 | 3.99 |
| -3 | -1 | 1 | 12.39 | 0.74 |
| -3 | -1 | -1 | 8.52 | 0.36 |
| -3 | -1 | -1 | 11.94 | 0.43 |
| 3 | -1 | 1 | 10.39 | 0.53 |
| 3 | -1 | 1 | 16.62 | 0.44 |
| 3 | -1 | -1 | 14.86 | 0.80 |
| -3 | 1 | 1 | 10.28 | 0.37 |
| -3 | 1 | 1 | 18.10 | 0.59 |
| -3 | 1 | -1 | 41.48 | 0.76 |
| -3 | 1 | -1 | 14.59 | 0.52 |
| 3 | 1 | 1 | 7.70 | 0.37 |
| 3 | 1 | -1 | 11.09 | 0.41 |
| -3 | -1 | 2 | 718.80 | 4.16 |
| -3 | -1 | 2 | 424.18 | 2.70 |
| -3 | -1 | 2 | 318.22 | 2.67 |
| -3 | -1 | 2 | 554.69 | 2.17 |
| -3 | -1 | -2 | 471.37 | 2.49 |
| -3 | -1 | -2 | 551.01 | 2.46 |
| 3 | -1 | 2 | 506.58 | 3.04 |
| 3 | -1 | 2 | 583.45 | 2.53 |
| 3 | -1 | -2 | 804.78 | 5.47 |
| -3 | 1 | 2 | 609.79 | 2.67 |
| -3 | 1 | -2 | 671.36 | 2.90 |
| 3 | 1 | -2 | 797.71 | 3.32 |
| -3 | -1 | 3 | 894.87 | 4.09 |
| -3 | -1 | 3 | 582.41 | 2.66 |
| -3 | -1 | -3 | 494.07 | 2.93 |
| 3 | -1 | 3 | 442.75 | 2.92 |
| 3 | -1 | 3 | 615.03 | 3.08 |
| 3 | -1 | -3 | 760.11 | 4.47 |
| 3 | -1 | -3 | 689.82 | 5.48 |
| -3 | 1 | 3 | 662.39 | 3.11 |
| -3 | 1 | -3 | 616.82 | 3.61 |
| 3 | 1 | 3 | 593.08 | 2.99 |
| 3 | 1 | -3 | 993.49 | 4.31 |
| -3 | -1 | 4 | 230.03 | 1.95 |
| -3 | -1 | -4 | 205.23 | 2.16 |
| 3 | -1 | 4 | 245.39 | 2.26 |
| 3 | -1 | -4 | 294.42 | 2.98 |
| 3 | -1 | -4 | 275.40 | 3.68 |
| -3 | 1 | 4 | 256.38 | 2.19 |
| -3 | 1 | -4 | 235.50 | 2.41 |
| 3 | 1 | 4 | 238.07 | 2.22 |
| 3 | 1 | -4 | 401.06 | 3.17 |
| 3 | 1 | -4 | 301.97 | 3.31 |
| -3 | -1 | 5 | 6.23 | 0.44 |
| -3 | -1 | 5 | 8.06 | 0.67 |
| -3 | -1 | -5 | 7.20 | 0.74 |
| -3 | -1 | -5 | 4.13 | 0.52 |
| 3 | -1 | 5 | 8.67 | 0.69 |
| 3 | -1 | -5 | 9.65 | 1.17 |
| 3 | -1 | -5 | 4.73 | 0.88 |
| 3 | -1 | -5 | 4.62 | 0.77 |
| -3 | 1 | 5 | 5.44 | 0.53 |
| -3 | 1 | -5 | 8.33 | 0.76 |
| 3 | 1 | 5 | 4.55 | 0.69 |
| 3 | 1 | -5 | 9.03 | 1.01 |
| 3 | 1 | -5 | 11.95 | 1.22 |
| -3 | -1 | -6 | 192.16 | 2.56 |
| 3 | -1 | 6 | 201.26 | 2.63 |
| 3 | -1 | -6 | 224.45 | 2.88 |
| 3 | -1 | -6 | 200.45 | 3.61 |
| 3 | -1 | -6 | 202.91 | 3.53 |
| -3 | 1 | 6 | 176.22 | 3.50 |
| -3 | 1 | -6 | 188.20 | 3.89 |
| 3 | 1 | 6 | 169.23 | 2.47 |
| 3 | 1 | -6 | 289.81 | 3.29 |
| 3 | 1 | -6 | 216.35 | 3.36 |
| -3 | -1 | 7 | 0.12 | 0.67 |
| -3 | -1 | -7 | 1.29 | 0.57 |
| -3 | -1 | -7 | 1.02 | 0.61 |
| 3 | -1 | 7 | 0.07 | 0.95 |
| 3 | -1 | -7 | -0.71 | 0.82 |
| 3 | -1 | -7 | 1.51 | 1.03 |
| 3 | -1 | -7 | 4.28 | 1.19 |
| -3 | 1 | 7 | 3.11 | 0.69 |
| -3 | 1 | -7 | 1.11 | 0.66 |
| 3 | 1 | -7 | 5.56 | 0.88 |
| 3 | 1 | -7 | 3.17 | 0.88 |
| -3 | -1 | 8 | 37.67 | 1.59 |
| -3 | -1 | -8 | 38.00 | 1.60 |
| -3 | -1 | -8 | 37.81 | 1.74 |
| 3 | -1 | 8 | 38.69 | 2.47 |
| 3 | -1 | -8 | 55.00 | 1.99 |
| 3 | -1 | -8 | 52.82 | 2.95 |
| 3 | -1 | -8 | 42.40 | 2.40 |
| -3 | 1 | 8 | 40.79 | 1.74 |
| -3 | 1 | -8 | 39.78 | 1.80 |
| 3 | 1 | 8 | 30.07 | 2.65 |
| 3 | 1 | -8 | 70.54 | 2.30 |
| 3 | 1 | -8 | 44.35 | 2.58 |
| -3 | -1 | 9 | 72.58 | 2.30 |
| -3 | -1 | -9 | 83.07 | 2.40 |
| -3 | -1 | -9 | 87.56 | 2.67 |
| 3 | -1 | 9 | 77.54 | 2.99 |
| 3 | -1 | -9 | 110.75 | 3.54 |
| 3 | -1 | -9 | 96.34 | 3.40 |
| -3 | 1 | 9 | 67.02 | 2.41 |
| -3 | 1 | -9 | 76.18 | 2.16 |
| -3 | 1 | -9 | 81.26 | 2.62 |
| 3 | 1 | 9 | 56.19 | 2.90 |
| 3 | 1 | -9 | 125.76 | 2.99 |
| 3 | 1 | -9 | 113.16 | 3.43 |
| -3 | -1 | 10 | 21.94 | 2.26 |
| -3 | -1 | -10 | 24.21 | 2.15 |
| -3 | -1 | -10 | 23.56 | 2.38 |
| 3 | -1 | 10 | 19.46 | 2.27 |
| 3 | -1 | -10 | 36.42 | 3.17 |
| 3 | -1 | -10 | 29.83 | 2.94 |
| -3 | 1 | 10 | 14.91 | 1.75 |
| -3 | 1 | -10 | 22.86 | 1.99 |
| -3 | 1 | -10 | 22.88 | 2.29 |
| 3 | 1 | 10 | 17.72 | 2.46 |
| 3 | 1 | -10 | 3.07 | 1.80 |
| 3 | 1 | -10 | 37.69 | 3.03 |
| -3 | -1 | 11 | 3.76 | 1.83 |
| -3 | -1 | -11 | 3.45 | 1.54 |
| -3 | -1 | -11 | 2.43 | 1.66 |
| 3 | -1 | 11 | -1.75 | 2.94 |
| 3 | -1 | -11 | -0.78 | 1.94 |
| 3 | -1 | -11 | 2.36 | 1.75 |
| -3 | 1 | 11 | -2.30 | 1.89 |
| -3 | 1 | -11 | 4.06 | 1.47 |
| -3 | 1 | -11 | 2.77 | 1.62 |
| 3 | 1 | -11 | 5.89 | 1.86 |
| -3 | -1 | 12 | 30.71 | 3.75 |
| -3 | -1 | -12 | 35.55 | 4.12 |
| -3 | -1 | -12 | 44.56 | 3.77 |
| 3 | -1 | -12 | 51.85 | 4.66 |
| -3 | 1 | 12 | 13.37 | 2.90 |
| -3 | 1 | -12 | 60.08 | 3.90 |
| -3 | 1 | -12 | 38.48 | 3.74 |
| 3 | 1 | 12 | 36.54 | 5.44 |
| -3 | -1 | 13 | -2.71 | 3.32 |
| -3 | -1 | -13 | -5.66 | 2.97 |
| -3 | -1 | -13 | -6.14 | 2.36 |
| 3 | -1 | 13 | 0.50 | 5.89 |
| -3 | 1 | 13 | 12.79 | 4.05 |
| -3 | 1 | -13 | 7.95 | 2.51 |
| -3 | -1 | -14 | 16.03 | 4.99 |
| -3 | 1 | -14 | 20.21 | 5.16 |
| 5 | -1 | 0 | 16.46 | 0.92 |
| -5 | 1 | 0 | 2.06 | 0.42 |
| 5 | 1 | 0 | 14.87 | 0.67 |
| 5 | -1 | 1 | 23.94 | 0.99 |
| 5 | 1 | 1 | 30.21 | 1.23 |
| 5 | 1 | 1 | 17.93 | 0.72 |
| 5 | 1 | -1 | 21.23 | 0.78 |
| -5 | -1 | 2 | 19.77 | 0.98 |
| -5 | -1 | -2 | 46.41 | 1.01 |
| -5 | -1 | -2 | 53.52 | 0.90 |
| 5 | -1 | 2 | 46.81 | 1.28 |
| 5 | -1 | -2 | 68.09 | 1.83 |
| -5 | 1 | 2 | 82.18 | 1.26 |
| -5 | 1 | -2 | 63.16 | 1.22 |
| 5 | 1 | -2 | 58.63 | 1.20 |
| -5 | -1 | 3 | 153.30 | 1.32 |
| -5 | -1 | -3 | 146.39 | 1.89 |
| -5 | -1 | -3 | 150.47 | 1.80 |
| 5 | -1 | 3 | 130.11 | 2.08 |
| 5 | -1 | 3 | 61.71 | 1.80 |
| 5 | -1 | -3 | 213.78 | 3.31 |
| -5 | 1 | 3 | 194.42 | 1.90 |
| -5 | 1 | -3 | 194.75 | 2.59 |
| 5 | 1 | -3 | 218.34 | 2.34 |
| -5 | -1 | 4 | 33.93 | 0.87 |
| -5 | -1 | -4 | 26.15 | 1.30 |
| 5 | -1 | 4 | 33.51 | 1.22 |
| 5 | -1 | -4 | 36.24 | 1.79 |
| -5 | 1 | 4 | 38.46 | 1.11 |
| -5 | 1 | -4 | 36.26 | 1.22 |
| 5 | 1 | 4 | 23.87 | 1.36 |
| 5 | 1 | -4 | 43.79 | 1.24 |
| -5 | -1 | 5 | 12.69 | 0.78 |
| -5 | -1 | -5 | 10.11 | 1.09 |
| 5 | -1 | 5 | 19.25 | 1.37 |
| 5 | -1 | -5 | 18.21 | 1.62 |
| -5 | 1 | -5 | 21.25 | 1.42 |
| 5 | 1 | 5 | 10.09 | 1.07 |
| 5 | 1 | -5 | 14.75 | 1.12 |
| -5 | -1 | 6 | 9.99 | 0.89 |
| -5 | -1 | -6 | 2.67 | 0.69 |
| -5 | -1 | -6 | -0.18 | 0.62 |
| 5 | -1 | 6 | 1.47 | 0.99 |
| 5 | -1 | -6 | 0.95 | 1.30 |
| -5 | 1 | 6 | 1.50 | 0.78 |
| -5 | 1 | -6 | 1.88 | 0.63 |
| 5 | 1 | 6 | -2.69 | 1.22 |
| 5 | 1 | -6 | 0.63 | 0.69 |
| -5 | -1 | 7 | 3.81 | 0.82 |
| -5 | -1 | -7 | 1.78 | 0.65 |
| -5 | -1 | -7 | 0.02 | 0.77 |
| 5 | -1 | 7 | 1.39 | 1.22 |
| 5 | -1 | -7 | 7.89 | 1.42 |
| -5 | 1 | 7 | 3.04 | 0.89 |
| -5 | 1 | -7 | 3.85 | 0.85 |
| 5 | 1 | 7 | -0.53 | 1.51 |
| 5 | 1 | -7 | 3.78 | 0.90 |
| -5 | -1 | 8 | 6.45 | 1.07 |
| -5 | -1 | -8 | 5.51 | 0.92 |
| -5 | -1 | -8 | 5.81 | 1.00 |
| 5 | -1 | 8 | 10.26 | 2.02 |
| 5 | -1 | -8 | 4.32 | 1.36 |
| -5 | 1 | 8 | 3.18 | 1.14 |
| -5 | 1 | -8 | 4.01 | 1.04 |
| 5 | 1 | 8 | 4.92 | 2.19 |
| 5 | 1 | -8 | 8.00 | 1.14 |
| -5 | -1 | 9 | 38.33 | 2.24 |
| -5 | -1 | -9 | 51.91 | 2.29 |
| -5 | -1 | -9 | 44.20 | 2.46 |
| 5 | -1 | 9 | 37.65 | 2.85 |
| 5 | -1 | -9 | 55.55 | 3.02 |
| -5 | 1 | 9 | 38.49 | 2.40 |
| -5 | 1 | -9 | 57.72 | 2.47 |
| -5 | 1 | -9 | 53.61 | 2.59 |
| 5 | 1 | 9 | -4.30 | 2.83 |
| -5 | -1 | 10 | 1.50 | 1.67 |
| -5 | -1 | -10 | 2.88 | 1.47 |
| -5 | -1 | -10 | 2.24 | 1.40 |
| 5 | -1 | 10 | -0.41 | 3.35 |
| 5 | -1 | -10 | 3.96 | 1.85 |
| -5 | 1 | 10 | 3.35 | 1.74 |
| -5 | 1 | -10 | 4.10 | 1.35 |
| -5 | 1 | -10 | 0.26 | 1.43 |
| -5 | -1 | 11 | 1.75 | 2.96 |
| -5 | -1 | -11 | 2.36 | 1.73 |
| -5 | -1 | -11 | 3.65 | 2.62 |
| 5 | -1 | -11 | -0.53 | 2.36 |
| -5 | 1 | 11 | 3.52 | 2.61 |
| -5 | 1 | -11 | 8.07 | 1.81 |
| -5 | 1 | -11 | 3.58 | 2.72 |
| -5 | -1 | 12 | 4.71 | 2.96 |
| -5 | -1 | -12 | -0.51 | 2.22 |
| -5 | -1 | -12 | 8.43 | 2.46 |
| -5 | 1 | -12 | 1.59 | 2.34 |
| -5 | 1 | -12 | 6.42 | 2.28 |
| -5 | -1 | 13 | 3.23 | 4.12 |
| -5 | -1 | -13 | 0.59 | 2.95 |
| -5 | -1 | -13 | 2.31 | 2.83 |
| -5 | 1 | 13 | 3.03 | 3.79 |
| -5 | 1 | -13 | 0.55 | 2.90 |
| 7 | -1 | 0 | 2.76 | 0.84 |
| 7 | 1 | 0 | 0.83 | 0.48 |
| -7 | -1 | -1 | 5.64 | 0.68 |
| 7 | -1 | 1 | 6.99 | 1.05 |
| 7 | -1 | -1 | 7.32 | 1.15 |
| 7 | 1 | 1 | 7.04 | 0.87 |
| 7 | 1 | -1 | 8.87 | 0.92 |
| -7 | -1 | 2 | 3.54 | 0.80 |
| -7 | -1 | -2 | 15.10 | 1.70 |
| 7 | -1 | 2 | 10.03 | 1.11 |
| 7 | -1 | -2 | 9.04 | 1.20 |
| 7 | 1 | -2 | 10.71 | 1.00 |
| 7 | -1 | 3 | 55.12 | 1.89 |
| 7 | -1 | -3 | 66.63 | 2.55 |
| -7 | 1 | -3 | 76.97 | 2.32 |
| 7 | 1 | -3 | 82.68 | 1.88 |
| -7 | -1 | -4 | 5.23 | 0.81 |
| 7 | -1 | -4 | 15.43 | 1.54 |
| -7 | 1 | 4 | 8.64 | 1.35 |
| -7 | 1 | -4 | 5.59 | 0.93 |
| 7 | 1 | -4 | 6.98 | 0.78 |
| -7 | -1 | 5 | 7.26 | 0.93 |
| -7 | -1 | -5 | 5.93 | 0.88 |
| 7 | -1 | -5 | 6.43 | 1.35 |
| -7 | 1 | 5 | 5.68 | 1.16 |
| -7 | 1 | -5 | 5.11 | 0.88 |
| 7 | 1 | -5 | 5.72 | 0.77 |
| -7 | -1 | 6 | 0.98 | 0.95 |
| -7 | -1 | -6 | 0.99 | 0.87 |
| -7 | -1 | -6 | 5.34 | 1.02 |
| 7 | -1 | -6 | 0.95 | 1.15 |
| -7 | 1 | 6 | 5.82 | 1.19 |
| -7 | 1 | -6 | 6.19 | 1.08 |
| 7 | 1 | -6 | 5.24 | 1.26 |
| -7 | -1 | 7 | -0.02 | 1.16 |
| -7 | -1 | -7 | 4.99 | 1.00 |
| -7 | -1 | -7 | 1.50 | 1.06 |
| 7 | -1 | -7 | 0.26 | 1.32 |
| -7 | 1 | 7 | 0.95 | 1.36 |
| -7 | 1 | -7 | 1.18 | 1.17 |
| 7 | 1 | -7 | 3.38 | 1.23 |
| -7 | -1 | 8 | -0.13 | 1.43 |
| -7 | -1 | -8 | 2.24 | 1.08 |
| -7 | -1 | -8 | 2.71 | 1.41 |
| 7 | -1 | -8 | -2.73 | 1.49 |
| -7 | 1 | 8 | -1.31 | 1.61 |
| -7 | 1 | -8 | 1.48 | 1.32 |
| -7 | -1 | 9 | 32.46 | 2.85 |
| -7 | -1 | -9 | 40.58 | 2.83 |
| -7 | -1 | -9 | 33.21 | 3.07 |
| 7 | -1 | -9 | 29.78 | 3.16 |
| -7 | 1 | 9 | 31.27 | 3.03 |
| -7 | 1 | -9 | 41.13 | 2.99 |
| -7 | 1 | -9 | 40.05 | 3.08 |
| -7 | -1 | -10 | 1.28 | 1.88 |
| -7 | -1 | -10 | 4.20 | 2.05 |
| -7 | 1 | 10 | 0.20 | 2.57 |
| -7 | 1 | -10 | 0.45 | 2.23 |
| -7 | 1 | -10 | 3.01 | 2.09 |
| -7 | -1 | 11 | -6.91 | 2.81 |
| -7 | -1 | -11 | 0.00 | 2.33 |
| -7 | -1 | -11 | 0.98 | 2.52 |
| -7 | 1 | 11 | -5.45 | 2.95 |
| -7 | 1 | -11 | -1.37 | 2.25 |
| -7 | 1 | -11 | -6.10 | 2.26 |
| -7 | -1 | 12 | -7.36 | 3.65 |
| -7 | -1 | -12 | 4.82 | 2.74 |
| -7 | -1 | -12 | 4.60 | 2.88 |
| -7 | 1 | 12 | -6.29 | 3.54 |
| -7 | 1 | -12 | -4.71 | 2.94 |
| -7 | 1 | -12 | 8.00 | 3.10 |
| 9 | -1 | 0 | 82.55 | 2.62 |
| 9 | 1 | 0 | 76.97 | 2.26 |
| -9 | -1 | -1 | 2.00 | 1.10 |
| 9 | -1 | 1 | 6.24 | 1.18 |
| 9 | -1 | -1 | 1.42 | 1.10 |
| 9 | 1 | 1 | 0.67 | 0.85 |
| 9 | 1 | -1 | 2.81 | 0.86 |
| -9 | -1 | 2 | 9.65 | 1.34 |
| -9 | -1 | -2 | 44.46 | 2.01 |
| 9 | -1 | 2 | 35.90 | 2.07 |
| 9 | -1 | -2 | 39.81 | 2.23 |
| -9 | 1 | -2 | 63.91 | 2.63 |
| 9 | 1 | -2 | 46.34 | 1.88 |
| -9 | -1 | -3 | 41.53 | 2.01 |
| 9 | -1 | 3 | 28.20 | 2.05 |
| 9 | -1 | -3 | 37.39 | 2.32 |
| -9 | 1 | -3 | 50.92 | 2.46 |
| 9 | 1 | -3 | 39.48 | 1.85 |
| -9 | -1 | -4 | 41.88 | 2.13 |
| 9 | -1 | -4 | 33.63 | 2.39 |
| -9 | 1 | -4 | 39.87 | 2.39 |
| 9 | 1 | -4 | 33.31 | 1.94 |
| -9 | -1 | -5 | -0.65 | 1.26 |
| -9 | -1 | -5 | 1.10 | 1.24 |
| 9 | -1 | -5 | 0.35 | 1.37 |
| -9 | 1 | -5 | 1.08 | 1.31 |
| 9 | 1 | -5 | -2.21 | 1.04 |
| -9 | -1 | -6 | 51.74 | 2.79 |
| -9 | -1 | -6 | 46.29 | 2.72 |
| 9 | -1 | -6 | 33.38 | 2.69 |
| -9 | 1 | 6 | 25.42 | 2.97 |
| -9 | 1 | -6 | 52.99 | 3.02 |
| 9 | 1 | -6 | 41.78 | 2.44 |
| -9 | -1 | 7 | -1.78 | 2.03 |
| -9 | -1 | -7 | 1.22 | 1.55 |
| -9 | -1 | -7 | -1.73 | 1.52 |
| 9 | -1 | -7 | 1.26 | 1.71 |
| -9 | 1 | 7 | 3.82 | 2.10 |
| -9 | 1 | -7 | 2.62 | 1.73 |
| -9 | -1 | 8 | 15.21 | 2.58 |
| -9 | -1 | -8 | 17.31 | 2.29 |
| 9 | -1 | -8 | 12.21 | 2.16 |
| -9 | 1 | 8 | 8.47 | 2.56 |
| -9 | 1 | -8 | 18.61 | 3.63 |
| -9 | -1 | 9 | 8.64 | 2.97 |
| -9 | -1 | -9 | 36.59 | 3.76 |
| -9 | -1 | -9 | 25.31 | 2.69 |
| -9 | 1 | 9 | 9.43 | 3.05 |
| -9 | 1 | -9 | 23.54 | 2.60 |
| -9 | 1 | -9 | 26.09 | 2.69 |
| -9 | -1 | 10 | -8.65 | 3.39 |
| -9 | -1 | -10 | 18.31 | 2.79 |
| -9 | -1 | -10 | 12.80 | 3.01 |
| -9 | 1 | 10 | 8.97 | 3.30 |
| -9 | 1 | -10 | 2.20 | 2.56 |
| -9 | 1 | -10 | 8.45 | 2.98 |
| -9 | -1 | -11 | 5.11 | 3.23 |
| -9 | -1 | -11 | -0.39 | 3.53 |
| -9 | 1 | -11 | 2.52 | 3.13 |
| -9 | 1 | -11 | -8.08 | 3.50 |
| -11 | -1 | 0 | 1.67 | 1.65 |
| -11 | -1 | 0 | -1.89 | 1.64 |
| 11 | -1 | 0 | 0.30 | 1.54 |
| -11 | -1 | 1 | 4.63 | 1.63 |
| -11 | -1 | 1 | 2.69 | 1.79 |
| -11 | -1 | -1 | -1.04 | 1.59 |
| -11 | -1 | -1 | -0.80 | 1.65 |
| 11 | -1 | 1 | 0.82 | 1.63 |
| 11 | -1 | -1 | 3.79 | 1.68 |
| -11 | 1 | -1 | -1.09 | 1.92 |
| 11 | 1 | -1 | -1.36 | 1.40 |
| -11 | -1 | -2 | 11.79 | 1.78 |
| -11 | -1 | -2 | 8.35 | 1.80 |
| 11 | -1 | 2 | 10.03 | 2.03 |
| 11 | -1 | -2 | 7.45 | 1.73 |
| -11 | 1 | -2 | 10.98 | 2.03 |
| 11 | 1 | -2 | 5.43 | 1.54 |
| -11 | -1 | -3 | 16.44 | 1.94 |
| -11 | -1 | -3 | 12.52 | 1.86 |
| 11 | -1 | -3 | 9.41 | 1.89 |
| -11 | 1 | -3 | 12.33 | 2.16 |
| 11 | 1 | -3 | 8.36 | 1.63 |
| -11 | -1 | -4 | 10.01 | 1.89 |
| -11 | -1 | -4 | 9.27 | 1.98 |
| 11 | -1 | -4 | 3.72 | 1.90 |
| -11 | 1 | -4 | 10.32 | 2.18 |
| 11 | 1 | -4 | 2.28 | 1.64 |
| -11 | -1 | -5 | 1.41 | 2.42 |
| -11 | -1 | -5 | -0.82 | 2.10 |
| 11 | -1 | -5 | 0.91 | 1.96 |
| -11 | 1 | -5 | -0.47 | 2.23 |
| 11 | 1 | -5 | -3.35 | 1.81 |
| -11 | -1 | -6 | 2.99 | 2.13 |
| -11 | -1 | -6 | 3.72 | 2.44 |
| 11 | -1 | -6 | -1.52 | 3.30 |
| -11 | 1 | -6 | 3.86 | 2.42 |
| -11 | -1 | -7 | -2.03 | 2.74 |
| 11 | -1 | -7 | 3.25 | 2.79 |
| -11 | 1 | -7 | -1.70 | 2.72 |
| -11 | 1 | -7 | 6.25 | 2.76 |
| -11 | -1 | -8 | 8.85 | 2.70 |
| -11 | -1 | -8 | 3.11 | 3.12 |
| -11 | 1 | -8 | 6.88 | 2.82 |
| -11 | 1 | -8 | 6.01 | 3.16 |
| -11 | -1 | -9 | 28.01 | 3.74 |
| -11 | -1 | -9 | 17.68 | 4.40 |
| -11 | 1 | -9 | 9.08 | 3.65 |
| -11 | 1 | -9 | 0.03 | 3.55 |
| 13 | -1 | 0 | 3.33 | 2.67 |
| -13 | 1 | 0 | 0.06 | 3.33 |
| 13 | -1 | -1 | 1.90 | 2.52 |
| -13 | 1 | -1 | 5.05 | 3.22 |
| 13 | -1 | -2 | -1.68 | 2.62 |
| -13 | 1 | -2 | -5.30 | 3.24 |
| -13 | -1 | -3 | 11.83 | 3.19 |
| 13 | -1 | -3 | 6.68 | 2.83 |
| -13 | 1 | -3 | 13.50 | 3.40 |
| -13 | -1 | -4 | -4.62 | 3.08 |
| 13 | -1 | -4 | 1.83 | 2.84 |
| -13 | 1 | -4 | -5.77 | 3.41 |
| -13 | -1 | -5 | -2.87 | 3.46 |
| -13 | 1 | -5 | 8.30 | 3.33 |
| -13 | 1 | -5 | 13.56 | 4.34 |
| -13 | 1 | -6 | -3.92 | 5.62 |
| 0 | -2 | 0 | 404.27 | 3.16 |
| 0 | -2 | 0 | 175.71 | 1.65 |
| 0 | 2 | 0 | 661.69 | 3.98 |
| 0 | 2 | 0 | 644.67 | 4.06 |
| 0 | 2 | 0 | 258.72 | 1.94 |
| 0 | -2 | 1 | 2379.85 | 5.59 |
| 0 | -2 | -1 | 3061.58 | 8.58 |
| 0 | -2 | -1 | 2578.94 | 7.00 |
| 0 | -2 | -1 | 2774.80 | 7.62 |
| 0 | -2 | -1 | 3448.84 | 7.35 |
| 0 | 2 | 1 | 4232.00 | 11.67 |
| 0 | 2 | 1 | 3423.02 | 8.04 |
| 0 | 2 | 1 | 1559.27 | 4.94 |
| 0 | 2 | -1 | 7303.30 | 16.75 |
| 0 | -2 | 2 | 369.93 | 3.22 |
| 0 | -2 | 2 | 397.75 | 2.78 |
| 0 | -2 | -2 | 342.61 | 2.86 |
| 0 | -2 | -2 | 315.58 | 2.62 |
| 0 | 2 | 2 | 306.91 | 2.27 |
| 0 | 2 | -2 | 733.12 | 3.62 |
| 0 | 2 | -2 | 271.47 | 1.52 |
| 0 | -2 | 3 | 4266.66 | 11.00 |
| 0 | -2 | -3 | 3759.41 | 9.85 |
| 0 | -2 | -3 | 4012.45 | 10.24 |
| 0 | -2 | 4 | 56.21 | 1.29 |
| 0 | -2 | 4 | 49.59 | 1.15 |
| 0 | -2 | -4 | 49.73 | 1.24 |
| 0 | -2 | -4 | 54.52 | 1.35 |
| 0 | 2 | 4 | 66.76 | 1.28 |
| 0 | 2 | -4 | 39.73 | 0.86 |
| 0 | -2 | 5 | 1429.33 | 6.56 |
| 0 | -2 | 5 | 1432.49 | 6.37 |
| 0 | -2 | -5 | 1383.28 | 6.68 |
| 0 | -2 | -5 | 1382.11 | 7.16 |
| 0 | 2 | 5 | 1448.56 | 6.26 |
| 0 | 2 | -5 | 1163.09 | 5.25 |
| 0 | -2 | 6 | 179.91 | 2.64 |
| 0 | -2 | -6 | 157.10 | 2.47 |
| 0 | -2 | -6 | 168.05 | 2.77 |
| 0 | 2 | 6 | 155.51 | 3.53 |
| 0 | 2 | -6 | 144.91 | 2.17 |
| 0 | -2 | 7 | 881.12 | 6.00 |
| 0 | -2 | -7 | 938.43 | 6.35 |
| 0 | -2 | -7 | 883.60 | 6.90 |
| 0 | 2 | 7 | 892.37 | 6.00 |
| 0 | 2 | -7 | 785.40 | 3.99 |
| 0 | 2 | -7 | 754.04 | 5.98 |
| 0 | -2 | 8 | 4.45 | 0.92 |
| 0 | -2 | -8 | 12.13 | 1.27 |
| 0 | -2 | -8 | 9.71 | 1.09 |
| 0 | 2 | 8 | 1.95 | 0.99 |
| 0 | 2 | -8 | 0.56 | 0.50 |
| 0 | 2 | -8 | 5.27 | 0.91 |
| 0 | -2 | 9 | 807.28 | 6.75 |
| 0 | -2 | -9 | 1217.11 | 8.34 |
| 0 | 2 | 9 | 761.58 | 6.56 |
| 0 | 2 | -9 | 984.51 | 7.63 |
| 0 | -2 | 10 | 5.68 | 1.55 |
| 0 | -2 | -10 | 13.67 | 1.58 |
| 0 | -2 | -10 | 19.86 | 2.40 |
| 0 | 2 | 10 | 7.19 | 1.75 |
| 0 | 2 | -10 | 1.58 | 1.68 |
| 0 | 2 | -10 | 7.01 | 1.37 |
| 0 | -2 | 11 | 283.94 | 5.24 |
| 0 | -2 | -11 | 649.43 | 6.92 |
| 0 | 2 | 11 | 278.31 | 5.27 |
| 0 | 2 | -11 | 460.66 | 6.42 |
| 0 | -2 | 12 | 41.77 | 5.63 |
| 0 | -2 | -12 | 118.54 | 4.78 |
| 0 | 2 | 12 | 17.31 | 5.39 |
| 0 | 2 | -12 | 92.35 | 4.76 |
| 0 | -2 | 13 | 126.23 | 6.05 |
| 0 | -2 | -13 | 346.12 | 6.88 |
| 0 | 2 | 13 | 132.60 | 6.64 |
| 0 | 2 | -13 | 350.92 | 6.95 |
| 0 | -2 | 14 | -3.95 | 4.90 |
| 0 | -2 | -14 | 7.39 | 2.44 |
| 0 | 2 | 14 | 5.89 | 8.49 |
| 0 | 2 | -14 | 9.34 | 3.35 |
| -2 | -2 | 0 | 3474.80 | 6.99 |
| -2 | -2 | 0 | 1516.63 | 4.62 |
| 2 | -2 | 0 | 4012.33 | 10.67 |
| 2 | -2 | 0 | 2585.40 | 6.99 |
| -2 | 2 | 0 | 2046.64 | 6.20 |
| 2 | 2 | 0 | 2602.07 | 5.34 |
| -2 | -2 | 1 | 1625.61 | 4.78 |
| -2 | -2 | 1 | 1307.19 | 6.27 |
| -2 | -2 | 1 | 1980.49 | 4.74 |
| -2 | -2 | -1 | 2291.39 | 5.96 |
| -2 | -2 | -1 | 2064.47 | 4.21 |
| -2 | -2 | -1 | 2511.19 | 6.32 |
| 2 | -2 | 1 | 2310.41 | 6.84 |
| 2 | -2 | 1 | 1691.09 | 6.02 |
| 2 | -2 | 1 | 2733.38 | 7.27 |
| 2 | -2 | -1 | 2471.16 | 9.12 |
| 2 | -2 | -1 | 1907.57 | 5.65 |
| -2 | 2 | 1 | 4753.24 | 9.83 |
| -2 | 2 | 1 | 1179.08 | 4.66 |
| -2 | 2 | -1 | 2481.77 | 6.69 |
| 2 | 2 | 1 | 1443.86 | 3.32 |
| 2 | 2 | 1 | 850.03 | 3.29 |
| 2 | 2 | -1 | 1728.36 | 3.66 |
| -2 | -2 | 2 | 324.85 | 3.30 |
| -2 | -2 | 2 | 601.02 | 3.21 |
| -2 | -2 | -2 | 543.69 | 3.12 |
| -2 | -2 | -2 | 491.87 | 2.55 |
| -2 | -2 | -2 | 664.57 | 3.40 |
| 2 | -2 | 2 | 444.21 | 2.65 |
| 2 | -2 | 2 | 456.37 | 3.32 |
| 2 | -2 | 2 | 560.97 | 3.29 |
| 2 | -2 | -2 | 597.44 | 4.67 |
| 2 | -2 | -2 | 546.20 | 4.80 |
| 2 | -2 | -2 | 574.82 | 2.97 |
| -2 | 2 | 2 | 476.83 | 2.95 |
| 2 | 2 | 2 | 310.27 | 2.25 |
| -2 | -2 | 3 | 2954.99 | 4.91 |
| -2 | -2 | 3 | 3322.42 | 10.33 |
| -2 | -2 | 3 | 2924.43 | 7.44 |
| -2 | -2 | -3 | 3195.81 | 8.35 |
| -2 | -2 | -3 | 2923.06 | 7.42 |
| 2 | -2 | 3 | 1973.00 | 5.51 |
| 2 | -2 | 3 | 2685.10 | 8.66 |
| 2 | -2 | 3 | 3107.14 | 8.25 |
| 2 | -2 | -3 | 3361.46 | 11.38 |
| -2 | 2 | 3 | 3051.95 | 7.89 |
| -2 | 2 | -3 | 3259.57 | 8.09 |
| 2 | 2 | 3 | 2723.05 | 7.27 |
| 2 | 2 | -3 | 3421.56 | 7.57 |
| -2 | -2 | 4 | 327.36 | 3.05 |
| -2 | -2 | -4 | 357.35 | 3.00 |
| -2 | -2 | -4 | 311.88 | 2.75 |
| 2 | -2 | 4 | 327.74 | 2.89 |
| 2 | -2 | -4 | 399.27 | 3.98 |
| 2 | -2 | -4 | 306.29 | 3.79 |
| -2 | 2 | 4 | 328.39 | 2.76 |
| -2 | 2 | -4 | 314.54 | 2.67 |
| 2 | 2 | 4 | 299.38 | 2.64 |
| 2 | 2 | -4 | 367.17 | 2.89 |
| -2 | -2 | 5 | 1066.55 | 5.32 |
| -2 | -2 | 5 | 994.19 | 5.08 |
| -2 | -2 | -5 | 1124.12 | 5.83 |
| -2 | -2 | -5 | 1000.70 | 5.57 |
| 2 | -2 | 5 | 1088.73 | 5.74 |
| 2 | -2 | -5 | 1223.12 | 7.16 |
| 2 | -2 | -5 | 999.18 | 7.19 |
| -2 | 2 | 5 | 1194.97 | 5.79 |
| -2 | 2 | -5 | 1020.91 | 5.31 |
| 2 | 2 | 5 | 1015.28 | 5.33 |
| 2 | 2 | -5 | 1178.82 | 5.89 |
| -2 | -2 | 6 | 651.23 | 4.11 |
| -2 | -2 | -6 | 706.09 | 5.02 |
| -2 | -2 | -6 | 648.32 | 5.38 |
| 2 | -2 | 6 | 684.04 | 4.98 |
| 2 | -2 | -6 | 743.08 | 5.78 |
| 2 | -2 | -6 | 573.68 | 7.45 |
| -2 | 2 | 6 | 597.45 | 6.92 |
| -2 | 2 | -6 | 657.24 | 4.69 |
| 2 | 2 | 6 | 643.13 | 4.69 |
| 2 | 2 | -6 | 738.13 | 5.22 |
| -2 | -2 | 7 | 640.80 | 4.92 |
| -2 | -2 | -7 | 731.95 | 5.64 |
| -2 | -2 | -7 | 731.14 | 5.93 |
| 2 | -2 | 7 | 662.97 | 8.70 |
| 2 | -2 | -7 | 945.95 | 6.83 |
| 2 | -2 | -7 | 640.36 | 6.55 |
| -2 | 2 | 7 | 730.22 | 5.38 |
| -2 | 2 | -7 | 738.83 | 5.58 |
| 2 | 2 | 7 | 682.58 | 5.47 |
| 2 | 2 | -7 | 822.86 | 6.74 |
| -2 | -2 | 8 | 69.65 | 2.06 |
| -2 | -2 | -8 | 109.58 | 2.55 |
| -2 | -2 | -8 | 101.05 | 2.63 |
| 2 | -2 | 8 | 87.90 | 2.61 |
| 2 | -2 | -8 | 151.53 | 3.03 |
| 2 | -2 | -8 | 97.55 | 2.96 |
| -2 | 2 | 8 | 82.16 | 2.25 |
| -2 | 2 | -8 | 124.90 | 2.67 |
| 2 | 2 | 8 | 76.03 | 2.67 |
| 2 | 2 | -8 | 106.70 | 3.03 |
| -2 | -2 | 9 | 657.49 | 5.92 |
| -2 | -2 | -9 | 858.72 | 7.07 |
| -2 | -2 | -9 | 852.36 | 7.65 |
| 2 | -2 | 9 | 716.22 | 6.64 |
| 2 | -2 | -9 | 1451.04 | 9.25 |
| -2 | 2 | 9 | 709.94 | 6.30 |
| -2 | 2 | -9 | 784.79 | 5.61 |
| -2 | 2 | -9 | 802.51 | 6.97 |
| 2 | 2 | 9 | 624.63 | 6.21 |
| 2 | 2 | -9 | 1038.81 | 8.23 |
| -2 | -2 | 10 | 53.74 | 2.62 |
| -2 | -2 | -10 | 84.95 | 2.93 |
| -2 | -2 | -10 | 82.56 | 3.20 |
| 2 | -2 | 10 | 52.73 | 3.23 |
| 2 | -2 | -10 | 147.22 | 3.58 |
| 2 | -2 | -10 | 91.21 | 3.61 |
| -2 | 2 | 10 | 49.29 | 2.79 |
| -2 | 2 | -10 | 72.30 | 2.32 |
| -2 | 2 | -10 | 75.63 | 2.88 |
| 2 | 2 | 10 | 50.78 | 3.38 |
| 2 | 2 | -10 | 93.43 | 3.55 |
| -2 | -2 | 11 | 230.45 | 4.67 |
| -2 | -2 | -11 | 435.03 | 5.92 |
| -2 | -2 | -11 | 383.28 | 6.26 |
| 2 | -2 | 11 | 244.88 | 5.50 |
| 2 | -2 | -11 | 562.92 | 7.59 |
| 2 | -2 | -11 | 435.41 | 6.97 |
| -2 | 2 | 11 | 233.13 | 4.83 |
| -2 | 2 | -11 | 332.56 | 4.51 |
| -2 | 2 | -11 | 368.84 | 5.87 |
| 2 | 2 | 11 | 229.33 | 5.54 |
| 2 | 2 | -11 | 479.49 | 6.73 |
| -2 | -2 | 12 | 101.70 | 4.41 |
| -2 | -2 | -12 | 240.73 | 5.26 |
| -2 | -2 | -12 | 204.92 | 5.41 |
| 2 | -2 | 12 | 74.45 | 8.63 |
| 2 | -2 | -12 | 281.71 | 6.33 |
| 2 | -2 | -12 | 222.52 | 5.64 |
| -2 | 2 | 12 | 99.24 | 4.81 |
| -2 | 2 | -12 | 216.08 | 5.35 |
| 2 | 2 | 12 | 119.13 | 10.66 |
| 2 | 2 | -12 | 240.66 | 6.80 |
| -2 | -2 | 13 | 113.60 | 5.40 |
| -2 | -2 | -13 | 292.53 | 6.52 |
| 2 | -2 | 13 | 55.17 | 5.37 |
| 2 | -2 | -13 | 376.29 | 9.42 |
| -2 | 2 | 13 | 108.41 | 5.88 |
| -2 | 2 | -13 | 311.36 | 6.91 |
| 2 | 2 | 13 | 95.50 | 8.40 |
| 2 | 2 | -13 | 322.06 | 7.41 |
| -2 | -2 | -14 | 20.30 | 3.60 |
| -2 | 2 | -14 | 26.95 | 5.51 |
| -4 | -2 | 0 | 1729.95 | 5.08 |
| 4 | -2 | 0 | 1840.73 | 8.44 |
| 4 | -2 | 0 | 1382.25 | 4.57 |
| -4 | 2 | 0 | 1328.40 | 5.63 |
| -4 | -2 | 1 | 1748.52 | 4.65 |
| -4 | -2 | 1 | 1024.34 | 5.75 |
| -4 | -2 | -1 | 1638.09 | 5.41 |
| -4 | -2 | -1 | 1326.75 | 3.84 |
| -4 | -2 | -1 | 1994.84 | 5.48 |
| 4 | -2 | 1 | 1480.36 | 7.00 |
| 4 | -2 | 1 | 1445.47 | 4.50 |
| 4 | -2 | -1 | 1809.75 | 9.06 |
| -4 | 2 | 1 | 3334.07 | 9.01 |
| -4 | 2 | 1 | 969.90 | 4.78 |
| -4 | 2 | -1 | 1974.83 | 6.74 |
| 4 | 2 | -1 | 1326.47 | 4.12 |
| -4 | -2 | 2 | 111.70 | 2.05 |
| -4 | -2 | 2 | 291.12 | 2.00 |
| -4 | -2 | -2 | 317.96 | 2.59 |
| -4 | -2 | -2 | 259.12 | 1.97 |
| -4 | -2 | -2 | 339.62 | 2.45 |
| 4 | -2 | 2 | 249.45 | 2.72 |
| 4 | -2 | 2 | 275.75 | 2.16 |
| 4 | -2 | -2 | 327.99 | 4.14 |
| -4 | 2 | 2 | 267.72 | 2.52 |
| 4 | 2 | -2 | 280.21 | 1.80 |
| -4 | -2 | 3 | 1356.83 | 7.00 |
| -4 | -2 | 3 | 1804.52 | 5.36 |
| -4 | -2 | -3 | 2187.04 | 7.46 |
| -4 | -2 | -3 | 1923.70 | 6.22 |
| 4 | -2 | 3 | 1554.24 | 6.68 |
| 4 | -2 | 3 | 2007.13 | 6.42 |
| 4 | -2 | -3 | 2254.26 | 11.47 |
| -4 | 2 | 3 | 2226.95 | 7.30 |
| 4 | 2 | 3 | 970.58 | 4.13 |
| -4 | -2 | 4 | 231.04 | 2.66 |
| -4 | -2 | 4 | 174.91 | 1.87 |
| -4 | -2 | -4 | 199.93 | 2.27 |
| 4 | -2 | 4 | 128.15 | 1.96 |
| 4 | -2 | 4 | 205.34 | 2.31 |
| 4 | -2 | -4 | 243.98 | 3.19 |
| 4 | -2 | -4 | 219.68 | 3.85 |
| -4 | 2 | 4 | 235.86 | 2.56 |
| 4 | 2 | 4 | 157.42 | 2.09 |
| -4 | -2 | 5 | 720.30 | 4.12 |
| -4 | -2 | 5 | 787.35 | 4.32 |
| -4 | -2 | -5 | 937.82 | 8.58 |
| -4 | -2 | -5 | 861.76 | 5.25 |
| 4 | -2 | 5 | 900.44 | 5.28 |
| 4 | -2 | -5 | 879.09 | 6.10 |
| 4 | -2 | -5 | 868.17 | 7.71 |
| -4 | 2 | -5 | 1037.65 | 9.48 |
| 4 | 2 | 5 | 745.04 | 4.64 |
| -4 | -2 | 6 | 416.39 | 3.57 |
| -4 | -2 | -6 | 487.16 | 4.50 |
| -4 | -2 | -6 | 465.14 | 4.40 |
| 4 | -2 | 6 | 457.98 | 4.18 |
| 4 | -2 | -6 | 439.67 | 4.40 |
| 4 | -2 | -6 | 448.97 | 6.07 |
| -4 | 2 | 6 | 509.52 | 4.32 |
| -4 | 2 | -6 | 504.21 | 4.69 |
| 4 | 2 | 6 | 377.48 | 3.76 |
| -4 | -2 | 7 | 530.22 | 4.37 |
| -4 | -2 | -7 | 650.58 | 5.59 |
| -4 | -2 | -7 | 621.54 | 5.50 |
| 4 | -2 | 7 | 630.20 | 5.35 |
| 4 | -2 | -7 | 603.80 | 5.13 |
| 4 | -2 | -7 | 551.26 | 6.71 |
| -4 | 2 | 7 | 642.44 | 5.18 |
| -4 | 2 | -7 | 719.81 | 5.99 |
| 4 | 2 | 7 | 493.01 | 4.75 |
| -4 | -2 | 8 | 62.29 | 2.01 |
| -4 | -2 | -8 | 73.88 | 2.26 |
| -4 | -2 | -8 | 86.69 | 2.53 |
| 4 | -2 | 8 | 69.91 | 3.13 |
| 4 | -2 | -8 | 113.38 | 2.69 |
| 4 | -2 | -8 | 74.96 | 3.11 |
| -4 | 2 | 8 | 68.34 | 2.28 |
| -4 | 2 | -8 | 84.92 | 2.51 |
| 4 | 2 | 8 | 43.17 | 4.03 |
| -4 | -2 | 9 | 521.12 | 5.20 |
| -4 | -2 | -9 | 702.08 | 6.76 |
| -4 | -2 | -9 | 706.35 | 7.06 |
| 4 | -2 | 9 | 589.18 | 6.48 |
| 4 | -2 | -9 | -0.39 | 1.66 |
| 4 | -2 | -9 | 820.65 | 8.62 |
| -4 | 2 | 9 | 616.34 | 5.97 |
| -4 | 2 | -9 | 797.21 | 7.38 |
| 4 | 2 | 9 | 492.10 | 6.13 |
| -4 | -2 | 10 | 46.03 | 2.70 |
| -4 | -2 | -10 | 73.00 | 2.98 |
| -4 | -2 | -10 | 77.29 | 3.28 |
| 4 | -2 | 10 | 38.84 | 4.18 |
| 4 | -2 | -10 | 77.97 | 3.60 |
| -4 | 2 | 10 | 51.47 | 2.94 |
| -4 | 2 | -10 | 57.08 | 2.60 |
| -4 | 2 | -10 | 65.86 | 3.07 |
| 4 | 2 | 10 | 31.97 | 3.44 |
| -4 | -2 | 11 | 201.16 | 4.79 |
| -4 | -2 | -11 | 332.15 | 5.66 |
| -4 | -2 | -11 | 338.66 | 6.08 |
| 4 | -2 | 11 | 228.89 | 8.20 |
| 4 | -2 | -11 | 378.70 | 6.43 |
| -4 | 2 | 11 | 251.48 | 6.63 |
| -4 | 2 | -11 | 297.82 | 4.90 |
| -4 | 2 | -11 | 315.38 | 5.75 |
| 4 | 2 | 11 | 232.03 | 8.56 |
| -4 | -2 | 12 | 52.44 | 4.36 |
| -4 | -2 | -12 | 172.81 | 4.91 |
| -4 | -2 | -12 | 173.72 | 5.33 |
| 4 | -2 | -12 | 180.04 | 5.50 |
| -4 | 2 | 12 | 77.12 | 4.80 |
| -4 | 2 | -12 | 150.43 | 4.76 |
| -4 | 2 | -12 | 148.45 | 5.00 |
| -4 | -2 | 13 | 47.70 | 5.18 |
| -4 | -2 | -13 | 271.01 | 6.14 |
| -4 | -2 | -13 | 244.42 | 6.42 |
| -4 | 2 | 13 | 101.47 | 5.56 |
| -4 | 2 | -13 | 221.73 | 6.04 |
| -6 | -2 | 0 | 131.05 | 1.67 |
| -6 | -2 | 0 | 112.65 | 1.38 |
| 6 | -2 | 0 | 232.31 | 3.48 |
| -6 | 2 | 0 | 125.97 | 2.00 |
| 6 | 2 | 0 | 100.80 | 1.63 |
| -6 | -2 | 1 | 1268.31 | 4.72 |
| -6 | -2 | -1 | 1035.24 | 4.34 |
| 6 | -2 | 1 | 1058.54 | 7.13 |
| 6 | -2 | -1 | 1220.58 | 8.27 |
| -6 | 2 | 1 | 1098.54 | 5.74 |
| -6 | 2 | -1 | 1593.90 | 6.57 |
| 6 | 2 | -1 | 999.43 | 4.80 |
| -6 | -2 | 2 | -0.02 | 0.72 |
| -6 | -2 | -2 | 6.99 | 0.68 |
| -6 | -2 | -2 | 20.87 | 1.08 |
| 6 | -2 | 2 | 5.41 | 0.92 |
| 6 | -2 | -2 | 10.99 | 1.33 |
| -6 | 2 | 2 | 4.01 | 0.86 |
| 6 | 2 | -2 | 3.06 | 0.43 |
| -6 | -2 | 3 | 501.15 | 4.44 |
| -6 | -2 | -3 | 1432.33 | 10.06 |
| -6 | -2 | -3 | 1336.74 | 5.92 |
| 6 | -2 | 3 | 1109.70 | 7.17 |
| 6 | -2 | -3 | 1316.68 | 10.02 |
| -6 | 2 | 3 | 1218.98 | 6.74 |
| 6 | 2 | -3 | 1455.63 | 5.53 |
| -6 | -2 | 4 | 13.09 | 1.26 |
| -6 | -2 | 4 | 1.46 | 0.55 |
| -6 | -2 | -4 | 4.70 | 0.78 |
| -6 | -2 | -4 | 3.96 | 0.56 |
| 6 | -2 | 4 | 0.48 | 0.72 |
| 6 | -2 | 4 | -0.20 | 0.94 |
| 6 | -2 | -4 | 3.57 | 1.19 |
| -6 | 2 | 4 | 22.83 | 1.92 |
| -6 | 2 | -4 | 1.48 | 0.69 |
| 6 | 2 | -4 | 1.54 | 0.45 |
| -6 | -2 | 5 | 676.85 | 4.01 |
| -6 | -2 | -5 | 784.42 | 5.90 |
| -6 | -2 | -5 | 742.33 | 5.39 |
| 6 | -2 | 5 | 672.71 | 4.45 |
| 6 | -2 | -5 | 676.36 | 7.21 |
| -6 | 2 | 5 | 813.71 | 5.53 |
| -6 | 2 | -5 | 873.53 | 6.73 |
| 6 | 2 | -5 | 779.49 | 3.97 |
| -6 | -2 | 6 | 92.04 | 1.90 |
| -6 | -2 | -6 | 91.77 | 2.31 |
| -6 | -2 | -6 | 88.11 | 2.24 |
| 6 | -2 | 6 | 67.25 | 2.45 |
| 6 | -2 | -6 | 72.19 | 2.86 |
| -6 | 2 | 6 | 99.11 | 2.40 |
| -6 | 2 | -6 | 108.14 | 2.66 |
| -6 | -2 | 7 | 457.88 | 4.07 |
| -6 | -2 | -7 | 530.01 | 5.60 |
| -6 | -2 | -7 | 552.37 | 5.56 |
| 6 | -2 | 7 | 93.12 | 3.16 |
| 6 | -2 | -7 | 441.28 | 6.11 |
| -6 | 2 | 7 | 557.53 | 5.21 |
| -6 | 2 | -7 | 625.79 | 6.35 |
| -6 | -2 | 8 | 2.28 | 1.21 |
| -6 | -2 | -8 | 5.02 | 1.04 |
| -6 | -2 | -8 | 3.66 | 1.17 |
| 6 | -2 | -8 | 4.86 | 1.65 |
| -6 | 2 | 8 | 2.83 | 1.49 |
| -6 | 2 | -8 | 2.35 | 1.15 |
| -6 | -2 | 9 | 434.81 | 4.91 |
| -6 | -2 | -9 | 593.34 | 6.82 |
| -6 | -2 | -9 | 599.36 | 6.87 |
| 6 | -2 | -9 | 491.06 | 6.64 |
| -6 | 2 | 9 | 507.05 | 5.84 |
| -6 | 2 | -9 | 652.28 | 7.36 |
| -6 | -2 | 10 | 4.63 | 1.98 |
| -6 | -2 | -10 | 7.15 | 1.76 |
| -6 | -2 | -10 | 6.05 | 1.96 |
| 6 | -2 | -10 | 7.78 | 2.19 |
| -6 | 2 | 10 | 3.34 | 2.50 |
| -6 | 2 | -10 | 10.73 | 1.90 |
| -6 | 2 | -10 | 9.92 | 1.99 |
| -6 | -2 | 11 | 148.81 | 4.46 |
| -6 | -2 | -11 | 277.36 | 5.74 |
| -6 | -2 | -11 | 307.94 | 6.19 |
| 6 | -2 | -11 | 224.19 | 5.65 |
| -6 | 2 | 11 | 182.30 | 5.24 |
| -6 | 2 | -11 | 268.94 | 5.50 |
| -6 | 2 | -11 | 295.07 | 6.09 |
| -6 | -2 | 12 | 12.77 | 3.17 |
| -6 | -2 | -12 | 50.26 | 4.61 |
| -6 | -2 | -12 | 49.51 | 4.68 |
| -6 | 2 | 12 | 19.22 | 3.38 |
| -6 | 2 | -12 | 60.46 | 4.66 |
| -6 | 2 | -12 | 55.72 | 4.87 |
| -8 | -2 | 0 | 477.58 | 3.77 |
| -8 | -2 | 0 | 414.94 | 3.21 |
| 8 | -2 | 0 | 371.38 | 5.10 |
| -8 | -2 | 1 | 627.38 | 3.99 |
| -8 | -2 | 1 | 319.18 | 4.01 |
| -8 | -2 | -1 | 662.16 | 4.95 |
| -8 | -2 | -1 | 540.38 | 4.23 |
| 8 | -2 | 1 | 467.08 | 5.63 |
| 8 | -2 | -1 | 508.84 | 6.03 |
| 8 | 2 | -1 | 527.18 | 4.46 |
| -8 | -2 | 2 | 29.66 | 1.92 |
| -8 | -2 | -2 | 176.13 | 2.84 |
| -8 | -2 | -2 | 134.37 | 2.23 |
| 8 | -2 | 2 | 112.35 | 2.89 |
| 8 | -2 | -2 | 161.26 | 3.62 |
| -8 | 2 | 2 | 196.23 | 3.56 |
| 8 | 2 | -2 | 140.59 | 2.55 |
| -8 | -2 | 3 | 245.94 | 3.31 |
| -8 | -2 | -3 | 856.39 | 6.21 |
| -8 | -2 | -3 | 724.75 | 5.20 |
| 8 | -2 | 3 | 517.19 | 5.85 |
| 8 | -2 | -3 | 654.00 | 7.06 |
| -8 | 2 | 3 | 530.65 | 4.92 |
| -8 | 2 | -3 | 999.88 | 8.10 |
| 8 | 2 | -3 | 721.14 | 5.06 |
| -8 | -2 | -4 | 139.83 | 2.85 |
| -8 | -2 | -4 | 127.26 | 2.56 |
| 8 | -2 | 4 | 78.42 | 2.60 |
| 8 | -2 | -4 | 113.14 | 3.35 |
| -8 | 2 | 4 | 97.86 | 2.74 |
| -8 | 2 | -4 | 167.27 | 3.58 |
| -8 | -2 | -5 | 453.05 | 5.18 |
| -8 | -2 | -5 | 417.49 | 4.55 |
| 8 | -2 | -5 | 340.34 | 5.38 |
| -8 | 2 | 5 | 330.61 | 4.22 |
| -8 | 2 | -5 | 522.76 | 6.14 |
| 8 | 2 | -5 | 418.56 | 4.75 |
| -8 | -2 | 6 | 200.51 | 2.82 |
| -8 | -2 | -6 | 248.80 | 4.18 |
| -8 | -2 | -6 | 242.17 | 3.94 |
| 8 | -2 | -6 | 177.96 | 4.20 |
| -8 | 2 | 6 | 197.67 | 3.77 |
| -8 | 2 | -6 | 282.32 | 4.81 |
| -8 | -2 | 7 | 272.51 | 3.56 |
| -8 | -2 | -7 | 342.69 | 5.15 |
| -8 | -2 | -7 | 344.82 | 4.96 |
| 8 | -2 | -7 | 228.27 | 4.76 |
| -8 | 2 | 7 | 283.61 | 4.55 |
| -8 | 2 | -7 | 383.94 | 5.80 |
| -8 | -2 | 8 | 44.87 | 2.81 |
| -8 | -2 | -8 | 65.71 | 3.19 |
| -8 | -2 | -8 | 59.48 | 3.31 |
| 8 | -2 | -8 | 35.33 | 3.25 |
| -8 | 2 | 8 | 42.47 | 3.52 |
| -8 | 2 | -8 | 65.56 | 3.57 |
| -8 | -2 | -9 | 367.18 | 6.34 |
| 8 | -2 | -9 | 270.67 | 5.46 |
| -8 | 2 | 9 | 143.02 | 5.54 |
| -8 | 2 | -9 | 417.44 | 7.02 |
| -8 | -2 | 10 | 30.52 | 3.01 |
| -8 | -2 | -10 | 49.83 | 4.09 |
| -8 | -2 | -10 | 52.10 | 4.43 |
| 8 | -2 | -10 | 38.57 | 4.30 |
| -8 | 2 | 10 | 22.00 | 3.11 |
| -8 | 2 | -10 | 47.07 | 4.22 |
| -8 | 2 | -10 | 59.16 | 4.37 |
| -8 | -2 | 11 | 52.52 | 5.02 |
| -8 | -2 | -11 | 179.46 | 5.60 |
| -8 | -2 | -11 | 158.34 | 6.01 |
| -8 | 2 | 11 | 100.55 | 5.25 |
| -8 | 2 | -11 | 180.26 | 5.63 |
| -8 | 2 | -11 | 183.53 | 5.90 |
| -10 | -2 | 0 | 197.07 | 3.17 |
| -10 | -2 | 0 | 164.17 | 3.12 |
| 10 | -2 | 0 | 133.84 | 3.80 |
| -10 | -2 | 1 | 439.97 | 4.37 |
| -10 | -2 | 1 | 110.97 | 3.36 |
| -10 | -2 | -1 | 435.36 | 4.63 |
| -10 | -2 | -1 | 400.13 | 4.09 |
| 10 | -2 | 1 | 278.87 | 5.12 |
| 10 | -2 | -1 | 284.15 | 5.15 |
| -10 | -2 | 2 | 11.91 | 2.08 |
| -10 | -2 | -2 | 82.45 | 2.68 |
| -10 | -2 | -2 | 60.28 | 2.39 |
| 10 | -2 | 2 | 52.77 | 2.79 |
| 10 | -2 | -2 | 51.57 | 2.90 |
| 10 | 2 | -2 | 57.04 | 2.27 |
| -10 | -2 | -3 | 533.83 | 5.78 |
| -10 | -2 | -3 | 452.79 | 4.87 |
| 10 | -2 | 3 | 319.29 | 5.50 |
| 10 | -2 | -3 | 367.58 | 5.85 |
| -10 | 2 | -3 | 600.55 | 7.27 |
| 10 | 2 | -3 | 411.35 | 4.38 |
| -10 | -2 | -4 | 59.88 | 2.75 |
| -10 | -2 | -4 | 48.04 | 2.82 |
| 10 | -2 | -4 | 40.39 | 3.00 |
| -10 | 2 | -4 | 72.29 | 3.47 |
| 10 | 2 | -4 | 53.19 | 2.31 |
| -10 | -2 | -5 | 313.20 | 5.13 |
| -10 | -2 | -5 | 288.56 | 4.61 |
| 10 | -2 | -5 | 195.30 | 4.63 |
| -10 | 2 | -5 | 354.63 | 6.06 |
| -10 | -2 | -6 | 125.89 | 4.01 |
| -10 | -2 | -6 | 119.71 | 3.97 |
| 10 | -2 | -6 | 65.91 | 3.66 |
| -10 | 2 | -6 | 141.31 | 4.68 |
| -10 | -2 | -7 | 225.71 | 5.21 |
| -10 | -2 | -7 | 232.46 | 7.27 |
| -10 | 2 | -7 | 287.69 | 6.08 |
| -10 | -2 | -8 | 32.99 | 4.06 |
| -10 | -2 | -8 | 19.08 | 2.97 |
| 10 | -2 | -8 | 17.61 | 2.75 |
| -10 | 2 | 8 | 23.53 | 3.27 |
| -10 | 2 | -8 | 39.16 | 4.56 |
| -10 | -2 | -9 | 239.44 | 5.97 |
| -10 | -2 | -9 | 232.18 | 6.27 |
| -10 | 2 | 9 | 107.59 | 5.65 |
| -10 | 2 | -9 | 236.72 | 6.17 |
| -10 | 2 | -9 | 286.90 | 7.60 |
| -10 | -2 | -10 | 21.03 | 4.66 |
| -10 | -2 | -10 | 26.15 | 6.22 |
| -10 | 2 | -10 | 16.27 | 4.02 |
| -10 | 2 | -10 | 12.81 | 4.31 |
| 12 | -2 | 0 | 8.64 | 2.73 |
| -12 | -2 | -1 | 249.88 | 4.94 |
| -12 | -2 | -1 | 100.63 | 4.34 |
| 12 | -2 | 1 | 184.78 | 5.19 |
| 12 | -2 | -1 | 175.40 | 5.30 |
| -12 | 2 | -1 | 210.78 | 7.03 |
| -12 | -2 | -2 | -1.84 | 2.35 |
| 12 | -2 | 2 | 2.38 | 2.29 |
| 12 | -2 | -2 | 6.21 | 2.20 |
| -12 | -2 | -3 | 295.14 | 6.25 |
| 12 | -2 | -3 | 198.66 | 5.18 |
| -12 | 2 | -3 | 298.74 | 6.46 |
| -12 | -2 | -4 | -5.84 | 2.42 |
| 12 | -2 | -4 | -0.99 | 2.34 |
| -12 | 2 | -4 | 8.12 | 2.78 |
| -12 | -2 | -5 | 195.49 | 5.46 |
| 12 | -2 | -5 | 109.42 | 4.82 |
| -12 | 2 | -5 | 199.57 | 6.22 |
| -12 | -2 | -6 | 18.24 | 2.98 |
| 12 | -2 | -6 | 4.69 | 2.79 |
| -12 | 2 | -6 | 10.91 | 3.30 |
| -12 | -2 | -7 | 128.02 | 5.88 |
| -12 | 2 | -7 | 167.37 | 6.63 |
| 14 | -2 | -1 | 73.30 | 6.52 |
| -14 | 2 | -1 | 74.39 | 8.45 |
| 14 | -2 | -2 | 11.52 | 5.18 |
| -1 | -3 | 0 | 1208.50 | 6.12 |
| -1 | -3 | 0 | 1120.60 | 6.52 |
| 1 | -3 | 0 | 1206.47 | 7.00 |
| 1 | -3 | 0 | 1014.94 | 5.75 |
| 1 | -3 | 0 | 1407.72 | 7.18 |
| -1 | 3 | 0 | 1599.15 | 6.24 |
| 1 | 3 | 0 | 1186.46 | 4.73 |
| -1 | -3 | 1 | 30.71 | 1.16 |
| -1 | -3 | -1 | 57.28 | 1.36 |
| -1 | -3 | -1 | 55.20 | 1.12 |
| -1 | -3 | -1 | 54.36 | 1.37 |
| 1 | -3 | 1 | 63.58 | 1.56 |
| 1 | -3 | 1 | 54.13 | 1.17 |
| 1 | -3 | 1 | 53.53 | 1.48 |
| 1 | -3 | -1 | 59.47 | 1.69 |
| 1 | -3 | -1 | 67.54 | 1.57 |
| -1 | 3 | 1 | 45.64 | 1.16 |
| -1 | 3 | -1 | 74.79 | 1.35 |
| 1 | 3 | 1 | 67.45 | 1.26 |
| 1 | 3 | 1 | 47.72 | 1.12 |
| 1 | 3 | -1 | 70.80 | 1.14 |
| -1 | -3 | 2 | 66.54 | 1.31 |
| -1 | -3 | 2 | 48.94 | 1.54 |
| -1 | -3 | -2 | 67.16 | 1.46 |
| -1 | -3 | -2 | 71.67 | 1.33 |
| 1 | -3 | 2 | 54.15 | 0.91 |
| 1 | -3 | 2 | 52.18 | 1.47 |
| 1 | -3 | 2 | 75.44 | 1.60 |
| 1 | -3 | -2 | 68.61 | 1.80 |
| 1 | -3 | -2 | 74.79 | 1.57 |
| -1 | 3 | 2 | 50.80 | 1.23 |
| -1 | 3 | -2 | 155.92 | 1.91 |
| -1 | 3 | -2 | 90.18 | 1.46 |
| 1 | 3 | 2 | 43.84 | 1.14 |
| 1 | 3 | -2 | 126.50 | 2.27 |
| -1 | -3 | 3 | 116.87 | 2.30 |
| -1 | -3 | 3 | 128.05 | 1.96 |
| -1 | -3 | -3 | 135.02 | 2.10 |
| -1 | -3 | -3 | 145.59 | 2.02 |
| 1 | -3 | 3 | 71.96 | 0.91 |
| 1 | -3 | 3 | 114.16 | 2.15 |
| 1 | -3 | 3 | 133.89 | 2.08 |
| 1 | -3 | -3 | 143.56 | 2.52 |
| 1 | -3 | -3 | 140.11 | 1.98 |
| -1 | 3 | 3 | 136.46 | 1.96 |
| -1 | 3 | -3 | 241.47 | 2.12 |
| -1 | 3 | -3 | 140.56 | 1.58 |
| 1 | 3 | 3 | 116.27 | 1.78 |
| 1 | 3 | -3 | 292.11 | 2.95 |
| 1 | 3 | -3 | 98.47 | 1.06 |
| -1 | -3 | 4 | 25.25 | 1.17 |
| -1 | -3 | 4 | 27.13 | 1.04 |
| -1 | -3 | -4 | 28.68 | 1.11 |
| -1 | -3 | -4 | 40.85 | 1.26 |
| 1 | -3 | 4 | 28.02 | 1.15 |
| 1 | -3 | 4 | 28.55 | 1.18 |
| 1 | -3 | -4 | 28.87 | 1.36 |
| 1 | -3 | -4 | 29.68 | 1.33 |
| 1 | -3 | -4 | 29.24 | 1.00 |
| -1 | 3 | 4 | 28.42 | 1.15 |
| -1 | 3 | -4 | 25.87 | 0.81 |
| 1 | 3 | 4 | 36.76 | 1.27 |
| 1 | 3 | -4 | 84.12 | 2.55 |
| 1 | 3 | -4 | 21.94 | 0.73 |
| -1 | -3 | 5 | 21.96 | 1.10 |
| -1 | -3 | -5 | 19.04 | 1.15 |
| -1 | -3 | -5 | 23.04 | 1.23 |
| 1 | -3 | 5 | 20.78 | 1.14 |
| 1 | -3 | 5 | 17.26 | 1.15 |
| 1 | -3 | -5 | 33.96 | 1.54 |
| -1 | 3 | -5 | 19.17 | 0.86 |
| 1 | 3 | 5 | 22.22 | 1.25 |
| 1 | 3 | -5 | 16.02 | 0.82 |
| -1 | -3 | 6 | 196.61 | 2.90 |
| -1 | -3 | 6 | 194.41 | 2.95 |
| -1 | -3 | -6 | 194.96 | 3.52 |
| -1 | -3 | -6 | 207.18 | 3.99 |
| 1 | -3 | 6 | 199.11 | 3.10 |
| 1 | -3 | 6 | 208.22 | 3.03 |
| 1 | -3 | -6 | 220.01 | 4.84 |
| 1 | -3 | -6 | 199.66 | 3.61 |
| -1 | 3 | 6 | 185.38 | 3.08 |
| -1 | 3 | -6 | 231.45 | 4.20 |
| 1 | 3 | 6 | 188.68 | 3.00 |
| 1 | 3 | -6 | 200.80 | 2.51 |
| -1 | -3 | 7 | 4.49 | 0.67 |
| -1 | -3 | 7 | 9.93 | 1.00 |
| -1 | -3 | -7 | 11.62 | 1.23 |
| -1 | -3 | -7 | 7.00 | 1.04 |
| 1 | -3 | 7 | 9.86 | 1.19 |
| 1 | -3 | -7 | 13.91 | 1.60 |
| 1 | -3 | -7 | 6.10 | 1.00 |
| -1 | 3 | 7 | 7.39 | 1.11 |
| -1 | 3 | -7 | 6.15 | 0.94 |
| 1 | 3 | 7 | 10.06 | 1.49 |
| 1 | 3 | -7 | 5.49 | 0.74 |
| -1 | -3 | 8 | 5.33 | 0.96 |
| -1 | -3 | -8 | 1.73 | 0.88 |
| -1 | -3 | -8 | 2.06 | 0.88 |
| 1 | -3 | 8 | 1.85 | 1.03 |
| 1 | -3 | -8 | 2.75 | 1.12 |
| 1 | -3 | -8 | 3.91 | 1.03 |
| -1 | 3 | 8 | 6.77 | 1.16 |
| -1 | 3 | -8 | 1.75 | 0.78 |
| 1 | 3 | 8 | 1.93 | 1.32 |
| 1 | 3 | -8 | 2.92 | 1.03 |
| -1 | -3 | 9 | 10.95 | 1.39 |
| -1 | -3 | -9 | 15.09 | 1.53 |
| -1 | -3 | -9 | 13.06 | 1.47 |
| 1 | -3 | 9 | 10.50 | 1.57 |
| 1 | -3 | -9 | 25.06 | 2.37 |
| 1 | -3 | -9 | 9.91 | 1.50 |
| -1 | 3 | 9 | 12.73 | 1.63 |
| -1 | 3 | -9 | 14.89 | 1.23 |
| -1 | 3 | -9 | 11.79 | 1.27 |
| 1 | 3 | 9 | 17.34 | 2.04 |
| 1 | 3 | -9 | 14.21 | 1.44 |
| -1 | -3 | 10 | 2.61 | 1.47 |
| -1 | -3 | -10 | 5.98 | 1.41 |
| -1 | -3 | -10 | 7.42 | 1.49 |
| 1 | -3 | 10 | 2.93 | 1.75 |
| 1 | -3 | -10 | 4.84 | 1.61 |
| 1 | -3 | -10 | 5.16 | 1.50 |
| -1 | 3 | 10 | 2.07 | 1.96 |
| -1 | 3 | -10 | 1.26 | 1.10 |
| -1 | 3 | -10 | 7.86 | 1.39 |
| 1 | 3 | 10 | 2.99 | 2.24 |
| 1 | 3 | -10 | 12.30 | 1.62 |
| -1 | -3 | 11 | 3.87 | 2.00 |
| -1 | -3 | -11 | 14.93 | 1.99 |
| -1 | -3 | -11 | 13.06 | 2.04 |
| 1 | -3 | 11 | 6.32 | 2.55 |
| 1 | -3 | -11 | 26.27 | 3.20 |
| 1 | -3 | -11 | 14.27 | 1.99 |
| -1 | 3 | 11 | 7.27 | 2.66 |
| -1 | 3 | -11 | 8.83 | 1.80 |
| 1 | 3 | 11 | 7.46 | 3.03 |
| 1 | 3 | -11 | 22.95 | 2.23 |
| -1 | -3 | 12 | 38.94 | 4.21 |
| -1 | -3 | -12 | 96.09 | 6.22 |
| -1 | -3 | -12 | 93.92 | 4.36 |
| 1 | -3 | 12 | 41.22 | 3.63 |
| 1 | -3 | -12 | 94.74 | 4.35 |
| -1 | 3 | 12 | 40.59 | 4.97 |
| -1 | 3 | -12 | 82.84 | 4.57 |
| 1 | 3 | 12 | 27.10 | 4.06 |
| 1 | 3 | -12 | 75.92 | 7.75 |
| -1 | -3 | 13 | -2.00 | 3.51 |
| -1 | -3 | -13 | 4.48 | 2.34 |
| 1 | -3 | -13 | 14.61 | 2.91 |
| -1 | 3 | 13 | 0.43 | 4.44 |
| -1 | 3 | -13 | 4.46 | 2.69 |
| 1 | 3 | 13 | 2.08 | 4.97 |
| 1 | 3 | -13 | 5.83 | 2.99 |
| -3 | -3 | 0 | 2.53 | 0.41 |
| -3 | -3 | 0 | 4.03 | 0.54 |
| -3 | -3 | 0 | 3.02 | 0.71 |
| 3 | -3 | 0 | 4.59 | 0.67 |
| 3 | -3 | 0 | 8.73 | 0.63 |
| -3 | 3 | 0 | 57.49 | 1.41 |
| 3 | 3 | 0 | 3.84 | 0.55 |
| -3 | -3 | 1 | 176.55 | 2.03 |
| -3 | -3 | 1 | 132.66 | 1.25 |
| -3 | -3 | 1 | 128.34 | 2.38 |
| -3 | -3 | 1 | 146.19 | 1.89 |
| -3 | -3 | -1 | 234.03 | 2.37 |
| -3 | -3 | -1 | 228.80 | 1.61 |
| -3 | -3 | -1 | 215.94 | 2.60 |
| 3 | -3 | 1 | 179.66 | 2.49 |
| 3 | -3 | 1 | 134.14 | 2.04 |
| 3 | -3 | 1 | 225.46 | 2.63 |
| 3 | -3 | -1 | 207.91 | 3.16 |
| 3 | -3 | -1 | 176.23 | 2.98 |
| 3 | -3 | -1 | 130.72 | 1.77 |
| -3 | 3 | 1 | 176.67 | 2.33 |
| -3 | 3 | -1 | 247.21 | 2.68 |
| 3 | 3 | 1 | 150.85 | 1.58 |
| -3 | -3 | 2 | 121.44 | 2.43 |
| -3 | -3 | 2 | 291.98 | 2.54 |
| -3 | -3 | -2 | 362.92 | 3.08 |
| -3 | -3 | -2 | 342.93 | 2.35 |
| 3 | -3 | 2 | 244.97 | 2.63 |
| 3 | -3 | 2 | 233.97 | 2.85 |
| 3 | -3 | 2 | 318.62 | 3.06 |
| 3 | -3 | -2 | 327.89 | 4.08 |
| 3 | -3 | -2 | 286.02 | 4.07 |
| 3 | -3 | -2 | 210.32 | 2.06 |
| -3 | 3 | 2 | 230.56 | 2.64 |
| -3 | 3 | -2 | 453.24 | 3.60 |
| 3 | 3 | 2 | 189.12 | 2.04 |
| -3 | -3 | 3 | 2.98 | 0.66 |
| -3 | -3 | 3 | 9.23 | 0.81 |
| -3 | -3 | -3 | 7.89 | 0.68 |
| -3 | -3 | -3 | 33.99 | 0.94 |
| 3 | -3 | 3 | 4.75 | 0.50 |
| 3 | -3 | 3 | 7.81 | 0.75 |
| 3 | -3 | 3 | 12.59 | 0.99 |
| 3 | -3 | -3 | 7.38 | 0.86 |
| 3 | -3 | -3 | 7.12 | 0.94 |
| -3 | 3 | 3 | 9.39 | 1.00 |
| 3 | 3 | 3 | 13.18 | 1.05 |
| -3 | -3 | 4 | 133.76 | 2.48 |
| -3 | -3 | 4 | 127.47 | 1.92 |
| -3 | -3 | -4 | 163.10 | 2.94 |
| -3 | -3 | -4 | 172.63 | 2.19 |
| 3 | -3 | 4 | 76.76 | 1.38 |
| 3 | -3 | 4 | 136.73 | 3.48 |
| 3 | -3 | 4 | 159.63 | 2.38 |
| 3 | -3 | -4 | 180.92 | 3.21 |
| 3 | -3 | -4 | 154.29 | 3.70 |
| -3 | 3 | -4 | 194.37 | 2.51 |
| 3 | 3 | 4 | 125.91 | 2.06 |
| -3 | -3 | 5 | 52.57 | 1.56 |
| -3 | -3 | 5 | 53.77 | 1.53 |
| -3 | -3 | -5 | 75.60 | 2.69 |
| 3 | -3 | 5 | 60.39 | 1.75 |
| 3 | -3 | -5 | 78.48 | 2.39 |
| 3 | -3 | -5 | 63.84 | 2.51 |
| -3 | 3 | 5 | 55.34 | 2.54 |
| -3 | 3 | -5 | 75.59 | 1.67 |
| 3 | 3 | 5 | 58.46 | 1.80 |
| -3 | -3 | 6 | 3.94 | 0.62 |
| -3 | -3 | 6 | 4.04 | 0.80 |
| -3 | -3 | -6 | 2.67 | 0.70 |
| -3 | -3 | -6 | 3.51 | 0.86 |
| 3 | -3 | 6 | 2.20 | 1.50 |
| 3 | -3 | -6 | 13.68 | 2.31 |
| 3 | -3 | -6 | 4.63 | 1.29 |
| -3 | 3 | 6 | 13.69 | 1.48 |
| -3 | 3 | -6 | 14.11 | 1.29 |
| 3 | 3 | 6 | 7.33 | 1.16 |
| 3 | 3 | -6 | 3.78 | 0.88 |
| -3 | -3 | 7 | 19.24 | 1.45 |
| -3 | -3 | 7 | 21.43 | 1.41 |
| -3 | -3 | -7 | 29.20 | 1.65 |
| -3 | -3 | -7 | 32.43 | 1.73 |
| 3 | -3 | 7 | 34.10 | 2.38 |
| 3 | -3 | -7 | 27.02 | 2.89 |
| 3 | -3 | -7 | 23.65 | 2.32 |
| -3 | 3 | 7 | 26.05 | 1.69 |
| -3 | 3 | -7 | 27.26 | 1.57 |
| 3 | 3 | -7 | 26.05 | 2.03 |
| -3 | -3 | 8 | 29.28 | 1.67 |
| -3 | -3 | -8 | 40.38 | 2.01 |
| -3 | -3 | -8 | 37.57 | 2.10 |
| 3 | -3 | 8 | 31.75 | 2.37 |
| 3 | -3 | -8 | 46.94 | 2.84 |
| 3 | -3 | -8 | 45.84 | 2.82 |
| -3 | 3 | 8 | 34.00 | 2.10 |
| -3 | 3 | -8 | 41.67 | 1.96 |
| 3 | 3 | 8 | 39.17 | 3.01 |
| 3 | 3 | -8 | 57.40 | 3.22 |
| -3 | -3 | 9 | 2.74 | 1.17 |
| -3 | -3 | -9 | 1.98 | 1.15 |
| -3 | -3 | -9 | 3.05 | 1.18 |
| 3 | -3 | 9 | 12.68 | 1.83 |
| 3 | -3 | -9 | 0.24 | 1.27 |
| 3 | -3 | -9 | 3.61 | 1.55 |
| -3 | 3 | 9 | 0.29 | 1.51 |
| -3 | 3 | -9 | 2.74 | 1.14 |
| 3 | 3 | -9 | 1.51 | 1.58 |
| -3 | -3 | 10 | 5.80 | 1.55 |
| -3 | -3 | -10 | 14.62 | 1.71 |
| -3 | -3 | -10 | 15.13 | 1.92 |
| 3 | -3 | 10 | 11.02 | 2.37 |
| 3 | -3 | -10 | 8.32 | 1.67 |
| 3 | -3 | -10 | 16.11 | 2.15 |
| -3 | 3 | 10 | 6.05 | 2.06 |
| -3 | 3 | -10 | 6.20 | 1.48 |
| -3 | 3 | -10 | 12.03 | 1.60 |
| 3 | 3 | 10 | 0.80 | 2.92 |
| 3 | 3 | -10 | 19.47 | 2.14 |
| -3 | -3 | 11 | 12.71 | 2.10 |
| -3 | -3 | -11 | 19.51 | 2.08 |
| -3 | -3 | -11 | 28.14 | 3.35 |
| 3 | -3 | 11 | 9.07 | 3.39 |
| 3 | -3 | -11 | 33.37 | 3.71 |
| -3 | 3 | 11 | 17.13 | 3.30 |
| -3 | 3 | -11 | 19.99 | 1.82 |
| -3 | 3 | -11 | 20.03 | 2.21 |
| 3 | 3 | 11 | 14.02 | 4.00 |
| 3 | 3 | -11 | 27.74 | 2.63 |
| -3 | -3 | 12 | 15.28 | 2.82 |
| -3 | -3 | -12 | 30.82 | 3.84 |
| -3 | -3 | -12 | 23.21 | 2.54 |
| 3 | -3 | 12 | 1.30 | 4.24 |
| 3 | -3 | -12 | 37.33 | 4.26 |
| -3 | 3 | 12 | 8.50 | 3.29 |
| -3 | 3 | -12 | 16.82 | 2.48 |
| 3 | 3 | 12 | -9.41 | 6.66 |
| -3 | -3 | 13 | 4.60 | 3.45 |
| -3 | -3 | -13 | 7.99 | 2.74 |
| -3 | -3 | -13 | 9.05 | 2.68 |
| 3 | -3 | 13 | -9.30 | 7.33 |
| 3 | -3 | -13 | 7.48 | 3.16 |
| -3 | 3 | 13 | -9.75 | 3.67 |
| -3 | 3 | -13 | 7.61 | 2.78 |
| -5 | -3 | 0 | 409.69 | 3.08 |
| 5 | -3 | 0 | 318.56 | 4.72 |
| 5 | -3 | 0 | 312.92 | 2.70 |
| -5 | 3 | 0 | 429.73 | 3.98 |
| -5 | -3 | 1 | 11.84 | 0.86 |
| -5 | -3 | 1 | 9.95 | 1.06 |
| -5 | -3 | 1 | 25.29 | 1.62 |
| -5 | -3 | -1 | 14.34 | 0.80 |
| -5 | -3 | -1 | 25.02 | 1.69 |
| 5 | -3 | 1 | 10.67 | 1.15 |
| 5 | -3 | 1 | 12.32 | 1.08 |
| 5 | -3 | -1 | 12.81 | 0.98 |
| -5 | 3 | 1 | 35.09 | 1.53 |
| -5 | 3 | -1 | 18.55 | 1.21 |
| -5 | -3 | 2 | 7.84 | 0.90 |
| -5 | -3 | 2 | 19.40 | 1.48 |
| -5 | -3 | -2 | 24.97 | 1.03 |
| -5 | -3 | -2 | 19.63 | 0.79 |
| -5 | -3 | -2 | 35.29 | 1.86 |
| 5 | -3 | 2 | 15.74 | 1.16 |
| 5 | -3 | -2 | 19.96 | 1.61 |
| -5 | 3 | 2 | 25.45 | 1.82 |
| -5 | -3 | 3 | 14.67 | 1.32 |
| -5 | -3 | 3 | 39.68 | 1.81 |
| -5 | -3 | 3 | 29.75 | 1.06 |
| -5 | -3 | -3 | 46.44 | 1.35 |
| -5 | -3 | -3 | 42.94 | 1.10 |
| 5 | -3 | 3 | 26.36 | 1.32 |
| 5 | -3 | 3 | 31.72 | 1.34 |
| -5 | 3 | 3 | 34.76 | 1.49 |
| -5 | -3 | 4 | 6.94 | 0.90 |
| -5 | -3 | 4 | 12.53 | 0.88 |
| -5 | -3 | -4 | 18.88 | 1.32 |
| -5 | -3 | -4 | 18.87 | 1.09 |
| 5 | -3 | 4 | 6.99 | 0.87 |
| 5 | -3 | 4 | 13.37 | 1.30 |
| 5 | -3 | -4 | 14.96 | 1.78 |
| -5 | 3 | 4 | 16.50 | 1.85 |
| -5 | 3 | -4 | 17.49 | 1.61 |
| 5 | 3 | 4 | 9.01 | 1.04 |
| -5 | -3 | 5 | 7.58 | 0.81 |
| -5 | -3 | 5 | 10.10 | 1.26 |
| -5 | -3 | -5 | 4.94 | 0.79 |
| -5 | -3 | -5 | 21.35 | 1.57 |
| 5 | -3 | 5 | 6.52 | 0.95 |
| 5 | -3 | -5 | 17.84 | 2.21 |
| -5 | 3 | 5 | 10.46 | 1.25 |
| -5 | 3 | -5 | 6.21 | 0.78 |
| 5 | 3 | 5 | 2.40 | 1.20 |
| -5 | -3 | 6 | 86.66 | 1.78 |
| -5 | -3 | 6 | 89.00 | 1.95 |
| -5 | -3 | -6 | 134.08 | 2.74 |
| -5 | -3 | -6 | 131.03 | 2.59 |
| 5 | -3 | 6 | 103.03 | 2.52 |
| 5 | -3 | -6 | 112.31 | 3.48 |
| -5 | 3 | 6 | 127.80 | 2.77 |
| -5 | 3 | -6 | 155.65 | 3.00 |
| 5 | 3 | 6 | 90.57 | 2.57 |
| -5 | -3 | 7 | 3.45 | 0.90 |
| -5 | -3 | -7 | 2.54 | 0.86 |
| -5 | -3 | -7 | 3.35 | 0.95 |
| 5 | -3 | 7 | 3.85 | 1.38 |
| 5 | -3 | -7 | 1.53 | 1.24 |
| -5 | 3 | 7 | 1.43 | 1.30 |
| -5 | 3 | -7 | 3.67 | 0.96 |
| 5 | 3 | 7 | 2.50 | 1.92 |
| -5 | -3 | 8 | 3.61 | 1.09 |
| -5 | -3 | -8 | 3.92 | 1.12 |
| -5 | -3 | -8 | 5.56 | 1.18 |
| 5 | -3 | 8 | 10.60 | 2.10 |
| 5 | -3 | -8 | 2.57 | 1.45 |
| -5 | 3 | 8 | 8.83 | 1.58 |
| -5 | 3 | -8 | 12.16 | 1.45 |
| 5 | 3 | 8 | 1.68 | 3.08 |
| -5 | -3 | 9 | 11.11 | 1.53 |
| -5 | -3 | -9 | 9.65 | 1.48 |
| -5 | -3 | -9 | 10.64 | 1.57 |
| 5 | -3 | 9 | 9.03 | 2.85 |
| 5 | -3 | -9 | 4.33 | 2.02 |
| -5 | 3 | 9 | 4.92 | 1.87 |
| -5 | 3 | -9 | 9.96 | 1.49 |
| -5 | -3 | 10 | 5.36 | 1.86 |
| -5 | -3 | -10 | 8.03 | 1.76 |
| -5 | -3 | -10 | 8.22 | 1.82 |
| 5 | -3 | 10 | -8.25 | 4.01 |
| 5 | -3 | -10 | 10.75 | 2.22 |
| -5 | 3 | 10 | 5.74 | 2.32 |
| -5 | 3 | -10 | 8.28 | 1.80 |
| -5 | -3 | 11 | 4.33 | 2.37 |
| -5 | -3 | -11 | 5.24 | 1.91 |
| -5 | -3 | -11 | 9.64 | 2.31 |
| 5 | -3 | -11 | 6.36 | 2.36 |
| -5 | 3 | 11 | 0.38 | 2.72 |
| -5 | 3 | -11 | 5.25 | 2.26 |
| -5 | 3 | -11 | 8.63 | 2.06 |
| -5 | -3 | 12 | 25.19 | 3.22 |
| -5 | -3 | -12 | 71.24 | 4.58 |
| -5 | -3 | -12 | 69.84 | 4.66 |
| 5 | -3 | -12 | 47.24 | 4.85 |
| -5 | 3 | 12 | 10.69 | 3.37 |
| -5 | 3 | -12 | 49.56 | 4.79 |
| -5 | 3 | -12 | 71.61 | 4.76 |
| -5 | -3 | -13 | 12.09 | 3.92 |
| -5 | 3 | -13 | 10.88 | 4.40 |
| -7 | -3 | 0 | 124.66 | 2.02 |
| 7 | -3 | 0 | 84.44 | 2.69 |
| -7 | 3 | 0 | 144.25 | 3.15 |
| -7 | -3 | 1 | 10.66 | 1.10 |
| -7 | -3 | 1 | 6.79 | 0.95 |
| -7 | -3 | -1 | 19.73 | 1.12 |
| -7 | -3 | -1 | 8.99 | 0.87 |
| 7 | -3 | 1 | 8.54 | 1.18 |
| 7 | -3 | -1 | 7.79 | 1.22 |
| -7 | 3 | 1 | 26.92 | 2.22 |
| -7 | 3 | -1 | 20.71 | 2.04 |
| -7 | -3 | 2 | 2.87 | 1.24 |
| -7 | -3 | -2 | 3.14 | 0.80 |
| -7 | -3 | -2 | 4.93 | 0.58 |
| 7 | -3 | 2 | 3.25 | 0.89 |
| 7 | -3 | -2 | 4.54 | 1.16 |
| -7 | 3 | 2 | 11.66 | 1.55 |
| -7 | -3 | 3 | 0.70 | 1.05 |
| -7 | -3 | -3 | 12.00 | 1.29 |
| -7 | -3 | -3 | 11.03 | 1.16 |
| 7 | -3 | 3 | 17.43 | 1.69 |
| 7 | -3 | -3 | 4.80 | 1.27 |
| -7 | 3 | 3 | 12.38 | 1.52 |
| -7 | -3 | 4 | 2.93 | 1.12 |
| -7 | -3 | 4 | 5.08 | 0.95 |
| -7 | -3 | -4 | 2.02 | 0.77 |
| -7 | -3 | -4 | 2.60 | 0.99 |
| 7 | -3 | 4 | 1.70 | 0.85 |
| 7 | -3 | -4 | 1.85 | 1.20 |
| -7 | 3 | 4 | 2.10 | 1.26 |
| -7 | 3 | -4 | 5.14 | 1.01 |
| -7 | -3 | 5 | 5.61 | 0.95 |
| -7 | -3 | -5 | 7.84 | 0.96 |
| -7 | -3 | -5 | 8.41 | 1.03 |
| 7 | -3 | 5 | 3.53 | 1.09 |
| 7 | -3 | -5 | 5.12 | 1.42 |
| -7 | 3 | 5 | 5.32 | 1.34 |
| -7 | 3 | -5 | 9.37 | 1.25 |
| -7 | -3 | 6 | 48.36 | 1.80 |
| -7 | -3 | -6 | 66.59 | 2.41 |
| -7 | -3 | -6 | 61.73 | 2.26 |
| 7 | -3 | -6 | 46.86 | 2.92 |
| -7 | 3 | 6 | 57.41 | 2.68 |
| -7 | 3 | -6 | 79.65 | 2.86 |
| -7 | -3 | 7 | 3.21 | 1.24 |
| -7 | -3 | -7 | 3.51 | 1.10 |
| -7 | -3 | -7 | 0.88 | 1.32 |
| 7 | -3 | -7 | 8.36 | 1.78 |
| -7 | 3 | 7 | 2.55 | 1.65 |
| -7 | 3 | -7 | 4.02 | 1.29 |
| -7 | -3 | 8 | 3.42 | 1.44 |
| -7 | -3 | -8 | 3.48 | 1.46 |
| -7 | -3 | -8 | 3.27 | 1.56 |
| 7 | -3 | -8 | 6.61 | 1.89 |
| -7 | 3 | 8 | 5.52 | 1.99 |
| -7 | 3 | -8 | -0.56 | 1.60 |
| -7 | -3 | 9 | 0.71 | 1.90 |
| -7 | -3 | -9 | 6.25 | 1.84 |
| -7 | -3 | -9 | 5.96 | 1.92 |
| 7 | -3 | -9 | 5.88 | 2.09 |
| -7 | 3 | 9 | 4.46 | 2.50 |
| -7 | 3 | -9 | 0.69 | 1.81 |
| -7 | -3 | 10 | 1.88 | 2.34 |
| -7 | -3 | -10 | 2.79 | 2.05 |
| -7 | -3 | -10 | 4.37 | 2.33 |
| 7 | -3 | -10 | 3.67 | 2.43 |
| -7 | 3 | 10 | 1.62 | 2.56 |
| -7 | 3 | -10 | 5.67 | 2.14 |
| -7 | -3 | 11 | -1.78 | 2.90 |
| -7 | -3 | -11 | 11.87 | 2.63 |
| -7 | -3 | -11 | 11.69 | 2.82 |
| 7 | -3 | -11 | 1.72 | 2.90 |
| -7 | 3 | 11 | 5.11 | 3.00 |
| -7 | 3 | -11 | -2.07 | 2.56 |
| -7 | 3 | -11 | 8.50 | 2.81 |
| -7 | -3 | 12 | 19.84 | 5.21 |
| -7 | -3 | -12 | 50.29 | 5.85 |
| -7 | -3 | -12 | 35.04 | 4.27 |
| -7 | 3 | -12 | 34.41 | 5.90 |
| -7 | 3 | -12 | 33.36 | 3.81 |
| -9 | -3 | 0 | 3.54 | 1.24 |
| -9 | -3 | 0 | -0.86 | 1.15 |
| 9 | -3 | 0 | 3.16 | 1.35 |
| -9 | -3 | 1 | 20.69 | 1.89 |
| -9 | -3 | 1 | 8.92 | 2.03 |
| -9 | -3 | -1 | 22.87 | 1.87 |
| 9 | -3 | 1 | 13.76 | 1.60 |
| 9 | -3 | -1 | 11.78 | 1.68 |
| -9 | 3 | 1 | 20.20 | 1.75 |
| -9 | -3 | 2 | 11.67 | 1.74 |
| -9 | -3 | -2 | 50.86 | 2.13 |
| -9 | -3 | -2 | 52.25 | 2.46 |
| 9 | -3 | 2 | 24.63 | 2.38 |
| 9 | -3 | -2 | 33.26 | 2.62 |
| -9 | 3 | 2 | 69.04 | 2.83 |
| -9 | -3 | 3 | 0.87 | 1.94 |
| -9 | -3 | -3 | 2.15 | 1.12 |
| -9 | -3 | -3 | 1.42 | 1.10 |
| 9 | -3 | 3 | 1.32 | 1.37 |
| 9 | -3 | -3 | 1.25 | 1.37 |
| -9 | 3 | 3 | 1.38 | 1.99 |
| -9 | 3 | -3 | 0.11 | 1.46 |
| -9 | -3 | -4 | 36.03 | 2.27 |
| -9 | -3 | -4 | 34.02 | 2.23 |
| 9 | -3 | 4 | 16.52 | 1.82 |
| 9 | -3 | -4 | 21.90 | 2.63 |
| -9 | 3 | 4 | 35.71 | 3.02 |
| -9 | 3 | -4 | 41.75 | 2.92 |
| -9 | -3 | -5 | 15.56 | 1.68 |
| -9 | -3 | -5 | 13.96 | 1.56 |
| 9 | -3 | -5 | 5.23 | 1.79 |
| -9 | 3 | 5 | 11.43 | 2.03 |
| -9 | 3 | -5 | 14.19 | 1.76 |
| -9 | -3 | -6 | 3.16 | 1.66 |
| 9 | -3 | -6 | -0.57 | 1.81 |
| -9 | 3 | 6 | 1.40 | 2.14 |
| -9 | 3 | -6 | 0.28 | 1.81 |
| -9 | -3 | 7 | 7.66 | 1.85 |
| -9 | -3 | -7 | 11.24 | 1.86 |
| -9 | -3 | -7 | 6.13 | 1.98 |
| 9 | -3 | -7 | 6.64 | 2.19 |
| -9 | 3 | 7 | 14.04 | 2.47 |
| -9 | 3 | -7 | 13.04 | 2.27 |
| -9 | -3 | 8 | -5.80 | 4.46 |
| -9 | -3 | -8 | 4.40 | 2.02 |
| -9 | -3 | -8 | 4.48 | 3.00 |
| 9 | -3 | -8 | 9.71 | 2.47 |
| -9 | 3 | 8 | 9.41 | 2.63 |
| -9 | 3 | -8 | 9.52 | 2.38 |
| -9 | -3 | 9 | -0.30 | 3.00 |
| -9 | -3 | -9 | 0.75 | 2.40 |
| -9 | -3 | -9 | -3.48 | 2.80 |
| 9 | -3 | -9 | -1.31 | 2.64 |
| -9 | 3 | 9 | -3.14 | 2.86 |
| -9 | 3 | -9 | -4.33 | 2.43 |
| -9 | -3 | 10 | 5.15 | 3.98 |
| -9 | -3 | -10 | 0.30 | 2.69 |
| -9 | -3 | -10 | 2.45 | 3.20 |
| -9 | 3 | 10 | -2.10 | 3.26 |
| -9 | 3 | -10 | -2.64 | 2.82 |
| -9 | 3 | -10 | 13.87 | 2.96 |
| -11 | -3 | 0 | 20.50 | 2.31 |
| 11 | -3 | 0 | 32.87 | 3.30 |
| -11 | -3 | 1 | -0.62 | 3.20 |
| -11 | -3 | -1 | 0.91 | 1.88 |
| -11 | -3 | -1 | 0.23 | 1.88 |
| 11 | -3 | 1 | -1.18 | 1.87 |
| 11 | -3 | -1 | -0.71 | 1.83 |
| -11 | -3 | -2 | 11.57 | 1.95 |
| 11 | -3 | 2 | 5.00 | 2.24 |
| 11 | -3 | -2 | 6.11 | 2.00 |
| -11 | 3 | -2 | 11.20 | 2.43 |
| -11 | -3 | -3 | 3.79 | 2.00 |
| 11 | -3 | 3 | 2.61 | 2.25 |
| 11 | -3 | -3 | 5.70 | 2.09 |
| -11 | 3 | -3 | 6.81 | 2.39 |
| -11 | -3 | -4 | 10.86 | 2.19 |
| 11 | -3 | -4 | 0.15 | 3.63 |
| -11 | 3 | -4 | 13.62 | 4.24 |
| -11 | -3 | -5 | 3.22 | 2.26 |
| 11 | -3 | -5 | 5.34 | 2.40 |
| -11 | 3 | -5 | -0.26 | 2.50 |
| -11 | -3 | -6 | 54.07 | 4.28 |
| 11 | -3 | -6 | 15.48 | 2.91 |
| -11 | 3 | -6 | 37.37 | 3.51 |
| -11 | -3 | -7 | 2.28 | 2.78 |
| 11 | -3 | -7 | 0.95 | 2.87 |
| -11 | -3 | -8 | -4.43 | 3.14 |
| 11 | -3 | -8 | -0.48 | 3.27 |
| -11 | 3 | -8 | 10.42 | 3.46 |
| 13 | -3 | 0 | 12.40 | 3.18 |
| -13 | 3 | 0 | 11.97 | 3.85 |
| 13 | -3 | 1 | 8.58 | 3.01 |
| 13 | -3 | -1 | -0.39 | 2.93 |
| -13 | 3 | -1 | -0.17 | 3.61 |
| 13 | -3 | 2 | -2.29 | 3.03 |
| 13 | -3 | -2 | -4.87 | 2.77 |
| -13 | 3 | -2 | -3.38 | 3.60 |
| -13 | -3 | -3 | 0.96 | 3.13 |
| 13 | -3 | -3 | 2.46 | 2.92 |
| -13 | 3 | -3 | -8.05 | 3.49 |
| -13 | -3 | -4 | 0.76 | 3.95 |
| -13 | -3 | -4 | -1.42 | 3.25 |
| 13 | -3 | -4 | 2.65 | 3.43 |
| -13 | 3 | -4 | 1.12 | 3.76 |
| -13 | 3 | -5 | -1.40 | 5.75 |
| 0 | -4 | 0 | 141.76 | 2.68 |
| 0 | -4 | 0 | 166.07 | 2.91 |
| 0 | 4 | 0 | 285.94 | 2.89 |
| 0 | -4 | 1 | 147.21 | 2.64 |
| 0 | -4 | 1 | 116.40 | 1.80 |
| 0 | -4 | 1 | 236.48 | 3.58 |
| 0 | -4 | -1 | 291.06 | 3.78 |
| 0 | -4 | -1 | 293.75 | 3.69 |
| 0 | 4 | 1 | 432.82 | 3.67 |
| 0 | 4 | -1 | 423.78 | 3.40 |
| 0 | -4 | 2 | 414.98 | 4.82 |
| 0 | -4 | -2 | 582.46 | 5.34 |
| 0 | -4 | -2 | 584.70 | 4.96 |
| 0 | 4 | 2 | 557.22 | 4.23 |
| 0 | 4 | -2 | 765.73 | 4.50 |
| 0 | -4 | 3 | 828.91 | 6.84 |
| 0 | -4 | 3 | 1136.62 | 6.80 |
| 0 | -4 | -3 | 1356.57 | 8.20 |
| 0 | -4 | -3 | 1483.15 | 7.80 |
| 0 | -4 | -3 | 1387.95 | 7.36 |
| 0 | 4 | 3 | 1155.37 | 6.25 |
| 0 | 4 | -3 | 1305.85 | 4.49 |
| 0 | -4 | 4 | 303.94 | 4.19 |
| 0 | -4 | 4 | 383.20 | 4.09 |
| 0 | -4 | -4 | 620.47 | 5.70 |
| 0 | -4 | -4 | 533.18 | 4.94 |
| 0 | -4 | -4 | 515.11 | 4.38 |
| 0 | 4 | 4 | 484.43 | 4.25 |
| 0 | 4 | -4 | 404.38 | 2.75 |
| 0 | -4 | 5 | 166.08 | 3.11 |
| 0 | -4 | 5 | 185.78 | 3.17 |
| 0 | -4 | -5 | 232.58 | 3.54 |
| 0 | -4 | -5 | 255.64 | 3.64 |
| 0 | 4 | 5 | 253.69 | 3.27 |
| 0 | 4 | -5 | 189.56 | 2.15 |
| 0 | -4 | 6 | 180.28 | 3.44 |
| 0 | -4 | 6 | 183.66 | 3.25 |
| 0 | -4 | -6 | 227.18 | 3.93 |
| 0 | -4 | -6 | 223.58 | 3.83 |
| 0 | 4 | 6 | 272.29 | 3.92 |
| 0 | 4 | -6 | 189.78 | 2.49 |
| 0 | -4 | 7 | 156.53 | 3.16 |
| 0 | -4 | 7 | 155.30 | 3.11 |
| 0 | -4 | -7 | 222.60 | 3.96 |
| 0 | -4 | -7 | 209.25 | 3.84 |
| 0 | 4 | 7 | 224.19 | 3.68 |
| 0 | 4 | -7 | 199.26 | 4.32 |
| 0 | -4 | 8 | 228.95 | 3.92 |
| 0 | -4 | -8 | 433.42 | 5.61 |
| 0 | 4 | 8 | 359.18 | 4.82 |
| 0 | 4 | -8 | 311.09 | 3.99 |
| 0 | -4 | 9 | 288.42 | 4.70 |
| 0 | -4 | -9 | 617.16 | 6.95 |
| 0 | -4 | -9 | 573.22 | 7.04 |
| 0 | 4 | 9 | 454.87 | 5.80 |
| 0 | 4 | -9 | 457.26 | 5.18 |
| 0 | -4 | 10 | 125.60 | 3.86 |
| 0 | -4 | -10 | 317.32 | 5.44 |
| 0 | -4 | -10 | 243.88 | 5.14 |
| 0 | -4 | -10 | 0.02 | 1.33 |
| 0 | 4 | 10 | 188.34 | 4.66 |
| 0 | 4 | -10 | 231.95 | 4.39 |
| 0 | -4 | 11 | 61.81 | 3.97 |
| 0 | -4 | -11 | 149.94 | 4.66 |
| 0 | -4 | -11 | 121.09 | 4.27 |
| 0 | 4 | 11 | 77.91 | 4.72 |
| 0 | 4 | -11 | 110.99 | 4.26 |
| 0 | -4 | 12 | 41.98 | 4.91 |
| 0 | -4 | -12 | 173.45 | 5.24 |
| 0 | -4 | -12 | 139.76 | 4.86 |
| 0 | 4 | 12 | 59.25 | 5.71 |
| 0 | 4 | -12 | 130.90 | 5.14 |
| 0 | -4 | 13 | 28.48 | 4.34 |
| 0 | 4 | 13 | 1.01 | 4.29 |
| -2 | -4 | 0 | 1261.25 | 6.97 |
| -2 | -4 | 0 | 1061.04 | 7.33 |
| 2 | -4 | 0 | 1068.38 | 7.68 |
| 2 | -4 | 0 | 760.60 | 5.89 |
| 2 | -4 | 0 | 1095.04 | 6.99 |
| -2 | 4 | 0 | 2398.86 | 9.23 |
| 2 | 4 | 0 | 1394.47 | 5.55 |
| -2 | -4 | 1 | 478.38 | 4.28 |
| -2 | -4 | -1 | 829.24 | 5.62 |
| -2 | -4 | -1 | 756.21 | 5.88 |
| 2 | -4 | 1 | 470.02 | 4.10 |
| 2 | -4 | 1 | 516.33 | 4.93 |
| 2 | -4 | -1 | 623.05 | 6.04 |
| 2 | -4 | -1 | 536.32 | 5.34 |
| 2 | -4 | -1 | 705.56 | 5.44 |
| -2 | 4 | 1 | 1061.76 | 6.18 |
| -2 | 4 | -1 | 919.21 | 5.61 |
| 2 | 4 | 1 | 788.55 | 3.84 |
| 2 | 4 | 1 | 1308.94 | 5.78 |
| 2 | 4 | -1 | 856.40 | 4.13 |
| -2 | -4 | 2 | 576.15 | 4.35 |
| -2 | -4 | 2 | 792.69 | 6.85 |
| -2 | -4 | -2 | 2034.16 | 8.92 |
| -2 | -4 | -2 | 2210.91 | 7.50 |
| -2 | -4 | -2 | 1982.51 | 9.19 |
| 2 | -4 | 2 | 1237.32 | 5.77 |
| 2 | -4 | 2 | 1100.09 | 7.39 |
| 2 | -4 | 2 | 1630.58 | 8.33 |
| 2 | -4 | -2 | 1695.15 | 10.11 |
| 2 | -4 | -2 | 1656.94 | 9.89 |
| 2 | -4 | -2 | 1881.61 | 8.60 |
| -2 | 4 | 2 | 1852.15 | 8.20 |
| -2 | 4 | -2 | 2542.31 | 9.29 |
| 2 | 4 | 2 | 1733.84 | 7.06 |
| -2 | -4 | 3 | 803.07 | 6.94 |
| -2 | -4 | 3 | 1448.60 | 7.36 |
| -2 | -4 | -3 | 1953.62 | 8.98 |
| -2 | -4 | -3 | 2177.74 | 8.10 |
| 2 | -4 | 3 | 918.47 | 4.38 |
| 2 | -4 | 3 | 1201.35 | 7.94 |
| 2 | -4 | 3 | 1554.56 | 8.13 |
| 2 | -4 | -3 | 1786.57 | 10.48 |
| 2 | -4 | -3 | 1688.84 | 10.39 |
| 2 | -4 | -3 | 1925.03 | 8.40 |
| -2 | 4 | 3 | 1589.97 | 7.71 |
| 2 | 4 | 3 | 1520.21 | 6.95 |
| -2 | -4 | 4 | 713.38 | 6.49 |
| -2 | -4 | 4 | 971.02 | 6.07 |
| -2 | -4 | -4 | 1374.83 | 7.82 |
| -2 | -4 | -4 | 1550.43 | 7.41 |
| 2 | -4 | 4 | 460.20 | 2.94 |
| 2 | -4 | 4 | 913.34 | 7.09 |
| 2 | -4 | 4 | 1094.12 | 6.95 |
| 2 | -4 | -4 | 1345.76 | 10.27 |
| 2 | -4 | -4 | 1400.91 | 6.93 |
| -2 | 4 | 4 | 1347.25 | 7.24 |
| -2 | 4 | -4 | 1489.03 | 6.56 |
| 2 | 4 | 4 | 1103.74 | 6.21 |
| -2 | -4 | 5 | 317.54 | 3.77 |
| -2 | -4 | -5 | 466.24 | 4.90 |
| -2 | -4 | -5 | 475.81 | 4.48 |
| 2 | -4 | 5 | 335.32 | 4.56 |
| 2 | -4 | 5 | 389.61 | 4.68 |
| 2 | -4 | -5 | 438.15 | 5.31 |
| 2 | -4 | -5 | 468.26 | 3.92 |
| -2 | 4 | 5 | 466.45 | 4.63 |
| -2 | 4 | -5 | 449.20 | 4.29 |
| 2 | 4 | 5 | 407.17 | 4.08 |
| -2 | -4 | 6 | 409.75 | 6.09 |
| -2 | -4 | 6 | 465.20 | 4.59 |
| -2 | -4 | -6 | 716.02 | 6.28 |
| -2 | -4 | -6 | 700.57 | 5.99 |
| 2 | -4 | 6 | 506.53 | 5.65 |
| 2 | -4 | 6 | 533.83 | 5.52 |
| 2 | -4 | -6 | 592.50 | 6.56 |
| 2 | -4 | -6 | 620.68 | 4.38 |
| -2 | 4 | 6 | 657.59 | 5.62 |
| -2 | 4 | -6 | 665.34 | 5.07 |
| 2 | 4 | 6 | 632.63 | 5.31 |
| 2 | 4 | -6 | 690.55 | 4.61 |
| -2 | -4 | 7 | 231.39 | 3.68 |
| -2 | -4 | 7 | 220.28 | 3.43 |
| -2 | -4 | -7 | 342.61 | 4.65 |
| -2 | -4 | -7 | 356.14 | 4.65 |
| 2 | -4 | 7 | 261.13 | 4.12 |
| 2 | -4 | -7 | 325.61 | 5.03 |
| -2 | 4 | 7 | 321.43 | 4.26 |
| -2 | 4 | -7 | 323.38 | 3.86 |
| 2 | 4 | 7 | 313.60 | 4.52 |
| 2 | 4 | -7 | 380.57 | 3.97 |
| -2 | -4 | 8 | 420.27 | 4.64 |
| -2 | -4 | 8 | 386.04 | 4.76 |
| -2 | -4 | -8 | 642.32 | 6.68 |
| -2 | -4 | -8 | 700.20 | 6.94 |
| 2 | -4 | 8 | 458.37 | 5.62 |
| 2 | -4 | -8 | 790.46 | 7.66 |
| -2 | 4 | 8 | 601.27 | 6.09 |
| -2 | 4 | -8 | 630.28 | 5.74 |
| 2 | 4 | 8 | 561.44 | 5.96 |
| 2 | 4 | -8 | 764.76 | 6.53 |
| -2 | -4 | 9 | 340.90 | 4.82 |
| -2 | -4 | -9 | 645.20 | 7.09 |
| -2 | -4 | -9 | 631.46 | 7.11 |
| 2 | -4 | 9 | 365.49 | 5.54 |
| 2 | -4 | -9 | 807.55 | 7.93 |
| -2 | 4 | 9 | 518.68 | 6.12 |
| -2 | 4 | -9 | 564.89 | 5.95 |
| 2 | 4 | 9 | 495.87 | 6.19 |
| 2 | 4 | -9 | 751.87 | 6.69 |
| -2 | -4 | 10 | 183.89 | 4.17 |
| -2 | -4 | -10 | 462.98 | 6.45 |
| -2 | -4 | -10 | 446.60 | 6.56 |
| 2 | -4 | 10 | 237.35 | 5.25 |
| 2 | -4 | -10 | 555.12 | 6.78 |
| 2 | -4 | -10 | 575.88 | 8.06 |
| -2 | 4 | 10 | 314.12 | 5.43 |
| -2 | 4 | -10 | 343.91 | 4.12 |
| -2 | 4 | -10 | 380.39 | 5.41 |
| 2 | 4 | 10 | 304.58 | 5.69 |
| 2 | 4 | -10 | 539.67 | 6.41 |
| -2 | -4 | -11 | 167.10 | 4.77 |
| -2 | -4 | -11 | 127.61 | 5.84 |
| 2 | -4 | 11 | 68.50 | 4.89 |
| 2 | -4 | -11 | 209.08 | 4.93 |
| 2 | -4 | -11 | 191.70 | 5.42 |
| -2 | 4 | 11 | 67.67 | 6.26 |
| -2 | 4 | -11 | 138.53 | 4.36 |
| 2 | 4 | 11 | 100.08 | 5.50 |
| 2 | 4 | -11 | 185.25 | 5.16 |
| -2 | -4 | 12 | 58.17 | 4.74 |
| -2 | -4 | -12 | 269.71 | 6.09 |
| -2 | -4 | -12 | 245.94 | 6.29 |
| 2 | -4 | 12 | 67.32 | 6.17 |
| 2 | -4 | -12 | 232.18 | 6.18 |
| -2 | 4 | 12 | 109.31 | 5.50 |
| -2 | 4 | -12 | 200.94 | 5.61 |
| -2 | -4 | 13 | 36.56 | 4.00 |
| -2 | -4 | -13 | 163.27 | 5.84 |
| -2 | -4 | -13 | 109.44 | 4.95 |
| 2 | -4 | 13 | 16.63 | 5.19 |
| 2 | -4 | -13 | 106.75 | 5.50 |
| -2 | 4 | 13 | 20.68 | 3.82 |
| -2 | 4 | -13 | 75.45 | 5.61 |
| -4 | -4 | 0 | 1063.12 | 5.91 |
| -4 | -4 | 0 | 833.25 | 6.68 |
| 4 | -4 | 0 | 616.71 | 5.96 |
| 4 | -4 | 0 | 554.64 | 4.39 |
| -4 | 4 | 0 | 1148.05 | 7.80 |
| -4 | -4 | 1 | 524.76 | 4.12 |
| -4 | -4 | 1 | 352.67 | 4.54 |
| -4 | -4 | -1 | 637.54 | 4.65 |
| -4 | -4 | -1 | 507.18 | 6.16 |
| 4 | -4 | 1 | 347.02 | 4.17 |
| 4 | -4 | 1 | 322.46 | 3.59 |
| 4 | -4 | 1 | 495.49 | 4.46 |
| 4 | -4 | -1 | 378.55 | 5.45 |
| 4 | -4 | -1 | 303.10 | 3.01 |
| -4 | 4 | 1 | 1051.67 | 10.37 |
| -4 | 4 | -1 | 711.30 | 8.56 |
| 4 | 4 | 1 | 447.71 | 3.03 |
| -4 | -4 | 2 | 450.60 | 5.28 |
| -4 | -4 | -2 | 1509.78 | 7.46 |
| -4 | -4 | -2 | 1572.12 | 5.55 |
| 4 | -4 | 2 | 752.62 | 5.65 |
| 4 | -4 | 2 | 814.64 | 6.14 |
| 4 | -4 | 2 | 1085.19 | 6.50 |
| 4 | -4 | -2 | 987.50 | 8.40 |
| 4 | -4 | -2 | 688.34 | 4.12 |
| -4 | 4 | 2 | 1280.32 | 7.38 |
| -4 | 4 | -2 | 1914.11 | 8.90 |
| 4 | 4 | 2 | 1878.39 | 6.25 |
| -4 | -4 | 3 | 281.26 | 4.26 |
| -4 | -4 | 3 | 873.97 | 5.51 |
| -4 | -4 | -3 | 1465.20 | 7.65 |
| -4 | -4 | -3 | 1468.88 | 6.09 |
| 4 | -4 | 3 | 652.71 | 4.96 |
| 4 | -4 | 3 | 864.17 | 6.58 |
| 4 | -4 | 3 | 1060.33 | 6.53 |
| 4 | -4 | -3 | 1113.81 | 9.34 |
| 4 | 4 | 3 | 1091.83 | 5.38 |
| -4 | -4 | 4 | 427.64 | 5.11 |
| -4 | -4 | 4 | 565.38 | 4.65 |
| -4 | -4 | -4 | 1123.25 | 7.12 |
| -4 | -4 | -4 | 1119.45 | 5.95 |
| 4 | -4 | 4 | 424.32 | 4.24 |
| 4 | -4 | 4 | 814.44 | 5.92 |
| 4 | -4 | -4 | 861.40 | 7.55 |
| 4 | -4 | -4 | 787.72 | 8.18 |
| -4 | 4 | 4 | 930.66 | 6.53 |
| -4 | 4 | -4 | 1287.66 | 7.56 |
| 4 | 4 | 4 | 729.90 | 4.84 |
| -4 | -4 | 5 | 184.37 | 3.99 |
| -4 | -4 | 5 | 225.38 | 3.00 |
| -4 | -4 | -5 | 371.35 | 4.35 |
| 4 | -4 | 5 | 101.09 | 1.95 |
| 4 | -4 | 5 | 276.01 | 3.79 |
| 4 | -4 | -5 | 269.97 | 5.29 |
| 4 | -4 | -5 | 282.07 | 5.10 |
| -4 | 4 | 5 | 342.23 | 4.15 |
| -4 | 4 | -5 | 413.51 | 4.42 |
| 4 | 4 | 5 | 270.17 | 3.33 |
| -4 | -4 | 6 | 280.26 | 3.76 |
| -4 | -4 | 6 | 297.94 | 3.56 |
| -4 | -4 | -6 | 539.11 | 5.52 |
| -4 | -4 | -6 | 585.98 | 5.32 |
| 4 | -4 | 6 | 467.33 | 6.15 |
| 4 | -4 | -6 | 419.26 | 5.72 |
| 4 | -4 | -6 | 385.05 | 6.02 |
| -4 | 4 | 6 | 531.27 | 5.26 |
| -4 | 4 | -6 | 593.64 | 5.45 |
| 4 | 4 | 6 | 447.16 | 4.46 |
| -4 | -4 | 7 | 155.70 | 3.08 |
| -4 | -4 | 7 | 158.47 | 2.94 |
| -4 | -4 | -7 | 278.81 | 4.32 |
| -4 | -4 | -7 | 286.22 | 4.13 |
| 4 | -4 | 7 | 207.11 | 4.10 |
| 4 | -4 | -7 | 205.00 | 4.15 |
| 4 | -4 | -7 | 187.19 | 4.52 |
| -4 | 4 | 7 | 275.62 | 4.19 |
| -4 | 4 | -7 | 286.20 | 4.07 |
| 4 | 4 | 7 | 238.01 | 3.97 |
| -4 | -4 | 8 | 311.52 | 3.74 |
| -4 | -4 | 8 | 290.62 | 4.05 |
| -4 | -4 | -8 | 513.31 | 6.15 |
| -4 | -4 | -8 | 570.86 | 6.23 |
| 4 | -4 | 8 | 352.40 | 5.36 |
| 4 | -4 | -8 | 477.86 | 5.90 |
| 4 | -4 | -8 | 426.35 | 6.75 |
| -4 | 4 | 8 | 490.91 | 5.73 |
| -4 | 4 | -8 | 543.63 | 5.91 |
| 4 | 4 | 8 | 428.17 | 5.72 |
| -4 | -4 | 9 | 231.39 | 4.04 |
| -4 | -4 | -9 | 481.70 | 6.36 |
| -4 | -4 | -9 | 508.64 | 6.40 |
| 4 | -4 | 9 | 295.42 | 5.68 |
| 4 | -4 | -9 | 503.16 | 6.04 |
| 4 | -4 | -9 | 448.34 | 7.11 |
| -4 | 4 | 9 | 408.49 | 5.69 |
| -4 | 4 | -9 | 500.78 | 6.11 |
| 4 | 4 | 9 | 111.34 | 4.94 |
| -4 | -4 | 10 | 158.35 | 4.01 |
| -4 | -4 | -10 | 347.62 | 5.89 |
| -4 | -4 | -10 | 368.29 | 6.08 |
| 4 | -4 | 10 | 188.49 | 5.93 |
| 4 | -4 | -10 | 348.12 | 6.60 |
| -4 | 4 | 10 | 272.51 | 5.34 |
| -4 | 4 | -10 | 338.79 | 5.61 |
| -4 | -4 | 11 | 56.19 | 3.72 |
| -4 | -4 | -11 | 131.09 | 4.52 |
| -4 | -4 | -11 | 147.81 | 4.81 |
| 4 | -4 | 11 | 73.51 | 5.60 |
| 4 | -4 | -11 | 113.70 | 4.79 |
| -4 | 4 | 11 | 89.70 | 4.52 |
| -4 | 4 | -11 | 99.02 | 4.40 |
| -4 | 4 | -11 | 120.77 | 4.47 |
| -4 | -4 | 12 | 50.75 | 4.44 |
| -4 | -4 | -12 | 205.14 | 5.77 |
| -4 | -4 | -12 | 178.35 | 5.61 |
| 4 | -4 | -12 | 152.57 | 5.55 |
| -4 | 4 | 12 | 86.74 | 5.10 |
| -4 | 4 | -12 | 157.23 | 5.58 |
| -4 | -4 | 13 | 24.33 | 4.84 |
| -4 | -4 | -13 | 122.80 | 7.58 |
| -4 | 4 | -13 | 91.77 | 8.56 |
| -6 | -4 | 0 | 205.53 | 2.71 |
| -6 | -4 | 0 | 151.33 | 3.38 |
| 6 | -4 | 0 | 118.22 | 3.24 |
| 6 | -4 | 0 | 128.43 | 2.23 |
| -6 | 4 | 0 | 275.00 | 4.06 |
| -6 | -4 | 1 | 251.06 | 2.86 |
| -6 | -4 | -1 | 329.53 | 3.70 |
| 6 | -4 | 1 | 145.75 | 3.38 |
| 6 | -4 | 1 | 179.80 | 2.46 |
| 6 | -4 | -1 | 156.73 | 3.76 |
| -6 | 4 | 1 | 504.52 | 5.35 |
| -6 | 4 | -1 | 332.92 | 4.34 |
| -6 | -4 | 2 | 105.55 | 2.87 |
| -6 | -4 | 2 | 203.09 | 2.43 |
| -6 | -4 | -2 | 379.14 | 4.13 |
| -6 | -4 | -2 | 352.63 | 2.65 |
| 6 | -4 | 2 | 169.34 | 3.43 |
| 6 | -4 | 2 | 220.18 | 2.76 |
| 6 | -4 | -2 | 255.73 | 4.76 |
| -6 | 4 | 2 | 425.48 | 4.94 |
| -6 | -4 | 3 | 110.90 | 2.82 |
| -6 | -4 | 3 | 382.23 | 3.45 |
| -6 | -4 | -3 | 728.13 | 5.79 |
| -6 | -4 | -3 | 721.15 | 4.31 |
| 6 | -4 | 3 | 292.72 | 4.23 |
| 6 | -4 | 3 | 432.47 | 3.85 |
| 6 | -4 | -3 | 459.79 | 6.43 |
| -6 | 4 | 3 | 577.26 | 5.59 |
| -6 | -4 | 4 | 98.79 | 2.55 |
| -6 | -4 | 4 | 164.03 | 2.42 |
| -6 | -4 | -4 | 310.44 | 4.03 |
| -6 | -4 | -4 | 308.86 | 3.18 |
| 6 | -4 | 4 | 108.85 | 2.69 |
| 6 | -4 | 4 | 196.98 | 2.99 |
| 6 | -4 | -4 | 210.41 | 4.57 |
| -6 | 4 | 4 | 256.91 | 3.89 |
| -6 | 4 | -4 | 379.89 | 4.94 |
| -6 | -4 | 5 | 76.33 | 2.27 |
| -6 | -4 | 5 | 93.95 | 2.04 |
| -6 | -4 | -5 | 188.33 | 3.42 |
| -6 | -4 | -5 | 197.68 | 3.15 |
| 6 | -4 | 5 | 56.93 | 2.09 |
| 6 | -4 | 5 | 113.13 | 2.78 |
| 6 | -4 | -5 | 146.70 | 4.06 |
| -6 | 4 | 5 | 165.02 | 3.32 |
| -6 | 4 | -5 | 218.72 | 3.84 |
| -6 | -4 | 6 | 61.52 | 3.28 |
| -6 | -4 | 6 | 89.37 | 2.23 |
| -6 | -4 | -6 | 160.42 | 3.46 |
| -6 | -4 | -6 | 166.39 | 3.12 |
| 6 | -4 | -6 | 107.38 | 3.69 |
| -6 | 4 | 6 | 138.28 | 3.30 |
| -6 | 4 | -6 | 181.09 | 3.68 |
| -6 | -4 | 7 | 83.04 | 2.43 |
| -6 | -4 | -7 | 164.45 | 3.69 |
| -6 | -4 | -7 | 157.19 | 3.43 |
| 6 | -4 | 7 | -5.04 | 3.97 |
| 6 | -4 | -7 | 88.41 | 3.64 |
| -6 | 4 | 7 | 126.62 | 3.42 |
| -6 | 4 | -7 | 163.32 | 3.75 |
| -6 | -4 | 8 | 114.26 | 2.95 |
| -6 | -4 | -8 | 196.40 | 4.32 |
| -6 | -4 | -8 | 214.72 | 4.23 |
| 6 | -4 | -8 | 120.59 | 4.28 |
| -6 | 4 | 8 | 180.74 | 4.21 |
| -6 | 4 | -8 | 213.90 | 4.48 |
| -6 | -4 | 9 | 161.77 | 3.75 |
| -6 | -4 | -9 | 293.84 | 5.52 |
| -6 | -4 | -9 | 313.02 | 5.42 |
| 6 | -4 | -9 | 184.00 | 5.13 |
| -6 | 4 | 9 | 227.35 | 5.02 |
| -6 | 4 | -9 | 321.15 | 5.70 |
| -6 | -4 | 10 | 67.15 | 3.62 |
| -6 | -4 | -10 | 148.14 | 4.73 |
| -6 | -4 | -10 | 154.94 | 4.75 |
| 6 | -4 | -10 | 103.23 | 4.49 |
| -6 | 4 | 10 | 105.84 | 4.47 |
| -6 | 4 | -10 | 139.81 | 4.63 |
| -6 | -4 | 11 | 38.74 | 4.20 |
| -6 | -4 | -11 | 64.02 | 4.71 |
| -6 | -4 | -11 | 80.94 | 4.74 |
| 6 | -4 | -11 | 59.16 | 4.81 |
| -6 | 4 | 11 | 36.60 | 4.55 |
| -6 | 4 | -11 | 64.32 | 4.62 |
| -6 | 4 | -11 | 76.21 | 4.75 |
| -6 | -4 | 12 | 15.40 | 3.57 |
| -6 | -4 | -12 | 78.23 | 5.32 |
| -6 | -4 | -12 | 77.95 | 5.53 |
| 6 | -4 | -12 | 43.85 | 4.09 |
| -6 | 4 | 12 | 57.36 | 5.51 |
| -6 | 4 | -12 | 53.08 | 5.47 |
| -8 | -4 | 0 | 600.44 | 5.64 |
| -8 | -4 | 0 | 375.12 | 5.27 |
| 8 | -4 | 0 | 288.94 | 5.14 |
| -8 | 4 | 0 | 700.38 | 6.37 |
| -8 | -4 | 1 | 309.45 | 3.57 |
| -8 | -4 | 1 | 177.85 | 3.82 |
| -8 | -4 | -1 | 365.36 | 4.27 |
| -8 | -4 | -1 | 289.99 | 4.69 |
| 8 | -4 | 1 | 166.19 | 3.97 |
| 8 | -4 | -1 | 200.66 | 4.47 |
| -8 | 4 | 1 | 507.04 | 5.57 |
| -8 | -4 | 2 | 186.04 | 3.99 |
| -8 | -4 | -2 | 717.59 | 5.98 |
| 8 | -4 | 2 | 275.93 | 4.89 |
| 8 | -4 | -2 | 414.88 | 6.38 |
| -8 | 4 | 2 | 1307.30 | 8.74 |
| -8 | -4 | 3 | 97.89 | 3.30 |
| -8 | -4 | -3 | 667.21 | 6.09 |
| -8 | -4 | -3 | 614.54 | 5.27 |
| 8 | -4 | 3 | 239.30 | 4.54 |
| 8 | -4 | -3 | 375.07 | 6.23 |
| -8 | 4 | 3 | 1032.32 | 7.93 |
| -8 | -4 | 4 | 153.77 | 3.76 |
| -8 | -4 | -4 | 569.52 | 5.97 |
| -8 | -4 | -4 | 558.88 | 4.78 |
| 8 | -4 | 4 | 177.27 | 3.95 |
| 8 | -4 | -4 | 329.37 | 5.94 |
| -8 | 4 | 4 | 526.45 | 5.94 |
| -8 | 4 | -4 | 678.45 | 7.61 |
| -8 | -4 | 5 | 88.44 | 3.32 |
| -8 | -4 | -5 | 237.91 | 4.24 |
| -8 | -4 | -5 | 249.16 | 3.71 |
| 8 | -4 | 5 | 69.36 | 2.94 |
| 8 | -4 | -5 | 131.93 | 4.19 |
| -8 | 4 | 5 | 192.21 | 4.18 |
| -8 | 4 | -5 | 268.97 | 5.04 |
| -8 | -4 | 6 | 176.52 | 2.97 |
| -8 | -4 | -6 | 330.82 | 5.24 |
| -8 | -4 | -6 | 317.17 | 4.56 |
| 8 | -4 | -6 | 182.38 | 4.80 |
| -8 | 4 | 6 | 295.77 | 4.90 |
| -8 | 4 | -6 | 393.43 | 6.14 |
| -8 | -4 | 7 | 88.72 | 2.95 |
| -8 | -4 | -7 | 180.24 | 4.50 |
| -8 | -4 | -7 | 174.55 | 4.12 |
| 8 | -4 | -7 | 91.86 | 4.12 |
| -8 | 4 | 7 | 146.51 | 4.26 |
| -8 | 4 | -7 | 201.94 | 4.90 |
| -8 | -4 | 8 | 166.45 | 3.81 |
| -8 | -4 | -8 | 316.92 | 7.31 |
| -8 | -4 | -8 | 329.52 | 5.58 |
| 8 | -4 | -8 | 181.39 | 7.27 |
| -8 | 4 | 8 | 229.14 | 6.14 |
| -8 | 4 | -8 | 383.22 | 8.07 |
| -8 | -4 | 9 | 117.85 | 4.14 |
| -8 | -4 | -9 | 265.28 | 5.90 |
| -8 | -4 | -9 | 270.29 | 5.80 |
| 8 | -4 | -9 | 130.35 | 5.06 |
| -8 | 4 | 9 | 186.44 | 5.18 |
| -8 | 4 | -9 | 301.68 | 6.40 |
| -8 | -4 | 10 | 50.07 | 4.54 |
| -8 | -4 | -10 | 216.54 | 5.79 |
| -8 | -4 | -10 | 188.41 | 5.83 |
| 8 | -4 | -10 | 108.97 | 5.03 |
| -8 | 4 | 10 | 131.04 | 5.11 |
| -8 | 4 | -10 | 218.42 | 6.08 |
| -8 | -4 | -11 | 78.31 | 6.33 |
| -8 | -4 | -11 | 120.96 | 10.51 |
| -8 | 4 | -11 | 99.67 | 7.79 |
| -8 | 4 | -11 | 76.04 | 6.43 |
| -10 | -4 | 0 | 275.25 | 4.24 |
| -10 | -4 | 0 | 146.88 | 4.11 |
| 10 | -4 | 0 | 159.12 | 4.60 |
| -10 | -4 | 1 | 59.95 | 3.28 |
| -10 | -4 | -1 | 161.34 | 3.73 |
| -10 | -4 | -1 | 112.54 | 3.68 |
| 10 | -4 | 1 | 84.80 | 3.73 |
| 10 | -4 | -1 | 94.39 | 3.89 |
| -10 | 4 | 1 | 157.88 | 4.16 |
| -10 | -4 | 2 | 67.94 | 3.55 |
| -10 | -4 | -2 | 370.94 | 5.08 |
| -10 | -4 | -2 | 272.80 | 5.19 |
| 10 | -4 | 2 | 158.23 | 4.58 |
| 10 | -4 | -2 | 209.33 | 5.20 |
| -10 | 4 | 2 | 450.38 | 5.66 |
| -10 | -4 | -3 | 323.02 | 5.05 |
| 10 | -4 | 3 | 122.14 | 4.24 |
| 10 | -4 | -3 | 183.92 | 5.03 |
| -10 | 4 | 3 | 486.74 | 5.99 |
| -10 | 4 | -3 | 370.04 | 6.78 |
| -10 | -4 | -4 | 299.89 | 5.22 |
| 10 | -4 | 4 | 107.96 | 4.24 |
| 10 | -4 | -4 | 166.07 | 4.94 |
| -10 | 4 | 4 | 528.49 | 6.32 |
| -10 | 4 | -4 | 361.12 | 6.66 |
| -10 | -4 | -5 | 133.47 | 4.25 |
| 10 | -4 | -5 | 60.49 | 4.19 |
| -10 | 4 | 5 | 240.68 | 5.32 |
| -10 | 4 | -5 | 134.43 | 6.72 |
| -10 | -4 | -6 | 175.69 | 4.96 |
| 10 | -4 | -6 | 83.52 | 4.30 |
| -10 | 4 | 6 | 285.04 | 5.98 |
| -10 | 4 | -6 | 216.23 | 5.85 |
| -10 | -4 | -7 | 92.14 | 4.61 |
| 10 | -4 | -7 | 35.49 | 4.15 |
| -10 | 4 | 7 | 101.24 | 5.21 |
| -10 | 4 | -7 | 123.13 | 5.37 |
| -10 | -4 | -8 | 163.51 | 5.58 |
| 10 | -4 | -8 | 75.12 | 4.61 |
| -10 | 4 | 8 | 154.11 | 5.57 |
| -10 | 4 | -8 | 182.88 | 6.09 |
| -10 | -4 | -9 | 154.24 | 6.35 |
| 10 | -4 | -9 | 80.76 | 5.43 |
| -10 | 4 | 9 | 162.21 | 9.08 |
| -10 | 4 | -9 | 150.92 | 6.49 |
| 12 | -4 | 0 | 38.52 | 4.30 |
| -12 | -4 | -1 | 35.00 | 4.30 |
| 12 | -4 | 1 | 44.08 | 4.37 |
| 12 | -4 | -1 | 37.99 | 4.33 |
| -12 | -4 | -2 | 50.13 | 4.29 |
| 12 | -4 | 2 | 48.77 | 4.61 |
| 12 | -4 | -2 | 37.09 | 4.33 |
| -12 | 4 | -2 | 88.35 | 5.75 |
| -12 | -4 | -3 | 106.40 | 5.51 |
| -12 | -4 | -3 | 78.50 | 4.58 |
| 12 | -4 | 3 | 61.76 | 4.92 |
| 12 | -4 | -3 | 56.90 | 4.46 |
| -12 | 4 | -3 | 155.32 | 6.22 |
| -12 | -4 | -4 | 47.94 | 3.80 |
| -12 | -4 | -4 | 43.81 | 4.27 |
| 12 | -4 | -4 | 21.36 | 3.05 |
| -12 | 4 | -4 | 60.40 | 5.71 |
| -12 | -4 | -5 | 30.60 | 3.62 |
| -12 | -4 | -5 | 38.51 | 4.35 |
| 12 | -4 | -5 | 23.82 | 3.28 |
| -12 | 4 | -5 | 45.60 | 5.50 |
| -12 | -4 | -6 | 38.05 | 3.82 |
| 12 | -4 | -6 | 14.88 | 3.40 |
| -12 | 4 | -6 | 36.19 | 4.11 |
| -1 | -5 | 0 | 30.93 | 1.65 |
| -1 | -5 | 0 | 28.64 | 1.57 |
| 1 | -5 | 0 | 21.90 | 1.62 |
| 1 | -5 | 0 | 19.93 | 1.34 |
| 1 | -5 | 0 | 30.66 | 1.52 |
| -1 | 5 | 0 | 127.00 | 2.34 |
| 1 | 5 | 0 | 55.57 | 1.50 |
| -1 | -5 | 1 | 23.58 | 1.56 |
| -1 | -5 | 1 | 27.19 | 1.23 |
| -1 | -5 | 1 | 37.85 | 1.86 |
| -1 | -5 | -1 | 66.47 | 2.14 |
| -1 | -5 | -1 | 61.41 | 2.03 |
| 1 | -5 | 1 | 27.91 | 1.66 |
| 1 | -5 | 1 | 19.36 | 1.19 |
| 1 | -5 | 1 | 46.38 | 1.89 |
| 1 | -5 | -1 | 50.33 | 2.16 |
| 1 | -5 | -1 | 46.36 | 1.84 |
| 1 | -5 | -1 | 57.96 | 1.93 |
| -1 | 5 | 1 | 128.75 | 2.43 |
| -1 | 5 | -1 | 83.46 | 1.91 |
| 1 | 5 | 1 | 126.88 | 2.23 |
| 1 | 5 | -1 | 134.40 | 2.12 |
| -1 | -5 | 2 | 52.68 | 1.78 |
| -1 | -5 | 2 | 77.41 | 2.77 |
| -1 | -5 | -2 | 190.68 | 3.46 |
| -1 | -5 | -2 | 187.51 | 3.27 |
| 1 | -5 | 2 | 78.60 | 1.49 |
| 1 | -5 | 2 | 94.37 | 3.19 |
| 1 | -5 | -2 | 168.51 | 5.00 |
| 1 | -5 | -2 | 170.37 | 3.04 |
| -1 | 5 | 2 | 254.10 | 3.40 |
| -1 | 5 | -2 | 237.45 | 3.09 |
| 1 | 5 | 2 | 264.30 | 3.24 |
| -1 | -5 | 3 | 9.44 | 1.10 |
| -1 | -5 | 3 | 12.56 | 1.26 |
| -1 | -5 | -3 | 24.14 | 1.47 |
| -1 | -5 | -3 | 31.85 | 1.67 |
| 1 | -5 | 3 | 8.68 | 0.66 |
| 1 | -5 | 3 | 13.83 | 1.32 |
| 1 | -5 | -3 | 38.52 | 1.90 |
| 1 | -5 | -3 | 32.84 | 1.54 |
| -1 | 5 | 3 | 28.85 | 1.47 |
| 1 | 5 | 3 | 27.70 | 1.47 |
| -1 | -5 | 4 | 62.39 | 2.33 |
| -1 | -5 | -4 | 117.00 | 3.04 |
| -1 | -5 | -4 | 124.81 | 3.13 |
| -1 | -5 | -4 | 108.67 | 2.42 |
| 1 | -5 | 4 | 50.36 | 2.38 |
| 1 | -5 | 4 | 68.52 | 2.27 |
| 1 | -5 | -4 | 98.85 | 3.24 |
| 1 | -5 | -4 | 106.11 | 2.34 |
| -1 | 5 | 4 | 130.93 | 3.58 |
| -1 | 5 | -4 | 106.27 | 1.70 |
| 1 | 5 | 4 | 94.11 | 2.88 |
| -1 | -5 | 5 | 13.85 | 1.56 |
| -1 | -5 | 5 | 15.70 | 1.60 |
| -1 | -5 | -5 | 29.67 | 1.95 |
| -1 | -5 | -5 | 34.71 | 2.12 |
| 1 | -5 | 5 | 14.33 | 1.59 |
| 1 | -5 | 5 | 21.56 | 2.14 |
| 1 | -5 | -5 | 25.98 | 2.24 |
| 1 | -5 | -5 | 27.65 | 1.54 |
| -1 | 5 | 5 | 32.53 | 2.00 |
| -1 | 5 | -5 | 28.21 | 1.16 |
| 1 | 5 | 5 | 19.34 | 2.03 |
| 1 | 5 | -5 | 20.22 | 1.00 |
| -1 | -5 | 6 | 9.64 | 1.29 |
| -1 | -5 | 6 | 9.85 | 1.22 |
| -1 | -5 | -6 | 20.57 | 1.84 |
| -1 | -5 | -6 | 21.05 | 1.80 |
| -1 | -5 | -6 | 20.56 | 1.82 |
| 1 | -5 | 6 | 10.66 | 1.35 |
| 1 | -5 | 6 | 15.48 | 1.90 |
| 1 | -5 | -6 | 17.20 | 2.05 |
| 1 | -5 | -6 | 20.13 | 1.56 |
| -1 | 5 | 6 | 18.67 | 1.90 |
| 1 | 5 | 6 | 18.96 | 2.25 |
| 1 | 5 | -6 | 18.92 | 1.22 |
| -1 | -5 | 7 | 8.37 | 1.21 |
| -1 | -5 | 7 | 5.84 | 1.21 |
| -1 | -5 | -7 | 13.16 | 1.45 |
| -1 | -5 | -7 | 10.97 | 1.25 |
| -1 | -5 | -7 | 15.92 | 1.41 |
| 1 | -5 | 7 | 9.51 | 1.28 |
| 1 | -5 | 7 | 8.86 | 1.51 |
| 1 | -5 | -7 | 11.98 | 1.42 |
| 1 | -5 | -7 | 12.73 | 1.56 |
| -1 | 5 | 7 | 12.27 | 1.61 |
| -1 | 5 | -7 | 10.48 | 1.11 |
| 1 | 5 | 7 | 8.08 | 1.50 |
| 1 | 5 | -7 | 11.99 | 1.02 |
| -1 | -5 | 8 | 28.25 | 2.05 |
| -1 | -5 | 8 | 29.40 | 2.34 |
| -1 | -5 | -8 | 54.64 | 2.79 |
| -1 | -5 | -8 | 54.96 | 2.70 |
| -1 | -5 | -8 | 56.46 | 2.23 |
| 1 | -5 | 8 | 30.07 | 2.65 |
| 1 | -5 | -8 | 57.44 | 2.91 |
| 1 | -5 | -8 | 66.56 | 3.08 |
| -1 | 5 | 8 | 44.76 | 2.80 |
| -1 | 5 | -8 | 45.26 | 2.11 |
| 1 | 5 | 8 | 53.75 | 3.45 |
| 1 | 5 | -8 | 54.35 | 2.59 |
| -1 | -5 | 9 | 5.18 | 1.67 |
| -1 | -5 | -9 | 11.59 | 1.66 |
| -1 | -5 | -9 | 6.96 | 1.55 |
| -1 | -5 | -9 | 15.07 | 1.53 |
| 1 | -5 | 9 | 5.76 | 1.92 |
| 1 | -5 | -9 | 18.40 | 1.94 |
| -1 | 5 | 9 | 3.32 | 2.00 |
| -1 | 5 | -9 | 8.87 | 1.37 |
| 1 | 5 | 9 | 6.86 | 2.27 |
| 1 | 5 | -9 | 12.71 | 1.53 |
| -1 | -5 | 10 | 5.45 | 1.98 |
| -1 | -5 | -10 | 36.48 | 3.82 |
| -1 | -5 | -10 | 26.65 | 2.35 |
| -1 | -5 | -10 | 23.12 | 2.58 |
| 1 | -5 | 10 | 3.57 | 2.16 |
| 1 | -5 | -10 | 30.27 | 3.19 |
| -1 | 5 | 10 | 15.10 | 2.50 |
| -1 | 5 | -10 | 30.04 | 2.92 |
| 1 | 5 | 10 | 11.61 | 2.82 |
| 1 | 5 | -10 | 30.13 | 2.88 |
| -1 | -5 | 11 | 7.24 | 2.36 |
| -1 | -5 | -11 | 33.93 | 3.78 |
| -1 | -5 | -11 | 28.51 | 3.38 |
| 1 | -5 | 11 | 0.24 | 3.13 |
| 1 | -5 | -11 | 32.23 | 2.80 |
| -1 | 5 | 11 | 4.62 | 2.89 |
| -1 | 5 | -11 | 19.40 | 2.38 |
| 1 | 5 | 11 | 10.31 | 3.71 |
| -1 | -5 | 12 | 13.10 | 3.24 |
| -1 | -5 | -12 | 22.80 | 2.66 |
| -1 | -5 | -12 | 9.89 | 2.24 |
| 1 | -5 | 12 | 0.89 | 3.73 |
| 1 | -5 | -12 | 19.23 | 3.33 |
| 1 | -5 | -12 | 21.85 | 2.87 |
| -1 | 5 | 12 | 9.28 | 3.81 |
| -1 | 5 | -12 | 9.59 | 3.25 |
| 1 | 5 | 12 | -4.78 | 4.23 |
| -1 | -5 | 13 | -5.16 | 4.17 |
| 1 | -5 | 13 | -6.32 | 5.18 |
| 1 | -5 | -13 | 6.91 | 3.04 |
| -3 | -5 | 0 | 449.17 | 4.61 |
| -3 | -5 | 0 | 351.43 | 4.78 |
| 3 | -5 | 0 | 166.91 | 3.25 |
| 3 | -5 | 0 | 228.30 | 3.45 |
| -3 | 5 | 0 | 795.89 | 6.11 |
| 3 | 5 | 0 | 401.87 | 3.15 |
| -3 | -5 | 1 | 5.50 | 0.83 |
| -3 | -5 | 1 | 0.45 | 0.72 |
| -3 | -5 | -1 | 4.68 | 0.77 |
| -3 | -5 | -1 | 2.87 | 0.79 |
| 3 | -5 | 1 | 0.01 | 0.67 |
| 3 | -5 | 1 | -0.86 | 0.63 |
| 3 | -5 | -1 | 7.23 | 1.01 |
| 3 | -5 | -1 | 3.47 | 1.42 |
| 3 | -5 | -1 | 1.28 | 0.66 |
| -3 | 5 | 1 | 7.71 | 0.86 |
| -3 | 5 | -1 | 32.50 | 1.50 |
| 3 | 5 | 1 | 5.84 | 0.73 |
| -3 | -5 | 2 | 1.65 | 0.78 |
| -3 | -5 | 2 | 3.16 | 0.97 |
| -3 | -5 | -2 | 10.86 | 1.28 |
| -3 | -5 | -2 | 9.75 | 1.09 |
| -3 | -5 | -2 | 7.65 | 1.10 |
| 3 | -5 | 2 | 3.01 | 0.62 |
| 3 | -5 | 2 | 3.31 | 0.68 |
| 3 | -5 | 2 | 4.36 | 1.05 |
| 3 | -5 | -2 | 4.49 | 0.89 |
| 3 | -5 | -2 | 5.51 | 1.02 |
| 3 | -5 | -2 | 4.94 | 0.74 |
| -3 | 5 | -2 | 7.08 | 1.21 |
| 3 | 5 | 2 | 11.00 | 0.97 |
| -3 | -5 | 3 | 2.50 | 0.81 |
| -3 | -5 | 3 | 2.34 | 1.00 |
| -3 | -5 | -3 | 9.10 | 1.26 |
| -3 | -5 | -3 | 9.78 | 0.82 |
| -3 | -5 | -3 | 8.92 | 1.18 |
| 3 | -5 | 3 | 2.88 | 0.60 |
| 3 | -5 | 3 | 5.49 | 1.13 |
| 3 | -5 | 3 | 8.67 | 1.27 |
| 3 | -5 | -3 | 8.27 | 1.46 |
| 3 | -5 | -3 | 6.35 | 1.35 |
| 3 | -5 | -3 | 9.02 | 0.88 |
| -3 | 5 | 3 | 9.09 | 1.28 |
| 3 | 5 | 3 | 10.78 | 1.02 |
| -3 | -5 | 4 | -0.29 | 1.09 |
| -3 | -5 | 4 | 7.25 | 1.04 |
| -3 | -5 | -4 | 6.29 | 0.99 |
| -3 | -5 | -4 | 4.68 | 0.83 |
| 3 | -5 | 4 | 3.34 | 0.59 |
| 3 | -5 | 4 | 2.67 | 0.85 |
| 3 | -5 | 4 | 4.80 | 1.06 |
| 3 | -5 | -4 | 4.81 | 1.30 |
| 3 | -5 | -4 | 4.83 | 1.34 |
| 3 | -5 | -4 | 5.85 | 0.78 |
| -3 | 5 | 4 | 7.51 | 1.19 |
| -3 | 5 | -4 | 6.91 | 1.20 |
| 3 | 5 | 4 | 6.47 | 1.05 |
| -3 | -5 | 5 | -0.18 | 1.12 |
| -3 | -5 | 5 | 3.12 | 0.88 |
| -3 | -5 | -5 | 7.99 | 1.00 |
| -3 | -5 | -5 | 3.02 | 1.11 |
| 3 | -5 | 5 | 1.60 | 0.89 |
| 3 | -5 | 5 | 6.96 | 1.67 |
| 3 | -5 | -5 | 1.68 | 1.16 |
| 3 | -5 | -5 | 0.46 | 1.23 |
| -3 | 5 | 5 | 1.74 | 1.04 |
| -3 | 5 | -5 | 0.05 | 0.67 |
| 3 | 5 | 5 | -1.43 | 1.17 |
| -3 | -5 | 6 | 59.96 | 2.52 |
| -3 | -5 | 6 | 75.08 | 2.32 |
| -3 | -5 | -6 | 157.94 | 3.41 |
| -3 | -5 | -6 | 178.13 | 3.36 |
| 3 | -5 | 6 | 91.32 | 2.81 |
| 3 | -5 | 6 | 96.68 | 3.22 |
| 3 | -5 | -6 | 111.31 | 3.33 |
| 3 | -5 | -6 | 108.32 | 3.53 |
| -3 | 5 | 6 | 153.13 | 3.25 |
| -3 | 5 | -6 | 164.07 | 2.91 |
| 3 | 5 | 6 | 157.23 | 4.67 |
| -3 | -5 | 7 | 2.11 | 1.04 |
| -3 | -5 | 7 | -1.24 | 1.03 |
| -3 | -5 | -7 | 1.51 | 1.08 |
| -3 | -5 | -7 | 0.06 | 1.07 |
| 3 | -5 | 7 | -3.47 | 1.42 |
| 3 | -5 | -7 | -0.33 | 1.18 |
| 3 | -5 | -7 | 1.95 | 1.35 |
| -3 | 5 | 7 | 2.46 | 1.32 |
| -3 | 5 | -7 | 2.99 | 0.90 |
| 3 | 5 | 7 | 4.21 | 1.91 |
| -3 | -5 | 8 | 0.98 | 1.13 |
| -3 | -5 | 8 | 1.60 | 1.35 |
| -3 | -5 | -8 | 5.88 | 1.33 |
| -3 | -5 | -8 | 1.23 | 1.26 |
| 3 | -5 | 8 | 1.12 | 1.95 |
| 3 | -5 | -8 | 12.92 | 1.68 |
| 3 | -5 | -8 | -1.88 | 1.36 |
| -3 | 5 | 8 | 0.85 | 1.61 |
| -3 | 5 | -8 | 0.72 | 1.08 |
| 3 | 5 | 8 | 3.20 | 2.06 |
| -3 | -5 | 9 | -2.45 | 1.58 |
| -3 | -5 | -9 | -1.93 | 1.41 |
| -3 | -5 | -9 | -1.73 | 1.50 |
| 3 | -5 | 9 | -3.86 | 2.09 |
| 3 | -5 | -9 | -0.90 | 1.39 |
| 3 | -5 | -9 | 0.35 | 1.56 |
| -3 | 5 | 9 | 1.52 | 2.04 |
| -3 | 5 | -9 | 0.64 | 1.30 |
| 3 | 5 | 9 | 6.77 | 2.64 |
| -3 | -5 | 10 | 3.01 | 1.89 |
| -3 | -5 | -10 | 0.37 | 1.76 |
| -3 | -5 | -10 | 1.51 | 1.84 |
| 3 | -5 | 10 | -2.05 | 2.68 |
| 3 | -5 | -10 | -2.21 | 1.65 |
| 3 | -5 | -10 | 9.42 | 2.07 |
| -3 | 5 | 10 | 2.92 | 2.64 |
| -3 | 5 | -10 | 1.78 | 1.70 |
| 3 | 5 | 10 | 5.70 | 3.50 |
| -3 | -5 | 11 | 3.45 | 2.41 |
| -3 | -5 | -11 | 3.02 | 2.14 |
| -3 | -5 | -11 | 9.27 | 2.19 |
| 3 | -5 | 11 | -2.57 | 3.71 |
| 3 | -5 | -11 | 7.71 | 2.22 |
| -3 | 5 | 11 | 3.25 | 2.80 |
| -3 | 5 | -11 | -1.41 | 2.10 |
| -3 | -5 | 12 | 18.41 | 3.04 |
| -3 | -5 | -12 | 52.62 | 4.49 |
| -3 | -5 | -12 | 38.45 | 4.14 |
| 3 | -5 | 12 | 12.84 | 4.95 |
| 3 | -5 | -12 | 43.52 | 4.63 |
| -3 | 5 | 12 | 18.92 | 3.41 |
| -3 | 5 | -12 | 48.35 | 4.94 |
| -5 | -5 | 0 | 2.84 | 0.87 |
| -5 | -5 | 0 | 4.64 | 1.12 |
| 5 | -5 | 0 | -0.21 | 1.14 |
| 5 | -5 | 0 | 2.13 | 1.12 |
| 5 | -5 | 0 | 8.86 | 1.31 |
| -5 | 5 | 0 | 11.80 | 1.39 |
| -5 | -5 | 1 | 31.80 | 1.68 |
| -5 | -5 | 1 | 17.49 | 2.00 |
| -5 | -5 | -1 | 41.16 | 1.71 |
| -5 | -5 | -1 | 38.99 | 2.06 |
| 5 | -5 | 1 | 12.53 | 1.50 |
| 5 | -5 | 1 | 16.10 | 1.24 |
| 5 | -5 | 1 | 22.61 | 2.03 |
| 5 | -5 | -1 | 18.68 | 2.25 |
| 5 | -5 | -1 | 12.90 | 1.08 |
| -5 | 5 | 1 | 79.28 | 2.66 |
| -5 | 5 | -1 | 45.11 | 2.20 |
| -5 | -5 | 2 | 48.31 | 2.58 |
| -5 | -5 | -2 | 130.53 | 2.93 |
| -5 | -5 | -2 | 116.22 | 2.94 |
| 5 | -5 | 2 | 38.15 | 2.17 |
| 5 | -5 | 2 | 61.07 | 2.23 |
| 5 | -5 | 2 | 71.38 | 2.28 |
| 5 | -5 | -2 | 67.60 | 2.91 |
| 5 | -5 | -2 | 51.39 | 1.35 |
| -5 | 5 | 2 | 180.65 | 3.49 |
| -5 | 5 | -2 | 151.06 | 3.22 |
| -5 | -5 | 3 | 2.47 | 1.35 |
| -5 | -5 | 3 | 7.38 | 1.10 |
| -5 | -5 | -3 | 21.93 | 1.74 |
| -5 | -5 | -3 | 18.14 | 1.39 |
| 5 | -5 | 3 | 12.25 | 1.36 |
| 5 | -5 | 3 | 14.75 | 1.53 |
| 5 | -5 | 3 | 16.43 | 1.95 |
| 5 | -5 | -3 | 12.04 | 1.54 |
| -5 | 5 | 3 | 17.24 | 1.93 |
| -5 | -5 | 4 | 12.43 | 1.55 |
| -5 | -5 | 4 | 33.86 | 1.74 |
| -5 | -5 | -4 | 81.77 | 2.43 |
| -5 | -5 | -4 | 85.80 | 2.49 |
| 5 | -5 | 4 | 22.83 | 1.88 |
| 5 | -5 | 4 | 42.19 | 2.33 |
| 5 | -5 | -4 | 56.85 | 2.85 |
| -5 | 5 | 4 | 78.18 | 2.54 |
| 5 | 5 | 4 | 45.84 | 2.20 |
| -5 | -5 | 5 | 8.76 | 1.46 |
| -5 | -5 | 5 | 11.50 | 1.25 |
| -5 | -5 | -5 | 24.92 | 1.85 |
| -5 | -5 | -5 | 27.02 | 2.20 |
| 5 | -5 | 5 | 5.48 | 1.06 |
| 5 | -5 | 5 | 16.33 | 1.58 |
| 5 | -5 | -5 | 14.98 | 1.57 |
| 5 | -5 | -5 | 20.78 | 2.45 |
| -5 | 5 | 5 | 18.87 | 2.13 |
| -5 | 5 | -5 | 27.34 | 1.82 |
| 5 | 5 | 5 | -3.09 | 1.70 |
| -5 | -5 | 6 | 3.00 | 1.28 |
| -5 | -5 | 6 | 3.82 | 1.17 |
| -5 | -5 | -6 | 4.06 | 1.16 |
| -5 | -5 | -6 | 8.25 | 1.27 |
| 5 | -5 | 6 | 4.43 | 2.14 |
| 5 | -5 | -6 | 5.88 | 1.39 |
| 5 | -5 | -6 | 1.75 | 1.50 |
| -5 | 5 | 6 | 7.14 | 1.53 |
| -5 | 5 | -6 | 6.85 | 1.14 |
| -5 | -5 | 7 | 5.64 | 1.27 |
| -5 | -5 | 7 | 6.21 | 1.38 |
| -5 | -5 | -7 | 18.90 | 2.18 |
| -5 | -5 | -7 | 14.55 | 1.57 |
| 5 | -5 | 7 | 4.32 | 2.04 |
| 5 | -5 | -7 | 8.29 | 1.55 |
| 5 | -5 | -7 | 7.92 | 1.71 |
| -5 | 5 | 7 | 11.53 | 1.78 |
| -5 | 5 | -7 | 15.44 | 1.52 |
| -5 | -5 | 8 | 17.57 | 2.16 |
| -5 | -5 | 8 | 9.35 | 1.62 |
| -5 | -5 | -8 | 35.13 | 2.80 |
| -5 | -5 | -8 | 52.30 | 2.99 |
| 5 | -5 | 8 | 9.89 | 2.57 |
| 5 | -5 | -8 | 19.49 | 1.96 |
| 5 | -5 | -8 | 24.53 | 3.04 |
| -5 | 5 | 8 | 41.74 | 3.01 |
| -5 | 5 | -8 | 37.55 | 2.74 |
| -5 | -5 | 9 | 3.86 | 1.74 |
| -5 | -5 | -9 | 15.95 | 2.03 |
| -5 | -5 | -9 | 20.62 | 2.21 |
| 5 | -5 | 9 | -0.46 | 3.24 |
| 5 | -5 | -9 | 7.92 | 2.06 |
| -5 | 5 | 9 | 7.48 | 2.26 |
| -5 | 5 | -9 | 12.74 | 1.91 |
| -5 | -5 | 10 | 4.77 | 2.02 |
| -5 | -5 | -10 | 35.63 | 3.51 |
| -5 | -5 | -10 | 21.23 | 2.57 |
| 5 | -5 | -10 | 15.60 | 2.35 |
| -5 | 5 | 10 | 14.27 | 2.57 |
| -5 | 5 | -10 | 14.74 | 2.23 |
| -5 | -5 | 11 | 8.22 | 2.60 |
| -5 | -5 | -11 | 20.04 | 2.56 |
| -5 | -5 | -11 | 29.07 | 3.09 |
| 5 | -5 | -11 | 23.44 | 2.97 |
| -5 | 5 | 11 | 17.91 | 3.01 |
| -5 | 5 | -11 | 10.92 | 2.75 |
| -5 | 5 | -11 | 15.37 | 2.71 |
| -5 | -5 | 12 | 5.23 | 3.64 |
| -5 | -5 | -12 | 6.48 | 2.80 |
| -5 | -5 | -12 | 8.24 | 3.02 |
| 5 | -5 | -12 | -2.23 | 2.90 |
| -5 | 5 | 12 | 10.73 | 3.74 |
| -5 | 5 | -12 | -1.26 | 3.11 |
| -7 | -5 | 0 | 7.43 | 2.28 |
| -7 | -5 | 0 | 4.03 | 1.11 |
| 7 | -5 | 0 | -0.06 | 1.19 |
| -7 | 5 | 0 | 6.76 | 1.33 |
| -7 | -5 | 1 | 5.78 | 1.24 |
| -7 | -5 | 1 | 4.28 | 1.29 |
| -7 | -5 | -1 | 7.48 | 1.30 |
| -7 | -5 | -1 | 7.25 | 1.17 |
| 7 | -5 | 1 | 1.37 | 1.13 |
| 7 | -5 | -1 | 2.56 | 1.36 |
| -7 | 5 | 1 | 12.27 | 1.48 |
| -7 | 5 | -1 | 12.46 | 1.54 |
| -7 | -5 | 2 | 18.23 | 2.17 |

| -7 | -5 | 2 | 19.73 | 1.86 |
|----|----|----|--------|------|
| -7 | -5 | -2 | 55.31 | 2.26 |
| -7 | -5 | -2 | 55.37 | 2.46 |
| 7 | -5 | 2 | 19.86 | 2.14 |
| 7 | -5 | -2 | 29.05 | 2.57 |
| -7 | 5 | 2 | 110.43 | 3.32 |
| -7 | -5 | 3 | 0.71 | 1.34 |
| -7 | -5 | 3 | 0.15 | 1.11 |
| -7 | -5 | -3 | 6.61 | 1.13 |
| 7 | -5 | 3 | 2.51 | 1.22 |
| 7 | -5 | -3 | 2.16 | 1.37 |
| -7 | 5 | 3 | 11.67 | 1.66 |
| -7 | -5 | 4 | 8.48 | 1.52 |
| -7 | -5 | 4 | 17.28 | 1.84 |
| -7 | -5 | -4 | 38.75 | 2.28 |
| -7 | -5 | -4 | 44.24 | 3.00 |
| 7 | -5 | 4 | 8.30 | 1.44 |
| 7 | -5 | -4 | 17.96 | 2.09 |
| -7 | 5 | 4 | 43.02 | 2.77 |
| -7 | 5 | -4 | 47.84 | 2.72 |
| -7 | -5 | 5 | 5.28 | 1.51 |
| -7 | -5 | 5 | 5.05 | 1.37 |
| -7 | -5 | -5 | 7.62 | 1.39 |
| -7 | -5 | -5 | 7.99 | 1.42 |
| 7 | -5 | 5 | 2.45 | 1.33 |
| 7 | -5 | -5 | 6.27 | 1.64 |
| -7 | 5 | 5 | 12.19 | 1.92 |
| -7 | 5 | -5 | 10.72 | 1.47 |
| -7 | -5 | 6 | 6.77 | 1.57 |
| -7 | -5 | 6 | 0.64 | 1.46 |
| -7 | -5 | -6 | 1.20 | 1.32 |
| -7 | -5 | -6 | 6.41 | 2.06 |
| 7 | -5 | -6 | 3.16 | 1.76 |
| -7 | 5 | 6 | 2.95 | 1.82 |
| -7 | 5 | -6 | 4.46 | 1.41 |
| -7 | -5 | 7 | 5.95 | 1.64 |
| -7 | -5 | -7 | 7.05 | 1.74 |
| -7 | -5 | -7 | 10.46 | 1.92 |
| 7 | -5 | -7 | -0.43 | 1.80 |
| -7 | 5 | 7 | 6.34 | 2.15 |
| -7 | 5 | -7 | 8.33 | 1.79 |
| -7 | -5 | 8 | 8.99 | 1.80 |
| -7 | -5 | -8 | 13.01 | 2.46 |
| -7 | -5 | -8 | 23.90 | 2.47 |
| 7 | -5 | -8 | 11.86 | 2.42 |
| -7 | 5 | -8 | 28.87 | 3.58 |
| -7 | -5 | 9 | 2.79 | 2.23 |
| -7 | -5 | -9 | 3.57 | 2.24 |
| -7 | -5 | -9 | 5.95 | 2.54 |
| 7 | -5 | -9 | 2.40 | 2.38 |
| -7 | 5 | 9 | 6.26 | 2.65 |
| -7 | 5 | -9 | 4.18 | 2.24 |
| -7 | -5 | 10 | 11.53 | 2.74 |
| -7 | -5 | -10 | 12.94 | 2.72 |
| -7 | -5 | -10 | 13.60 | 3.16 |
| 7 | -5 | -10 | 7.45 | 2.65 |
| -7 | 5 | 10 | 14.82 | 3.09 |
| -7 | 5 | -10 | 15.12 | 2.63 |
| -7 | -5 | 11 | 6.82 | 3.54 |
| -7 | -5 | -11 | 17.53 | 3.10 |
| -7 | -5 | -11 | 13.02 | 3.78 |
| 7 | -5 | -11 | 9.05 | 3.35 |
| -7 | 5 | 11 | 20.56 | 6.01 |
| -7 | 5 | -11 | 5.69 | 3.70 |
| -7 | 5 | -11 | 11.54 | 3.18 |
| -9 | -5 | 0 | 79.24 | 3.20 |
| -9 | -5 | 0 | 55.56 | 3.03 |
| 9 | -5 | 0 | 31.63 | 3.02 |
| -9 | 5 | 0 | 110.95 | 3.82 |
| -9 | -5 | 1 | 3.34 | 1.66 |
| -9 | -5 | -1 | 3.12 | 1.62 |
| -9 | -5 | -1 | 2.97 | 1.51 |
| 9 | -5 | 1 | -2.21 | 1.54 |
| 9 | -5 | -1 | 1.14 | 1.50 |
| -9 | 5 | 1 | 2.78 | 1.77 |
| -9 | -5 | 2 | 2.58 | 1.75 |
| -9 | -5 | -2 | 4.79 | 1.59 |
| -9 | -5 | -2 | 5.80 | 1.54 |
| 9 | -5 | 2 | 3.70 | 1.84 |
| 9 | -5 | -2 | 2.69 | 1.80 |
| -9 | 5 | 2 | 4.76 | 1.78 |
| -9 | -5 | 3 | 0.14 | 1.84 |
| -9 | -5 | -3 | 6.08 | 1.60 |
| -9 | -5 | -3 | 0.84 | 1.43 |
| 9 | -5 | 3 | 0.35 | 1.75 |
| 9 | -5 | -3 | 7.98 | 1.93 |
| -9 | 5 | 3 | 11.24 | 1.98 |
| -9 | -5 | 4 | -1.77 | 2.03 |
| -9 | -5 | -4 | 1.00 | 1.68 |
| 9 | -5 | 4 | -2.79 | 1.86 |
| 9 | -5 | -4 | -0.40 | 1.84 |
| -9 | 5 | 4 | 10.55 | 2.20 |
| -9 | 5 | -4 | 4.17 | 2.03 |
| -9 | -5 | -5 | 1.79 | 1.81 |
| 9 | -5 | 5 | 0.87 | 2.03 |
| 9 | -5 | -5 | -2.59 | 2.05 |
| -9 | 5 | 5 | 0.79 | 2.38 |
| -9 | 5 | -5 | -2.62 | 1.92 |
| -9 | -5 | -6 | 65.44 | 6.03 |
| 9 | -5 | -6 | 17.99 | 3.11 |
| -9 | 5 | 6 | 46.60 | 5.45 |
| -9 | 5 | -6 | 48.00 | 4.47 |
| -9 | -5 | -7 | -4.04 | 2.09 |
| 9 | -5 | -7 | -1.45 | 2.35 |
| -9 | 5 | 7 | 8.73 | 2.73 |
| -9 | 5 | -7 | 0.93 | 2.45 |
| -9 | -5 | 8 | 4.00 | 2.88 |
| -9 | -5 | -8 | 2.38 | 2.63 |
| 9 | -5 | -8 | -3.97 | 2.49 |
| -9 | 5 | 8 | -3.89 | 3.00 |
| -9 | 5 | -8 | -0.77 | 2.63 |
| -9 | -5 | 9 | -1.88 | 3.40 |
| -9 | -5 | -9 | -3.00 | 2.78 |
| 9 | -5 | -9 | 0.24 | 2.84 |
| -9 | 5 | 9 | -0.54 | 3.25 |
| -9 | 5 | -9 | 3.93 | 3.10 |
| -11 | -5 | 0 | -1.10 | 2.31 |
| 11 | -5 | 0 | 4.21 | 2.42 |
| -11 | -5 | 1 | 4.12 | 2.56 |
| -11 | -5 | -1 | 16.92 | 3.97 |
| 11 | -5 | 1 | 3.44 | 2.49 |
| 11 | -5 | -1 | 11.35 | 2.55 |
| -11 | -5 | -2 | 16.65 | 3.03 |
| -11 | -5 | -2 | 14.75 | 2.41 |
| 11 | -5 | 2 | 10.00 | 2.79 |
| 11 | -5 | -2 | 9.47 | 2.65 |
| -11 | 5 | 2 | 19.96 | 5.73 |
| -11 | -5 | -3 | 7.05 | 3.07 |
| -11 | -5 | -3 | 8.85 | 2.43 |
| 11 | -5 | 3 | 2.99 | 2.69 |
| 11 | -5 | -3 | 11.73 | 2.77 |
| -11 | 5 | 3 | 3.57 | 3.08 |
| -11 | 5 | -3 | 20.30 | 3.27 |
| -11 | -5 | -4 | 10.69 | 2.88 |
| -11 | -5 | -4 | 21.68 | 2.67 |
| 11 | -5 | 4 | 5.87 | 2.92 |
| 11 | -5 | -4 | 8.75 | 2.70 |
| -11 | 5 | 4 | 32.64 | 3.13 |
| -11 | 5 | -4 | 17.48 | 3.30 |
| -11 | -5 | -5 | 10.84 | 3.04 |
| -11 | -5 | -5 | 6.47 | 2.38 |
| 11 | -5 | -5 | 7.45 | 2.74 |
| -11 | 5 | 5 | 9.22 | 3.17 |
| -11 | 5 | -5 | 11.09 | 3.23 |
| -11 | -5 | -6 | -6.76 | 3.02 |
| -11 | -5 | -6 | 2.06 | 2.39 |
| 11 | -5 | -6 | 1.58 | 2.65 |
| -11 | 5 | 6 | 4.32 | 3.15 |
| -11 | 5 | -6 | -6.11 | 3.25 |
| -11 | -5 | -7 | 4.54 | 3.46 |
| -11 | -5 | -7 | 9.82 | 2.84 |
| 11 | -5 | -7 | -2.35 | 3.21 |
| -11 | 5 | 7 | 12.89 | 5.06 |
| -11 | 5 | -7 | 11.67 | 3.61 |
| 13 | -5 | 0 | 10.88 | 3.61 |
| 13 | -5 | 1 | -2.36 | 3.73 |
| 13 | -5 | -1 | -6.97 | 3.38 |
| -13 | 5 | -1 | 6.27 | 4.34 |
| 13 | -5 | 2 | 5.79 | 5.29 |
| 13 | -5 | -2 | -2.64 | 3.68 |
| -13 | 5 | -2 | -6.34 | 4.66 |
| 0 | -6 | 0 | 1178.09 | 9.14 |
| 0 | -6 | 0 | 1228.86 | 7.66 |
| 0 | -6 | 0 | 1387.09 | 10.68 |
| 0 | 6 | 0 | 3404.90 | 12.58 |
| 0 | -6 | 1 | 112.35 | 3.80 |
| 0 | -6 | 1 | 78.33 | 1.87 |
| 0 | -6 | 1 | 169.08 | 5.61 |
| 0 | -6 | -1 | 218.96 | 4.91 |
| 0 | -6 | -1 | 263.41 | 3.89 |
| 0 | -6 | -1 | 245.01 | 4.13 |
| 0 | 6 | -1 | 361.98 | 4.24 |
| 0 | -6 | 2 | 808.38 | 8.19 |
| 0 | -6 | -2 | 1596.05 | 10.97 |
| 0 | -6 | -2 | 1469.66 | 9.84 |
| 0 | 6 | 2 | 3164.70 | 12.44 |
| 0 | -6 | 3 | 2.97 | 1.14 |
| 0 | -6 | 3 | 6.79 | 1.35 |
| 0 | -6 | -3 | 10.10 | 1.30 |
| 0 | -6 | -3 | 41.36 | 2.19 |
| 0 | 6 | 3 | 4.57 | 2.16 |
| 0 | 6 | -3 | 8.66 | 0.81 |
| 0 | -6 | 4 | 383.12 | 5.78 |
| 0 | -6 | 4 | 495.22 | 5.80 |
| 0 | -6 | -4 | 1265.13 | 9.87 |
| 0 | -6 | -4 | 1166.95 | 9.08 |
| 0 | 6 | 4 | 1419.47 | 8.79 |
| 0 | 6 | -4 | 1057.21 | 4.78 |
| 0 | -6 | 5 | 67.47 | 2.70 |
| 0 | -6 | 5 | 109.56 | 3.06 |
| 0 | -6 | -5 | 243.50 | 4.54 |
| 0 | -6 | -5 | 237.05 | 4.09 |
| 0 | 6 | 5 | 281.30 | 4.26 |
| 0 | 6 | -5 | 187.02 | 2.23 |
| 0 | -6 | 6 | 360.83 | 5.61 |
| 0 | -6 | 6 | 485.97 | 5.92 |
| 0 | -6 | -6 | 1049.07 | 9.34 |
| 0 | -6 | -6 | 1014.54 | 7.78 |
| 0 | 6 | 6 | 1211.21 | 8.57 |
| 0 | 6 | -6 | 855.12 | 5.62 |
| 0 | -6 | 7 | 65.26 | 2.78 |
| 0 | -6 | 7 | 68.37 | 3.01 |
| 0 | -6 | -7 | 170.29 | 4.18 |
| 0 | -6 | -7 | 151.42 | 3.29 |
| 0 | 6 | 7 | 172.16 | 3.94 |
| 0 | 6 | -7 | 134.27 | 2.90 |
| 0 | -6 | 8 | 264.60 | 4.96 |
| 0 | -6 | 8 | 225.48 | 4.79 |
| 0 | -6 | -8 | 738.20 | 8.31 |
| 0 | -6 | -8 | 736.92 | 6.63 |
| 0 | 6 | 8 | 702.09 | 7.27 |
| 0 | 6 | -8 | 631.64 | 5.43 |
| 0 | -6 | 9 | 8.85 | 2.28 |
| 0 | -6 | -9 | 9.41 | 1.63 |
| 0 | -6 | -9 | 2.08 | 1.54 |
| 0 | 6 | 9 | 23.12 | 2.68 |
| 0 | 6 | -9 | 4.87 | 1.63 |
| 0 | -6 | 10 | 144.16 | 6.33 |
| 0 | -6 | -10 | 552.75 | 7.70 |
| 0 | 6 | 10 | 503.94 | 8.15 |
| 0 | -6 | 11 | 17.55 | 2.99 |
| 0 | -6 | -11 | 153.12 | 5.09 |
| 0 | 6 | 11 | 116.78 | 5.64 |
| 0 | -6 | 12 | 94.26 | 5.98 |
| 0 | -6 | -12 | 445.66 | 7.38 |
| 0 | -6 | -12 | 462.15 | 7.98 |
| 0 | 6 | 12 | 244.61 | 7.02 |
| -2 | -6 | 0 | 1032.42 | 10.59 |
| -2 | -6 | 0 | 766.31 | 7.74 |
| 2 | -6 | 0 | 483.47 | 6.41 |
| 2 | -6 | 0 | 328.16 | 4.46 |
| 2 | -6 | 0 | 640.14 | 6.52 |
| -2 | 6 | 0 | 1940.83 | 10.58 |
| 2 | 6 | 0 | 1182.22 | 6.49 |
| -2 | -6 | 1 | 369.45 | 4.81 |
| -2 | -6 | 1 | 407.23 | 5.89 |
| -2 | -6 | -1 | 660.52 | 6.91 |
| -2 | -6 | -1 | 610.68 | 6.75 |
| 2 | -6 | 1 | 128.88 | 2.55 |
| 2 | -6 | 1 | 335.53 | 6.03 |
| 2 | -6 | -1 | 406.73 | 6.00 |
| 2 | -6 | -1 | 343.23 | 4.86 |
| 2 | -6 | -1 | 460.82 | 5.42 |
| -2 | 6 | 1 | 1488.62 | 9.00 |
| -2 | 6 | -1 | 990.92 | 7.35 |
| 2 | 6 | 1 | 1157.14 | 6.68 |
| 2 | 6 | -1 | 699.36 | 4.88 |
| -2 | -6 | 2 | 165.02 | 3.10 |
| -2 | -6 | 2 | 313.53 | 5.30 |
| -2 | -6 | -2 | 858.64 | 7.65 |
| -2 | -6 | -2 | 726.08 | 7.15 |
| 2 | -6 | 2 | 180.43 | 2.63 |
| 2 | -6 | 2 | 329.32 | 5.19 |
| 2 | -6 | 2 | 271.08 | 4.35 |
| 2 | -6 | -2 | 524.20 | 6.74 |
| 2 | -6 | -2 | 515.33 | 7.62 |
| 2 | -6 | -2 | 615.79 | 9.05 |
| -2 | 6 | 2 | 1429.36 | 8.90 |
| -2 | 6 | -2 | 993.84 | 7.24 |
| 2 | 6 | 2 | 1584.71 | 8.15 |
| -2 | -6 | 3 | 25.75 | 2.10 |
| -2 | -6 | 3 | 30.31 | 2.05 |
| -2 | -6 | -3 | 212.56 | 4.03 |
| -2 | -6 | -3 | 113.68 | 3.02 |
| 2 | -6 | 3 | 25.27 | 1.13 |
| 2 | -6 | 3 | 63.72 | 2.58 |
| 2 | -6 | 3 | 39.23 | 2.87 |
| 2 | -6 | -3 | 83.66 | 3.11 |
| 2 | -6 | -3 | 90.58 | 3.20 |
| -2 | 6 | 3 | 152.93 | 3.30 |
| 2 | 6 | 3 | 181.69 | 3.16 |
| -2 | -6 | 4 | 131.79 | 3.62 |
| -2 | -6 | 4 | 221.62 | 3.82 |
| -2 | -6 | -4 | 643.61 | 6.76 |
| -2 | -6 | -4 | 634.94 | 6.47 |
| 2 | -6 | 4 | 257.77 | 4.69 |
| 2 | -6 | 4 | 274.19 | 4.74 |
| 2 | -6 | -4 | 483.25 | 6.49 |
| 2 | -6 | -4 | 508.12 | 6.63 |
| -2 | 6 | 4 | 747.80 | 6.59 |
| -2 | 6 | -4 | 678.13 | 6.30 |
| 2 | 6 | 4 | 636.69 | 5.67 |
| -2 | -6 | 5 | 92.18 | 3.10 |
| -2 | -6 | 5 | 165.45 | 3.39 |
| -2 | -6 | -5 | 443.59 | 5.78 |
| -2 | -6 | -5 | 543.28 | 5.65 |
| -2 | -6 | -5 | 472.57 | 5.58 |
| 2 | -6 | 5 | 190.07 | 4.13 |
| 2 | -6 | 5 | 191.25 | 4.06 |
| 2 | -6 | -5 | 367.08 | 5.75 |
| 2 | -6 | -5 | 376.65 | 5.93 |
| 2 | -6 | -5 | 371.93 | 5.29 |
| -2 | 6 | 5 | 494.13 | 5.55 |
| -2 | 6 | -5 | 465.04 | 4.37 |
| 2 | 6 | 5 | 519.70 | 5.76 |
| -2 | -6 | 6 | 158.29 | 3.80 |
| -2 | -6 | 6 | 249.15 | 4.21 |
| -2 | -6 | -6 | 605.26 | 6.97 |
| -2 | -6 | -6 | 719.29 | 6.76 |
| -2 | -6 | -6 | 607.49 | 6.35 |
| 2 | -6 | 6 | 264.23 | 4.95 |
| 2 | -6 | 6 | 306.61 | 5.06 |
| 2 | -6 | -6 | 469.92 | 6.56 |
| 2 | -6 | -6 | 478.06 | 6.83 |
| 2 | -6 | -6 | 484.19 | 5.08 |
| -2 | 6 | 6 | 660.76 | 6.58 |
| -2 | 6 | -6 | 639.10 | 5.23 |
| 2 | 6 | 6 | 715.39 | 6.63 |
| -2 | -6 | 7 | 90.99 | 3.10 |
| -2 | -6 | 7 | 118.68 | 3.34 |
| -2 | -6 | -7 | 310.74 | 5.30 |
| -2 | -6 | -7 | 322.74 | 4.97 |
| -2 | -6 | -7 | 278.56 | 4.55 |
| 2 | -6 | 7 | 115.06 | 3.59 |
| 2 | -6 | 7 | 108.15 | 3.80 |
| 2 | -6 | -7 | 230.89 | 4.87 |
| 2 | -6 | -7 | 250.54 | 5.33 |
| -2 | 6 | 7 | 312.96 | 4.95 |
| -2 | 6 | -7 | 271.72 | 3.74 |
| 2 | 6 | 7 | 320.62 | 4.91 |
| -2 | -6 | 8 | 137.41 | 3.67 |
| -2 | -6 | 8 | 107.18 | 3.53 |
| -2 | -6 | -8 | 398.43 | 6.23 |
| -2 | -6 | -8 | 437.66 | 6.05 |
| -2 | -6 | -8 | 423.73 | 5.56 |
| 2 | -6 | 8 | 135.38 | 4.31 |
| 2 | -6 | -8 | 366.88 | 6.07 |
| 2 | -6 | -8 | 381.62 | 6.65 |
| -2 | 6 | 8 | 386.50 | 5.68 |
| -2 | 6 | -8 | 353.31 | 4.56 |
| 2 | 6 | 8 | 393.71 | 5.78 |
| -2 | -6 | 9 | 13.44 | 2.12 |
| -2 | -6 | -9 | 43.74 | 3.28 |
| -2 | -6 | -9 | 51.91 | 3.19 |
| -2 | -6 | -9 | 56.73 | 2.84 |
| 2 | -6 | 9 | 12.70 | 2.76 |
| 2 | -6 | -9 | 45.06 | 3.20 |
| 2 | -6 | -9 | 46.09 | 3.53 |
| -2 | 6 | 9 | 38.91 | 3.58 |
| -2 | 6 | -9 | 32.26 | 2.83 |
| 2 | 6 | 9 | 25.74 | 2.78 |
| -2 | -6 | 10 | 65.27 | 3.94 |
| -2 | -6 | -10 | 397.65 | 8.65 |
| -2 | -6 | -10 | 313.67 | 5.29 |
| 2 | -6 | 10 | 68.83 | 5.79 |
| 2 | -6 | -10 | 292.53 | 5.78 |
| 2 | -6 | -10 | 286.76 | 6.41 |
| -2 | 6 | 10 | 228.98 | 5.56 |
| -2 | 6 | -10 | 231.69 | 4.86 |
| 2 | 6 | 10 | 236.64 | 6.01 |
| -2 | -6 | 11 | 17.86 | 2.74 |
| -2 | -6 | -11 | 225.43 | 5.71 |
| -2 | -6 | -11 | 190.20 | 5.34 |
| 2 | -6 | 11 | 49.87 | 5.77 |
| 2 | -6 | -11 | 191.29 | 5.28 |
| 2 | -6 | -11 | 194.26 | 5.57 |
| -2 | 6 | 11 | 136.31 | 5.56 |
| -2 | 6 | -11 | 145.05 | 5.18 |
| -2 | -6 | 12 | 60.16 | 5.21 |
| -2 | -6 | -12 | 334.29 | 6.78 |
| 2 | -6 | 12 | 68.48 | 5.36 |
| 2 | -6 | -12 | 240.05 | 6.24 |
| -2 | 6 | 12 | 196.59 | 6.37 |
| -4 | -6 | 0 | 888.92 | 7.12 |
| -4 | -6 | 0 | 672.82 | 7.38 |
| 4 | -6 | 0 | 255.50 | 4.65 |
| 4 | -6 | 0 | 333.92 | 5.46 |
| -4 | 6 | 0 | 1547.35 | 9.66 |
| -4 | -6 | 1 | 354.46 | 4.51 |
| -4 | -6 | 1 | 320.60 | 5.27 |
| -4 | -6 | -1 | 537.89 | 5.95 |
| -4 | -6 | -1 | 429.43 | 5.77 |
| 4 | -6 | 1 | 110.73 | 3.23 |
| 4 | -6 | 1 | 212.59 | 3.90 |
| 4 | -6 | 1 | 225.99 | 4.21 |
| 4 | -6 | -1 | 189.53 | 4.21 |
| 4 | -6 | -1 | 201.55 | 3.24 |
| -4 | 6 | 1 | 997.80 | 7.83 |
| -4 | 6 | -1 | 713.89 | 6.62 |
| 4 | 6 | 1 | 404.52 | 3.48 |
| -4 | -6 | 2 | 213.27 | 3.29 |
| -4 | -6 | 2 | 297.81 | 5.19 |
| -4 | -6 | -2 | 759.00 | 6.93 |
| -4 | -6 | -2 | 665.26 | 7.02 |
| 4 | -6 | 2 | 126.00 | 3.35 |
| 4 | -6 | 2 | 280.59 | 4.48 |
| 4 | -6 | 2 | 338.01 | 6.64 |
| 4 | -6 | -2 | 372.23 | 5.76 |
| 4 | -6 | -2 | 279.92 | 4.36 |
| -4 | 6 | 2 | 1265.68 | 8.81 |
| -4 | 6 | -2 | 872.23 | 7.31 |
| 4 | 6 | 2 | 911.28 | 5.41 |
| -4 | -6 | 3 | 17.40 | 1.98 |
| -4 | -6 | 3 | 24.78 | 1.84 |
| -4 | -6 | -3 | 99.17 | 2.87 |
| -4 | -6 | -3 | 90.71 | 2.75 |
| 4 | -6 | 3 | 11.21 | 1.62 |
| 4 | -6 | 3 | 35.11 | 1.98 |
| 4 | -6 | 3 | 38.30 | 2.36 |
| 4 | -6 | -3 | 58.44 | 2.86 |
| 4 | -6 | -3 | 47.33 | 2.77 |
| 4 | -6 | -3 | 36.32 | 1.59 |
| -4 | 6 | 3 | 91.17 | 2.89 |
| 4 | 6 | 3 | 141.60 | 2.71 |
| -4 | -6 | 4 | 89.53 | 3.04 |
| -4 | -6 | 4 | 168.85 | 3.28 |
| -4 | -6 | -4 | 580.83 | 6.34 |
| -4 | -6 | -4 | 680.44 | 5.41 |
| -4 | -6 | -4 | 550.93 | 6.35 |
| 4 | -6 | 4 | 265.17 | 4.67 |
| 4 | -6 | 4 | 286.09 | 5.17 |
| 4 | -6 | -4 | 377.72 | 5.95 |
| 4 | -6 | -4 | 343.61 | 5.89 |
| 4 | -6 | -4 | 247.45 | 2.90 |
| -4 | 6 | 4 | 548.61 | 5.99 |
| 4 | 6 | 4 | 1013.65 | 6.53 |
| -4 | -6 | 5 | 63.13 | 2.73 |
| -4 | -6 | 5 | 108.52 | 2.78 |
| -4 | -6 | -5 | 358.09 | 5.23 |
| -4 | -6 | -5 | 416.04 | 4.85 |
| -4 | -6 | -5 | 347.50 | 5.13 |
| 4 | -6 | 5 | 45.23 | 2.18 |
| 4 | -6 | 5 | 170.55 | 3.99 |
| 4 | -6 | 5 | 181.88 | 4.25 |
| 4 | -6 | -5 | 232.18 | 4.82 |
| 4 | -6 | -5 | 217.39 | 4.92 |
| -4 | 6 | 5 | 373.95 | 5.14 |
| -4 | 6 | -5 | 382.85 | 4.63 |
| 4 | 6 | 5 | 270.82 | 4.04 |
| -4 | -6 | 6 | 112.46 | 3.30 |
| -4 | -6 | 6 | 193.14 | 3.68 |
| -4 | -6 | -6 | 538.58 | 6.60 |
| -4 | -6 | -6 | 617.62 | 6.05 |
| -4 | -6 | -6 | 538.88 | 6.45 |
| 4 | -6 | 6 | 236.20 | 4.89 |
| 4 | -6 | -6 | 336.76 | 5.70 |
| 4 | -6 | -6 | 337.79 | 6.16 |
| -4 | 6 | 6 | 563.43 | 6.34 |
| -4 | 6 | -6 | 567.95 | 5.75 |
| 4 | 6 | 6 | 84.06 | 5.32 |
| -4 | -6 | 7 | 58.40 | 2.79 |
| -4 | -6 | 7 | 58.16 | 2.75 |
| -4 | -6 | -7 | 240.17 | 4.79 |
| -4 | -6 | -7 | 272.66 | 4.58 |
| -4 | -6 | -7 | 231.58 | 4.49 |
| 4 | -6 | 7 | 88.41 | 4.07 |
| 4 | -6 | -7 | 143.90 | 4.06 |
| 4 | -6 | -7 | 141.36 | 4.43 |
| -4 | 6 | 7 | 212.50 | 4.40 |
| -4 | 6 | -7 | 253.95 | 4.19 |
| 4 | 6 | 7 | 2.96 | 2.29 |
| -4 | -6 | 8 | 100.31 | 3.49 |
| -4 | -6 | 8 | 83.35 | 3.36 |
| -4 | -6 | -8 | 351.45 | 6.02 |
| -4 | -6 | -8 | 395.63 | 5.75 |
| -4 | -6 | -8 | 372.51 | 5.69 |
| 4 | -6 | 8 | 113.45 | 4.82 |
| 4 | -6 | -8 | 260.00 | 5.21 |
| 4 | -6 | -8 | 233.39 | 5.61 |
| -4 | 6 | 8 | 317.78 | 5.48 |
| -4 | 6 | -8 | 331.27 | 5.00 |
| -4 | -6 | 9 | 15.10 | 1.90 |
| -4 | -6 | 9 | 6.07 | 2.36 |
| -4 | -6 | -9 | 35.40 | 3.31 |
| -4 | -6 | -9 | 42.24 | 3.38 |
| -4 | -6 | -9 | 34.39 | 2.91 |
| 4 | -6 | 9 | 10.14 | 3.15 |
| 4 | -6 | -9 | 34.10 | 3.61 |
| 4 | -6 | -9 | 19.42 | 2.38 |
| -4 | 6 | 9 | 30.45 | 4.03 |
| -4 | 6 | -9 | 37.39 | 3.21 |
| -4 | -6 | 10 | 32.41 | 3.58 |
| -4 | -6 | -10 | 267.56 | 6.02 |
| -4 | -6 | -10 | 276.68 | 5.79 |
| -4 | -6 | -10 | 278.16 | 5.36 |
| 4 | -6 | 10 | 67.12 | 6.47 |
| 4 | -6 | -10 | 186.74 | 6.21 |
| 4 | -6 | -10 | 190.67 | 5.51 |
| -4 | 6 | 10 | 186.35 | 5.32 |
| -4 | 6 | -10 | 215.68 | 5.18 |
| -4 | -6 | 11 | 37.21 | 3.09 |
| -4 | -6 | -11 | 178.14 | 5.51 |
| -4 | -6 | -11 | 166.29 | 5.37 |
| -4 | -6 | -11 | 149.31 | 4.77 |
| 4 | -6 | -11 | 101.21 | 4.86 |
| -4 | 6 | 11 | 105.90 | 4.98 |
| -4 | 6 | -11 | 118.84 | 5.39 |
| -4 | -6 | 12 | 50.11 | 5.36 |
| -4 | -6 | -12 | 334.50 | 10.25 |
| -4 | -6 | -12 | 312.33 | 10.47 |
| 4 | -6 | -12 | 226.71 | 9.00 |
| -6 | -6 | 0 | 1262.37 | 8.68 |
| -6 | -6 | 0 | 958.17 | 9.08 |
| 6 | -6 | 0 | 366.83 | 5.72 |
| 6 | -6 | 0 | 547.61 | 5.26 |
| 6 | -6 | 0 | 603.23 | 5.87 |
| -6 | 6 | 0 | 2052.47 | 11.88 |
| -6 | -6 | 1 | 158.03 | 3.49 |
| -6 | -6 | 1 | 113.01 | 3.40 |
| -6 | -6 | -1 | 187.41 | 3.71 |
| -6 | -6 | -1 | 150.70 | 3.78 |
| 6 | -6 | 1 | 38.27 | 2.37 |
| 6 | -6 | 1 | 70.90 | 2.94 |
| 6 | -6 | -1 | 58.80 | 2.99 |
| 6 | -6 | -1 | 83.77 | 3.66 |
| -6 | 6 | 1 | 307.79 | 4.93 |
| -6 | 6 | -1 | 227.45 | 4.36 |
| -6 | -6 | 2 | 437.92 | 6.31 |
| -6 | -6 | -2 | 1097.03 | 8.29 |
| -6 | -6 | -2 | 921.96 | 8.76 |
| 6 | -6 | 2 | 198.99 | 3.98 |
| 6 | -6 | 2 | 428.47 | 5.02 |
| 6 | -6 | -2 | 451.51 | 6.77 |
| -6 | 6 | 2 | 2109.27 | 12.10 |
| -6 | 6 | -2 | 1279.85 | 9.53 |
| -6 | -6 | 3 | 4.95 | 1.34 |
| -6 | -6 | 3 | 4.38 | 1.38 |
| -6 | -6 | -3 | 12.41 | 1.44 |
| -6 | -6 | -3 | 23.64 | 2.01 |
| 6 | -6 | 3 | 3.82 | 1.28 |
| 6 | -6 | 3 | 7.69 | 1.80 |
| 6 | -6 | -3 | 2.48 | 1.42 |
| -6 | 6 | 3 | 16.97 | 1.89 |
| -6 | -6 | 4 | 118.35 | 3.51 |
| -6 | -6 | 4 | 237.78 | 3.60 |
| -6 | -6 | -4 | 851.25 | 7.83 |
| -6 | -6 | -4 | 925.22 | 5.94 |
| -6 | -6 | -4 | 774.17 | 8.09 |
| 6 | -6 | 4 | 125.46 | 3.13 |
| 6 | -6 | 4 | 126.70 | 4.38 |
| 6 | -6 | -4 | 419.62 | 6.80 |
| -6 | 6 | 4 | 956.68 | 8.30 |
| -6 | -6 | 5 | 15.67 | 1.82 |
| -6 | -6 | 5 | 41.66 | 2.39 |
| -6 | -6 | -5 | 152.33 | 3.83 |
| -6 | -6 | -5 | 157.69 | 3.46 |
| -6 | -6 | -5 | 141.27 | 3.78 |
| 6 | -6 | 5 | 21.95 | 2.23 |
| 6 | -6 | 5 | 3.55 | 2.63 |
| 6 | -6 | -5 | 72.03 | 3.60 |
| -6 | 6 | 5 | 137.28 | 3.81 |
| -6 | 6 | -5 | 172.55 | 3.89 |
| -6 | -6 | 6 | 128.48 | 3.76 |
| -6 | -6 | 6 | 157.76 | 3.43 |
| -6 | -6 | -6 | 697.35 | 7.76 |
| -6 | -6 | -6 | 815.28 | 6.90 |
| 6 | -6 | -6 | 380.20 | 6.78 |
| -6 | 6 | 6 | 655.25 | 7.22 |
| -6 | 6 | -6 | 797.58 | 7.85 |
| -6 | -6 | 7 | 24.59 | 2.81 |
| -6 | -6 | 7 | 27.12 | 2.84 |
| -6 | -6 | -7 | 106.10 | 3.87 |
| -6 | -6 | -7 | 108.47 | 3.80 |
| 6 | -6 | -7 | 49.37 | 3.56 |
| -6 | 6 | 7 | 110.47 | 3.99 |
| -6 | 6 | -7 | 100.57 | 3.73 |
| -6 | -6 | 8 | 139.82 | 4.92 |
| -6 | -6 | -8 | 437.90 | 9.62 |
| -6 | -6 | -8 | 513.55 | 6.70 |
| -6 | 6 | 8 | 461.34 | 7.92 |
| -6 | 6 | -8 | 465.81 | 6.68 |
| -6 | -6 | 9 | 5.24 | 2.12 |
| -6 | -6 | -9 | 8.50 | 2.12 |
| -6 | -6 | -9 | 20.06 | 2.73 |
| -6 | -6 | -9 | 10.22 | 1.98 |
| 6 | -6 | -9 | 5.04 | 2.15 |
| -6 | 6 | 9 | 7.81 | 2.67 |
| -6 | 6 | -9 | 4.80 | 2.32 |
| -6 | -6 | 10 | 62.52 | 4.05 |
| -6 | -6 | -10 | 303.29 | 6.52 |
| -6 | -6 | -10 | 356.32 | 6.82 |
| -6 | -6 | -10 | 336.41 | 6.39 |
| 6 | -6 | -10 | 184.25 | 5.51 |
| -6 | 6 | 10 | 280.34 | 5.89 |
| -6 | 6 | -10 | 287.44 | 6.23 |
| -6 | -6 | 11 | 23.00 | 3.41 |
| -6 | -6 | -11 | 80.45 | 5.51 |
| -6 | -6 | -11 | 79.36 | 8.97 |
| -6 | -6 | -11 | 85.23 | 4.98 |
| 6 | -6 | -11 | 41.45 | 3.79 |
| -6 | 6 | 11 | 90.96 | 8.29 |
| -6 | 6 | -11 | 52.89 | 5.82 |
| -8 | -6 | 0 | 386.36 | 5.18 |
| -8 | -6 | 0 | 316.22 | 5.65 |
| 8 | -6 | 0 | 127.69 | 4.15 |
| -8 | 6 | 0 | 596.21 | 7.00 |
| -8 | -6 | 1 | 164.84 | 4.45 |
| -8 | -6 | -1 | 295.87 | 4.75 |
| -8 | -6 | -1 | 223.21 | 4.84 |
| 8 | -6 | 1 | 70.32 | 3.40 |
| 8 | -6 | -1 | 92.99 | 3.78 |
| -8 | 6 | 1 | 470.06 | 6.31 |
| -8 | 6 | -1 | 352.68 | 5.76 |
| -8 | -6 | 2 | 151.94 | 4.23 |
| -8 | -6 | 2 | 65.54 | 2.79 |
| -8 | -6 | -2 | 353.01 | 5.22 |
| -8 | -6 | -2 | 284.62 | 5.41 |
| 8 | -6 | 2 | 74.46 | 3.45 |
| 8 | -6 | -2 | 137.63 | 4.45 |
| -8 | 6 | 2 | 593.69 | 6.99 |
| -8 | -6 | 3 | 13.64 | 2.02 |
| -8 | -6 | 3 | 21.04 | 2.56 |
| -8 | -6 | -3 | 96.03 | 3.46 |
| -8 | -6 | -3 | 69.75 | 3.18 |
| 8 | -6 | 3 | 8.18 | 1.84 |
| 8 | -6 | -3 | 37.28 | 3.32 |
| -8 | 6 | 3 | 154.86 | 4.35 |
| -8 | -6 | 4 | 41.14 | 3.28 |
| -8 | -6 | 4 | 83.64 | 2.88 |
| -8 | -6 | -4 | 295.62 | 5.23 |
| -8 | -6 | -4 | 259.98 | 5.32 |
| 8 | -6 | 4 | 39.48 | 3.08 |
| 8 | -6 | -4 | 120.84 | 4.48 |
| -8 | 6 | 4 | 599.89 | 7.20 |
| -8 | 6 | -4 | 355.42 | 6.23 |
| -8 | -6 | 5 | 21.52 | 2.28 |
| -8 | -6 | 5 | 46.63 | 2.88 |
| -8 | -6 | -5 | 208.36 | 4.86 |
| -8 | -6 | -5 | 196.01 | 4.80 |
| 8 | -6 | 5 | 26.00 | 3.19 |
| 8 | -6 | -5 | 97.42 | 4.25 |
| -8 | 6 | 5 | 345.82 | 5.81 |
| -8 | 6 | -5 | 251.48 | 5.49 |
| -8 | -6 | 6 | 68.55 | 3.23 |
| -8 | -6 | -6 | 271.26 | 5.69 |
| -8 | 6 | 6 | 296.16 | 8.36 |
| -8 | 6 | -6 | 308.96 | 7.74 |
| -8 | -6 | 7 | 16.74 | 2.45 |
| -8 | -6 | -7 | 154.83 | 4.91 |
| -8 | -6 | -7 | 137.87 | 4.45 |
| 8 | -6 | -7 | 73.28 | 4.30 |
| -8 | 6 | 7 | 152.58 | 4.97 |
| -8 | 6 | -7 | 155.98 | 5.04 |
| -8 | -6 | 8 | 38.68 | 3.60 |
| -8 | -6 | -8 | 177.46 | 5.44 |
| -8 | -6 | -8 | 185.51 | 5.21 |
| 8 | -6 | -8 | 87.32 | 4.61 |
| -8 | 6 | 8 | 150.55 | 5.37 |
| -8 | 6 | -8 | 188.88 | 5.65 |
| -8 | -6 | 9 | 4.03 | 2.75 |
| -8 | -6 | -9 | 44.82 | 4.55 |
| -8 | -6 | -9 | 36.37 | 3.80 |
| 8 | -6 | -9 | 9.99 | 2.66 |
| -8 | 6 | 9 | 29.44 | 3.30 |
| -8 | 6 | -9 | 24.12 | 3.11 |
| -10 | -6 | 0 | 169.40 | 4.93 |
| -10 | -6 | 1 | 74.40 | 4.08 |
| -10 | -6 | -1 | 91.61 | 4.04 |
| 10 | -6 | -1 | 47.45 | 5.62 |
| -10 | 6 | 1 | 183.55 | 5.30 |
| -10 | -6 | 2 | 70.61 | 4.17 |
| -10 | -6 | -2 | 210.56 | 5.31 |
| 10 | -6 | 2 | 53.85 | 5.39 |
| 10 | -6 | -2 | 96.33 | 4.53 |
| -10 | 6 | 2 | 315.41 | 8.53 |
| -10 | -6 | 3 | 3.28 | 2.88 |
| -10 | -6 | -3 | 41.83 | 4.56 |
| -10 | -6 | -3 | 24.75 | 2.62 |
| 10 | -6 | 3 | 7.14 | 2.54 |
| 10 | -6 | -3 | 10.29 | 2.44 |
| -10 | 6 | 3 | 68.44 | 4.85 |
| -10 | 6 | -3 | 49.37 | 4.57 |
| -10 | -6 | -4 | 193.48 | 5.46 |
| -10 | -6 | -4 | 161.36 | 5.00 |
| 10 | -6 | 4 | 38.79 | 4.17 |
| 10 | -6 | -4 | 79.44 | 4.45 |
| -10 | 6 | 4 | 331.85 | 6.13 |
| -10 | 6 | -4 | 240.00 | 6.48 |
| -10 | -6 | -5 | 107.67 | 5.01 |
| -10 | -6 | -5 | 86.21 | 4.28 |
| 10 | -6 | -5 | 42.47 | 4.21 |
| -10 | 6 | 5 | 177.53 | 5.38 |
| -10 | 6 | -5 | 114.78 | 5.51 |
| -10 | -6 | -6 | 188.22 | 5.86 |
| -10 | -6 | -6 | 163.41 | 5.17 |
| 10 | -6 | -6 | 70.33 | 4.59 |
| -10 | 6 | 6 | 318.35 | 6.32 |
| -10 | 6 | -6 | 204.38 | 6.15 |
| -10 | -6 | -7 | 73.30 | 5.30 |
| -10 | -6 | -7 | 60.99 | 4.34 |
| 10 | -6 | -7 | 54.18 | 4.66 |
| -10 | 6 | 7 | 111.77 | 5.66 |
| -10 | 6 | -7 | 97.87 | 5.69 |
| 12 | -6 | 0 | 120.80 | 5.51 |
| -12 | -6 | -1 | 24.29 | 3.41 |
| 12 | -6 | 1 | 7.50 | 3.26 |
| 12 | -6 | -1 | 16.34 | 3.15 |
| -12 | -6 | -2 | 145.46 | 5.91 |
| 12 | -6 | 2 | 85.06 | 6.08 |
| 12 | -6 | -2 | 99.64 | 5.37 |
| -12 | 6 | -2 | 310.80 | 7.82 |
| -12 | -6 | -3 | -2.73 | 3.10 |
| 12 | -6 | 3 | 1.71 | 3.65 |
| 12 | -6 | -3 | -0.38 | 3.26 |
| -12 | 6 | -3 | 14.28 | 3.94 |
| 12 | -6 | -4 | 113.43 | 6.94 |
| -12 | 6 | -4 | 264.07 | 9.36 |
| -1 | -7 | 0 | 50.50 | 3.40 |
| -1 | -7 | 0 | 53.31 | 1.99 |
| -1 | -7 | 0 | 47.51 | 2.42 |
| 1 | -7 | 0 | 36.41 | 2.58 |
| 1 | -7 | 0 | 45.46 | 2.33 |
| -1 | 7 | 0 | 117.02 | 2.96 |
| 1 | 7 | 0 | 93.58 | 2.37 |
| -1 | -7 | 1 | 15.27 | 1.75 |
| -1 | -7 | 1 | 24.40 | 2.02 |
| -1 | -7 | -1 | 32.67 | 2.57 |
| -1 | -7 | -1 | 34.66 | 2.16 |
| 1 | -7 | 1 | 7.54 | 1.00 |
| 1 | -7 | 1 | 23.85 | 2.04 |
| 1 | -7 | -1 | 29.14 | 2.42 |
| 1 | -7 | -1 | 28.03 | 2.03 |
| -1 | 7 | 1 | 80.96 | 2.66 |
| -1 | 7 | -1 | 49.92 | 2.38 |
| 1 | 7 | 1 | 53.25 | 2.10 |
| 1 | 7 | -1 | 40.85 | 1.91 |
| -1 | -7 | 2 | 5.72 | 1.19 |
| -1 | -7 | -2 | 19.97 | 2.22 |
| -1 | -7 | -2 | 18.61 | 1.83 |
| 1 | -7 | 2 | 0.82 | 0.75 |
| 1 | -7 | 2 | 5.09 | 1.19 |
| 1 | -7 | -2 | 12.92 | 1.65 |
| 1 | -7 | -2 | 13.39 | 1.68 |
| 1 | -7 | -2 | 23.95 | 1.98 |
| -1 | 7 | 2 | 41.24 | 2.30 |
| -1 | 7 | -2 | 18.63 | 2.12 |
| 1 | 7 | 2 | 50.24 | 3.04 |
| -1 | -7 | 3 | 1.15 | 1.08 |
| -1 | -7 | 3 | -1.56 | 1.02 |
| -1 | -7 | -3 | 3.29 | 1.11 |
| -1 | -7 | -3 | 10.14 | 1.41 |
| 1 | -7 | 3 | 6.58 | 1.31 |
| 1 | -7 | 3 | 0.50 | 1.31 |
| 1 | -7 | -3 | 0.46 | 1.11 |
| 1 | -7 | -3 | 2.54 | 1.25 |
| 1 | -7 | -3 | 1.22 | 1.15 |
| -1 | 7 | 3 | 2.86 | 1.32 |
| 1 | 7 | 3 | 4.44 | 1.59 |
| -1 | -7 | 4 | 0.61 | 1.13 |
| -1 | -7 | 4 | 5.50 | 1.32 |
| -1 | -7 | -4 | 7.95 | 1.23 |
| -1 | -7 | -4 | 17.05 | 1.79 |
| 1 | -7 | 4 | 1.71 | 1.15 |
| 1 | -7 | 4 | 0.86 | 1.38 |
| 1 | -7 | -4 | 6.04 | 1.48 |
| 1 | -7 | -4 | 11.91 | 1.52 |
| 1 | -7 | -4 | 8.54 | 1.36 |
| -1 | 7 | 4 | 13.14 | 1.61 |
| -1 | 7 | -4 | 11.94 | 1.22 |
| 1 | 7 | 4 | 22.18 | 2.44 |
| -1 | -7 | 5 | 1.16 | 1.30 |
| -1 | -7 | 5 | 0.86 | 1.41 |
| -1 | -7 | -5 | 13.75 | 1.47 |
| -1 | -7 | -5 | 15.81 | 1.89 |
| 1 | -7 | 5 | 2.34 | 1.35 |
| 1 | -7 | 5 | 4.58 | 1.61 |
| 1 | -7 | -5 | 15.30 | 1.65 |
| 1 | -7 | -5 | 14.29 | 1.72 |
| 1 | -7 | -5 | 13.78 | 1.51 |
| -1 | 7 | 5 | 25.20 | 2.43 |
| -1 | 7 | -5 | 18.51 | 1.78 |
| 1 | 7 | 5 | 31.26 | 2.55 |
| -1 | -7 | 6 | 5.31 | 1.38 |
| -1 | -7 | 6 | 9.61 | 1.78 |
| -1 | -7 | -6 | 24.66 | 2.53 |
| -1 | -7 | -6 | 23.40 | 2.16 |
| 1 | -7 | 6 | 3.30 | 1.52 |
| 1 | -7 | 6 | 8.22 | 1.91 |
| 1 | -7 | -6 | 21.37 | 2.62 |
| 1 | -7 | -6 | 22.62 | 2.16 |
| -1 | 7 | 6 | 24.61 | 2.59 |
| -1 | 7 | -6 | 23.26 | 1.89 |
| 1 | 7 | 6 | 29.55 | 2.83 |
| -1 | -7 | 7 | 2.40 | 1.41 |
| -1 | -7 | 7 | 4.01 | 1.83 |
| -1 | -7 | -7 | 20.71 | 2.22 |
| -1 | -7 | -7 | 13.77 | 1.57 |
| 1 | -7 | 7 | -2.08 | 1.44 |
| 1 | -7 | 7 | 2.55 | 2.02 |
| 1 | -7 | -7 | 10.01 | 1.89 |
| 1 | -7 | -7 | 16.60 | 1.61 |
| -1 | 7 | 7 | 10.27 | 1.90 |
| -1 | 7 | -7 | 17.58 | 2.08 |
| 1 | 7 | 7 | 14.38 | 2.09 |
| -1 | -7 | 8 | 2.75 | 1.63 |
| -1 | -7 | 8 | 0.55 | 2.13 |
| -1 | -7 | -8 | 23.03 | 2.87 |
| -1 | -7 | -8 | 22.24 | 2.58 |
| 1 | -7 | 8 | 3.42 | 1.64 |
| 1 | -7 | 8 | 2.03 | 2.39 |
| 1 | -7 | -8 | 16.27 | 2.09 |
| 1 | -7 | -8 | 22.28 | 2.36 |
| -1 | 7 | 8 | 19.96 | 2.27 |
| -1 | 7 | -8 | 21.54 | 1.71 |
| 1 | 7 | 8 | 17.54 | 2.41 |
| -1 | -7 | 9 | 2.39 | 1.95 |
| -1 | -7 | -9 | 2.66 | 1.65 |
| -1 | -7 | -9 | 2.50 | 1.89 |
| 1 | -7 | 9 | 5.58 | 2.78 |
| 1 | -7 | -9 | 6.63 | 1.83 |
| 1 | -7 | -9 | 8.41 | 2.48 |
| -1 | 7 | 9 | 5.13 | 2.61 |
| 1 | 7 | 9 | -0.17 | 2.59 |
| -1 | -7 | 10 | 9.04 | 2.96 |
| -1 | -7 | -10 | 19.28 | 2.29 |
| 1 | -7 | 10 | 8.37 | 3.31 |
| 1 | -7 | -10 | 14.50 | 2.15 |
| 1 | -7 | -10 | 19.97 | 2.59 |
| -1 | 7 | 10 | 12.88 | 2.92 |
| 1 | 7 | 10 | -2.12 | 3.21 |
| -1 | -7 | 11 | -3.14 | 3.05 |
| -1 | -7 | -11 | 1.85 | 2.17 |
| 1 | -7 | 11 | 11.24 | 3.49 |
| 1 | -7 | -11 | 4.63 | 2.14 |
| 1 | -7 | -11 | 3.00 | 2.41 |
| -1 | 7 | 11 | -5.56 | 3.15 |
| 1 | 7 | 11 | -9.40 | 3.84 |
| -3 | -7 | 0 | 7.44 | 1.49 |
| -3 | -7 | 0 | 7.52 | 1.34 |
| 3 | -7 | 0 | 1.41 | 1.25 |
| 3 | -7 | 0 | 3.69 | 1.11 |
| -3 | 7 | 0 | 15.82 | 1.77 |
| 3 | 7 | 0 | 4.20 | 1.14 |
| -3 | -7 | 1 | 4.05 | 1.55 |
| -3 | -7 | 1 | 4.47 | 1.25 |
| -3 | -7 | -1 | 7.19 | 1.31 |
| -3 | -7 | -1 | 5.79 | 1.13 |
| 3 | -7 | 1 | 0.29 | 1.34 |
| 3 | -7 | 1 | 1.58 | 1.01 |
| 3 | -7 | -1 | 2.39 | 1.31 |
| 3 | -7 | -1 | 4.00 | 1.19 |
| -3 | 7 | 1 | 15.46 | 1.87 |
| -3 | 7 | -1 | 8.74 | 1.39 |
| 3 | 7 | 1 | 9.45 | 1.21 |
| -3 | -7 | 2 | 47.46 | 2.29 |
| -3 | -7 | 2 | 82.78 | 3.14 |
| -3 | -7 | -2 | 182.64 | 3.99 |
| -3 | -7 | -2 | 171.89 | 3.92 |
| 3 | -7 | 2 | 29.22 | 2.98 |
| 3 | -7 | 2 | 58.32 | 2.53 |
| 3 | -7 | 2 | 54.03 | 3.38 |
| 3 | -7 | -2 | 96.80 | 3.50 |
| 3 | -7 | -2 | 95.56 | 3.28 |
| 3 | -7 | -2 | 91.32 | 2.90 |
| -3 | 7 | 2 | 369.41 | 5.09 |
| -3 | 7 | -2 | 226.98 | 4.09 |
| 3 | 7 | 2 | 273.03 | 3.63 |
| -3 | -7 | 3 | 19.67 | 2.12 |
| -3 | -7 | 3 | 12.12 | 1.46 |
| -3 | -7 | -3 | 66.84 | 2.81 |
| -3 | -7 | -3 | 70.38 | 2.69 |
| 3 | -7 | 3 | 6.65 | 0.93 |
| 3 | -7 | 3 | 20.16 | 1.97 |
| 3 | -7 | 3 | 16.16 | 2.02 |
| 3 | -7 | -3 | 34.90 | 2.66 |
| 3 | -7 | -3 | 35.55 | 2.54 |
| 3 | -7 | -3 | 34.19 | 2.37 |
| -3 | 7 | 3 | 110.51 | 3.23 |
| -3 | -7 | 4 | 28.49 | 2.43 |
| -3 | -7 | 4 | 22.24 | 2.27 |
| -3 | -7 | -4 | 123.81 | 3.52 |
| -3 | -7 | -4 | 122.89 | 3.35 |
| 3 | -7 | 4 | 20.31 | 2.28 |
| 3 | -7 | 4 | 39.53 | 2.45 |
| 3 | -7 | 4 | 37.02 | 3.05 |
| 3 | -7 | -4 | 82.95 | 3.34 |
| 3 | -7 | -4 | 80.26 | 3.24 |
| 3 | -7 | -4 | 77.18 | 2.91 |
| -3 | 7 | 4 | 152.74 | 3.66 |
| -3 | 7 | -4 | 140.73 | 2.86 |
| -3 | -7 | 5 | 4.84 | 1.52 |
| -3 | -7 | -5 | 4.45 | 1.41 |
| -3 | -7 | -5 | 3.11 | 1.21 |
| 3 | -7 | 5 | 2.56 | 1.32 |
| 3 | -7 | 5 | 5.05 | 1.86 |
| 3 | -7 | -5 | -0.71 | 1.34 |
| 3 | -7 | -5 | 3.92 | 1.37 |
| 3 | -7 | -5 | 3.52 | 1.47 |
| -3 | 7 | 5 | 11.38 | 1.76 |
| -3 | 7 | -5 | 2.79 | 0.99 |
| 3 | 7 | 5 | 4.15 | 1.90 |
| -3 | -7 | 6 | 1.76 | 1.48 |
| -3 | -7 | 6 | 1.79 | 1.51 |
| -3 | -7 | -6 | 1.97 | 1.32 |
| -3 | -7 | -6 | 4.63 | 1.20 |
| 3 | -7 | 6 | 0.20 | 1.50 |
| 3 | -7 | 6 | 2.12 | 2.22 |
| 3 | -7 | -6 | 8.22 | 1.63 |
| 3 | -7 | -6 | 7.60 | 1.60 |
| -3 | 7 | 6 | 3.99 | 1.74 |
| -3 | 7 | -6 | 5.49 | 1.16 |
| 3 | 7 | 6 | 7.52 | 1.95 |
| -3 | -7 | 7 | 0.24 | 1.61 |
| -3 | -7 | 7 | -3.55 | 1.75 |
| -3 | -7 | -7 | -0.82 | 1.37 |
| -3 | -7 | -7 | 1.23 | 1.22 |
| 3 | -7 | 7 | -6.23 | 2.34 |
| 3 | -7 | -7 | -0.28 | 1.48 |
| 3 | -7 | -7 | 1.81 | 1.58 |
| -3 | 7 | 7 | 2.12 | 1.85 |
| -3 | 7 | -7 | -0.73 | 1.29 |
| 3 | 7 | 7 | 5.02 | 2.29 |
| -3 | -7 | 8 | 23.21 | 2.88 |
| -3 | -7 | 8 | 8.99 | 2.30 |
| -3 | -7 | -8 | 88.62 | 3.90 |
| -3 | -7 | -8 | 98.39 | 3.52 |
| 3 | -7 | 8 | 15.67 | 2.95 |
| 3 | -7 | -8 | 60.38 | 3.58 |
| 3 | -7 | -8 | 55.52 | 3.71 |
| -3 | 7 | 8 | 93.08 | 3.86 |
| -3 | 7 | -8 | 77.25 | 3.20 |
| 3 | 7 | 8 | 9.80 | 2.78 |
| -3 | -7 | 9 | 7.56 | 2.13 |
| -3 | -7 | 9 | 7.52 | 2.42 |
| -3 | -7 | -9 | 38.61 | 3.65 |
| -3 | -7 | -9 | 28.16 | 2.53 |
| 3 | -7 | 9 | 6.43 | 4.56 |
| 3 | -7 | -9 | 18.80 | 2.50 |
| -3 | 7 | 9 | 36.21 | 3.78 |
| -3 | 7 | -9 | 25.18 | 2.26 |
| -3 | -7 | 10 | 0.40 | 2.58 |
| -3 | -7 | -10 | 63.20 | 4.15 |
| -3 | -7 | -10 | 51.82 | 3.81 |
| 3 | -7 | 10 | -0.57 | 3.99 |
| 3 | -7 | -10 | 47.39 | 4.20 |
| 3 | -7 | -10 | 42.50 | 4.09 |
| -3 | 7 | 10 | 62.02 | 4.51 |
| -3 | 7 | -10 | 39.47 | 4.48 |
| -3 | -7 | 11 | 4.61 | 2.74 |
| -3 | -7 | -11 | -2.49 | 2.29 |
| 3 | -7 | 11 | -7.41 | 4.82 |
| 3 | -7 | -11 | 2.70 | 2.64 |
| -3 | 7 | 11 | -6.88 | 3.06 |
| -5 | -7 | 0 | 36.71 | 2.80 |
| -5 | -7 | 0 | 29.64 | 2.27 |
| 5 | -7 | 0 | 6.73 | 1.42 |
| 5 | -7 | 0 | 12.45 | 1.44 |
| 5 | -7 | 0 | 15.88 | 2.50 |
| -5 | 7 | 0 | 66.96 | 2.81 |
| -5 | -7 | 1 | 31.35 | 3.22 |
| -5 | -7 | 1 | 21.28 | 2.19 |
| -5 | -7 | -1 | 33.03 | 2.56 |
| -5 | -7 | -1 | 30.32 | 2.27 |
| 5 | -7 | 1 | 5.66 | 1.34 |
| 5 | -7 | 1 | 13.38 | 1.52 |
| 5 | -7 | 1 | 25.74 | 4.26 |
| 5 | -7 | -1 | 9.03 | 1.55 |
| 5 | -7 | -1 | 14.99 | 1.48 |
| -5 | 7 | 1 | 61.52 | 2.72 |
| -5 | 7 | -1 | 51.04 | 2.70 |
| -5 | -7 | 2 | 2.40 | 1.27 |
| -5 | -7 | 2 | 4.27 | 1.45 |
| -5 | -7 | -2 | 8.68 | 1.40 |
| -5 | -7 | -2 | 11.23 | 1.69 |
| 5 | -7 | 2 | 0.62 | 1.13 |
| 5 | -7 | 2 | 9.63 | 1.46 |
| 5 | -7 | 2 | 3.47 | 2.42 |
| 5 | -7 | -2 | 1.68 | 1.37 |
| 5 | -7 | -2 | 0.93 | 1.19 |
| -5 | 7 | 2 | 19.34 | 2.33 |
| -5 | 7 | -2 | 10.29 | 1.70 |
| -5 | -7 | 3 | 3.39 | 1.35 |
| -5 | -7 | 3 | -1.30 | 1.32 |
| -5 | -7 | -3 | -0.87 | 1.21 |
| -5 | -7 | -3 | 3.07 | 1.18 |
| 5 | -7 | 3 | 2.01 | 1.19 |
| 5 | -7 | 3 | 0.54 | 1.17 |
| 5 | -7 | 3 | 2.88 | 2.20 |
| 5 | -7 | -3 | 1.45 | 1.45 |
| 5 | -7 | -3 | 0.36 | 1.46 |
| -5 | 7 | 3 | 10.21 | 1.67 |
| -5 | -7 | 4 | 0.91 | 1.37 |
| -5 | -7 | 4 | 1.28 | 1.50 |
| -5 | -7 | -4 | 3.64 | 1.48 |
| -5 | -7 | -4 | 8.39 | 1.36 |
| 5 | -7 | 4 | -1.10 | 1.21 |
| 5 | -7 | 4 | 0.16 | 1.26 |
| 5 | -7 | 4 | 2.21 | 2.16 |
| 5 | -7 | -4 | 2.81 | 1.62 |
| -5 | 7 | 4 | 11.56 | 1.86 |
| -5 | 7 | -4 | 8.18 | 1.37 |
| -5 | -7 | 5 | 4.28 | 1.69 |
| -5 | -7 | 5 | 3.07 | 1.59 |
| -5 | -7 | -5 | 18.16 | 2.03 |
| -5 | -7 | -5 | 19.91 | 2.23 |
| 5 | -7 | 5 | 4.81 | 1.36 |
| 5 | -7 | 5 | 5.43 | 2.44 |
| 5 | -7 | -5 | 5.43 | 1.72 |
| -5 | 7 | 5 | 24.73 | 2.83 |
| -5 | 7 | -5 | 19.56 | 2.32 |
| -5 | -7 | 6 | 0.89 | 1.71 |
| -5 | -7 | 6 | 2.35 | 1.78 |
| -5 | -7 | -6 | 12.64 | 1.86 |
| -5 | -7 | -6 | 15.49 | 1.86 |
| 5 | -7 | 6 | 11.36 | 2.82 |
| 5 | -7 | -6 | 4.90 | 1.83 |
| -5 | 7 | 6 | 10.97 | 2.11 |
| -5 | 7 | -6 | 11.25 | 1.62 |
| -5 | -7 | 7 | 0.35 | 1.84 |
| -5 | -7 | 7 | 4.01 | 2.02 |
| -5 | -7 | -7 | 15.36 | 2.04 |
| -5 | -7 | -7 | 13.53 | 2.01 |
| -5 | -7 | -7 | 13.19 | 1.86 |
| 5 | -7 | 7 | 0.16 | 3.14 |
| 5 | -7 | -7 | 2.45 | 1.83 |
| -5 | 7 | 7 | 12.91 | 2.32 |
| -5 | 7 | -7 | 11.07 | 1.83 |
| -5 | -7 | 8 | 9.33 | 2.14 |
| -5 | -7 | 8 | 1.47 | 2.14 |
| -5 | -7 | -8 | 17.73 | 2.28 |
| -5 | -7 | -8 | 14.65 | 2.33 |
| -5 | -7 | -8 | 18.88 | 2.52 |
| 5 | -7 | -8 | 6.43 | 3.12 |
| -5 | 7 | 8 | 20.67 | 2.58 |
| -5 | 7 | -8 | 11.73 | 2.62 |
| -5 | -7 | 9 | 4.41 | 2.40 |
| -5 | -7 | -9 | -0.64 | 2.13 |
| -5 | -7 | -9 | -0.54 | 2.41 |
| -5 | -7 | -9 | 0.92 | 2.01 |
| 5 | -7 | -9 | -2.87 | 2.06 |
| -5 | 7 | 9 | 1.01 | 2.55 |
| -5 | 7 | -9 | -6.36 | 2.32 |
| -5 | -7 | 10 | -1.21 | 2.44 |
| -5 | -7 | -10 | 11.53 | 2.46 |
| -5 | -7 | -10 | 10.72 | 2.36 |
| 5 | -7 | -10 | 3.09 | 2.21 |
| -5 | 7 | 10 | 7.73 | 2.90 |
| -5 | 7 | -10 | 2.42 | 2.77 |
| -7 | -7 | 0 | 30.35 | 3.07 |
| -7 | -7 | 0 | 14.12 | 1.78 |
| 7 | -7 | 0 | 0.17 | 1.69 |
| -7 | 7 | 0 | 34.30 | 3.05 |
| -7 | -7 | 1 | 8.76 | 1.83 |
| -7 | -7 | 1 | 6.24 | 1.60 |
| -7 | -7 | -1 | 17.75 | 1.96 |
| -7 | -7 | -1 | 6.29 | 1.46 |
| 7 | -7 | 1 | 0.73 | 1.59 |
| 7 | -7 | -1 | 1.85 | 1.61 |
| -7 | 7 | 1 | 14.80 | 1.90 |
| -7 | 7 | -1 | 12.58 | 2.06 |
| -7 | -7 | 2 | 2.27 | 1.56 |
| -7 | -7 | 2 | 4.96 | 1.88 |
| -7 | -7 | -2 | 8.18 | 1.82 |
| -7 | -7 | -2 | 7.19 | 1.73 |
| 7 | -7 | 2 | -0.05 | 1.59 |
| 7 | -7 | -2 | 3.57 | 1.81 |
| -7 | 7 | 2 | 13.62 | 1.84 |
| -7 | -7 | 3 | 3.71 | 1.69 |
| -7 | -7 | 3 | 1.02 | 1.85 |
| -7 | -7 | -3 | 3.55 | 1.59 |
| -7 | -7 | -3 | -0.42 | 1.49 |
| 7 | -7 | 3 | 0.19 | 1.68 |
| 7 | -7 | -3 | 3.09 | 1.77 |
| -7 | 7 | 3 | 2.68 | 1.77 |
| -7 | -7 | 4 | 1.31 | 1.76 |
| -7 | -7 | 4 | 1.04 | 1.84 |
| -7 | -7 | -4 | 3.75 | 1.73 |
| -7 | -7 | -4 | 3.01 | 1.55 |
| 7 | -7 | 4 | -0.43 | 1.67 |
| 7 | -7 | -4 | 0.06 | 1.99 |
| -7 | 7 | 4 | 11.37 | 2.24 |
| -7 | 7 | -4 | 6.08 | 1.81 |
| -7 | -7 | 5 | -3.38 | 1.88 |
| -7 | -7 | 5 | -2.91 | 1.88 |
| -7 | -7 | -5 | 4.49 | 1.89 |
| -7 | -7 | -5 | 4.83 | 1.77 |
| 7 | -7 | 5 | -1.55 | 1.88 |
| 7 | -7 | -5 | 4.26 | 2.08 |
| -7 | 7 | 5 | 13.02 | 2.99 |
| -7 | 7 | -5 | 8.16 | 1.90 |
| -7 | -7 | 6 | 1.73 | 2.27 |
| -7 | -7 | 6 | -2.01 | 2.25 |
| -7 | -7 | -6 | 4.65 | 2.69 |
| -7 | -7 | -6 | 6.90 | 2.69 |
| 7 | -7 | -6 | 2.61 | 2.95 |
| -7 | 7 | 6 | 17.17 | 2.97 |
| -7 | 7 | -6 | 5.62 | 2.24 |
| -7 | -7 | 7 | -2.82 | 2.15 |
| -7 | -7 | -7 | 6.66 | 2.28 |
| -7 | -7 | -7 | 3.36 | 1.98 |
| 7 | -7 | -7 | -0.15 | 2.02 |
| -7 | 7 | 7 | -2.00 | 2.60 |
| -7 | 7 | -7 | 5.12 | 2.28 |
| -7 | -7 | 8 | -1.13 | 2.39 |
| -7 | -7 | -8 | 10.92 | 2.56 |
| -7 | -7 | -8 | 10.70 | 2.49 |
| 7 | -7 | -8 | 0.00 | 2.43 |
| -7 | 7 | 8 | 3.23 | 2.88 |
| -7 | 7 | -8 | 15.55 | 2.78 |
| -7 | -7 | 9 | -1.17 | 2.48 |
| -7 | -7 | -9 | 6.90 | 2.59 |
| -7 | -7 | -9 | 5.76 | 2.27 |
| 7 | -7 | -9 | 7.39 | 2.78 |
| -7 | 7 | 9 | -3.32 | 3.06 |
| -7 | 7 | -9 | 0.81 | 2.88 |
| -9 | -7 | 0 | 7.71 | 1.97 |
| 9 | -7 | 0 | 4.79 | 2.24 |
| -9 | 7 | 0 | 9.03 | 2.62 |
| -9 | -7 | 1 | -0.34 | 1.96 |
| -9 | -7 | -1 | -3.09 | 2.79 |
| -9 | -7 | -1 | 1.71 | 1.86 |
| 9 | -7 | 1 | 0.73 | 2.07 |
| 9 | -7 | -1 | 1.07 | 2.30 |
| -9 | 7 | 1 | 1.75 | 2.37 |
| -9 | -7 | 2 | 26.86 | 2.71 |
| -9 | -7 | -2 | 59.38 | 4.53 |
| -9 | -7 | -2 | 51.06 | 5.02 |
| 9 | -7 | 2 | 15.19 | 2.66 |
| 9 | -7 | -2 | 22.79 | 4.41 |
| -9 | 7 | 2 | 83.46 | 6.32 |
| -9 | -7 | 3 | 5.67 | 2.34 |
| -9 | -7 | -3 | 26.66 | 2.82 |
| -9 | -7 | -3 | 15.42 | 2.23 |
| 9 | -7 | 3 | 0.74 | 4.00 |
| 9 | -7 | -3 | 2.31 | 2.34 |
| -9 | 7 | 3 | 41.29 | 4.03 |
| -9 | -7 | 4 | 3.41 | 3.10 |
| -9 | -7 | -4 | 63.21 | 4.48 |
| -9 | -7 | -4 | 47.22 | 3.54 |
| 9 | -7 | 4 | 12.85 | 2.63 |
| 9 | -7 | -4 | 20.63 | 2.85 |
| -9 | 7 | 4 | 91.03 | 4.44 |
| -9 | 7 | -4 | 62.12 | 4.63 |
| -9 | -7 | -5 | 5.92 | 2.52 |
| -9 | -7 | -5 | -1.04 | 1.91 |
| 9 | -7 | -5 | 6.24 | 2.40 |
| -9 | 7 | 5 | 4.69 | 2.63 |
| -9 | 7 | -5 | 7.35 | 2.87 |
| -9 | -7 | 6 | -4.09 | 2.67 |
| -9 | -7 | -6 | 1.90 | 2.50 |
| -9 | -7 | -6 | 6.82 | 2.14 |
| 9 | -7 | -6 | -4.27 | 2.63 |
| -9 | 7 | 6 | 9.00 | 2.96 |
| -9 | 7 | -6 | 4.23 | 3.05 |
| -9 | -7 | 7 | 3.58 | 2.82 |
| -9 | -7 | -7 | -3.99 | 2.58 |
| -9 | -7 | -7 | 2.08 | 2.24 |
| 9 | -7 | -7 | 1.15 | 2.60 |
| -9 | 7 | 7 | 3.11 | 3.07 |
| -9 | 7 | -7 | 5.35 | 3.18 |
| -9 | -7 | -8 | 31.61 | 6.15 |
| -9 | -7 | -8 | 48.66 | 5.56 |
| -9 | 7 | -8 | 33.77 | 6.05 |
| -11 | -7 | 0 | 1.33 | 2.77 |
| 11 | -7 | 0 | 3.45 | 3.08 |
| -11 | -7 | 1 | -0.42 | 2.93 |
| -11 | -7 | -1 | 10.68 | 2.78 |
| 11 | -7 | 1 | -0.39 | 3.12 |
| 11 | -7 | -1 | 2.19 | 3.00 |
| -11 | -7 | -2 | 4.91 | 2.72 |
| 11 | -7 | 2 | 7.34 | 3.28 |
| 11 | -7 | -2 | -0.21 | 2.91 |
| -11 | 7 | -2 | 2.94 | 3.35 |
| -11 | -7 | -3 | -3.52 | 2.68 |
| 11 | -7 | 3 | -0.93 | 3.29 |
| 11 | -7 | -3 | 3.32 | 3.13 |
| -11 | 7 | 3 | -8.08 | 3.61 |
| -11 | 7 | -3 | -3.37 | 3.25 |
| -11 | -7 | -4 | 3.43 | 2.57 |
| 11 | -7 | 4 | -1.44 | 3.75 |
| 11 | -7 | -4 | -4.34 | 3.24 |
| -11 | 7 | 4 | -0.44 | 3.64 |
| -11 | 7 | -4 | -7.05 | 3.31 |
| 0 | -8 | 0 | 15.47 | 1.70 |
| 0 | -8 | 0 | 10.52 | 1.22 |
| 0 | -8 | 0 | 8.88 | 1.53 |
| 0 | 8 | 0 | 20.31 | 1.81 |
| 0 | -8 | 1 | 726.19 | 7.89 |
| 0 | -8 | 1 | 545.04 | 4.76 |
| 0 | -8 | 1 | 949.02 | 9.61 |
| 0 | -8 | -1 | 1184.85 | 10.46 |
| 0 | -8 | -1 | 1282.19 | 9.27 |
| 0 | -8 | -1 | 1164.54 | 10.20 |
| 0 | 8 | 1 | 2696.64 | 12.76 |
| 0 | 8 | -1 | 2022.75 | 10.82 |
| 0 | -8 | 2 | 82.37 | 3.24 |
| 0 | -8 | -2 | 122.55 | 3.84 |
| 0 | -8 | -2 | 122.09 | 3.48 |
| 0 | -8 | -2 | 118.64 | 3.45 |
| 0 | 8 | 2 | 252.50 | 4.31 |
| 0 | -8 | 3 | 177.94 | 4.40 |
| 0 | -8 | -3 | 802.29 | 8.88 |
| 0 | -8 | -3 | 893.61 | 8.48 |
| 0 | -8 | -3 | 745.91 | 7.89 |
| 0 | 8 | 3 | 1812.55 | 10.86 |
| 0 | -8 | 4 | 33.22 | 2.71 |
| 0 | -8 | 4 | 27.13 | 2.91 |
| 0 | -8 | -4 | 101.32 | 3.70 |
| 0 | -8 | -4 | 129.61 | 3.83 |
| 0 | -8 | -4 | 109.02 | 3.33 |
| 0 | 8 | 4 | 202.83 | 4.54 |
| 0 | -8 | 5 | 232.71 | 5.27 |
| 0 | -8 | 5 | 197.95 | 4.77 |
| 0 | -8 | -5 | 987.16 | 10.04 |
| 0 | -8 | -5 | 951.92 | 8.61 |
| 0 | 8 | 5 | 1739.09 | 11.15 |
| 0 | -8 | 6 | 1.02 | 1.65 |
| 0 | -8 | 6 | -1.31 | 2.18 |
| 0 | -8 | -6 | 4.55 | 1.43 |
| 0 | -8 | -6 | 1.88 | 1.39 |
| 0 | 8 | 6 | 15.42 | 2.21 |
| 0 | -8 | 7 | 100.62 | 3.98 |
| 0 | -8 | 7 | 157.95 | 5.00 |
| 0 | -8 | -7 | 656.01 | 8.54 |
| 0 | -8 | -7 | 677.36 | 7.13 |
| 0 | 8 | 7 | 834.15 | 8.43 |
| 0 | -8 | 8 | 8.66 | 2.37 |
| 0 | -8 | -8 | 55.10 | 4.28 |
| 0 | -8 | -8 | 57.50 | 3.31 |
| 0 | 8 | 8 | 52.88 | 4.17 |
| 0 | -8 | 9 | 31.99 | 3.58 |
| 0 | -8 | -9 | 363.61 | 6.80 |
| 0 | -8 | -9 | 392.01 | 7.21 |
| 0 | 8 | 9 | 321.69 | 6.48 |
| 0 | -8 | 10 | -6.54 | 3.41 |
| 0 | -8 | -10 | 45.49 | 4.08 |
| 0 | -8 | -10 | 42.11 | 4.12 |
| 0 | 8 | 10 | 30.68 | 3.69 |
| 0 | -8 | 11 | 8.54 | 4.67 |
| 0 | -8 | -11 | 444.05 | 7.54 |
| 0 | -8 | -11 | 429.54 | 7.89 |
| -2 | -8 | 0 | 71.66 | 3.15 |
| -2 | -8 | 0 | 53.09 | 2.37 |
| -2 | -8 | 0 | 21.86 | 2.15 |
| 2 | -8 | 0 | 10.85 | 1.57 |
| 2 | -8 | 0 | 17.88 | 2.12 |
| -2 | 8 | 0 | 48.03 | 2.56 |
| 2 | 8 | 0 | 27.93 | 3.40 |
| -2 | -8 | 1 | 518.56 | 6.47 |
| -2 | -8 | 1 | 584.48 | 7.72 |
| -2 | -8 | -1 | 830.80 | 8.52 |
| -2 | -8 | -1 | 968.05 | 7.78 |
| -2 | -8 | -1 | 768.07 | 8.56 |
| 2 | -8 | 1 | 272.30 | 4.38 |
| 2 | -8 | 1 | 500.06 | 6.82 |
| 2 | -8 | -1 | 526.40 | 7.34 |
| 2 | -8 | -1 | 530.92 | 6.41 |
| 2 | -8 | -1 | 538.98 | 6.66 |
| -2 | 8 | 1 | 1600.25 | 10.46 |
| -2 | 8 | -1 | 1270.57 | 9.25 |
| 2 | 8 | 1 | 1366.93 | 8.72 |
| -2 | -8 | 2 | 15.37 | 1.98 |
| -2 | -8 | -2 | 10.31 | 2.01 |
| -2 | -8 | -2 | 11.23 | 1.68 |
| 2 | -8 | 2 | 2.13 | 1.00 |
| 2 | -8 | 2 | 7.92 | 1.60 |
| 2 | -8 | 2 | 9.86 | 2.10 |
| 2 | -8 | -2 | 9.83 | 1.62 |
| 2 | -8 | -2 | 7.19 | 1.49 |
| 2 | -8 | -2 | 8.98 | 1.45 |
| -2 | 8 | 2 | 22.27 | 2.39 |
| -2 | 8 | -2 | 26.59 | 2.20 |
| 2 | 8 | 2 | 24.35 | 2.80 |
| -2 | -8 | 3 | 204.07 | 4.90 |
| -2 | -8 | 3 | 88.19 | 3.34 |
| -2 | -8 | -3 | 490.55 | 6.87 |
| -2 | -8 | -3 | 425.79 | 6.27 |
| 2 | -8 | 3 | 116.57 | 4.15 |
| 2 | -8 | -3 | 328.53 | 5.98 |
| 2 | -8 | -3 | 346.91 | 5.72 |
| 2 | -8 | -3 | 308.02 | 4.85 |
| -2 | 8 | 3 | 981.03 | 8.46 |
| 2 | 8 | 3 | 911.88 | 7.45 |
| -2 | -8 | 4 | 4.36 | 1.47 |
| -2 | -8 | 4 | 2.90 | 1.63 |
| -2 | -8 | -4 | 21.04 | 2.55 |
| -2 | -8 | -4 | 23.21 | 2.19 |
| 2 | -8 | 4 | 3.07 | 1.43 |
| 2 | -8 | 4 | -0.13 | 2.19 |
| 2 | -8 | -4 | 14.50 | 1.77 |
| 2 | -8 | -4 | 14.73 | 1.83 |
| 2 | -8 | -4 | 10.48 | 1.42 |
| -2 | 8 | 4 | 26.96 | 2.72 |
| -2 | 8 | -4 | 26.94 | 1.94 |
| 2 | 8 | 4 | 36.71 | 2.93 |
| -2 | -8 | 5 | 140.15 | 4.32 |
| -2 | -8 | 5 | 134.50 | 3.95 |
| -2 | -8 | -5 | 700.04 | 8.39 |
| -2 | -8 | -5 | 628.90 | 7.56 |
| 2 | -8 | 5 | 183.15 | 4.72 |
| 2 | -8 | 5 | 139.86 | 4.73 |
| 2 | -8 | -5 | 503.44 | 7.41 |
| 2 | -8 | -5 | 548.49 | 7.54 |
| 2 | -8 | -5 | 457.73 | 5.48 |
| -2 | 8 | 5 | 991.68 | 8.81 |
| -2 | 8 | -5 | 663.77 | 5.29 |
| 2 | 8 | 5 | 1376.08 | 9.61 |
| -2 | -8 | 6 | 1.47 | 1.56 |
| -2 | -8 | 6 | 1.81 | 1.93 |
| -2 | -8 | -6 | 14.84 | 2.01 |
| -2 | -8 | -6 | 10.87 | 1.82 |
| 2 | -8 | 6 | 3.61 | 1.81 |
| 2 | -8 | 6 | -2.73 | 2.47 |
| 2 | -8 | -6 | 14.28 | 2.04 |
| 2 | -8 | -6 | 10.46 | 1.81 |
| 2 | -8 | -6 | 8.37 | 1.71 |
| -2 | 8 | 6 | 12.22 | 2.19 |
| -2 | 8 | -6 | 8.54 | 1.36 |
| 2 | 8 | 6 | 25.22 | 2.54 |
| -2 | -8 | 7 | 60.94 | 3.42 |
| -2 | -8 | 7 | 98.07 | 4.32 |
| -2 | -8 | -7 | 494.77 | 7.49 |
| -2 | -8 | -7 | 477.90 | 6.58 |
| 2 | -8 | 7 | 100.29 | 4.17 |
| 2 | -8 | 7 | 54.38 | 4.46 |
| 2 | -8 | -7 | 350.00 | 6.42 |
| 2 | -8 | -7 | 352.37 | 6.70 |
| -2 | 8 | 7 | 497.02 | 6.83 |
| -2 | 8 | -7 | 413.38 | 6.67 |
| 2 | 8 | 7 | 643.39 | 7.42 |
| -2 | -8 | 8 | -1.33 | 2.61 |

```
-2   -8   -8     15.03    2.28
-2   -8   -8      9.11    1.90
 2   -8   -8     10.16    2.17
-2    8    8     16.64    2.78
 2    8    8      2.55    3.12
-2   -8    9     30.52    3.81
-2   -8    9     15.62    2.99
-2   -8   -9    246.13    5.85
-2   -8   -9    252.27    5.32
 2   -8    9     12.04    3.67
 2   -8   -9    191.89    5.32
 2   -8   -9    181.58    5.39
-2    8    9    196.76    5.50
-2   -8   10     -0.06    3.03
-2   -8  -10     12.60    2.19
 2   -8   10      5.68    4.10
 2   -8  -10      7.41    2.17
 2   -8  -10      9.25    2.53
-2    8   10     20.06    3.31
 2   -8  -11    297.83    8.58
-4   -8    0     18.24    1.88
-4   -8    0     20.50    2.28
 4   -8    0      4.66    1.40
 4   -8    0      7.69    1.42
-4    8    0     74.16    3.17
-4   -8    1    541.64    6.34
-4   -8    1    573.80    7.83
-4   -8   -1    859.37    8.44
-4   -8   -1    733.84    8.60
 4   -8    1    211.52    4.07
 4   -8    1    328.30    5.31
 4   -8    1    336.95    6.01
 4   -8   -1    329.33    5.52
 4   -8   -1    318.92    4.84
-4    8    1   1472.44   10.61
-4    8   -1   1209.92    9.66
-4   -8    2      6.28    1.55
-4   -8    2      9.45    1.67
-4   -8   -2     18.13    1.86
-4   -8   -2     15.29    1.82
 4   -8    2     -2.45    1.23
 4   -8    2      4.36    1.38
 4   -8    2      9.36    2.46
 4   -8   -2      6.08    1.58
 4   -8   -2      6.10    1.32
-4    8    2     16.35    1.85
-4    8   -2     22.85    2.28
-4   -8    3    187.74    4.77
-4   -8    3     73.80    3.25
-4   -8   -3    472.90    6.65
-4   -8   -3    441.28    6.62
 4   -8    3    167.43    4.22
 4   -8    3    145.92    4.68
 4   -8   -3    245.13    5.25
 4   -8   -3    170.62    3.34
-4    8    3    914.34    8.58
-4   -8    4      2.71    1.60
-4   -8    4      3.88    1.83
-4   -8   -4     10.01    1.58
-4   -8   -4      9.76    1.68
 4   -8    4      3.41    1.40
 4   -8    4      5.87    1.55
 4   -8    4      7.71    2.79
 4   -8   -4      5.19    1.53
 4   -8   -4      7.00    1.79
 4   -8   -4      2.62    1.37
-4    8    4     21.84    2.22
-4    8   -4     23.47    2.33
-4   -8    5     97.83    3.63
```

| | | | | |
|---|---|---|---|---|
| -4 | -8 | -5 | 647.77 | 8.05 |
| -4 | -8 | -5 | 601.81 | 7.67 |
| 4 | -8 | 5 | 203.52 | 5.01 |
| 4 | -8 | 5 | 179.31 | 6.31 |
| 4 | -8 | -5 | 359.51 | 6.42 |
| 4 | -8 | -5 | 371.84 | 6.57 |
| -4 | 8 | 5 | 1031.31 | 9.29 |
| -4 | 8 | -5 | 702.40 | 6.52 |
| -4 | -8 | 6 | 1.76 | 2.03 |
| -4 | -8 | 6 | -0.28 | 2.07 |
| -4 | -8 | -6 | 14.15 | 2.15 |
| -4 | -8 | -6 | 22.35 | 2.64 |
| 4 | -8 | 6 | 6.24 | 1.89 |
| 4 | -8 | 6 | -4.93 | 2.84 |
| 4 | -8 | -6 | 8.77 | 1.77 |
| 4 | -8 | -6 | 3.57 | 1.85 |
| -4 | 8 | 6 | 15.08 | 2.42 |
| -4 | 8 | -6 | 10.65 | 1.76 |
| -4 | -8 | 7 | 41.24 | 3.56 |
| -4 | -8 | 7 | 70.48 | 7.07 |
| -4 | -8 | -7 | 423.65 | 10.05 |
| -4 | -8 | -7 | 445.05 | 6.78 |
| 4 | -8 | 7 | 56.39 | 5.59 |
| 4 | -8 | -7 | 255.72 | 5.70 |
| -4 | 8 | 7 | 474.48 | 8.72 |
| -4 | 8 | -7 | 436.55 | 5.72 |
| -4 | -8 | 8 | 4.14 | 2.37 |
| -4 | -8 | 8 | 1.78 | 2.54 |
| -4 | -8 | -8 | 10.54 | 2.03 |
| -4 | -8 | -8 | 15.13 | 2.18 |
| 4 | -8 | 8 | 3.65 | 3.97 |
| 4 | -8 | -8 | 6.80 | 2.06 |
| 4 | -8 | -8 | 0.91 | 2.01 |
| -4 | 8 | 8 | 19.50 | 2.83 |
| -4 | 8 | -8 | 11.19 | 2.42 |
| -4 | -8 | 9 | 39.38 | 4.30 |
| -4 | -8 | 9 | 9.71 | 2.77 |
| -4 | -8 | -9 | 223.50 | 5.80 |
| -4 | -8 | -9 | 232.74 | 5.47 |
| 4 | -8 | -9 | 130.29 | 5.14 |
| 4 | -8 | -9 | 124.79 | 4.89 |
| -4 | 8 | 9 | 167.33 | 5.46 |
| -4 | 8 | -9 | 186.04 | 7.07 |
| -4 | -8 | 10 | 7.52 | 2.87 |
| -4 | -8 | -10 | 9.45 | 2.27 |
| -4 | -8 | -10 | 7.14 | 2.76 |
| 4 | -8 | -10 | 1.20 | 2.40 |
| -4 | 8 | 10 | 7.25 | 3.21 |
| -6 | -8 | 0 | 13.75 | 1.89 |
| -6 | -8 | 0 | 15.05 | 2.05 |
| 6 | -8 | 0 | 6.81 | 1.78 |
| 6 | -8 | 0 | 6.31 | 1.55 |
| -6 | 8 | 0 | 20.45 | 2.23 |
| -6 | -8 | 1 | 625.92 | 6.51 |
| -6 | -8 | 1 | 647.71 | 8.46 |
| -6 | -8 | -1 | 983.13 | 8.89 |
| -6 | -8 | -1 | 790.43 | 9.24 |
| 6 | -8 | 1 | 368.60 | 5.31 |
| 6 | -8 | -1 | 306.46 | 5.76 |
| 6 | -8 | -1 | 331.24 | 4.46 |
| -6 | 8 | 1 | 1615.91 | 11.69 |
| -6 | 8 | -1 | 1346.34 | 10.86 |
| -6 | -8 | 2 | 47.02 | 3.07 |
| -6 | -8 | 2 | 8.88 | 2.00 |
| -6 | -8 | -2 | 84.49 | 3.68 |
| -6 | -8 | -2 | 80.18 | 3.46 |
| 6 | -8 | 2 | 13.38 | 1.84 |
| 6 | -8 | 2 | 37.42 | 2.70 |
| 6 | -8 | -2 | 31.62 | 3.24 |
| 6 | -8 | -2 | 9.64 | 1.45 |
| -6 | 8 | 2 | 143.16 | 4.28 |
| -6 | -8 | 3 | 243.65 | 5.41 |
| -6 | -8 | 3 | 82.36 | 3.62 |
| -6 | -8 | -3 | 617.03 | 7.59 |
| -6 | -8 | -3 | 546.61 | 7.72 |
| 6 | -8 | -3 | 256.16 | 5.65 |
| -6 | 8 | 3 | 1023.42 | 9.71 |
| -6 | -8 | 4 | 19.25 | 2.38 |
| -6 | -8 | 4 | 10.65 | 2.16 |
| -6 | -8 | -4 | 73.91 | 3.75 |
| -6 | -8 | -4 | 77.56 | 3.55 |
| 6 | -8 | 4 | 26.29 | 3.04 |
| 6 | -8 | -4 | 32.04 | 3.43 |
| -6 | 8 | 4 | 146.44 | 4.69 |
| -6 | 8 | -4 | 89.65 | 3.90 |
| -6 | -8 | 5 | 79.66 | 3.76 |
| -6 | -8 | -5 | 732.95 | 8.71 |
| -6 | -8 | -5 | 667.91 | 8.53 |
| 6 | -8 | -5 | 339.05 | 6.75 |
| -6 | 8 | 5 | 860.06 | 10.83 |
| -6 | 8 | -5 | 854.78 | 8.38 |
| -6 | -8 | 6 | -0.03 | 2.26 |
| -6 | -8 | 6 | -0.89 | 2.25 |
| -6 | -8 | -6 | 11.35 | 2.08 |
| -6 | -8 | -6 | 9.79 | 2.03 |
| 6 | -8 | -6 | 2.35 | 2.31 |
| -6 | 8 | 6 | 16.49 | 2.52 |
| -6 | -8 | 7 | 46.91 | 4.13 |
| -6 | -8 | 7 | 15.95 | 2.59 |
| -6 | -8 | -7 | 511.20 | 7.94 |
| -6 | -8 | -7 | 470.86 | 7.37 |
| 6 | -8 | -7 | 215.33 | 5.71 |
| -6 | 8 | 7 | 714.60 | 8.39 |
| -6 | 8 | -7 | 540.22 | 7.12 |
| -6 | -8 | 8 | 3.22 | 2.58 |
| -6 | -8 | -8 | 41.99 | 4.34 |
| -6 | -8 | -8 | 33.60 | 3.61 |
| 6 | -8 | -8 | 11.88 | 2.60 |
| -6 | 8 | 8 | 37.11 | 4.60 |
| -6 | 8 | -8 | 38.44 | 4.65 |
| -6 | -8 | 9 | 16.71 | 2.87 |
| -6 | -8 | -9 | 264.35 | 6.55 |
| -6 | -8 | -9 | 260.79 | 5.99 |
| 6 | -8 | -9 | 116.32 | 5.10 |
| -6 | 8 | 9 | 250.26 | 7.07 |
| -6 | 8 | -9 | 213.16 | 6.27 |
| -8 | -8 | 0 | 5.21 | 2.12 |
| 8 | -8 | 0 | 4.36 | 2.19 |
| -8 | 8 | 0 | 20.24 | 4.41 |
| -8 | -8 | 1 | 271.86 | 5.90 |
| -8 | -8 | 1 | 33.93 | 4.01 |
| -8 | -8 | -1 | 351.84 | 7.21 |
| -8 | -8 | -1 | 322.54 | 7.95 |
| 8 | -8 | 1 | 99.87 | 4.33 |
| 8 | -8 | -1 | 104.12 | 4.58 |
| -8 | -8 | 2 | 8.31 | 2.27 |
| -8 | -8 | 2 | -4.05 | 2.59 |
| -8 | -8 | -2 | 5.85 | 2.29 |
| -8 | -8 | -2 | 6.02 | 2.53 |
| 8 | -8 | 2 | 4.69 | 2.28 |
| 8 | -8 | -2 | 3.48 | 2.97 |
| -8 | 8 | 2 | 12.80 | 2.61 |
| -8 | -8 | -3 | 199.60 | 7.28 |
| -8 | -8 | -3 | 187.71 | 5.19 |
| 8 | -8 | 3 | 34.92 | 4.15 |
| 8 | -8 | -3 | 85.35 | 4.47 |
| -8 | 8 | 3 | 368.21 | 6.39 |
| -8 | -8 | 4 | 7.95 | 2.51 |
| -8 | -8 | 4 | 9.15 | 2.78 |
| -8 | -8 | -4 | 6.95 | 2.63 |
| -8 | -8 | -4 | 6.38 | 2.14 |
| 8 | -8 | 4 | 1.92 | 2.65 |
| 8 | -8 | -4 | 0.77 | 2.35 |
| -8 | 8 | 4 | 6.95 | 2.37 |
| -8 | 8 | -4 | -0.73 | 2.40 |
| -8 | -8 | 5 | 62.19 | 4.54 |
| -8 | -8 | 5 | 3.60 | 2.50 |
| -8 | -8 | -5 | 312.97 | 6.56 |
| -8 | -8 | -5 | 268.37 | 5.99 |
| 8 | -8 | -5 | 108.77 | 4.80 |
| -8 | 8 | 5 | 504.14 | 7.31 |
| -8 | 8 | -5 | 345.52 | 6.63 |
| -8 | -8 | 6 | 1.55 | 2.64 |
| -8 | -8 | -6 | 14.88 | 2.91 |
| -8 | -8 | -6 | 4.05 | 2.25 |
| 8 | -8 | -6 | -3.48 | 2.45 |
| -8 | 8 | 6 | 9.78 | 2.85 |
| -8 | 8 | -6 | 9.99 | 3.14 |
| -8 | -8 | 7 | 11.22 | 2.88 |
| -8 | -8 | -7 | 228.95 | 6.37 |
| -8 | -8 | -7 | 201.82 | 5.55 |
| 8 | -8 | -7 | 79.55 | 4.75 |
| -8 | 8 | 7 | 369.21 | 6.77 |
| -8 | 8 | -7 | 222.61 | 6.56 |
| -10 | -8 | 0 | 12.49 | 2.69 |
| 10 | -8 | 0 | -1.97 | 2.76 |
| -10 | -8 | 1 | 165.87 | 5.50 |
| -10 | -8 | -1 | 206.09 | 5.65 |
| 10 | -8 | 1 | 70.65 | 4.97 |
| 10 | -8 | -1 | 92.35 | 5.03 |
| -10 | -8 | 2 | 1.92 | 2.86 |
| -10 | -8 | -2 | 10.71 | 2.51 |
| 10 | -8 | 2 | -1.87 | 2.98 |
| 10 | -8 | -2 | 5.95 | 2.94 |
| -10 | 8 | 2 | 15.36 | 3.32 |
| -10 | -8 | -3 | 153.84 | 5.22 |
| 10 | -8 | 3 | 67.81 | 5.20 |
| 10 | -8 | -3 | 54.12 | 4.84 |
| -10 | 8 | 3 | 224.92 | 6.45 |
| -10 | 8 | -3 | 206.15 | 6.45 |
| -10 | -8 | -4 | 6.18 | 2.54 |
| 10 | -8 | 4 | 6.28 | 3.84 |
| 10 | -8 | -4 | -0.25 | 2.89 |
| -10 | 8 | 4 | 8.54 | 3.00 |
| -10 | 8 | -4 | 6.65 | 3.46 |
| 10 | -8 | -5 | 83.18 | 6.01 |
| -10 | 8 | -5 | 264.78 | 8.86 |
| -1 | -9 | 0 | -0.30 | 1.33 |
| -1 | -9 | 0 | 0.58 | 1.68 |
| -1 | -9 | 0 | -0.96 | 1.22 |
| 1 | -9 | 0 | 0.51 | 1.44 |
| 1 | -9 | 0 | 2.23 | 1.44 |
| -1 | 9 | 0 | -2.08 | 1.50 |
| 1 | 9 | 0 | -1.06 | 1.67 |
| -1 | -9 | 1 | 8.58 | 1.64 |
| -1 | -9 | 1 | 11.74 | 1.82 |
| -1 | -9 | -1 | 15.71 | 1.86 |
| -1 | -9 | -1 | 23.16 | 2.63 |
| -1 | -9 | -1 | 21.84 | 2.28 |
| 1 | -9 | 1 | 9.25 | 1.71 |
| 1 | -9 | 1 | 14.27 | 1.95 |
| 1 | -9 | -1 | 8.16 | 1.77 |
| 1 | -9 | -1 | 13.84 | 1.63 |
| 1 | -9 | -1 | 8.75 | 1.50 |
| -1 | 9 | 1 | 25.72 | 2.65 |
| -1 | 9 | -1 | 24.00 | 2.81 |
| 1 | 9 | 1 | 27.37 | 2.15 |
| -1 | -9 | 2 | 8.29 | 1.70 |
| -1 | -9 | -2 | 13.91 | 1.88 |
| -1 | -9 | -2 | 13.69 | 1.78 |
| -1 | -9 | -2 | 14.60 | 1.81 |
| 1 | -9 | 2 | 4.93 | 1.64 |
| 1 | -9 | 2 | 5.16 | 2.29 |
| 1 | -9 | -2 | 11.23 | 1.68 |
| 1 | -9 | -2 | 11.95 | 1.72 |
| 1 | -9 | -2 | 13.50 | 1.71 |
| -1 | 9 | 2 | 28.44 | 2.65 |
| 1 | 9 | 2 | 27.44 | 2.77 |
| -1 | -9 | 3 | 6.84 | 1.73 |
| -1 | -9 | 3 | 2.94 | 2.00 |
| -1 | -9 | -3 | 16.38 | 2.02 |
| -1 | -9 | -3 | 17.66 | 1.95 |
| -1 | -9 | -3 | 19.58 | 2.41 |
| 1 | -9 | 3 | 9.70 | 1.94 |
| 1 | -9 | 3 | 3.54 | 2.48 |
| 1 | -9 | -3 | 13.11 | 1.78 |
| 1 | -9 | -3 | 19.51 | 2.03 |
| 1 | -9 | -3 | 14.55 | 1.73 |
| -1 | 9 | 3 | 37.10 | 2.90 |
| 1 | 9 | 3 | 34.60 | 3.02 |
| -1 | -9 | 4 | 9.06 | 2.02 |
| -1 | -9 | 4 | 3.15 | 2.31 |
| -1 | -9 | -4 | 7.40 | 1.73 |
| -1 | -9 | -4 | 11.96 | 1.86 |
| -1 | -9 | -4 | 7.50 | 1.58 |
| 1 | -9 | 4 | 6.04 | 1.82 |
| 1 | -9 | 4 | 1.24 | 2.44 |
| 1 | -9 | -4 | 11.61 | 1.82 |
| 1 | -9 | -4 | 7.56 | 1.70 |
| 1 | -9 | -4 | 9.84 | 1.60 |
| -1 | 9 | 4 | 17.91 | 2.31 |
| 1 | 9 | 4 | 42.17 | 4.14 |
| -1 | -9 | 5 | 4.79 | 1.82 |
| -1 | -9 | 5 | 3.48 | 2.48 |
| -1 | -9 | -5 | 14.85 | 2.18 |
| -1 | -9 | -5 | 22.15 | 2.27 |
| -1 | -9 | -5 | 17.53 | 2.05 |
| 1 | -9 | 5 | 4.05 | 2.01 |
| 1 | -9 | 5 | 12.73 | 2.72 |
| 1 | -9 | -5 | 17.22 | 2.28 |
| 1 | -9 | -5 | 13.43 | 1.90 |
| 1 | -9 | -5 | 9.83 | 1.72 |
| -1 | 9 | 5 | 34.55 | 3.54 |
| 1 | 9 | 5 | 25.63 | 2.51 |
| -1 | -9 | 6 | 1.77 | 2.11 |
| -1 | -9 | 6 | -3.70 | 2.50 |
| -1 | -9 | -6 | -0.55 | 1.55 |
| -1 | -9 | -6 | -3.78 | 1.77 |
| -1 | -9 | -6 | 0.96 | 1.66 |
| 1 | -9 | 6 | 0.58 | 1.81 |
| 1 | -9 | 6 | 3.54 | 2.80 |
| 1 | -9 | -6 | 4.16 | 1.69 |
| 1 | -9 | -6 | -2.31 | 1.76 |
| 1 | -9 | -6 | 2.74 | 1.73 |
| -1 | 9 | 6 | 3.49 | 2.44 |
| 1 | 9 | 6 | 6.74 | 2.70 |
| -1 | -9 | 7 | 2.78 | 2.12 |
| -1 | -9 | 7 | 5.19 | 3.00 |
| -1 | -9 | -7 | 18.28 | 2.41 |
| -1 | -9 | -7 | 20.34 | 2.49 |
| -1 | -9 | -7 | 15.24 | 2.13 |
| 1 | -9 | 7 | 1.07 | 2.14 |
| 1 | -9 | 7 | 3.27 | 4.07 |
| 1 | -9 | -7 | 11.39 | 1.84 |
| 1 | -9 | -7 | 11.56 | 2.12 |
| -1 | 9 | 7 | 29.11 | 4.00 |
| 1 | 9 | 7 | 43.26 | 4.66 |
| -1 | -9 | 8 | -0.13 | 2.37 |
| -1 | -9 | 8 | 5.39 | 3.29 |
| -1 | -9 | -8 | 11.32 | 2.01 |
| -1 | -9 | -8 | 17.42 | 2.66 |
| -1 | -9 | -8 | 12.32 | 2.32 |
| 1 | -9 | 8 | -4.74 | 2.59 |
| 1 | -9 | 8 | 18.79 | 3.40 |
| 1 | -9 | -8 | 16.54 | 2.22 |
| 1 | -9 | -8 | 5.48 | 2.26 |
| -1 | 9 | 8 | 24.88 | 3.25 |
| 1 | 9 | 8 | 10.13 | 3.68 |
| -1 | -9 | 9 | 2.86 | 2.73 |
| -1 | -9 | 9 | 4.30 | 3.36 |
| -1 | -9 | -9 | 3.40 | 1.94 |
| -1 | -9 | -9 | 8.94 | 2.43 |
| 1 | -9 | 9 | -0.91 | 3.75 |
| 1 | -9 | -9 | 7.23 | 2.23 |
| 1 | -9 | -9 | 1.90 | 2.24 |
| -1 | 9 | 9 | 13.81 | 3.97 |
| -1 | -9 | 10 | -1.86 | 5.69 |
| -1 | -9 | -10 | 2.97 | 2.43 |
| -1 | -9 | -10 | -1.10 | 2.62 |
| 1 | -9 | -10 | -1.10 | 2.54 |
| 1 | -9 | -10 | 2.49 | 2.58 |
| -3 | -9 | 0 | 37.57 | 3.09 |
| -3 | -9 | 0 | 30.95 | 2.69 |
| 3 | -9 | 0 | 15.09 | 1.81 |
| 3 | -9 | 0 | 19.81 | 2.44 |
| -3 | 9 | 0 | 63.03 | 3.29 |
| -3 | -9 | 1 | 73.76 | 3.46 |
| -3 | -9 | 1 | 94.72 | 3.65 |
| -3 | -9 | -1 | 115.62 | 4.47 |
| -3 | -9 | -1 | 108.92 | 3.47 |
| -3 | -9 | -1 | 105.89 | 3.77 |
| 3 | -9 | 1 | 47.87 | 2.80 |
| 3 | -9 | 1 | 49.74 | 2.87 |
| 3 | -9 | 1 | 57.49 | 4.32 |
| 3 | -9 | -1 | 54.07 | 3.11 |
| 3 | -9 | -1 | 57.60 | 2.91 |
| -3 | 9 | 1 | 205.90 | 4.52 |
| -3 | 9 | -1 | 163.92 | 4.33 |
| -3 | -9 | 2 | 4.50 | 1.49 |
| -3 | -9 | -2 | 6.24 | 1.64 |
| -3 | -9 | -2 | 3.58 | 1.49 |
| 3 | -9 | 2 | 1.79 | 1.44 |
| 3 | -9 | 2 | 1.41 | 1.48 |
| 3 | -9 | 2 | 6.74 | 2.78 |
| 3 | -9 | -2 | 4.35 | 1.64 |
| 3 | -9 | -2 | -2.40 | 1.37 |
| -3 | 9 | 2 | 9.43 | 1.71 |
| 3 | 9 | 2 | 3.83 | 2.50 |
| -3 | -9 | 3 | -0.26 | 1.54 |
| -3 | -9 | 3 | 4.55 | 2.09 |
| -3 | -9 | -3 | 4.99 | 1.63 |
| -3 | -9 | -3 | 7.71 | 1.77 |
| 3 | -9 | 3 | -1.24 | 1.50 |
| 3 | -9 | 3 | -3.80 | 2.76 |
| 3 | -9 | -3 | 4.10 | 1.63 |
| 3 | -9 | -3 | 8.18 | 1.74 |
| 3 | -9 | -3 | 2.69 | 1.56 |
| -3 | 9 | 3 | 10.64 | 1.92 |
| -3 | 9 | -3 | 4.91 | 1.45 |
| -3 | -9 | 4 | 3.93 | 1.79 |
| -3 | -9 | 4 | -0.77 | 2.26 |
| -3 | -9 | -4 | 8.79 | 1.87 |
| -3 | -9 | -4 | 2.11 | 1.55 |
| 3 | -9 | 4 | -0.24 | 1.88 |
| 3 | -9 | 4 | 6.34 | 2.96 |
| 3 | -9 | -4 | 6.33 | 1.89 |
| 3 | -9 | -4 | 2.44 | 1.70 |
| 3 | -9 | -4 | 0.15 | 1.58 |
| -3 | 9 | 4 | 12.43 | 2.07 |
| -3 | 9 | -4 | 3.78 | 1.52 |
| -3 | -9 | 5 | 30.55 | 3.37 |
| -3 | -9 | 5 | 20.07 | 2.60 |
| -3 | -9 | -5 | 92.31 | 4.14 |
| -3 | -9 | -5 | 84.33 | 3.62 |
| 3 | -9 | 5 | 33.28 | 3.27 |
| 3 | -9 | 5 | 25.90 | 3.26 |
| 3 | -9 | -5 | 54.49 | 3.73 |
| 3 | -9 | -5 | 61.12 | 3.66 |
| -3 | 9 | 5 | 184.46 | 4.97 |
| -3 | 9 | -5 | 89.32 | 3.20 |
| -3 | -9 | 6 | 4.64 | 2.26 |
| -3 | -9 | 6 | 7.50 | 3.68 |
| -3 | -9 | -6 | 23.17 | 3.29 |
| -3 | -9 | -6 | 18.72 | 2.34 |
| 3 | -9 | 6 | 10.52 | 3.03 |
| 3 | -9 | 6 | 9.05 | 3.39 |
| -3 | 9 | -6 | 18.87 | 2.67 |
| -3 | -9 | 7 | 12.30 | 2.53 |
| -3 | -9 | 7 | 7.19 | 2.71 |
| -3 | -9 | -7 | 68.08 | 4.08 |
| -3 | -9 | -7 | 76.62 | 3.74 |
| 3 | -9 | 7 | 16.40 | 2.73 |
| 3 | -9 | 7 | 5.19 | 3.64 |
| 3 | -9 | -7 | 41.57 | 3.83 |
| 3 | -9 | -7 | 40.41 | 3.76 |
| -3 | 9 | 7 | 92.93 | 4.71 |
| -3 | -9 | 8 | 0.15 | 2.74 |
| -3 | -9 | 8 | 1.44 | 2.90 |
| -3 | -9 | -8 | -0.08 | 2.14 |
| -3 | -9 | -8 | 2.05 | 2.11 |
| 3 | -9 | 8 | -10.87 | 4.30 |
| 3 | -9 | -8 | -2.64 | 1.89 |
| 3 | -9 | -8 | 0.26 | 2.18 |
| -3 | 9 | 8 | -1.92 | 3.02 |
| -3 | -9 | 9 | 7.54 | 3.13 |
| -3 | -9 | 9 | -1.14 | 3.15 |
| -3 | -9 | -9 | 3.46 | 2.23 |
| -3 | -9 | -9 | -2.08 | 2.54 |
| 3 | -9 | -9 | 2.13 | 2.24 |
| 3 | -9 | -9 | 3.05 | 2.26 |
| -3 | 9 | 9 | 2.61 | 3.28 |
| -5 | -9 | 0 | 2.62 | 1.80 |
| -5 | -9 | 0 | 0.00 | 1.52 |
| 5 | -9 | 0 | 0.10 | 1.68 |
| 5 | -9 | 0 | 0.82 | 1.57 |
| 5 | -9 | 0 | -1.44 | 3.23 |
| -5 | 9 | 0 | 2.23 | 2.04 |
| -5 | -9 | 1 | 11.34 | 2.00 |
| -5 | -9 | 1 | 3.86 | 1.90 |
| -5 | -9 | -1 | 7.64 | 1.83 |
| -5 | -9 | -1 | 6.84 | 2.00 |
| 5 | -9 | 1 | 0.46 | 1.77 |
| 5 | -9 | 1 | 3.03 | 1.73 |
| 5 | -9 | 1 | -0.35 | 3.23 |
| 5 | -9 | -1 | 0.58 | 1.76 |
| 5 | -9 | -1 | 1.28 | 1.63 |
| 5 | -9 | -1 | 4.52 | 3.22 |
| -5 | 9 | 1 | 13.63 | 2.03 |
| -5 | 9 | -1 | 8.18 | 2.28 |
| -5 | -9 | 2 | 5.59 | 1.69 |
| -5 | -9 | 2 | 4.71 | 2.29 |
| -5 | -9 | -2 | 7.82 | 1.85 |
| -5 | -9 | -2 | 13.12 | 1.91 |
| 5 | -9 | 2 | -0.18 | 1.73 |
| 5 | -9 | 2 | 5.87 | 1.82 |
| 5 | -9 | 2 | -0.26 | 3.98 |
| 5 | -9 | -2 | 3.25 | 1.80 |
| 5 | -9 | -2 | 0.57 | 1.62 |
| -5 | 9 | 2 | 15.79 | 2.16 |
| -5 | -9 | 3 | 7.59 | 2.02 |
| -5 | -9 | 3 | 0.62 | 2.45 |
| -5 | -9 | -3 | 22.69 | 2.35 |
| -5 | -9 | -3 | 17.86 | 2.14 |
| 5 | -9 | 3 | 9.39 | 1.97 |
| 5 | -9 | 3 | 2.77 | 1.86 |
| 5 | -9 | 3 | -0.05 | 3.38 |
| 5 | -9 | -3 | 6.76 | 1.95 |
| -5 | 9 | 3 | 38.83 | 3.35 |
| -5 | -9 | 4 | 5.86 | 2.10 |
| -5 | -9 | 4 | 1.19 | 2.62 |
| -5 | -9 | -4 | 8.04 | 1.83 |
| -5 | -9 | -4 | 5.78 | 1.88 |
| 5 | -9 | 4 | -0.55 | 2.01 |
| 5 | -9 | 4 | -1.48 | 3.55 |
| 5 | -9 | -4 | 3.78 | 1.92 |
| -5 | 9 | 4 | 11.02 | 2.69 |
| -5 | 9 | -4 | 1.27 | 1.90 |
| -5 | -9 | 5 | 16.36 | 3.83 |
| -5 | -9 | 5 | 0.38 | 2.54 |
| -5 | -9 | -5 | 14.67 | 2.23 |
| -5 | -9 | -5 | 8.25 | 2.06 |
| 5 | -9 | -5 | -3.71 | 1.92 |
| -5 | 9 | 5 | 13.18 | 2.57 |
| -5 | 9 | -5 | 11.09 | 2.13 |
| -5 | -9 | 6 | 1.41 | 2.50 |
| -5 | -9 | 6 | -4.51 | 2.51 |
| -5 | -9 | -6 | -3.87 | 1.72 |
| -5 | -9 | -6 | 6.54 | 1.94 |
| 5 | -9 | -6 | -0.41 | 2.00 |
| -5 | 9 | 6 | 1.40 | 2.74 |
| -5 | 9 | -6 | 3.87 | 2.38 |
| -5 | -9 | 7 | -1.22 | 2.77 |
| -5 | -9 | 7 | -2.49 | 2.71 |
| -5 | -9 | -7 | 9.72 | 2.19 |
| -5 | -9 | -7 | 20.17 | 2.49 |
| 5 | -9 | -7 | 5.55 | 2.46 |
| -5 | 9 | 7 | 12.19 | 3.45 |
| -5 | 9 | -7 | -1.20 | 2.64 |
| -5 | -9 | 8 | 3.72 | 3.27 |
| -5 | -9 | 8 | 8.13 | 2.87 |
| -5 | -9 | -8 | 12.40 | 2.18 |
| -5 | -9 | -8 | 6.43 | 2.46 |
| 5 | -9 | -8 | 3.37 | 2.26 |
| -5 | 9 | 8 | 6.38 | 3.04 |
| -7 | -9 | 0 | 1.65 | 2.03 |
| 7 | -9 | 0 | 4.21 | 3.17 |
| -7 | 9 | 0 | 2.99 | 2.53 |
| -7 | -9 | 1 | 11.18 | 2.05 |
| -7 | -9 | 1 | 1.03 | 3.75 |
| -7 | -9 | -1 | 11.92 | 2.46 |
| -7 | -9 | -1 | 12.83 | 2.05 |
| 7 | -9 | 1 | 3.69 | 2.55 |
| 7 | -9 | -1 | 3.90 | 3.23 |
| -7 | 9 | 1 | 23.82 | 2.72 |
| -7 | -9 | 2 | 0.13 | 1.97 |
| -7 | -9 | 2 | 8.07 | 2.93 |
| -7 | -9 | -2 | 11.24 | 2.63 |
| -7 | -9 | -2 | 11.72 | 2.26 |
| 7 | -9 | 2 | -2.11 | 3.24 |
| 7 | -9 | -2 | 4.33 | 2.22 |
| -7 | 9 | 2 | 23.44 | 2.60 |
| -7 | -9 | 3 | -1.53 | 2.39 |
| -7 | -9 | 3 | 3.81 | 2.77 |
| -7 | -9 | -3 | 8.46 | 2.33 |
| -7 | -9 | -3 | 4.65 | 2.18 |
| 7 | -9 | 3 | 0.01 | 3.47 |
| 7 | -9 | -3 | -2.25 | 2.28 |
| -7 | 9 | 3 | 10.19 | 2.31 |
| -7 | -9 | 4 | -2.49 | 2.47 |
| -7 | -9 | 4 | 10.80 | 2.72 |
| -7 | -9 | -4 | 10.58 | 2.26 |
| -7 | -9 | -4 | 4.49 | 2.08 |
| 7 | -9 | -4 | 5.12 | 2.43 |
| -7 | 9 | 4 | 13.58 | 2.56 |
| -7 | 9 | -4 | 10.99 | 2.65 |
| -7 | -9 | 5 | 3.52 | 2.98 |
| -7 | -9 | 5 | -3.82 | 2.62 |
| -7 | -9 | -5 | 5.25 | 2.27 |
| -7 | -9 | -5 | 13.07 | 2.51 |
| 7 | -9 | -5 | 3.47 | 2.46 |
| -7 | 9 | 5 | 14.59 | 2.70 |
| -7 | 9 | -5 | 11.28 | 2.79 |
| -7 | -9 | 6 | -3.05 | 3.13 |
| -7 | -9 | 6 | 2.07 | 2.77 |
| -7 | -9 | -6 | -0.35 | 2.35 |
| -7 | -9 | -6 | 2.30 | 2.26 |
| 7 | -9 | -6 | -3.66 | 2.53 |
| -7 | 9 | 6 | -3.38 | 2.59 |
| -7 | 9 | -6 | -7.02 | 2.97 |
| -7 | -9 | 7 | 3.31 | 3.10 |
| -7 | -9 | -7 | 10.08 | 2.70 |
| -7 | -9 | -7 | 2.10 | 2.40 |
| 7 | -9 | -7 | 8.05 | 3.20 |
| -7 | 9 | 7 | 19.14 | 3.21 |
| -7 | 9 | -7 | 25.40 | 3.52 |
| -9 | -9 | 0 | 12.01 | 2.51 |
| -9 | -9 | 0 | 1.49 | 3.81 |
| 9 | -9 | 0 | 0.48 | 2.74 |
| -9 | -9 | 1 | 40.29 | 4.17 |
| -9 | -9 | 1 | 3.03 | 3.45 |
| -9 | -9 | -1 | 44.75 | 3.94 |
| 9 | -9 | 1 | 19.68 | 3.07 |
| 9 | -9 | -1 | 10.54 | 2.83 |
| -9 | 9 | 1 | 48.06 | 5.34 |
| -9 | -9 | 2 | 0.78 | 2.57 |
| -9 | -9 | 2 | -0.90 | 3.14 |
| -9 | -9 | -2 | -1.30 | 2.22 |
| 9 | -9 | 2 | 7.20 | 3.22 |
| 9 | -9 | -2 | -0.46 | 2.64 |
| -9 | 9 | 2 | 0.77 | 3.01 |
| -9 | -9 | 3 | -1.81 | 3.11 |
| -9 | -9 | 3 | 8.90 | 3.16 |
| -9 | -9 | -3 | 0.05 | 2.30 |
| 9 | -9 | 3 | -2.44 | 3.46 |
| 9 | -9 | -3 | -1.45 | 2.68 |
| -9 | 9 | 3 | 6.61 | 3.42 |
| -9 | 9 | -3 | 14.04 | 3.34 |
| -9 | -9 | 4 | 7.22 | 3.89 |
| -9 | -9 | -4 | 9.76 | 5.46 |
| -9 | -9 | -4 | 4.55 | 2.68 |
| 9 | -9 | -4 | -5.95 | 2.94 |
| -9 | 9 | 4 | 9.44 | 2.96 |
| -9 | 9 | -4 | -8.88 | 3.26 |
| 0 | -10 | 0 | 173.65 | 4.66 |
| 0 | 10 | 0 | 298.43 | 5.38 |
| 0 | -10 | 1 | 2.96 | 1.70 |
| 0 | -10 | -1 | 1.80 | 1.64 |
| 0 | -10 | -1 | 7.57 | 1.80 |
| 0 | -10 | -1 | 5.72 | 1.81 |
| 0 | 10 | 1 | 9.27 | 2.13 |
| 0 | -10 | 2 | 310.32 | 6.22 |
| 0 | -10 | 2 | 271.93 | 6.28 |
| 0 | -10 | -2 | 443.37 | 7.17 |
| 0 | -10 | -2 | 466.84 | 6.21 |
| 0 | -10 | -2 | 423.52 | 6.65 |
| 0 | 10 | 2 | 833.01 | 8.10 |
| 0 | -10 | 3 | 5.08 | 2.08 |
| 0 | -10 | 3 | -6.06 | 2.84 |
| 0 | -10 | -3 | 2.64 | 1.68 |
| 0 | -10 | -3 | 1.86 | 1.78 |
| 0 | -10 | -3 | 3.98 | 1.72 |
| 0 | 10 | 3 | 4.98 | 2.06 |
| 0 | -10 | 4 | 185.27 | 5.36 |
| 0 | -10 | 4 | 253.71 | 6.37 |
| 0 | -10 | -4 | 412.38 | 7.16 |
| 0 | -10 | -4 | 484.11 | 6.95 |
| 0 | -10 | -4 | 382.72 | 6.17 |
| 0 | 10 | 4 | 782.92 | 8.17 |
| 0 | -10 | 5 | 7.68 | 2.39 |
| 0 | -10 | 5 | -1.48 | 3.39 |
| 0 | -10 | -5 | 11.41 | 2.20 |
| 0 | -10 | -5 | 13.53 | 2.67 |
| 0 | -10 | -5 | 6.85 | 1.97 |
| 0 | 10 | 5 | 2.36 | 2.55 |
| 0 | -10 | 6 | 35.90 | 3.97 |
| 0 | -10 | 6 | 69.02 | 5.48 |
| 0 | -10 | -6 | 135.99 | 4.83 |
| 0 | -10 | -6 | 144.50 | 4.76 |
| 0 | -10 | -6 | 132.28 | 4.29 |
| 0 | 10 | 6 | 266.35 | 6.05 |
| 0 | -10 | 7 | 3.97 | 2.56 |
| 0 | -10 | 7 | -2.29 | 3.54 |
| 0 | -10 | -7 | -1.03 | 1.90 |
| 0 | -10 | -7 | 4.34 | 2.23 |
| 0 | 10 | 7 | 7.53 | 3.34 |
| 0 | -10 | 8 | 61.32 | 4.53 |
| 0 | -10 | 8 | 54.02 | 5.88 |
| 0 | -10 | -8 | 219.66 | 5.67 |
| 0 | -10 | -8 | 271.66 | 6.11 |
| 0 | 10 | 8 | 465.16 | 7.80 |
| -2 | -10 | 0 | 300.84 | 5.74 |
| -2 | -10 | 0 | 298.47 | 6.01 |
| 2 | -10 | 0 | 212.92 | 4.31 |
| 2 | -10 | 0 | 191.21 | 4.84 |
| -2 | 10 | 0 | 523.20 | 7.13 |
| -2 | -10 | 1 | 79.62 | 3.77 |
| -2 | -10 | -1 | 71.24 | 3.79 |
| -2 | -10 | -1 | 61.88 | 3.32 |
| 2 | -10 | 1 | 49.48 | 3.34 |
| 2 | -10 | 1 | 48.03 | 4.82 |
| 2 | -10 | -1 | 49.67 | 3.29 |
| 2 | -10 | -1 | 37.89 | 3.10 |
| -2 | 10 | 1 | 122.21 | 4.86 |
| -2 | 10 | -1 | 109.62 | 4.39 |
| 2 | 10 | 1 | 15.10 | 2.89 |
| -2 | -10 | 2 | 492.14 | 7.77 |
| -2 | -10 | 2 | 321.56 | 6.37 |
| -2 | -10 | -2 | 686.19 | 8.56 |
| -2 | -10 | -2 | 692.33 | 8.05 |
| -2 | -10 | -2 | 641.94 | 8.39 |
| 2 | -10 | 2 | 321.97 | 6.22 |
| 2 | -10 | 2 | 418.87 | 7.68 |
| 2 | -10 | -2 | 518.80 | 6.75 |
| 2 | -10 | -2 | 429.14 | 6.36 |
| -2 | 10 | 2 | 1243.93 | 10.50 |
| 2 | 10 | 2 | 418.09 | 9.23 |
| -2 | -10 | 3 | 32.45 | 3.40 |
| -2 | -10 | 3 | 22.38 | 3.04 |
| -2 | -10 | -3 | 46.26 | 3.69 |
| -2 | -10 | -3 | 55.44 | 3.55 |
| -2 | -10 | -3 | 61.70 | 3.39 |
| 2 | -10 | 3 | 38.78 | 3.42 |
| 2 | -10 | 3 | 37.66 | 3.82 |
| 2 | -10 | -3 | 36.54 | 3.45 |
| 2 | -10 | -3 | 49.15 | 3.46 |
| 2 | -10 | -3 | 34.93 | 3.12 |
| -2 | 10 | 3 | 106.50 | 4.10 |
| 2 | 10 | 3 | 15.30 | 2.96 |
| -2 | -10 | -4 | 581.27 | 12.13 |
| -2 | -10 | -4 | 659.78 | 7.82 |
| -2 | -10 | -4 | 556.52 | 8.27 |
| 2 | -10 | 4 | 391.92 | 7.76 |
| 2 | -10 | -4 | 418.21 | 11.04 |
| 2 | -10 | -4 | 479.06 | 7.08 |
| 2 | -10 | -4 | 368.98 | 5.66 |
| -2 | 10 | 4 | 1013.17 | 9.81 |
| -2 | -10 | 5 | 22.33 | 2.75 |
| -2 | -10 | 5 | 21.38 | 3.24 |
| -2 | -10 | -5 | 63.21 | 4.04 |
| -2 | -10 | -5 | 56.66 | 3.51 |
| 2 | -10 | 5 | 20.16 | 2.60 |
| 2 | -10 | 5 | 39.24 | 4.24 |
| 2 | -10 | -5 | 45.48 | 3.70 |
| 2 | -10 | -5 | 53.15 | 3.75 |
| 2 | -10 | -5 | 39.29 | 3.65 |
| -2 | 10 | 5 | 108.51 | 4.67 |
| -2 | -10 | 6 | 59.00 | 4.22 |
| -2 | -10 | 6 | 100.93 | 5.24 |
| -2 | -10 | -6 | 219.59 | 5.80 |
| -2 | -10 | -6 | 233.39 | 5.70 |
| -2 | -10 | -6 | 196.14 | 5.05 |
| 2 | -10 | 6 | 90.44 | 4.59 |
| 2 | -10 | 6 | 93.79 | 6.04 |
| 2 | -10 | -6 | 156.76 | 5.01 |
| 2 | -10 | -6 | 187.87 | 5.23 |
| -2 | 10 | 6 | 405.28 | 6.86 |
| -2 | -10 | 7 | 4.84 | 2.76 |
| -2 | -10 | 7 | 8.68 | 3.52 |
| -2 | -10 | -7 | 31.03 | 3.84 |
| -2 | -10 | -7 | 38.15 | 4.21 |
| -2 | -10 | -7 | 31.23 | 2.84 |
| 2 | -10 | 7 | 6.96 | 2.77 |
| 2 | -10 | 7 | 1.83 | 4.28 |
| 2 | -10 | -7 | 13.02 | 2.20 |
| 2 | -10 | -7 | 31.08 | 3.87 |
| -2 | 10 | 7 | 68.86 | 4.90 |
| -2 | -10 | 8 | 79.48 | 5.47 |
| -2 | -10 | -8 | 348.60 | 6.86 |
| -2 | -10 | -8 | 378.67 | 6.92 |
| 2 | -10 | 8 | 19.49 | 5.00 |
| 2 | -10 | -8 | 249.45 | 5.91 |
| 2 | -10 | -8 | 276.45 | 6.14 |
| -2 | 10 | 8 | 588.42 | 8.08 |
| -4 | -10 | 0 | 271.88 | 5.57 |
| -4 | -10 | 0 | 270.47 | 6.06 |
| 4 | -10 | 0 | 174.99 | 4.56 |
| 4 | -10 | 0 | 143.34 | 4.17 |
| 4 | -10 | 0 | 199.23 | 7.16 |
| -4 | 10 | 0 | 469.67 | 7.05 |
| -4 | -10 | 1 | 37.12 | 3.85 |
| -4 | -10 | 1 | 46.46 | 3.64 |
| -4 | -10 | -1 | 49.24 | 3.82 |
| -4 | -10 | -1 | 51.18 | 3.72 |
| 4 | -10 | 1 | 27.59 | 2.97 |
| 4 | -10 | 1 | 20.23 | 2.32 |
| 4 | -10 | 1 | 22.33 | 4.16 |
| 4 | -10 | -1 | 23.56 | 2.40 |
| 4 | -10 | -1 | 26.29 | 3.16 |
| 4 | -10 | -1 | 25.87 | 3.99 |
| -4 | 10 | 1 | 73.09 | 4.10 |
| -4 | 10 | -1 | 70.63 | 4.52 |
| -4 | -10 | 2 | 247.41 | 9.79 |
| -4 | -10 | -2 | 617.36 | 8.11 |
| 4 | -10 | 2 | 296.08 | 5.83 |
| 4 | -10 | 2 | 426.68 | 8.05 |
| 4 | -10 | -2 | 364.69 | 6.17 |
| 4 | -10 | -2 | 252.76 | 4.88 |
| -4 | -10 | 3 | 14.53 | 2.42 |
| -4 | -10 | 3 | 21.92 | 3.21 |
| -4 | -10 | -3 | 43.81 | 3.29 |
| 4 | -10 | 3 | 18.92 | 2.47 |
| 4 | -10 | 3 | 17.31 | 4.18 |
| 4 | -10 | -3 | 21.43 | 3.50 |
| 4 | -10 | -3 | 11.64 | 2.38 |
| -4 | 10 | 3 | 63.23 | 3.94 |
| -4 | 10 | -3 | 38.31 | 3.66 |
| -4 | -10 | 4 | 278.21 | 6.21 |
| -4 | -10 | 4 | 236.82 | 5.90 |
| -4 | -10 | -4 | 565.44 | 8.09 |
| -4 | -10 | -4 | 496.94 | 7.57 |
| 4 | -10 | 4 | 279.80 | 6.12 |
| 4 | -10 | 4 | 288.73 | 11.35 |
| 4 | -10 | -4 | 338.62 | 6.28 |
| -4 | 10 | 4 | 903.55 | 9.17 |
| -4 | 10 | -4 | 673.35 | 8.44 |
| -4 | -10 | 5 | 12.43 | 2.76 |
| -4 | -10 | 5 | 15.97 | 3.17 |
| -4 | -10 | -5 | 47.73 | 4.23 |
| -4 | -10 | -5 | 41.84 | 3.47 |
| 4 | -10 | 5 | 19.44 | 2.86 |
| 4 | -10 | 5 | 18.39 | 4.26 |
| 4 | -10 | -5 | 32.42 | 3.81 |
| -4 | 10 | 5 | 63.23 | 4.38 |
| -4 | -10 | 6 | 50.82 | 4.34 |
| -4 | -10 | -6 | 197.16 | 5.68 |
| -4 | -10 | -6 | 191.70 | 5.19 |
| 4 | -10 | -6 | 126.82 | 4.80 |
| -4 | 10 | 6 | 333.66 | 6.42 |
| -4 | -10 | 7 | 10.55 | 3.07 |
| -4 | -10 | 7 | 0.08 | 3.20 |
| -4 | -10 | -7 | 18.03 | 2.50 |
| -4 | -10 | -7 | 24.95 | 2.71 |
| 4 | -10 | -7 | 23.32 | 2.82 |
| -4 | 10 | 7 | 33.01 | 3.17 |
| -4 | -10 | -8 | 371.37 | 8.53 |
| 4 | -10 | -8 | 261.61 | 8.43 |
| -6 | -10 | 0 | 129.96 | 4.73 |
| 6 | -10 | 0 | 62.04 | 4.32 |
| 6 | -10 | 0 | 71.98 | 3.80 |
| -6 | 10 | 0 | 210.95 | 5.76 |
| -6 | -10 | 1 | 6.71 | 2.24 |
| -6 | -10 | 1 | 2.71 | 3.13 |
| -6 | -10 | -1 | -1.29 | 2.84 |
| -6 | -10 | -1 | 2.76 | 1.91 |
| 6 | -10 | 1 | -0.80 | 2.49 |
| 6 | -10 | 1 | 3.13 | 2.31 |
| 6 | -10 | -1 | 1.37 | 2.36 |
| -6 | 10 | 1 | 7.59 | 2.48 |
| -6 | -10 | 2 | 220.54 | 5.69 |
| -6 | -10 | 2 | 112.88 | 5.26 |
| -6 | -10 | -2 | 327.62 | 6.54 |
| -6 | -10 | -2 | 292.31 | 6.28 |
| 6 | -10 | -2 | 161.72 | 5.03 |
| -6 | 10 | 2 | 488.59 | 7.33 |
| -6 | -10 | 3 | -1.78 | 2.27 |
| -6 | -10 | 3 | 3.29 | 3.24 |
| -6 | -10 | -3 | -0.29 | 2.28 |
| -6 | -10 | -3 | 5.31 | 2.05 |
| 6 | -10 | -3 | -0.49 | 2.41 |
| -6 | 10 | 3 | -1.70 | 2.33 |
| -6 | 10 | -3 | -0.81 | 2.59 |
| -6 | -10 | 4 | 135.04 | 5.07 |
| -6 | -10 | 4 | 94.43 | 4.92 |
| -6 | -10 | -4 | 294.76 | 6.42 |
| -6 | -10 | -4 | 241.27 | 5.78 |
| 6 | -10 | -4 | 146.48 | 5.05 |
| -6 | 10 | 4 | 443.76 | 6.96 |
| -6 | 10 | -4 | 325.88 | 6.64 |
| -6 | -10 | 5 | -4.60 | 2.83 |
| -6 | -10 | 5 | -0.91 | 2.91 |
| -6 | -10 | -5 | 9.49 | 2.30 |
| -6 | -10 | -5 | 5.82 | 2.05 |
| 6 | -10 | -5 | 2.28 | 2.41 |
| -6 | 10 | 5 | -2.67 | 2.52 |
| -6 | 10 | -5 | 6.17 | 3.01 |
| -6 | -10 | 6 | 20.53 | 3.47 |
| -6 | -10 | 6 | 11.08 | 3.07 |
| -6 | -10 | -6 | 98.55 | 5.16 |
| -6 | -10 | -6 | 92.36 | 4.61 |
| 6 | -10 | -6 | 54.43 | 4.50 |
| -6 | 10 | 6 | 170.78 | 5.45 |
| -8 | -10 | 0 | 145.53 | 5.05 |
| -8 | -10 | 0 | 57.50 | 4.34 |
| 8 | -10 | 0 | 78.86 | 5.13 |
| -8 | -10 | 1 | 29.04 | 2.92 |
| -8 | -10 | 1 | -3.32 | 3.69 |
| -8 | -10 | -1 | 24.31 | 2.80 |
| 8 | -10 | 1 | 21.15 | 3.30 |
| 8 | -10 | -1 | 8.06 | 2.98 |
| -8 | 10 | 1 | 42.00 | 3.75 |
| -8 | -10 | 2 | 184.58 | 5.61 |
| -8 | -10 | 2 | 101.00 | 5.28 |
| -8 | -10 | -2 | 272.89 | 6.26 |
| 8 | -10 | -2 | 130.73 | 5.29 |
| -8 | 10 | 2 | 445.67 | 7.70 |
| -8 | -10 | 3 | 11.88 | 3.18 |
| -8 | -10 | 3 | 12.49 | 3.50 |
| -8 | -10 | -3 | 22.92 | 2.91 |
| 8 | -10 | -3 | 10.08 | 2.94 |
| -8 | 10 | 3 | 38.10 | 4.58 |
| -8 | 10 | -3 | 33.46 | 3.61 |
| -8 | -10 | -4 | 317.90 | 9.13 |
| 8 | -10 | -4 | 151.28 | 8.63 |
| -1 | -11 | 0 | 6.99 | 2.05 |
| 1 | -11 | 0 | 6.02 | 2.25 |
| -1 | 11 | 0 | 12.13 | 5.04 |
| -1 | -11 | 1 | 18.40 | 2.31 |
| -1 | -11 | 1 | 31.28 | 3.57 |
| -1 | -11 | -1 | 31.83 | 3.63 |
| -1 | -11 | -1 | 24.23 | 2.55 |
| 1 | -11 | 1 | 20.54 | 2.45 |
| 1 | -11 | 1 | 24.30 | 5.11 |
| 1 | -11 | -1 | 21.53 | 3.31 |
| -1 | 11 | 1 | 46.28 | 6.04 |
| 1 | 11 | 1 | 85.04 | 6.23 |
| -1 | -11 | 2 | 11.77 | 2.23 |
| -1 | -11 | 2 | 2.55 | 3.51 |
| -1 | -11 | -2 | 14.94 | 2.14 |
| -1 | -11 | -2 | 16.89 | 2.44 |
| 1 | -11 | 2 | 11.42 | 2.44 |
| 1 | -11 | 2 | 13.88 | 4.38 |
| 1 | -11 | -2 | 15.41 | 2.90 |
| 1 | -11 | -2 | 10.90 | 2.14 |
| -1 | 11 | 2 | 24.09 | 3.03 |
| 1 | 11 | 2 | 30.83 | 4.47 |
| -1 | -11 | 3 | -0.28 | 2.29 |
| -1 | -11 | 3 | 0.38 | 3.47 |
| -1 | -11 | -3 | 7.32 | 2.07 |
| -1 | -11 | -3 | 2.36 | 1.93 |
| 1 | -11 | 3 | 2.48 | 2.47 |
| 1 | -11 | 3 | 10.64 | 3.88 |
| 1 | -11 | -3 | 4.26 | 1.88 |
| 1 | -11 | -3 | 3.20 | 2.31 |
| 1 | -11 | -3 | 1.30 | 2.08 |
| -1 | 11 | 3 | -1.34 | 2.63 |
| -1 | -11 | 4 | 3.64 | 2.38 |
| -1 | -11 | 4 | 18.90 | 3.64 |
| -1 | -11 | -4 | 17.77 | 2.21 |
| -1 | -11 | -4 | 18.14 | 2.69 |
| -1 | -11 | -4 | 16.17 | 2.40 |
| 1 | -11 | 4 | 6.87 | 2.34 |
| 1 | -11 | 4 | 8.82 | 4.16 |
| 1 | -11 | -4 | 18.05 | 2.38 |
| 1 | -11 | -4 | 19.80 | 2.63 |
| 1 | -11 | -4 | 16.85 | 2.46 |
| -1 | 11 | 4 | 30.58 | 3.08 |
| -1 | -11 | 5 | 7.70 | 2.71 |
| -1 | -11 | 5 | 19.54 | 3.73 |
| -1 | -11 | -5 | 25.96 | 2.43 |
| -1 | -11 | -5 | 32.68 | 3.05 |
| -1 | -11 | -5 | 19.29 | 2.53 |
| 1 | -11 | 5 | 12.05 | 2.69 |
| 1 | -11 | 5 | 15.78 | 4.01 |
| 1 | -11 | -5 | 20.48 | 2.29 |
| 1 | -11 | -5 | 25.67 | 2.82 |
| 1 | -11 | -5 | 12.02 | 3.78 |
| -1 | 11 | 5 | 47.64 | 4.65 |
| -1 | -11 | 6 | 1.67 | 2.90 |
| -1 | -11 | 6 | 10.77 | 3.70 |
| -1 | -11 | -6 | 8.95 | 2.17 |
| -1 | -11 | -6 | 9.69 | 2.71 |
| -1 | -11 | -6 | 0.25 | 2.49 |
| 1 | -11 | 6 | 11.77 | 4.09 |
| 1 | -11 | -6 | 4.87 | 2.06 |
| 1 | -11 | -6 | 5.42 | 2.50 |
| -1 | 11 | 6 | 0.04 | 3.08 |
| -1 | -11 | 7 | -1.23 | 3.26 |
| -1 | -11 | -7 | 18.11 | 2.55 |
| -1 | -11 | -7 | 20.33 | 5.57 |
| 1 | -11 | 7 | 12.06 | 3.75 |
| 1 | -11 | -7 | 5.39 | 2.32 |
| 1 | -11 | -7 | 15.78 | 3.05 |
| -3 | -11 | 0 | 2.06 | 2.15 |
| 3 | -11 | 0 | 1.61 | 2.34 |
| 3 | -11 | 0 | 2.23 | 4.47 |
| -3 | 11 | 0 | -4.07 | 2.78 |
| -3 | -11 | 1 | 5.62 | 2.10 |
| -3 | -11 | 1 | 2.16 | 3.15 |
| -3 | -11 | -1 | 6.30 | 2.17 |
| -3 | -11 | -1 | -0.59 | 2.06 |
| 3 | -11 | 1 | 0.90 | 2.27 |
| 3 | -11 | 1 | 7.52 | 4.63 |
| 3 | -11 | -1 | -2.36 | 2.29 |
| 3 | -11 | -1 | 8.64 | 4.15 |
| -3 | 11 | 1 | 5.63 | 2.74 |
| -3 | -11 | 2 | 1.56 | 2.21 |
| -3 | -11 | 2 | -2.35 | 3.44 |
| -3 | -11 | -2 | 2.21 | 2.00 |
| -3 | -11 | -2 | 3.58 | 2.18 |
| 3 | -11 | 2 | -10.59 | 4.41 |
| 3 | -11 | -2 | 1.69 | 2.40 |
| 3 | -11 | -2 | 4.98 | 2.38 |
| -3 | 11 | 2 | 3.01 | 2.55 |
| -3 | -11 | 3 | 24.54 | 2.83 |
| -3 | -11 | 3 | 23.11 | 3.66 |
| -3 | -11 | -3 | 52.20 | 4.15 |
| -3 | -11 | -3 | 47.31 | 3.65 |
| 3 | -11 | 3 | 23.16 | 2.90 |
| 3 | -11 | 3 | 17.04 | 4.73 |
| 3 | -11 | -3 | 26.87 | 2.71 |
| 3 | -11 | -3 | 14.35 | 2.71 |
| -3 | 11 | 3 | 62.72 | 4.42 |
| -3 | -11 | 4 | -2.57 | 2.38 |
| -3 | -11 | 4 | 7.41 | 3.50 |
| -3 | -11 | -4 | 3.10 | 2.33 |
| -3 | -11 | -4 | -0.90 | 2.17 |
| 3 | -11 | 4 | 1.58 | 2.62 |
| 3 | -11 | 4 | 11.23 | 4.87 |
| 3 | -11 | -4 | 4.87 | 2.42 |
| -3 | 11 | 4 | 8.97 | 3.48 |
| -3 | -11 | 5 | 2.86 | 2.65 |
| -3 | -11 | 5 | 10.85 | 3.37 |
| -3 | -11 | -5 | 1.37 | 2.30 |
| -3 | -11 | -5 | -1.62 | 2.25 |
| 3 | -11 | 5 | 0.75 | 2.63 |
| 3 | -11 | 5 | -9.44 | 4.79 |
| 3 | -11 | -5 | 0.35 | 2.41 |
| -3 | 11 | 5 | 4.27 | 2.53 |
| -3 | -11 | 6 | 4.75 | 3.53 |
| -3 | -11 | -6 | 3.69 | 2.39 |
| -3 | -11 | -6 | 0.74 | 3.14 |
| -3 | -11 | -6 | 1.60 | 2.48 |
| 3 | -11 | 6 | -2.39 | 3.10 |
| 3 | -11 | -6 | 2.85 | 2.57 |
| -3 | 11 | 6 | -2.54 | 2.81 |
| -5 | -11 | 0 | 10.00 | 2.19 |
| 5 | -11 | 0 | -1.17 | 2.62 |
| -5 | 11 | 0 | -4.25 | 2.98 |
| -5 | -11 | 1 | 10.77 | 2.37 |
| -5 | -11 | 1 | 23.57 | 3.71 |
| -5 | -11 | -1 | 20.83 | 2.95 |
| -5 | -11 | -1 | 36.70 | 3.65 |
| 5 | -11 | 1 | 18.52 | 2.88 |
| 5 | -11 | -1 | 11.50 | 2.92 |
| -5 | 11 | 1 | 38.69 | 3.63 |
| -5 | -11 | 2 | 8.12 | 2.41 |
| -5 | -11 | 2 | 11.08 | 3.64 |
| -5 | -11 | -2 | 6.88 | 2.63 |
| -5 | -11 | -2 | 9.39 | 2.32 |
| 5 | -11 | 2 | 0.90 | 2.75 |
| 5 | -11 | -2 | -1.80 | 2.72 |
| -5 | 11 | 2 | 15.97 | 2.94 |
| -5 | -11 | 3 | -0.27 | 2.46 |
| -5 | -11 | 3 | -1.58 | 3.65 |
| -5 | -11 | -3 | 2.50 | 2.41 |
| -5 | -11 | -3 | 0.76 | 2.16 |
| 5 | -11 | 3 | -2.02 | 2.70 |
| 5 | -11 | -3 | 5.60 | 2.66 |
| -5 | 11 | 3 | -2.74 | 2.65 |
| -5 | -11 | 4 | 11.07 | 2.89 |
| -5 | -11 | 4 | 1.64 | 3.32 |
| -5 | -11 | -4 | 15.36 | 2.68 |
| -5 | -11 | -4 | 10.65 | 2.35 |
| 5 | -11 | -4 | 13.01 | 2.87 |
| -5 | 11 | 4 | 34.48 | 3.10 |
| -5 | -11 | 5 | 12.22 | 3.28 |
| -5 | -11 | 5 | 16.94 | 3.67 |
| -5 | -11 | -5 | 27.22 | 3.08 |
| -5 | -11 | -5 | 12.09 | 2.60 |
| 5 | -11 | -5 | 14.13 | 3.16 |
| -5 | 11 | 5 | 40.99 | 4.92 |
| -7 | -11 | 0 | 11.46 | 2.55 |
| -7 | -11 | 0 | -10.43 | 4.31 |
| -7 | -11 | 1 | 17.13 | 3.01 |
| -7 | -11 | 1 | 7.57 | 4.40 |
| -7 | -11 | -1 | 8.21 | 2.70 |
| 7 | -11 | -1 | 9.20 | 4.59 |
| -7 | 11 | 1 | 20.21 | 3.73 |
| -7 | -11 | 2 | -0.41 | 4.39 |
| -7 | -11 | -2 | 10.36 | 3.22 |
| -7 | 11 | 2 | 22.64 | 5.20 |
| 0 | -12 | 0 | 33.00 | 3.82 |
| 0 | -12 | 0 | 49.45 | 4.61 |
| 0 | -12 | 1 | 87.85 | 4.32 |
| 0 | -12 | 1 | 89.42 | 6.40 |
| 0 | -12 | -1 | 80.13 | 4.70 |
| 0 | -12 | 2 | 82.62 | 4.42 |
| 0 | -12 | 2 | 100.52 | 6.52 |
| 0 | -12 | -2 | 118.79 | 4.45 |
| 0 | -12 | 3 | 49.23 | 4.26 |
| 0 | -12 | 3 | 44.13 | 4.64 |
| 0 | -12 | -3 | 46.77 | 4.13 |
| 0 | -12 | 4 | 66.86 | 4.60 |
| 0 | -12 | 4 | 100.59 | 6.84 |
| 0 | -12 | -4 | 115.96 | 4.85 |
| 0 | -12 | -4 | 96.72 | 4.76 |
| 0 | -12 | 5 | 56.73 | 7.04 |
| -2 | -12 | 0 | 21.89 | 2.83 |
| -2 | -12 | 0 | 43.02 | 5.34 |
| 2 | -12 | 0 | 12.89 | 2.83 |
| 2 | -12 | 0 | 21.52 | 4.92 |
| -2 | -12 | 1 | 203.82 | 5.39 |
| -2 | -12 | 1 | 174.93 | 6.72 |
| -2 | -12 | -1 | 254.44 | 5.73 |
| 2 | -12 | 1 | 141.11 | 4.96 |
| 2 | -12 | 1 | 214.08 | 7.83 |
| 2 | -12 | -1 | 159.22 | 5.54 |
| 2 | -12 | -1 | 183.71 | 7.66 |
| -2 | 12 | 1 | 358.22 | 7.10 |
| -2 | -12 | 2 | 68.41 | 4.36 |
| -2 | -12 | 2 | 55.60 | 6.01 |
| -2 | -12 | -2 | 123.35 | 4.96 |
| -2 | -12 | -2 | 105.27 | 4.42 |
| 2 | -12 | 2 | 49.49 | 4.18 |
| 2 | -12 | 2 | 75.09 | 7.24 |
| 2 | -12 | -2 | 50.81 | 4.22 |
| -2 | 12 | 2 | 110.84 | 6.02 |
| -2 | -12 | 3 | 101.07 | 4.82 |
| -2 | -12 | 3 | 89.88 | 6.25 |
| -2 | -12 | -3 | 192.25 | 5.56 |
| -2 | -12 | -3 | 165.09 | 4.99 |
| 2 | -12 | 3 | 94.50 | 4.77 |
| 2 | -12 | 3 | 104.57 | 7.83 |
| -2 | 12 | 3 | 270.90 | 6.66 |
| -2 | -12 | 4 | 45.20 | 4.70 |
| -2 | -12 | 4 | 50.05 | 6.01 |
| -2 | -12 | -4 | 96.11 | 5.00 |
| -2 | -12 | -4 | 82.69 | 4.65 |
| 2 | -12 | 4 | 51.00 | 4.62 |
| 2 | -12 | 4 | 93.65 | 9.27 |
| -2 | 12 | 4 | 136.48 | 7.28 |
| -4 | -12 | 0 | 17.88 | 2.75 |
| -4 | -12 | 0 | 14.19 | 4.10 |
| 4 | -12 | 0 | 0.31 | 3.07 |
| -4 | -12 | 1 | 187.16 | 5.51 |
| -4 | -12 | 1 | 140.44 | 6.61 |
| -4 | -12 | -1 | 194.05 | 5.42 |
| 4 | -12 | 1 | 111.16 | 4.95 |
| -4 | 12 | 1 | 305.31 | 7.02 |
| -4 | -12 | 2 | 58.08 | 4.59 |
| -4 | -12 | 2 | 30.62 | 4.36 |
| -4 | -12 | -2 | 86.58 | 6.00 |
| -4 | -12 | -2 | 77.53 | 4.53 |
| 4 | -12 | 2 | 53.30 | 4.72 |
| -4 | 12 | 2 | 127.74 | 6.10 |

```
  -4  -12   3  137.38    8.27
   0    0   0    0.00    0.00
;
_shelx_hkl_checksum   50276
```